\documentclass[a4paper,fleqn]{cas-sc}

\usepackage[T1]{fontenc}
\usepackage[polutonikogreek,italian,english]{babel}
\usepackage[page,toc,titletoc,title]{appendix}
\usepackage[numbers,sort&compress]{natbib}
\usepackage{graphicx}
\usepackage{graphics}
\usepackage{braket}
\usepackage{bbold}
\usepackage{amssymb}
\usepackage{amsmath}
\usepackage{nicefrac}
\usepackage{dcolumn}
\usepackage{bm}
\usepackage{slashed}
\usepackage{datetime}
\usepackage{mciteplus}
\usepackage{multirow}
\usepackage{bbold}
\usepackage{color}
\usepackage{xcolor}
\usepackage{float}
\usepackage[utf8]{inputenc}
\usepackage[normalem]{ulem}
\usepackage{acro}

\DeclareAcronym{AI}{
  short=AI,
  long=Artificial Intelligence,
}
\DeclareAcronym{BSM}{
  short=BSM,
  long=Beyond-the-Standard-Model,
}
\DeclareAcronym{QCD}{
  short=QCD,
  long=Quantum ChromoDynamics,
}
\DeclareAcronym{SM}{
  short=SM,
  long=Standard Model,
}
\DeclareAcronym{CP}{
  short=CP,
  long=Charge-Parity,
}
\DeclareAcronym{HF-NRevo}{
  short=HF-NRevo,
  long=Heavy-flavor NonRelativistic evolution,
}
\DeclareAcronym{SLAC}{
  short=SLAC,
  long=Stanford Linear Accelerator Center,
}
\DeclareAcronym{BNL}{
  short=BNL,
  long=Brookhaven National Laboratory,
}
\DeclareAcronym{FCNCs}{
  short=FCNCs,
  long=Flavor-Changing Neutral Currents,
}
\DeclareAcronym{GIM}{
  short=GIM,
  long=Glashow--Iliopoulos--Maiani,
}
\DeclareAcronym{CEM}{
  short=CEM,
  long= Color Evaporation Model,
}
\DeclareAcronym{CSM}{
  short=CSM,
  long= Color Singlet Mechanism,
}
\DeclareAcronym{CO}{
  short=CO,
  long=Color Octet,
}
\DeclareAcronym{NRQCD}{
  short=NRQCD,
  long=NonRelativistic QCD,
}
\DeclareAcronym{SDC}{
  short=SDC,
  long=Short-Distance Coefficient,
}
\DeclareAcronym{SDCs}{
  short=SDCs,
  long=Short-Distance Coefficients,
}
\DeclareAcronym{LDME}{
  short=LDME,
  long=Long-Distance Matrix Element,
}
\DeclareAcronym{LDMEs}{
  short=LDMEs,
  long=Long-Distance Matrix Elements,
}
\DeclareAcronym{LO}{
  short=LO,
  long=Leading Order,
}
\DeclareAcronym{NLO}{
  short=NLO,
  long=Next-to-Leading Order,
}
\DeclareAcronym{NNLO}{
  short=NNLO,
  long=Next-to-NLO,
}
\DeclareAcronym{MHOUs}{
  short=MHOUs,
  long=Missing Higher-Order Uncertainties,
}
\DeclareAcronym{DIS}{
  short=DIS,
  long=Deep Inelastic Scattering,
}
\DeclareAcronym{DGLAP}{
  short=DGLAP,
  long=Dokshitzer--Gribov--Lipatov--Altarelli--Parisi,
}
\DeclareAcronym{PDFs}{
  short=PDFs,
  long=Parton Distribution Functions,
}
\DeclareAcronym{FFs}{
  short=FFs,
  long=Fragmentation Functions,
}
\DeclareAcronym{MPIs}{
  short=MPIs,
  long=Multi-Parton Interactions,
}
\DeclareAcronym{DPS}{
  short=DPS,
  long=Double-Parton Scattering,
}
\DeclareAcronym{SCET}{
  short=SCET,
  long=Soft and Collinear Effective Theory,
}
\DeclareAcronym{TM}{
  short=TM,
  long=Transverse-Momentum,
}
\DeclareAcronym{TMD}{
  short=TMD,
  long=Transverse-Momentum-Dependent,
}
\DeclareAcronym{FFNS}{
  short=FFNS,
  long=Fixed-Flavor Number Scheme,
}
\DeclareAcronym{VFNS}{
  short=VFNS,
  long=Variable-Flavor Number Scheme,
}
\DeclareAcronym{ZM-VFNS}{
  short=ZM-VFNS,
  long=Zero-Mass-Variable-Flavor Number Scheme,
}
\DeclareAcronym{GM-VFNS}{
  short=GM-VFNS,
  long=General-Mass Variable-Flavor Number Scheme,
}
\DeclareAcronym{ABF}{
  short=ABF,
  long=Altarelli--Ball--Forte,
}
\DeclareAcronym{BFKL}{
  short=BFKL,
  long=Balitsky--Fadin--Kuraev--Lipatov,
}
\DeclareAcronym{LL}{
  short=LL,
  long=Leading Logarithmic,
}
\DeclareAcronym{NLL}{
  short=NLL,
  long=Next-to-Leading Logarithmic,
}
\DeclareAcronym{NNLL}{
  short=NNLL,
  long=Next-to-NLL,
}
\DeclareAcronym{LVM}{
  short=LVM,
  long=Light Vector Meson,
}
\DeclareAcronym{UGD}{
  short=UGD,
  long=Unintegrated Gluon Distribution,
}
\DeclareAcronym{LHC}{
  short=LHC,
  long=Large Hadron Collider,
}
\DeclareAcronym{HL-LHC}{
  short=HL-LHC,
  long=High-Luminosity Large Hadron Collider,
}
\DeclareAcronym{FCC}{
  short=FCC,
  long=Future Circular Collider,
}
\DeclareAcronym{EIC}{
  short=EIC,
  long=Electron-Ion Collider,
}
\DeclareAcronym{HFAG}{
  short=HFAG,
  long=Heavy Flavor Averaging Group,
}
\DeclareAcronym{SCA}{
  short=SCA,
  long=Small-Cone Algorithm
}
\DeclareAcronym{BLM}{
  short=BLM,
  long=Brodsky--Lepage--Mackenzie,
}
\DeclareAcronym{SNAJ}{
  short=SNAJ,
  long=Suzuki--Nejad--Amiri--Ji,
}
\DeclareAcronym{MSb}{
  short={$\boldsymbol{\overline{\rm MS}}$},
  long=Modified Minimal Subtraction,
}
\DeclareAcronym{MOM}{
  short=MOM,
  long=MOMentum,
}
\DeclareAcronym{HELL}{
  short={HELL},
  long=High Energy Large Logarithms,
}
\DeclareAcronym{PSSZ}{
  short={PSSZ},
  long=Peterson--Schlatter--Schmitt--Zerwas,
}

\newcommand{\deffont}[1]{\begin{otherlanguage*}{polutonikogreek}#1\end{otherlanguage*}}

\def\tsc#1{\csdef{#1}{\textsc{\lowercase{#1}}\xspace}}
\tsc{WGM}
\tsc{QE}

\newcommand{\drv}{{\rm d}}

\newcommand{\as}{\alpha_s}
\newcommand{\LQCD}{\Lambda_{\rm QCD}}
\newcommand{\MSb}{\overline{\rm MS}}

\newcommand{\NLO}{{\rm NLO}}

\newcommand{\LL}{{\rm LL/LO}}

\newcommand{\NLLp}{{\rm NLL/NLO^+}}
\newcommand{\NLLpp}{{\rm NLL/NLO^{(+)}}}
\newcommand{\HENLOp}{{\rm HE}\mbox{-}{\rm NLO^+}}

\newcommand{\CnLL}{{\cal C}_n^\LL}

\newcommand{\CnNLLp}{{\cal C}_n^\NLLp}

\newcommand{\CnHENLOp}{{\cal C}_n^{{\rm HE}\text{-}{\rm NLO}^+}}
\newcommand{\DY}{\Delta Y}

\newcommand{\vqTTa}{\langle {\vec q}_T^{\;2} \rangle}

\newcommand{\F}{{\cal F}}

\newcommand{\HQ}{{\cal H}_Q}

\newcommand{\Jpsi}{J/\psi}

\newcommand{\BCs}{B_c(^1S_0)}

\newcommand{\TQQ}{T_{4Q}}
\newcommand{\TQc}{T_{4c}}

\newcommand{\TQcTpp}{T_{4c}(2^{++})}
\newcommand{\TQb}{T_{4b}}

\newcommand{\TQbTpp}{T_{4b}(2^{++})}
\newcommand{\PQQ}{P_{5Q}}

\newcommand{{\HFNRevo}}{\textsc{HF-NRevo}}

\newcommand{{\Jethad}}{\textsc{Jethad}}
\newcommand{{\symJethad}}{\textsc{symJethad}}
\newcommand{{\psymJethad}}{\textsc{(sym)Jethad}}
\newcommand{{\Hell}}{\textsc{Hell}}
\newcommand{{\RadISH}}{\textsc{RadISH}}
\newcommand{{\Pegasus}}{\textsc{QCD-PEGASUS}}
\newcommand{{\HOPPET}}{\textsc{HOPPET}}
\newcommand{{\QCDNUM}}{\textsc{QCDNUM}}
\newcommand{{\APFEL}}{\textsc{APFEL}}
\newcommand{{\APFELpp}}{\textsc{APFEL++}}
\newcommand{{\APFELppp}}{\textsc{APFEL(++)}}
\newcommand{{\EKO}}{\textsc{EKO}}
\newcommand{{\FeynCalc}}{\textsc{FeynCalc}}

\newcommand{{\HCFF}}{{\tt HCFF1.0}}
\newcommand{{\NRFF}}{{\tt NRFF1.0}}

\begin{document}
\let\WriteBookmarks\relax
\def\floatpagepagefraction{1}
\def\textpagefraction{.001}

\shorttitle{Triply Heavy $\Omega$ Baryons with {\Jethad}: A High-Energy Viewpoint}

\shortauthors{Celiberto, Francesco Giovanni}  

\title []{\Huge Triply Heavy $\Omega$ Baryons with {\Jethad}: \\ A High-Energy Viewpoint}  

\author[1]{Francesco Giovanni Celiberto}[orcid=0000-0003-3299-2203]

\cormark[1]


\ead{francesco.celiberto@uah.es}


\affiliation[1]{organization={Universidad de Alcal\'a (UAH), Departamento de F\'isica y Matem\'aticas},
            addressline={Campus Universitario}, 
            city={Alcal\'a de Henares},
            postcode={E-28805}, 
            state={Madrid},
            country={Spain}}




\begin{abstract}
We investigate the leading-power fragmentation of triply heavy $\Omega$ baryons in high-energy hadronic collisions.
Extending our previous work on the $\Omega_{3c}$ sector, we release the full {\tt OMG3Q1.0} family of collinear fragmentation functions by completing the description of the charm channel and delivering the novel $\Omega_{3b}$ functions.
These hadron-structure-oriented functions are constructed from improved proxy-model calculations for heavy-quark and gluon fragmentation, matched to a flavor-aware DGLAP evolution based on the {\HFNRevo} scheme.
For phenomenological applications, we employ the \textsc{(sym)Jethad} multimodular interface to compute and analyze NLL/NLO$^+$ semi-inclusive $\Omega_{3Q}$ plus jet distributions at the HL-LHC and FCC.
This work consolidates the link between hadron structure, rare baryon production, and resummed QCD at the energy frontier.
\end{abstract}



\begin{keywords}
 {\tt OMG3Q1.0} FF release \sep
 Hadronic structure \sep
 Precision QCD \sep
 Rare baryons \sep
 Omega sector \sep
 Heavy flavor \sep
 Fragmentation \sep
 Resummation \sep
 HL-LHC \sep
 FCC
\end{keywords}

\maketitle

\newcounter{appcnt}


\tableofcontents
\clearpage

\setlength{\parskip}{3pt}%

\section{Overview}
\label{sec:introduction}

Understanding the mechanisms behind heavy-flavor production in high-energy hadronic environments is essential for probing the inner workings of the strong force.
Heavy quarks, due to their mass and unique interactions, offer a privileged window into both Standard Model (SM) dynamics and potential manifestations of \ac{BSM} physics, especially in scenarios involving new states coupling preferentially to heavy flavors.
Owing to their large masses, heavy quarks also permit accurate perturbative \ac{QCD} predictions, making them ideal tools for investigating strong dynamics at next-generation facilities such as the \ac{HL-LHC}~\cite{Apollinari:2015wtw,Apollinari:2015bam,Apollinari:2017lan,Chapon:2020heu}, the \ac{EIC}~\cite{AbdulKhalek:2021gbh,Khalek:2022bzd,Hentschinski:2022xnd,Amoroso:2022eow,Abir:2023fpo,Allaire:2023fgp}, and the \ac{FCC}~\cite{FCC:2018byv,FCC:2018evy,FCC:2018vvp,FCC:2018bvk,FCC:2025lpp,FCC:2025uan,FCC:2025jtd}.

Quantum Chromodynamics (QCD), the non-Abelian gauge theory describing the strong force, is a central pillar of the SM.
Formulated on the $SU(N_c)$ group with $N_c = 3$, it governs the dynamics of quarks---fermions in the fundamental triplet representation---and gluons---massless vector bosons in the adjoint octet representation~\cite{Gell-Mann:1962yej,Gell-Mann:1964ewy,Zweig:1964jf,Fritzsch:1973pi}.
Beyond its established role within the SM, QCD provides a rich theoretical ground for exploring \ac{BSM} physics, including axion models for solving the strong CP problem~\cite{Peccei:1977hh,Peccei:1977ur,Peccei:2006as,Duffy:2009ig}, non-Abelian dark sectors~\cite{Forestell:2017wov,Huang:2020crf}, quarkyonic matter~\cite{McLerran:2007qj,Hidaka:2008yy,McLerran:2018hbz}, and higher-dimensional operators~\cite{Buchmuller:1985jz,Witten:1979kh,Dudek:2010wm,Afonin:2019unu}.

A particularly relevant sector of QCD concerns the formation of hadrons containing two or more heavy quarks.
Among these, heavy quarkonium states---mesons composed of a $|Q\bar{Q}\rangle$ pair---hold a prominent position.
Their study traces back to the so-called ``November Revolution'' of 1974, marked by the independent discoveries of the $\Jpsi$ meson at SLAC~\cite{SLAC-SP-017:1974ind} and BNL~\cite{E598:1974sol}, later confirmed by the ADONE experiment in Frascati~\cite{Bacci:1974za}.
These milestones significantly deepened our understanding of the strong interaction and the quark content of matter.

While quarkonia are considered conventional hadrons, the color-neutrality of QCD allows for more intricate multi-quark structures to form, giving rise to exotic hadrons.
Such states are characterized by quantum numbers that cannot be accommodated within simple $|Q\bar{Q}\rangle$ or $qqq$ configurations.
They fall into two main classes: gluonic excitations, such as glueballs and hybrids~\cite{Close:1991pf,Close:1997qda,Close:1998zz,Minkowski:1998mf,Close:2000yg,Mathieu:2008me,Hsiao:2013dta,D0:2020tig,Csorgo:2019ewn}, and multi-quark states, including tetraquarks and pentaquarks~\cite{Gell-Mann:1964ewy,Jaffe:1976ig,Jaffe:1976ih,Ader:1981db}.

The experimental discovery of the $X(3872)$ at Belle in 2003~\cite{Belle:2003nnu} is widely regarded as the beginning of the ``Second Quarkonium Revolution.''
More recently, in 2021, the observation of the $X(2900)$ by LHCb~\cite{LHCb:2020bls} provided the first evidence for an exotic hadron carrying open-charm quantum numbers, further highlighting the richness of the heavy-flavor sector.

In parallel with the growing interest in exotic matter, the study of heavy-flavor dynamics continues to play a decisive role in unveiling the microscopic composition of hadrons.
Heavy-quark systems provide an exceptional bridge between the perturbative and nonperturbative regimes of QCD, serving as precision laboratories for testing confinement mechanisms and hadronization dynamics.
Within this landscape, triply heavy baryons, such as the $\Omega_{3c}$ and $\Omega_{3b}$, occupy a singular place in the hadronic spectrum predicted by QCD.
Composed exclusively of charm or bottom valence quarks, these baryons are free from light-quark contamination and thus offer a pristine environment for exploring color confinement in the heavy sector~\cite{Bjorken:1985ei,Fleck:1989mb,Martynenko:2007je,Martynenko:2013eoa,Karliner:2014gca,Yoshida:2015tia}.
As color-singlet bound states of three heavy quarks, their masses, decay properties, production mechanisms, and potential experimental signatures have been the focus of extensive theoretical investigations~\cite{Ebert:2002ig,Roberts:2007ni,Chen:2011mb,Padmanath:2013zfa,Brown:2014ena,Meinel:2012qz}.

Although most studies have traditionally concentrated on doubly heavy baryons, triply heavy systems such as $\Omega_{3c}$ have naturally emerged as their conceptual extension~\cite{Karliner:2014gca,Flynn:2003vz,Shah:2016vmd}.
The $\Omega_{3c}$, predicted to have a mass around $4.8\,\text{GeV}$~\cite{Shah:2017jkr}, is expected to be stable under strong interactions and to decay only weakly, with a lifetime comparable to that of singly charmed baryons~\cite{Bagan:1994dy}.
Its non-observation so far is largely attributed to the formidable experimental challenges posed by its high production threshold and complex multibody decay patterns~\cite{Chang:2006eu,GomshiNobary:2005ur}.

Fragmentation-based computations suggest extremely suppressed yields at current collider energies, with fragmentation probabilities in the range $10^{-5}$ to $10^{-7}$ and total cross sections at the nanobarn level~\cite{Chen:2011mb,GomshiNobary:2005ur}.
Nevertheless, indirect evidence for their existence could arise through the decay products of heavier multiquark states, as suggested by theoretical studies~\cite{Gershtein:2000nx} and by recent LHCb observations of pentaquark candidates containing open charm and strangeness~\cite{LHCb:2015yax,LHCb:2019kea,LHCb:2020jpq}.

Triply heavy baryons represent a cornerstone in the broader classification of hadrons within QCD~\cite{Brambilla:2019esw,Esposito:2016noz,Lebed:2016hpi,Faustov:2021qqf}.
Their internal structure---free from light degrees of freedom---renders them ideal benchmarks for testing potential models, effective field theories, and lattice-QCD predictions~\cite{Mathur:2018rwu,Padmanath:2013zfa,Francis:2018jyb}.
Comparative studies with quarkonia, particularly in terms of radial excitations and spin splittings, can constrain the parameters governing the heavy-quark potential, such as the effective coupling and confinement scale~\cite{Eichten:1994gt,Godfrey:1985xj}.
Moreover, analogies with fully heavy tetraquarks ($|Q\bar{Q}Q\bar{Q}\rangle$) and pentaquarks ($|QQQ\bar{Q}Q\rangle$)~\cite{Ali:2017jda,LHCb:2020bwg,Karliner:2020vsi} suggest that $\Omega_{3c}$ may be interpreted as the baryonic endpoint of a unified heavy-flavor spectrum, linking mesonic, baryonic, and multiquark configurations through a common fragmentation mechanism.
This conceptual continuity also manifests in effective-field-theory formulations, where the nonrelativistic nature of heavy constituents leads to compact color configurations analogous to atomic systems such as hydrogen or helium~\cite{Pineda:2011dg,Celiberto:2024mab,Celiberto:2024beg,Celiberto:2025dfe,Celiberto:2025ziy,Celiberto:2025vra}.

Triply heavy baryons containing charm and bottom quarks are particularly promising for disentangling the interplay of color and confinement at different distance scales.
Their Coulomb-like internal structure implies ground-state radii inversely proportional to the product of the heavy-quark mass and the strong coupling constant, resulting in significantly more compact states than ordinary baryons.
This compactness alters the relative importance of meson-exchange and color-exchange mechanisms, leading to dynamics qualitatively distinct from those in light-flavor systems.
Various theoretical approaches, ranging from nonrelativistic variational calculations and potential models~\cite{Llanes-Estrada:2011gwu,Wei:2016jyk,Yang:2019lsg,Gomez-Rocha:2023jfr,Najjar:2024deh} to recent quantum-computing simulations employing Cornell-type potentials~\cite{deArenaza:2024dhe}, have yielded consistent predictions for their mass spectra.
In particular, the $\Omega_{3c}$ baryon is expected to lie within the discovery reach of the HL-LHC~\cite{Apollinari:2015wtw}, with even greater prospects foreseen at the FCC~\cite{FCC:2025lpp,FCC:2025uan,FCC:2025jtd}, and may also be accessible at present and future lepton machines such as Belle~II~\cite{Belle-II:2010dht}.

The ATLAS, CMS, and LHCb experiments have already collected extensive datasets at $\sqrt{s} = 13$~TeV and 13.6~TeV, paving the way for dedicated searches for triply heavy baryons.
Future high-luminosity runs will substantially improve the feasibility of these studies, especially in boosted regimes where modern vertexing and tracking systems can be fully exploited~\cite{ATLAS:2016bek,CMS:2024aqx,LHCb:2020frr}.
In this respect, the forward coverage and excellent mass resolution of LHCb make it the most suitable facility for such analyses, while the FCC---with its projected 100~TeV energy and unprecedented luminosity---would open a completely new window onto the heavy-baryon sector, pushing experimental sensitivity far beyond current limits.

From a phenomenological perspective, triply heavy baryons offer a compelling case study due to the distinctive features of their formation mechanisms.
In general, the production of multiply heavy hadrons is expected to proceed predominantly via parton fragmentation into baryonic final states (see, \emph{e.g.}, Refs.~\cite{Braaten:1994bz,Braaten:1993rw,Braaten:1993mp,Braaten:1994xb,Braaten:1993jn,Kiselev:1994pu}).
This mechanism is inherently sensitive to both perturbative short-distance dynamics and the nonperturbative processes governing hadronization.
For heavy baryons, a widely adopted framework is the quark-diquark model~\cite{Anselmino:1992vg,Ebert:1995fp}, wherein a compact diquark, such as $|cc\rangle$, forms first and subsequently hadronizes into the full baryon through the capture of an additional heavy quark~\cite{MoosaviNejad:2016qdx,Chang:2006eu}.

This factorized description facilitates the construction of \ac{FFs}, which encode the probability for a parton to produce a specific baryon carrying a given momentum fraction.
Because heavy-quark masses exceed the QCD confinement scale, the initial conditions for FFs in the heavy sector can be computed by combining perturbative ingredients with nonperturbative components.
These are typically derived from model parametrizations or from wave function overlaps inspired by potential models and nonrelativistic effective theories~\cite{Caswell:1985ui,Bodwin:1994jh,Braaten:1993rw,Cho:1995vh,Cho:1995ce,Bodwin:2005hm}.

Fragmentation into triply heavy baryons has been explored at both leading and next-to-leading order in $\alpha_s$~\cite{Adamov:1997yk,Yang:2002gh,GomshiNobary:2004mq,MoosaviNejad:2017bda,MoosaviNejad:2017rvi,Delpasand:2019xpk}, with particular focus on the constituent--quark and gluon fragmentation channels.
These studies emphasize the importance of diquark correlations, binding effects, and color configurations in shaping the formation process.
Incorporating higher-order corrections and parton-evolution effects is essential for translating such theoretical constructs into realistic collider predictions~\cite{Cacciari:1993mq,Buza:1996wv,Cacciari:2001cw,Mitov:2006wy}.

Although no triply heavy baryon has yet been observed experimentally, indirect detection may occur through decay cascades of exotic multiquark states~\cite{An:2019idk,Ortiz-Pacheco:2023kjn,Liu:2024mwn}.
Tetraquarks and pentaquarks with hidden or open heavy flavor are routinely produced at the LHC (see~\cite{Chen:2016qju,Esposito:2016noz,Lebed:2016hpi} for reviews), and their decay topologies may feature final states with three heavy valence quarks.
Indeed, the $\Omega_{3c}$ can arise from hadronic transitions of exotic hadrons or as a decay product in sequential multibody chains~\cite{Chen:2011mb,Wang:2018utj,Yang:2019lsg}.

These motivations underscore the relevance of constructing realistic FFs for triply heavy baryons, not only to advance theoretical control over heavy-quark hadronization, but also to provide phenomenological inputs for searches at current and future colliders.
In this context, the $\Omega_{3c}$ is particularly significant as a possible daughter particle in exotic multiquark decays, and as a benchmark state for rare-baryon production in high-energy collisions.

To address this, in Refs.~\cite{Celiberto:2025ogy} we introduced the {\tt OMG3Q1.0} determinations~\cite{Celiberto:2025_OMG3Q10}, representing the first public release of collinear FFs for the $\Omega_{3c}$ baryon.
These functions expand on recent efforts to characterize rare and exotic hadron production via fragmentation, complementing studies on doubly and fully heavy tetraquarks~\cite{Celiberto:2024beg,Celiberto:2025dfe,Celiberto:2025ziy,Celiberto:2025vra} and fully heavy pentaquarks~\cite{Celiberto:2025ipt}.

The {\tt OMG3Q1.0} set is built from diquark-inspired NLO inputs for both heavy-quark~\cite{MoosaviNejad:2017rvi} and gluon~\cite{Delpasand:2019xpk} fragmentation channels, evaluated at the lowest perturbative scale.
Evolution to higher energy scales is performed using \ac{DGLAP} equations in a \ac{VFNS}~\cite{Mele:1990cw,Cacciari:1993mq,Buza:1996wv}, implemented within the \ac{HF-NRevo} scheme~\cite{Celiberto:2025euy,Celiberto:2024mex,Celiberto:2024bxu,Celiberto:2024rxa,Celiberto:2025xvy}.

The evolved FFs are provided in {\tt LHAPDF6} format~\cite{Buckley:2014ana}, enabling their direct use in high-energy simulations and cross section calculations.
They can be interfaced with resummation-based tools and Monte Carlo generators to deliver predictions for $\Omega_{3c}$ production across a wide range of collider settings.

In this review, we expand upon the study outlined in Ref.~\cite{Celiberto:2025ogy}, completing the release of the hadron-structure-oriented {\tt OMG3Q1.0} sets by fully detailing the $\Omega_{3c}$ set and introducing, for the first time, their bottom-flavored counterpart, $\Omega_{3b}$.

Our analysis is carried out within the $\NLLp$ hybrid-factorization framework, which embeds the resummation of leading (\ac{LL}), next-to-leading (\ac{NLL}), and selected higher-order (NLL$^+$) energy logarithms into a collinear structure defined at next-to-leading order.\footnote{The $\NLLp$ notation, originally introduced in the context of Mueller--Navelet jet studies~\cite{Celiberto:2022gji} and later extended to rare and exotic hadron production~\cite{Celiberto:2024mab,Celiberto:2024beg,Celiberto:2025ogy}, reflects the combined logarithmic and fixed-order accuracy of the approach, in line with current standards in QCD resummation.}

Among several hybrid schemes proposed for forward observables, we adopt a formulation in which high-energy logarithmic enhancement is consistently merged with fixed-order collinear elements.
This structure, successfully employed in earlier investigations~\cite{Bolognino:2021mrc,Celiberto:2022dyf}, provides a reliable description of large-rapidity processes while maintaining compatibility with established collinear inputs.

We apply the $\NLLp$ hybrid formalism to the inclusive production of fully heavy baryons carrying bottom flavor, $\Omega_{3b}$, focusing on their fragmentation dynamics and semi-inclusive features within this resummation-enhanced setting.

The review is structured as follows.
In Section~\ref{sec:HF_fragmentation}, we detail the construction of our collinear fragmentation model for triply heavy baryons.
Section~\ref{sec:HE_resummation} introduces the $\NLLpp$ scheme and its implementation within our hybrid framework.
Numerical predictions and phenomenological insights are discussed in Section~\ref{sec:phenomenology}, while final remarks and outlook are collected in Section~\ref{sec:conclusions}.

\section{Heavy-flavor fragmentation and the rare $\Omega$ sector}
\label{sec:HF_fragmentation}

In this section, we present the methodology employed to build the hadron-structure-oriented {\tt OMG3Q1.0} family of FFs.
These functions describe the collinear VFNS fragmentation of the rare $\Omega_{3c}$ and $\Omega_{3b}$ baryons, starting from initial-scale inputs for both the constituent heavy-quark and gluon channels.
These inputs are modeled within an effective diquark framework, which captures the dominant nonperturbative dynamics governing triply heavy baryon formation.

Section~\ref{ssec:FFs-intro} opens with a concise overview of heavy-flavor fragmentation across different sectors. from heavy-light hadrons and quarkonia to exotic configurations.
We then focus on the key aspects of the diquarklike proxy model adopted for triply heavy baryons in Section~\ref{ssec:FFs-diquark}.
Section~\ref{ssec:FFs-initial-scale} describes the modeling of the initial-scale heavy-quark and gluon inputs, while Section~\ref{ssec:FFs-OMG3Q10} presents the timelike DGLAP evolution of the {\tt OMG3Q1.0} functions via the {\HFNRevo} scheme.

The derivation of the {\tt OMG3Q1.0} set relied on extensive symbolic computations, performed with {\symJethad}---the \textsc{Mathematica}~\cite{Mathematica_V14-2} module embedded in the {\Jethad} environment~\cite{Celiberto:2020wpk,Celiberto:2022rfj,Celiberto:2023fzz,Celiberto:2024mrq,Celiberto:2024swu,Celiberto:2025_P5Q_review}---specifically developed to handle analytic manipulations in hadron-structure modeling and high-precision QCD.

\subsection{Unveiling the dynamics of heavy-flavor fragmentation}
\label{ssec:FFs-intro}

The fragmentation of heavy-flavored hadrons is more intricate than that of light ones due to the perturbative nature of heavy-quark masses in their lowest Fock states. While light-hadron FFs reflect only nonperturbative physics at their starting scale, heavy-flavor FFs require a mix of perturbative and nonperturbative components.

For singly heavy hadrons (\emph{e.g}, $D$, $B$, $\Lambda_Q$), fragmentation proceeds in two stages~\cite{Cacciari:1996wr,Cacciari:1993mq,Jaffe:1993ie,Kniehl:2005mk,Helenius:2018uul,Helenius:2023wkn,Generet:2023vte}. 
First, a parton $i$ with large transverse momentum from a hard process fragments into a heavy quark $Q$ via a perturbative cascade, governed by a \ac{SDC}. 
Since $\alpha_s(m_Q) < 1$, this step is calculable~\cite{Mele:1990yq,Mele:1990cw,Rijken:1996vr,Mitov:2006wy,Blumlein:2006rr,Melnikov:2004bm,Mitov:2004du,Biello:2024zti}.

At a later stage, $Q$ hadronizes into a bound state through a fully nonperturbative process modeled by either phenomenology~\cite{Kartvelishvili:1977pi,Bowler:1981sb,Peterson:1982ak,Andersson:1983jt,Collins:1984ms,Colangelo:1992kh} or effective field theory~\cite{Georgi:1990um,Eichten:1989zv,Grinstein:1992ss,Neubert:1993mb,Jaffe:1993ie}.
The resulting FF at initial scale $\mu_{F,0} \sim m_Q$ is given by~\cite{Cacciari:1996wr,Cacciari:1997du}:
\begin{equation}
\label{FFs_HF_initial}
 D_{i}^{\HQ} (z, \mu_{F,0}) \;=\;
 \int_z^1 \frac{\drv y}{y} D_i^Q (y, \mu_{F,0}) \, D_{\rm [np]}^{\HQ} \left( \frac{z}{y} \right) \;.
\end{equation}
Here, $D_i^Q$ is the perturbative input, and $D_{\rm [np]}^{\HQ}$ is a universal nonperturbative function, independent of $i$ and $\mu_{F,0}$. 
The variable $z$ represents the hadron's momentum fraction; $y$ is that of the intermediate heavy quark.

Evolution from $\mu_{F,0}$ to higher scales is performed using DGLAP timelike equations, yielding fully evolved VFNS FFs.
This two-stage logic applies also to quarkonia, where the lowest Fock state is $|Q\bar{Q}\rangle$. NRQCD~\cite{Caswell:1985ui,Thacker:1990bm,Bodwin:1994jh,Cho:1995vh,Cho:1995ce,Leibovich:1996pa,Bodwin:2005hm,Grinstein:1998xb,Kramer:2001hh,QuarkoniumWorkingGroup:2004kpm,Lansberg:2005aw,Lansberg:2019adr} enables factorization between perturbative $|Q\bar{Q}\rangle$ production and nonperturbative hadronization via LDMEs. 
Quarkonia are described as linear combinations of Fock states ordered by powers of $\alpha_s$ and the relative velocity $v_{\cal Q}$.

At low transverse momentum, quarkonium forms via direct $|Q\bar{Q}\rangle$ production and hadronization. 
At high transverse momentum, single-parton fragmentation into the quarkonium dominates. The former matches \ac{FFNS} two-parton fragmentation~\cite{Alekhin:2009ni,Fleming:2012wy,Kang:2014tta,Echevarria:2019ynx,Boer:2023zit,Celiberto:2024mex,Celiberto:2024bxu,Celiberto:2024rxa,Celiberto:2025xvy}; the latter follows a VFNS collinear approach with DGLAP evolution.

Initial-scale FFs for gluon and heavy-quark fragmentation into $S$-wave, color-singlet vector quarkonia were first computed at LO in~\cite{Braaten:1993rw,Braaten:1993mp}, and extended to NLO in~\cite{Zheng:2019gnb,Zheng:2021sdo}. These laid the ground for VFNS FF sets such as {\tt ZCW19$^+$}~\cite{Celiberto:2022dyf,Celiberto:2023fzz} and {\tt ZCFW22} for $B_c$ mesons~\cite{Celiberto:2022keu,Celiberto:2024omj}, which match LHCb measurements~\cite{LHCb:2014iah,LHCb:2016qpe,Celiberto:2024omj} indicating that $\BCs$ production rates stay below 0.1\% of singly bottomed $B$ mesons.

NRQCD also supports the description of di-$J/\psi$ states~\cite{LHCb:2020bwg,ATLAS:2023bft,CMS:2023owd} as exotic tetraquarks~\cite{Zhang:2020hoh,Zhu:2020xni}. Here, fully heavy $\TQQ$ formation begins from short-distance $|QQ\bar{Q}\bar{Q}\rangle$ production and follows the two-stage fragmentation scheme.

The gluon-initiated, color-singlet FF for $\TQQ$ was computed in~\cite{Feng:2020riv}. It was combined with a $Q$-initiated input, modeled using the Suzuki framework~\cite{Suzuki:1977km,Suzuki:1985up,Amiri:1986zv,Nejad:2021mmp}, to form the {\tt TQ4Q1.0} set.
The method was extended to heavy-light tetraquarks in {\tt TQHL1.0}~\cite{Celiberto:2023rzw,Celiberto:2024mrq}, and then improved in {\tt TQ4Q1.1} and {\tt TQHL1.1}~\cite{Celiberto:2024beg,Celiberto:2025dfe,Celiberto:2025ziy,Celiberto:2025vra} by refining the $[Q \to \TQQ]$ inputs using NRQCD modeling~\cite{Bai:2024ezn} (for applications, see Refs.~\cite{Ma:2025ryo,Nakhaei:2025zty}).

Recently, Ref.~\cite{Celiberto:2025ipt} introduced the {\tt PQ5Q1.0} FF set for fully heavy pentaquarks, combining DGLAP evolution with two modeling options: a compact $|ccc\bar{c}c\rangle$ state~\cite{Farashaeian:2024son} or a dicharm-charm-dicharm system~\cite{Farashaeian:2024cpd}.

\subsection{Rare-baryon fragmentation in a diquark picture}
\label{ssec:FFs-diquark}

The quark-diquark model offers a streamlined yet effective framework for describing baryons as systems where two quarks form a tightly bound diquark interacting with a third quark~\cite{Gell-Mann:1964ewy}.
This picture has seen extensive use in hadron spectroscopy and in modeling production and decay processes of baryons with heavy quarks~\cite{Maiani:2004vq,Jaffe:2003sg,Guo:2013xga,DeSanctis:2016zph}.
Diquarks are typically treated as either spin-0 (scalar) or spin-1 (axial-vector) constituents, with their internal nonperturbative structure encapsulated in phenomenological form factors.

Scalar diquarks, associated with simpler spin dynamics, require a single form factor, whereas axial-vector diquarks involve multiple ones to capture richer internal correlations. 
Both types have been widely used in modeling baryon formation and in parameterizing polarized parton distributions, especially in spectator models~\cite{Bacchetta:2008af,Bacchetta:2010si,Bacchetta:2020vty,Bacchetta:2024fci,Chakrabarti:2023djs,Banu:2024ywv}.

Early applications of this framework to fragmentation include studies on both light and heavy baryons~\cite{Nzar:1995wb,Ma:2001ri,Yang:2002gh,Falk:1993gb,Adamov:1997yk,MoosaviNejad:2017rvi,Delpasand:2019xpk}, and have been extended to exotic systems like pentaquarks~\cite{Maiani:2015vwa} and heavy tetraquarks via relativistic diquark-antidiquark models~\cite{Faustov:2020qfm,Faustov:2021hjs,Faustov:2022mvs}.

In this work, we follow a widely adopted strategy~\cite{Adamov:1997yk,Martynenko:1996bt,MoosaviNejad:2017rvi,Delpasand:2019xpk} that treats the diquark as a scalar. This choice simplifies spin algebra, reduces the number of form factors, and makes analytic calculations more tractable---an advantage evident in the modeling of unpolarized baryons with $J=1/2$~\cite{Adamov:1997yk,Martynenko:1996bt,GomshiNobary:2007xk}.
Axial-vector diquarks, while essential for describing spin-$3/2$ baryons and polarization effects, significantly increase theoretical complexity. Modeling the polarized fragmentation of the $\Omega_{3c}^*$, for instance, would necessarily require vector diquarks.

Here, the scalar diquark approximation serves as a practical starting point for describing fragmentation into unpolarized triply heavy baryons like $\Omega_{3c}$. More complete spin-sensitive treatments, including axial-vector components, may be developed in future work.
Notably, modeling $\Omega_{3c}$ and $\Omega_{3b}$ production using scalar diquarks remains compatible with spin-color symmetry constraints, provided a two-step mechanism is assumed: the heavy quark first fragments into a color-antitriplet diquark, which subsequently hadronizes into the baryon through nonperturbative dynamics.

Although a scalar diquark does not fully reproduce the symmetrization required for $J=1/2$ baryons composed of three identical fermions, this can be effectively restored through hadronization effects in the nonperturbative part of the FF.
This separation mirrors the NRQCD approach to quarkonia, where color-octet $[Q\bar{Q}]$ pairs are projected onto singlet states via LDMEs.
This interpretation, while not explicitly stated in~\cite{MoosaviNejad:2017rvi,Delpasand:2020mqv}, aligns with the modeling choices therein, where scalar diquarks are used to compute FFs into $\Omega_{3c}$ and $\Omega_{3b}$, yielding physically consistent and phenomenologically viable results.

\begin{figure}[!t]
\centering
\includegraphics[width=0.475\textwidth]{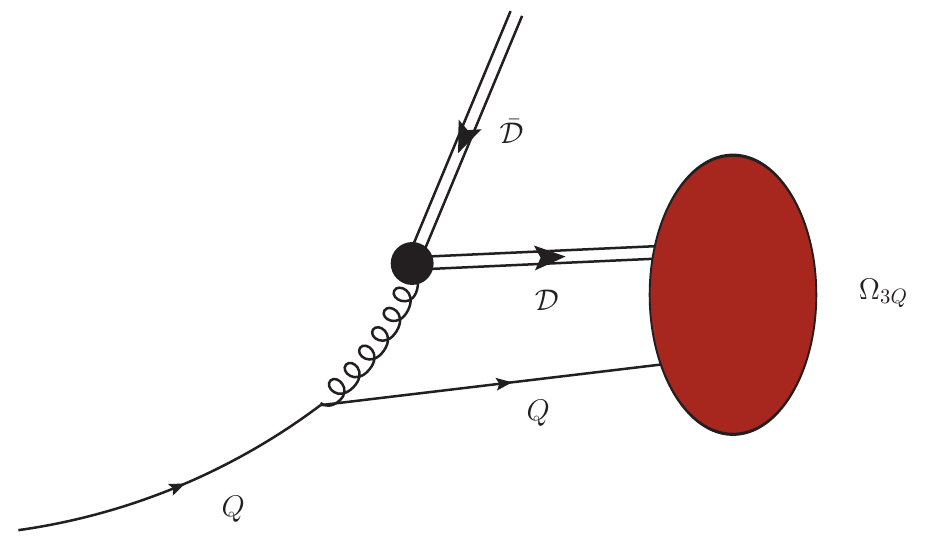}
\hspace{0.40cm}
\includegraphics[width=0.475\textwidth]{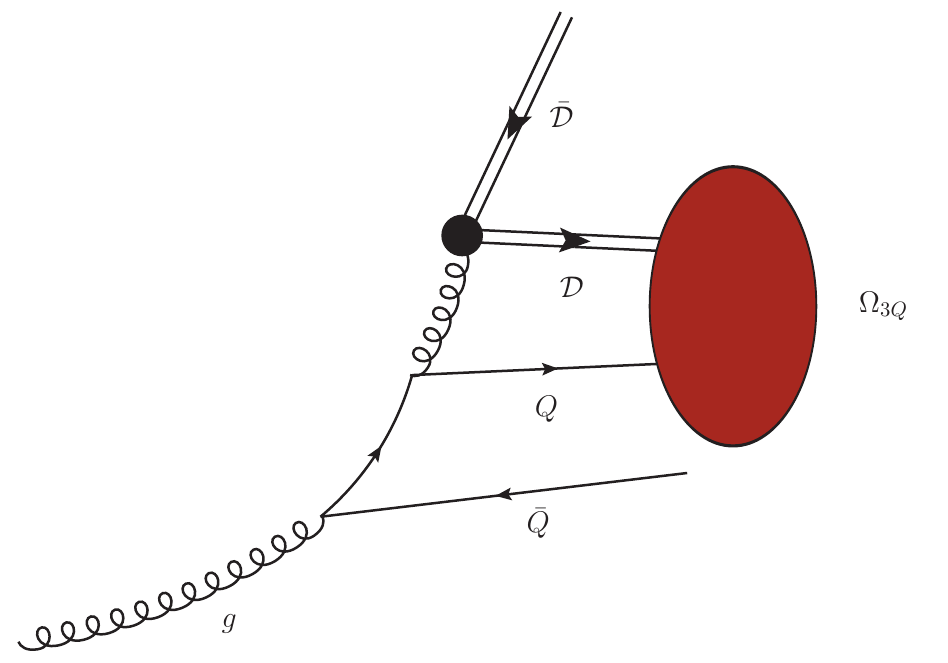}

\caption{
Representative leading-order diagrams for the diquarklike proxy model describing the initial-scale collinear fragmentation of a constituent heavy quark (left) and a gluon (right) $\Omega_{3Q}$ baryon.
Double lines indicate ${\cal D}$ or $\bar{{\cal D}}$ diquark states, orange blobs denote the nonperturbative hadronization component of the corresponding FFs, and black dots represent scalar diquark form-factor couplings.
Diagrams created with {\tt JaxoDraw 2.0}~\cite{Binosi:2008ig}.
}
\label{fig:OQQ_FF_diagrams}
\end{figure}

The initial-scale inputs for the fragmentation functions used in this work are derived from existing perturbative calculations targeting the production of triply heavy baryons in a color-singlet configuration. Specifically, we rely on quark-induced channels computed at LO~\cite{MoosaviNejad:2017bda} and NLO~\cite{MoosaviNejad:2017rvi}, and on the gluon-induced NLO result from~\cite{Delpasand:2019xpk}.
These works model the baryon as a spin-$1/2$ state composed of three charm or bottom quarks, with a spin-0 scalar diquark used to simplify the internal dynamics.

Two main production scenarios are considered in the literature. The first is the direct fragmentation mechanism, where the heavy quark forms the three-quark baryon via a full three-body interaction. The second, adopted here, is the diquark-based model, where two heavy quarks are grouped into a tightly bound scalar diquark that subsequently hadronizes with a third quark. This approach significantly reduces the complexity of spin and color algebra, while still capturing the essential features of unpolarized $J = 1/2$ baryons.

This diquark-based strategy corresponds to the so-called indirect or two-step mechanism discussed in Section~\ref{ssec:FFs-intro} and originally developed in Refs.~\cite{MoosaviNejad:2017rvi,Delpasand:2019xpk}. It separates the perturbative production of a colored diquark from its nonperturbative hadronization into a physical baryon, providing a clean and systematic framework for constructing quark and gluon FFs. Crucially, this setup enables consistent NLO calculations within a VFNS scheme.

The underlying formalism is based on the Suzuki model~\cite{Suzuki:1985up}, which plays a role analogous to NRQCD in quarkonium production. It defines a perturbative fragmentation amplitude where the initiating heavy quark produces a color-antitriplet diquark and a spectator, followed by hadronization into the baryon. The model uses simplifying assumptions, such as collinearity of final-state momenta, and introduces the nonperturbative content via a wave function at the origin and an effective diquark form factor.

While the Suzuki model can be used for both direct and diquark fragmentation, it is particularly effective in the latter, where it provides compact analytic expressions for the SDCs. This is especially useful when dealing with baryons like $\Omega_{3Q}$, where direct fragmentation becomes intractable due to the complexity of three-body spin-color structures. 
By modeling the diquark as an effective constituent, the Suzuki approach enables an analytically viable treatment of triply heavy baryon formation.

The method retains sensitivity to internal hadron structure through constituent-level vertices, spin-dependent terms, and color couplings. 
Unlike approaches that treat the baryon as pointlike, it preserves a direct link between the perturbative subprocess and the nonperturbative wave function.

In the specific implementations of Refs.~\cite{MoosaviNejad:2017rvi,Delpasand:2019xpk}, the hadronization step is modeled using a light-cone distribution amplitude in the spirit of Lepage and Brodsky~\cite{Lepage:1980fj}, approximated via a Dirac $\delta$-function to enforce collinearity. 
This simplification embeds the nonperturbative content directly into the perturbative expression, eliminating the need to treat the wave function as an external object. Similar strategies have been applied to heavy-flavor fragmentation into fully heavy tetraquarks~\cite{Nejad:2021mmp,Celiberto:2023rzw} and into pentaquark-quarkonium bound states~\cite{Farashaeian:2024son,Farashaeian:2024cpd,Celiberto:2025ipt}, where a separation between perturbative and nonperturbative dynamics is achieved through effective constituent modeling.

This diquark-based Suzuki framework is thus well suited for generating analytically controlled initial conditions for the fragmentation into $\Omega_{3Q}$ baryons in a VFNS context, while offering a solid foundation for future extensions to polarized or spin-$3/2$ states via axial-vector diquarks.

A key ingredient in building the initial-scale FFs for $\Omega_{3Q}$ baryons is the modeling of the nonperturbative transition from the constituent diquark ${\cal D}$ to the final hadron.
Following Refs.~\cite{MoosaviNejad:2017rvi,Delpasand:2019xpk}, we employ the phenomenological \ac{PSSZ} parametrization~\cite{Peterson:1982ak}, widely adopted in heavy-flavor studies and suited to multi-heavy hadrons. 
The functional form reads
\begin{equation}
\label{HFF_PSSZ}
 D_{\cal D}^{{\Omega}_{3Q}} (z) \;\equiv\; D_{\rm [np]}^{\rm [PSSZ]} (z) \;=\; {\cal N}_P \frac{z(1-z)^2}{\left[(1-z)^2 + \tau_P z \right]^2} \;,
\end{equation}
with normalization
\begin{equation}
\label{HFF_PSSZ_Nrm}
 {\cal N}_P \,=\, 
  \left\{
 \frac{\tau_P^2-6\tau_P+4}{(4-\tau_P)\sqrt{4\tau_P-\tau_P^2}}
  \left[ 
 \arctan \frac{2-\tau_P}{\sqrt{4 \tau_P - \tau_P^2}} + \arctan \frac{\tau_P}{\sqrt{4 \tau_P - \tau_P^2}} 
  \right]
 + \frac{1}{2} \ln \tau_P + \frac{1}{4-\tau_P} 
  \right\} 
  \;.
\end{equation}
derived by integrating over all hadrons containing the heavy quark~\cite{Cacciari:1997du,Peterson:1982ak}.
The shape parameter $\tau_P$ encodes nonperturbative effects and cannot be fixed from first principles.

In singly charmed systems like $D$ mesons, $\tau_P \approx \Lambda/m_c$, with $\Lambda \sim m_{\bar q}$, leading to $\langle z \rangle \simeq 1 - \sqrt{\tau_P}$, consistent with mass-dependent scaling predictions~\cite{Suzuki:1977km,Bjorken:1977md,Kinoshita:1985mh,Cacciari:1996wr}.
For $\Omega_{3c}$ or $\Omega_{3b}$ baryons, the heavy masses lie above the perturbative threshold.
In line with~\cite{Adamov:1997yk,MoosaviNejad:2017rvi,Delpasand:2019xpk}, we set $\tau_P = [m_Q/(m_Q + m_{\bar Q})]^2 = 1/4$.

This choice captures the essential features of the hadronization process in the diquark-based two-step framework~\cite{MoosaviNejad:2017rvi,Delpasand:2019xpk}, complementing the perturbative sector derived via the Suzuki model~\cite{Suzuki:1985up}.
The Peterson function reflects realistic momentum sharing in baryons with multiple heavy quarks and ensures a peaked behavior at large $z$, as expected when the final hadron inherits most of the parent parton's momentum.

Its simple analytical structure, governed by $\tau_P$, facilitates implementation within collinear convolutions.
It also aligns well with VFNS evolution schemes, ensuring smooth matching between perturbative and nonperturbative sectors.
These features make the Peterson form a physically motivated and computationally efficient choice for modeling $D_{\cal D}^{\Omega_{3Q}}(z)$.

The resulting initial-scale FF for a parton $i$ fragmenting into a $\Omega_{3Q}$ baryon is given by
\begin{equation}
\label{FFs_OQQ_initial}
D{i}^{\Omega_{3Q}} (z, \mu_{F,0}) \;=\;
\int_z^1 \frac{\drv y}{y} D_i^{\cal D} (y, \mu_{F,0}) , D_{\cal D}^{\Omega_{3Q}} \left( \frac{z}{y} \right) \;,
\end{equation}
mirroring the generic structure of heavy-hadron FFs in Eq.~\eqref{FFs_HF_initial}, with substitutions
\begin{subequations}
\begin{align}
\label{FFs_OQQ_vs_HF_initial_a}
D_i^Q (y, \mu_{F,0})
\quad &\longrightarrow \quad
D_i^{\cal D} (y, \mu_{F,0})
\;,
\\[0.25cm]
\label{FFs_OQQ_vs_HF_initial_b}
D_{\rm [np]}^{\HQ} \left( \frac{z}{y} \right)
\quad &\longrightarrow \quad
D_{\cal D}^{\Omega_{3Q}} \left( \frac{z}{y} \right)
\;.
\end{align}
\end{subequations}
Thus, the quark $Q$ in standard hadronization is replaced by a scalar diquark ${\cal D}$, which encapsulates the intermediate stage in the two-step transition from parton to baryon.
SDCs entering Eq.~\eqref{FFs_OQQ_initial} are discussed in the next section.

\subsection{Initial-scale partonic channels}
\label{ssec:FFs-initial-scale}

The diagram on the left of Fig.~\ref{fig:OQQ_FF_diagrams} depicts the initial-scale fragmentation process $[Q \to {\Omega}_{3Q}]$ in a diquark-based framework. 
Double lines represent heavy diquark states, ${\cal D} \equiv |QQ\rangle$, or their antidiquark counterparts, $\bar{{\cal D}} \equiv |\bar{Q}\bar{Q}\rangle$. 
The firebrick blob corresponds to the nonperturbative hadronization term $D{\rm [np]}^{\HQ} \equiv D_{\cal D}^{{\Omega}_{3Q}}$, while the black dot indicates the scalar diquark coupling vertex.

The calculation includes a scalar diquark form factor at this vertex. 
Its internal structure is modeled with a simple pole behavior in transverse-momentum space, governed by a typical scale of about 5~GeV. To streamline the expression of the initial FF, we omit the analytic form of this factor, but its impact is fully included via normalization and diagrammatic structure.

The diagram corresponds to a $[Q \to (Q, {\cal D}) + \bar{\cal D}]$ splitting. 
Exchanging $Q \leftrightarrow \bar{Q}$ yields the mirrored process $[\bar{Q} \to (\bar{Q}, \bar{\cal D}) + {\cal D}]$. 
While a dedicated analysis of possible production asymmetries between ${\Omega}_{3Q}$ and $\bar{\Omega}_{3Q}$ is beyond the present scope, we assume symmetric formation. 
Therefore, predictions focus on inclusive observables averaged over baryons and antibaryons, with quark and antiquark fragmentation treated equally.

The explicit LO and NLO expressions for the heavy-quark-induced initial-scale SDC were computed symbolically using {\symJethad}~\cite{Celiberto:2020wpk,Celiberto:2022rfj,Celiberto:2023fzz,Celiberto:2024mrq,Celiberto:2024swu,Celiberto:2025_P5Q_review} and {\FeynCalc}~\cite{Mertig:1990an,Shtabovenko:2016sxi,Shtabovenko:2020gxv}. 
Our results reproduce those found in earlier studies, confirming consistency with prior literature. 
The LO expression reads
\begin{equation}
\label{FFs_OQQ_c}
 D_{Q}^{\cal D} (z, \mu_{F,0}) \,=\, 
 \as^2
 \drv_{Q}^{\rm [LO]} (z; \mu_{F,0}) 
 \,+\,
 \as^3 \, \drv_{Q}^{\rm [NLO]} (z; \mu_{F,0})
 \;.
\end{equation}
with $\as \equiv \as(5m_Q)$ and
\begin{equation}
\label{FFs_OQQ_c_d_LO}
 \drv_{Q}^{\rm [LO]} (z; \mu_{F,0}) =
 {\cal N}_Q(z)
 \frac{{\cal S}_Q^{\rm [LO]}(z; {\cal R}_{q_T/Q})}{{\cal T}_Q^{\rm [LO]}(z; {\cal R}_{q_T/Q})}
 \;,
\end{equation}
where
\begin{equation}
\label{R_qTQ}
  {\cal R}_{q_T/Q}^2 \equiv \vqTTa/m_Q^2
 \;,
\end{equation}
\begin{equation}
\label{FFs_OQQ_c_d_norm}
 {\cal N}_Q(z) = \frac{\pi^2}{\sqrt{6}} \, f_{\cal B}^2 C_F^2 \, z^3(1-z)^3
 \;,
\end{equation}
\begin{equation}
\begin{split}
\label{FFs_OQQ_Q_d_num_LO}
  {\cal S}_Q^{\rm [LO]}(z; {\cal R}_{q_T/Q}) 
  \,=\, 
  4(41z^2 - 80z + 96 + 256/z^2) \,+\,
  {\cal R}_{q_T/Q}^2 (8z^3 + 5z^2 - 16z + 16)
  \,+\,
  {\cal R}_{q_T/Q}^4 \, 4z^2
  \;,
\end{split}
\end{equation}
and
\begin{equation}
\label{FFs_OQQ_Q_d_den_LO}
 {\cal T}_Q^{\rm [LO]}(z; {\cal R}_{q_T/Q}) 
  \,=\, 
  ({\cal R}_{q_T/Q}^2 + (4-3z)^2)^2
  \,
  ({\cal R}_{q_T/Q}^2 \, z^2 + z^2 - 16 z + 16)^2
 \;.
\end{equation}
The parameter $f_{\cal B}$ in Eq.~\eqref{FFs_OQQ_c_d_norm} is the decay constant for a triply heavy baryon.
We adopt the value $f_{\cal B} = 0.25$~GeV, in line with common assumptions in the literature~\cite{ParticleDataGroup:2020ssz}.
Furthermore, $C_F$ stands for the Casimir factor connected to emission of a gluon from a quark.
The next-to-leading order correction to the heavy-quark SDC was first derived for singly heavy mesons in Ref.~\cite{MoosaviNejad:2016qdx}, and subsequently applied to triply heavy baryons in Ref.~\cite{MoosaviNejad:2017rvi}.
It reads
\begin{equation}
\label{FFs_OQQ_c_d_NLO}
 \drv_{c}^{\rm [NLO]} (z; \mu_{F,0}) =
 {\cal N}_Q(z)
 \frac{{\cal S}_Q^{\rm [NLO]}(z; {\cal R}_{q_T/Q})}{{\cal T}_Q^{\rm [NLO]}(z; {\cal R}_{q_T/Q})}
 \;,
\end{equation}
where
\begin{equation}
\begin{split}
\label{FFs_OQQ_Q_d_num_NLO}
  {\cal S}_Q^{\rm [NLO]}&(z; {\cal R}_{q_T/Q}) 
  \,=\, 
  96 \pi z \,
\big[
  {\cal R}_{q_T/Q}^{14} \, (17-20 z) z^{14} \\
\,&+\,
  {\cal R}_{q_T/Q}^{12} \, (240 z^4-1664 z^3+3813 z^2-4130 z+1711) z^{12} \\
\,&+\,
  {\cal R}_{q_T/Q}^{10} \, (4368 z^6-39072 z^5+140635 z^4-269144 z^3+300490 z^2-189616 z+52243) z^{10} \\
\,&+\,
  {\cal R}_{q_T/Q}^8 \, (20976 z^8-208040 z^7+870151 z^6-2070634 z^5+3255115 z^4 \\
\,&-\,
  3533740 z^3+2600201 z^2-1202330 z+268205) z^8 \\
\,&+\,
  {\cal R}_{q_T/Q}^6 \, 4 (1-z)^2 (11256 z^8-75035 z^7-151779 z^6+2535886 z^5 \\
\,&-\,
  8913460 z^4+15864355 z^3-15844619 z^2+8395110 z-1823490) z^6 \\
\,&+\,
  {\cal R}_{q_T/Q}^4 \, 4 (1-z)^4 (12324 z^8-22830 z^7-2045973 z^6+14641206 z^5-42277675 z^4 \\
\,&+\,
  65338590 z^3-57387186 z^2+26330832 z-4933872) z^4 \\
  \,&+\,
  {\cal R}_{q_T/Q}^2 \, 48 (1-z)^6 (563 z^8+2272 z^7-252314 z^6+1752016 z^5-5068605 z^4 \\
\,&+\,
  7770780 z^3-6800544 z^2+3505032 z-849528) z^2 \\
\,&+\,
  144 (1-z)^8 (41 z^8+438 z^7-36063 z^6+258240 z^5 \\
\,&-\,
  833328 z^4+1614600 z^3-1793016 z^2+676512 z-198288)
\big]
  \;,
\end{split}
\end{equation}
and
\begin{equation}
\label{FFs_OQQ_Q_d_den_NLO}
\begin{split}
 {\cal T}_Q^{\rm [NLO]}(z; {\cal R}_{q_T/Q}) 
  \,&=\, 
 [{\cal R}_{q_T/Q}^2 \, z^2 + 4 (3-2 z)^2]^2 \,
 [{\cal R}_{q_T/Q}^2 \, z^2 + (6-z)^2]^2 \,
[{\cal R}_{q_T/Q}^2 \, z^2 + (1-z)^2] \\
\,&\times\, 
[{\cal R}_{q_T/Q}^2 \, z^2 + 36(1-z)^2]^2 \,
[{\cal R}_{q_T/Q}^2 \, z^2 + z^2-35 z+36]
  \;.
\end{split}
\end{equation}
As suggested by LO kinematics (Fig.~\ref{fig:OQQ_FF_diagrams}, left panel), the initial energy scale is set to $\mu_{F,0} = 5 m_Q$.

Our heavy-quark FF differs from that in Ref.~\cite{MoosaviNejad:2017rvi} in two key ways. First, the normalization factor ${\cal N}_Q(z)$ in Eq.~\eqref{FFs_OQQ_c_d_norm} was not computed there but imposed through a normalization condition. 
Second, the parameter $\vqTTa$, tied to the Suzuki model, warrants a more refined treatment.

The Suzuki framework incorporates spin effects and effectively mimics TMD fragmentation. 
In the collinear limit, instead of integrating over the squared transverse momentum of the outgoing charm quark, one replaces it with its average value, $\vqTTa$, treated as a free parameter to be fixed phenomenologically. 
A larger $\vqTTa$ shifts the FF peak to lower $z$, decreasing the overall normalization~\cite{GomshiNobary:1994eq}.

Reference~\cite{MoosaviNejad:2017rvi} used $\vqTTa = 1\,\text{GeV}^2$ as an upper bound. 
Here, we propose a more balanced and phenomenologically driven value. This improvement builds upon our earlier work on charm-to-$\TQQ$ FFs\cite{Celiberto:2024mab}, which drew insights from the fragmentation of light~\cite{Celiberto:2016hae,Celiberto:2017ptm,Bolognino:2018oth,Celiberto:2020wpk} and heavy-flavored hadrons~\cite{Celiberto:2021dzy,Celiberto:2021fdp,Celiberto:2022dyf,Celiberto:2022keu} in proton collisions. 
These studies typically probe average $z$ values above 0.4.

A numerical scan confirmed that $\vqTTa_{\TQc} = 70\,\text{GeV}^2$ ensures $\langle z \rangle > 0.4$ and comparable quark/gluon magnitudes. 
The same method was applied to $|QQQ\bar{Q}Q\rangle$ pentaquarks, for which we used $\vqTTa_{\PQQ} = 90\,\text{GeV}^2$~\cite{Celiberto:2025ipt}. 
For $\Omega_{3Q}$, we fix $\vqTTa_{{\Omega}_{3Q}} = 60\,\text{GeV}^2$, guided by the FF peak position.

This choice has theoretical backing. 
Early studies~\cite{Suzuki:1977km,Bjorken:1977md,Kinoshita:1985mh,Peterson:1982ak} showed that heavy-quark FFs peak at high $z$, with binding effects scaling with $m_Q$. Consider a $D$ meson with lowest Fock state $|c\bar{q}\rangle$ and momentum $k$. Assuming equal velocities $v \equiv v_c \simeq v_q$, one gets $k_c = z k = m_c v$ and $k_q = \Lambda_q v$. From $M_D \approx m_c$, it follows that $m_c v \approx z m_c v + \Lambda_q v$, yielding $\langle z \rangle_c \approx 1 - \Lambda_q / m_c$.

To better support the adopted hierarchy of $\vqTTa$ values across the $\Omega_{3Q}$, $\TQQ$, and $\PQQ$ families, we note that the fragmentation dynamics become increasingly rigid as the number of heavy constituents grows. 
This leads to more sharply peaked FFs at large $z$, and hence to a requirement of smaller $\vqTTa$ to reproduce the same peak location. 
The observed trend---from 90~GeV$^2$ in the five-quark case down to 60~GeV$^2$ in the baryonic one---is therefore consistent with the reduced phase-space compression in lighter fully-heavy systems. In this sense, our numerical scan over $\vqTTa$ values is not a blind tuning, but rather the final step of a physically motivated calibration.

This scaling behavior is also supported by theoretical arguments based on the equal-velocity condition. As discussed in Refs.~\cite{Suzuki:1977km,Bjorken:1977md,Kinoshita:1985mh,Peterson:1982ak}, the relation $\langle z \rangle_Q \approx 1 - \Lambda_q/m_Q$ captures the tendency of FFs to peak closer to $z \sim 1$ for larger heavy-quark masses. Although originally derived for heavy–light systems, this pattern survives qualitatively in fully heavy ones, where internal motion is suppressed and binding effects become comparatively less relevant. Our choices for $\vqTTa$ reflect this reasoning and are fully compatible with the observed peak positions and normalization patterns in the computed FFs. A dedicated study of the systematic uncertainty associated with this nonperturbative input is beyond the scope of the present work, but we are planning a future update of our FFs sets where this source will be explicitly assessed and included.

Now we turn our attention to the gluon case.
The initial-scale FF for the $[g \to {\Omega}_{3Q}]$ transition in the diquarklike picture is depicted in the right panel of Fig.~\ref{fig:OQQ_FF_diagrams}. 
Although both gluon- and quark-initiated channels are leading in their respective topologies, they differ in perturbative order: the gluon-induced process includes one extra QCD vertex and thus appears at higher order in $\alpha_s$.

As mentioned, the $[Q \to {\Omega}_{3Q}]$ FF is known at LO and NLO, corresponding to the left diagram in Fig.~\ref{fig:OQQ_FF_diagrams}. 
In contrast, the $[g \to {\Omega}_{3Q}]$ contribution starts at $\mathcal{O}(\alpha_s^3)$, as illustrated in the right diagram.

This hierarchy arises because gluons, lacking charm, must generate at least three charm quarks to form ${\Omega}_{3Q}$. 
This involves emitting a $[Q\bar{Q}]$ pair, forming a diquark-like $|QQ\rangle$ component, and adding a further QCD vertex for baryon formation. 
Accordingly, we refer to the gluon FF as NLO, keeping the LO label for the charm-induced one.

Using {\symJethad} and {\FeynCalc}, we obtained the explicit form of the NLO gluon SDC, originally derived in Ref.~\cite{Delpasand:2019xpk}. 
One has
\begin{equation}
\label{FFs_OQQ_g}
 D_{g}^{\cal D} (z, \mu_{F,0}) \,=\, 
 \as^3 \, \drv_{g}^{\rm [NLO]} (z; \mu_{F,0}) \;.
\end{equation}
with $\as \equiv \as(6m_Q)$ and
\begin{equation}
\label{FFs_OQQ_g_d}
 \drv_{g}^{\rm [NLO]} (z; \mu_{F,0}) =
 {\cal N}_g(z)
 \frac{{\cal S}_g^{\rm [NLO]}(z; {\cal R}_{q_T/Q})}{{\cal T}_g^{\rm [NLO]}(z; {\cal R}_{q_T/Q})}
 \;,
\end{equation}
where
\begin{equation}
\label{FFs_OQQ_g_d_norm}
 {\cal N}_g(z) = \frac{2 \pi^3}{3} \, f_{\cal B}^2 C_F^2 \, z^3(1-z)^2
 \;,
\end{equation}
\begin{equation}
\label{FFs_OQQ_g_d_num_NLO}
  {\cal S}_g^{\rm [NLO]}(z; {\cal R}_{q_T/Q}) 
  \,=\, 
  8(16z^2-32z+15) \,+\,
  {\cal R}_{q_T/Q}^2 \, 2z^2(4z^2-20z+17) \,+\,
  {\cal R}_{q_T/Q}^4 \, z^4
  \;,
\end{equation}
and
\begin{equation}
\label{FFs_OQQ_g_d_den_NLO}
\begin{split}
 {\cal T}_g^{\rm [NLO]}&(z; {\cal R}_{q_T/Q}) 
  \,=\, 
  (4 + z^2{\cal R}_{q_T/Q}^2)^5
 \;.
\end{split}
\end{equation}
As done in the heavy-quark case, we define ${\cal R}_{q_T/Q}^2 \equiv \vqTTa/m_Q^2$, with $\vqTTa \equiv \vqTTa_{{\rm \Omega}_{3Q}} = 60\,\text{GeV}^2$ (see Eq.~\eqref{R_qTQ}). 
However, here the gluon-induced LO kinematics set the starting scale at $\mu_{F,0} = 6 m_Q$, as shown in the right diagram of Fig.~\ref{fig:OQQ_FF_diagrams}.

\subsection{The {\tt OMG3Q1.0} functions from {\HFNRevo}}
\label{ssec:FFs-OMG3Q10}

To finalize the construction of our {\tt OMG3Q1.0} collinear FFs for fully heavy $\Omega$ baryons, we apply DGLAP evolution to the initial inputs.
Unlike light hadrons, both heavy-quark and gluon FFs here are subject to nontrivial thresholds, reflecting the perturbative origin of the underlying splittings (see Fig.~\ref{fig:OQQ_FF_diagrams}).

As previously noted, the quark channel activates at $5m_Q$, while the gluon one requires $6m_Q$.
To incorporate these scales, we adopt the {\HFNRevo} framework~\cite{Celiberto:2025euy,Celiberto:2024mex,Celiberto:2024bxu,Celiberto:2024rxa,Celiberto:2025xvy}, designed to evolve heavy-hadron FFs from nonrelativistic inputs.
This method involves three core elements: physical interpretation of the fragmentation process, DGLAP evolution, and estimation of MHOUs through threshold variations.

Originally developed for quarkonium, {\HFNRevo} has been extended to exotic states such as $\TQQ$ baryons~\cite{Celiberto:2024mab,Celiberto:2024beg} and now to ${\rm \Omega}_{3Q}$~\cite{Celiberto:2025ogy}.
Its dual-channel approach requires a threshold-sensitive evolution strategy, which we develop here.

We focus on the evolution procedure, leaving matching and uncertainty quantification to future work.
Following {\HFNRevo}, the heavy-quark FF is evolved first, from its starting scale $\mu_{F,0} = 5m_Q$ up to $6m_Q$, using only the quark-quark Altarelli-Parisi splitting kernel, $P_{qq}$.
This stage, being expanded and decoupled, is handled analytically with {\symJethad}~\cite{Celiberto:2020wpk,Celiberto:2022rfj,Celiberto:2023fzz,Celiberto:2024mrq,Celiberto:2024swu,Celiberto:2025_P5Q_review}.

Then, evolution proceeds numerically from $Q_0 \equiv 6m_Q$ using {\tt APFEL++}~\cite{Bertone:2013vaa,Carrazza:2014gfa,Bertone:2017gds}, leading to the NLO {\tt OMG3Q1.0} set, released in {\tt LHAPDF} format.
Future plans include linking with {\tt EKO}~\cite{Candido:2022tld,Hekhorn:2023gul}.

Though our approach omits light- and bottom-quark channels, current theory lacks initial FFs for nonconstituent-to-triply-heavy transitions.
These channels are thus generated dynamically by evolution and are expected to be strongly suppressed, as indicated by NRQCD studies of color-singlet pseudoscalar~\cite{Braaten:1993rw,Braaten:1993mp,Artoisenet:2014lpa,Zhang:2018mlo,Zheng:2021mqr,Zheng:2021ylc} and vector~\cite{Braaten:1993rw,Braaten:1993mp,Zheng:2019dfk} quarkonia.

\begin{figure}[!t]
\centering

   \includegraphics[scale=0.400,clip]{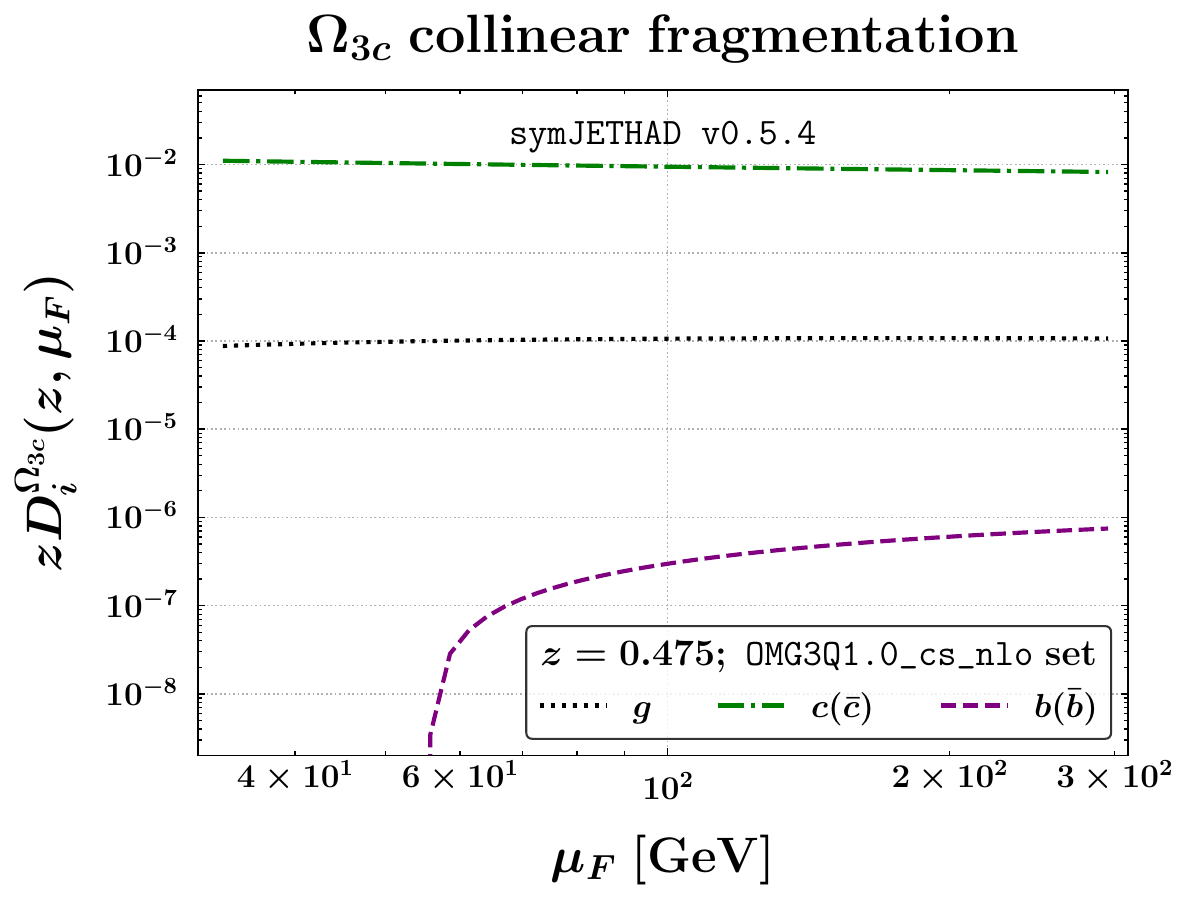}
   \includegraphics[scale=0.400,clip]{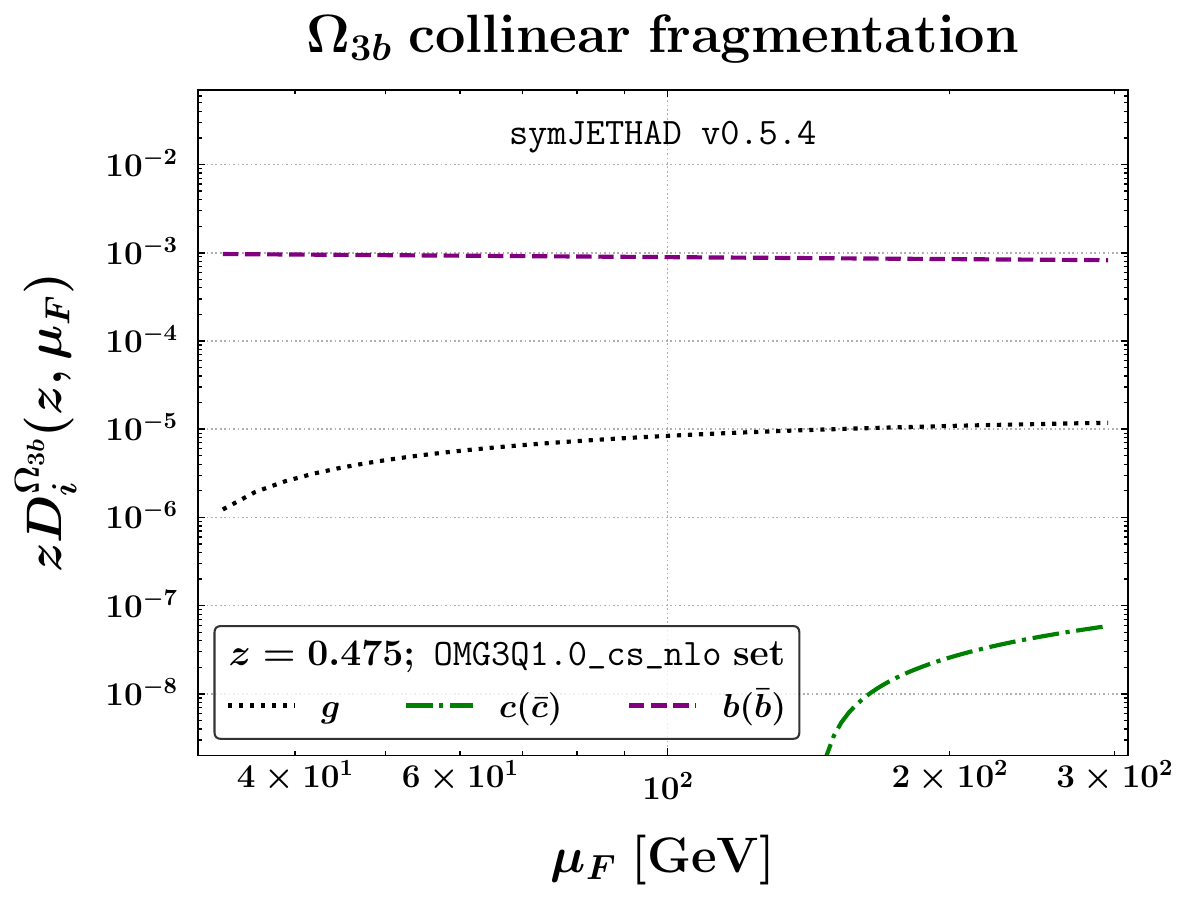}

\caption{Factorization-scale dependence of the {\tt OMG3Q1.0} NLO FFs describing the ZM-VFNS fragmentation of $\Omega_{3c}$ (left) and $\Omega_{3b}$ (right) rare baryons.
The hadron momentum fraction is set to $z = 0.475 \simeq \langle z \rangle$.}
\label{fig:NLO_FFs_OQQ}
\end{figure}

\begin{figure}[!t]
\centering

   \includegraphics[scale=0.400,clip]{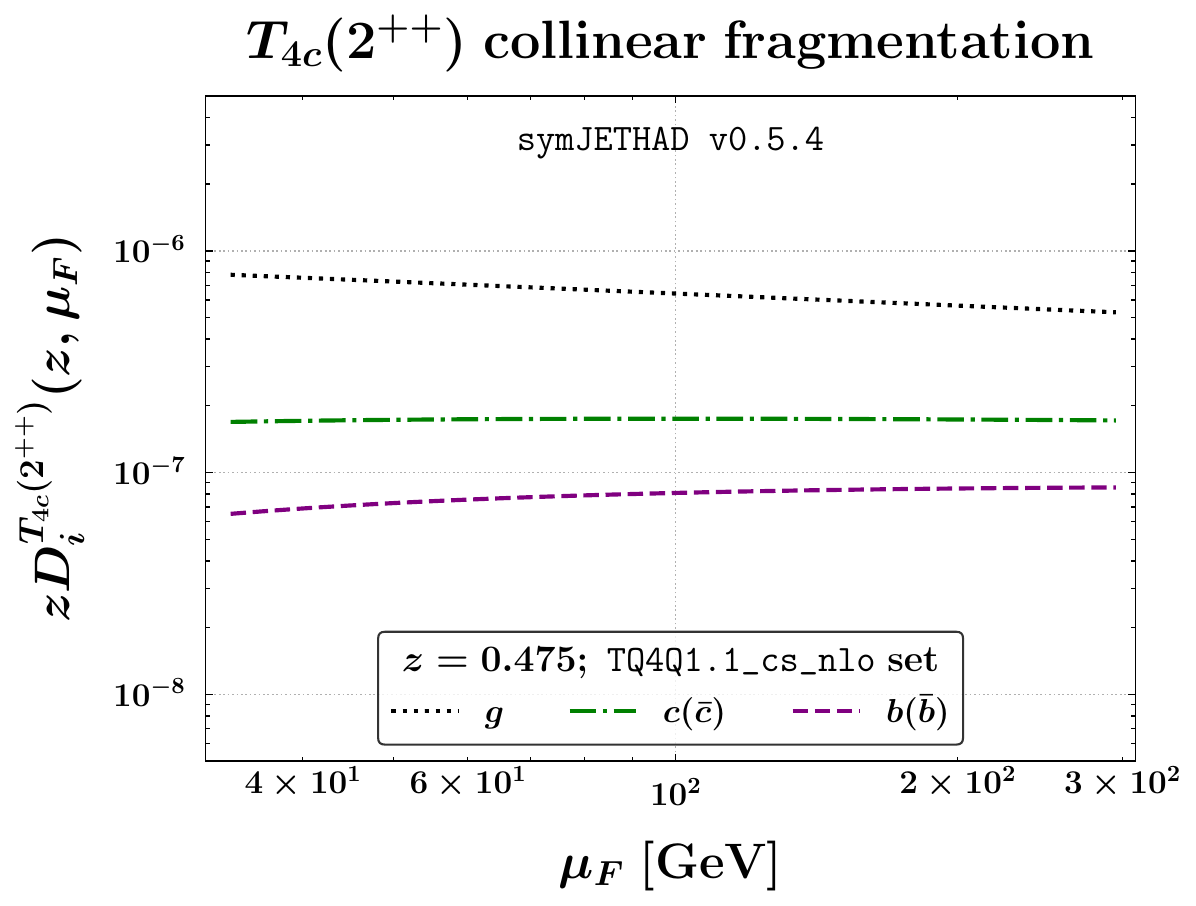}
   \includegraphics[scale=0.400,clip]{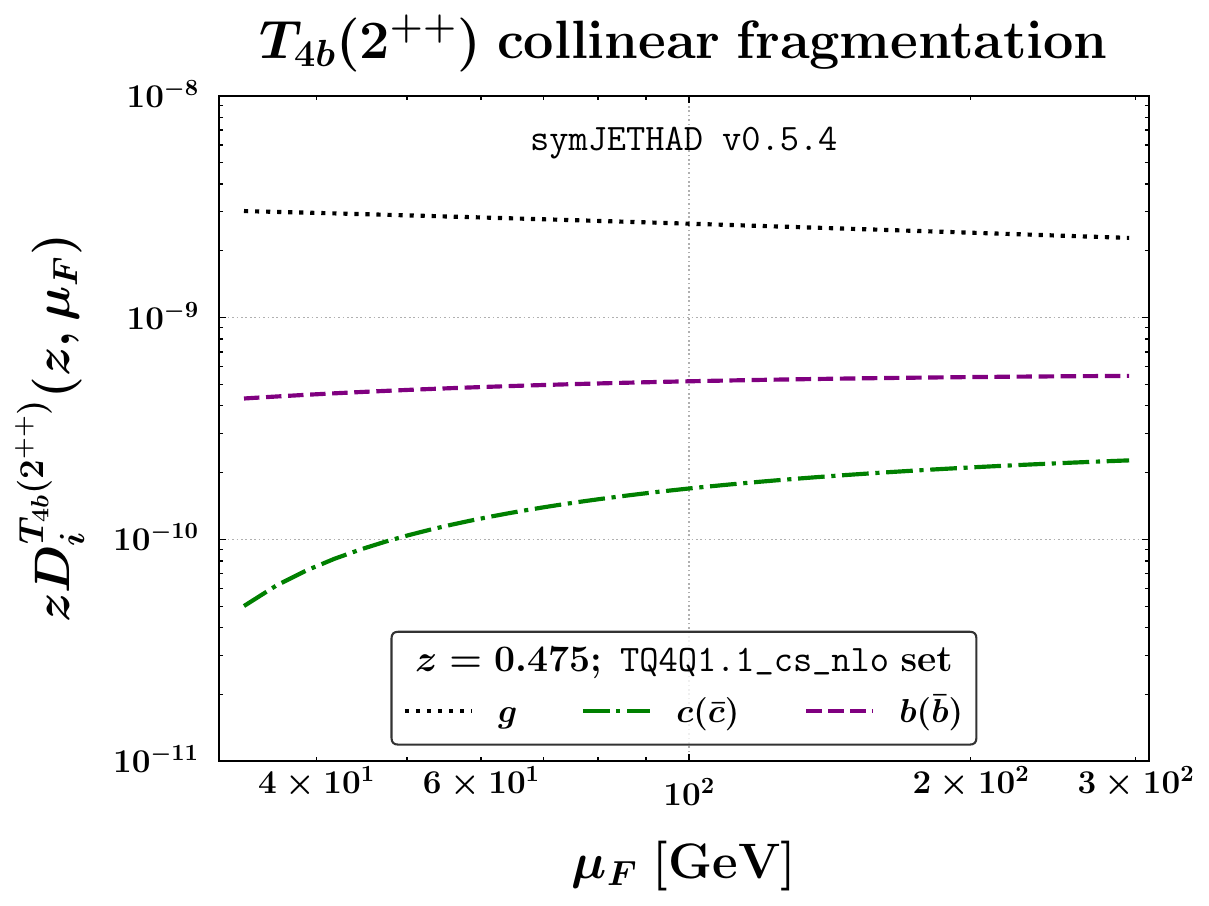}

\caption{Factorization-scale dependence of the {\tt TQ4Q1.1} NLO FFs describing the ZM-VFNS fragmentation of $\TQcTpp$ (left) and $\TQbTpp$ (right) tensor tetraquarks.
The hadron momentum fraction is set to $z = 0.475 \simeq \langle z \rangle$.}
\label{fig:NLO_FFs_TQQ}
\end{figure}

The left (right) panel of Fig.\ref{fig:NLO_FFs_OQQ} illustrates the dependence on the factorization scale $\mu_F$ of the NLO {\tt OMG3Q1.0} functions describing collinear VFNS fragmentation into ${\rm \Omega}_{3c}$ (${\rm \Omega}_{3b}$) baryons.
For comparison, the left (right) panel of Fig.\ref{fig:NLO_FFs_TQQ} shows the corresponding $\mu_F$ evolution for the NLO {\tt TQ4Q1.1} functions~\cite{Celiberto:2024beg,Celiberto:2025dfe,Celiberto:2025ziy,Celiberto:2025vra}, which describe collinear VFNS fragmentation into the tensor tetraquarks $\TQc$ ($\TQb$).
In both cases, the plots are evaluated at a representative value of the momentum fraction, $z = 0.475 \simeq \langle z \rangle$, which lies within the typical region where FFs yield dominant contributions to semihard final states~\cite{Celiberto:2016hae,Celiberto:2017ptm,Bolognino:2018oth,Celiberto:2020wpk,Celiberto:2021dzy,Celiberto:2021fdp,Celiberto:2022dyf,Celiberto:2022keu,Celiberto:2022kxx,Celiberto:2025euy}.

A direct inspection of the two figures reveals that the fragmentation channel initiated by the constituent heavy quark clearly dominates.
Its contribution substantially exceeds those from light partons and from nonconstituent heavy quarks across the entire range of $\mu_F$ values considered.
In the case of ${\rm \Omega}_{3Q}$ production, this dominance is even more pronounced, as the charm-initiated FF surpasses the gluon one by roughly two orders of magnitude.
This behavior stands in contrast with the case of the $\TQQ(2^{++})$ state, where the gluon FF dominates over the quark FF, with ratios ranging from a factor of 5 to nearly an order of magnitude.

The prevailing role of the heavy-quark FF in ${\rm \Omega}_{3Q}$ production reflects the initial-scale hierarchy between the two partonic channels.
Although the gluon FF is enhanced by DGLAP evolution (fed by timelike partonic splittings) the charm FF shows a smoother and more stable scale dependence, with no comparable enhancement mechanism.
As a result, the initial hierarchy of about five orders of magnitude between charm and gluon FFs is reduced to approximately two after evolution.

Despite its smaller magnitude, the gluon FF plays a crucial role in predicting (semi-)inclusive production rates of ${\rm \Omega}_{3Q}$ baryons at hadron colliders.
This is due to the dominant behavior of gluon PDFs over quark PDFs, which amplifies the contribution from the $[gg \to gg]$ partonic subprocess as well as from gluon-initiated fragmentation.
Accordingly, the simultaneous inclusion of both charm and gluon channels as initial-scale inputs in the {\tt OMG3Q1.0} set enhances the overall robustness and predictive power of our framework.

We stress that no initial-scale FF is provided for nonconstituent heavy quarks at $Q_0$; instead, these channels emerge dynamically via DGLAP evolution from gluon and charm contributions.
The resulting behavior and growth of this nonconstituent component with $\mu_F$ are therefore strongly influenced by the value of $Q_0$.
Notably, the higher $Q_0$ adopted in the ${\rm \Omega}_{3Q}$ case delays the onset of nonconstituent-heavy-quark fragmentation compared to the $T_{4Q}$ case, where evolution begins at a lower scale.

Finally, we observe that the gluon-to-${\rm \Omega}_{3Q}$ FF exhibits a gentle increase with $\mu_F$, whereas its gluon-to-$T_{4Q}$ counterpart shows a mild decrease.
In both scenarios, the dependence on the factorization scale is smooth, a feature with valuable phenomenological implications.
Recent analyses have shown that gluon FFs with such smooth $\mu_F$ behavior act as effective ``stabilizers'' for high-energy resummed observables involving the semi-inclusive production of singly~\cite{Celiberto:2021dzy,Celiberto:2021fdp} and multiply~\cite{Celiberto:2022dyf,Celiberto:2022keu,Celiberto:2023rzw} heavy-flavored hadrons.

This property has been termed the \emph{natural stability}~\cite{Celiberto:2022grc} of high-energy resummation (see Section~\ref{sec:HE_resummation}).
It constitutes one of the key ingredients underlying our forthcoming phenomenological analysis (see Section~\ref{sec:phenomenology}).

\section{NLL/NLO$^+$ hybrid factorization at work}
\label{sec:HE_resummation}

The first part of this section offers a brief overview of recent phenomenological advances in QCD high-energy resummation (see Section~\ref{ssec:HE_QCD}).
The second part presents the formal definition of our rare-baryon-sensitive observables, formulated within the hybrid high-energy and collinear factorization framework at $\NLLp$ accuracy (see Section~\ref{ssec:hybrid_factorization}).

\subsection{Phenomenology of the semi-hard QCD: A brief overview with open directions}
\label{ssec:HE_QCD}

Accurate predictions for high-energy observables rely on factorizing long-distance dynamics from short-distance effects in hadronic collisions. 
This is achieved via the collinear framework~\cite{Collins:1989gx,Sterman:1995fz}, where nonperturbative inputs are separated from perturbative computations.

However, in specific phase-space regions, large logarithmic corrections emerge, increasing with the perturbative order and jeopardizing the series' convergence. 
In such regimes, standard collinear factorization must be enhanced by all-order resummations.
In the semi-hard QCD regime~\cite{Gribov:1983ivg}, where $\sqrt{s} \gg Q \gg \LQCD$, logarithms like $\ln(s/Q^2)$ appear at each perturbative order and require resummation~\cite{Celiberto:2017ius,Bolognino:2021bjd,Mohammed:2022gbk,Gatto:2025kfl}.

The \ac{BFKL} formalism~\cite{Fadin:1975cb,Kuraev:1976ge,Kuraev:1977fs,Balitsky:1978ic} performs such resummations, accounting for all $(\alpha_s \ln s)^n$ terms (LL) and $\alpha_s(\alpha_s \ln s)^n$ terms (NLL). 
Within BFKL, scattering amplitudes are expressed as convolutions of a universal Green's function with two transverse-momentum-dependent emission vertices (impact factors), which describe forward-particle production.

The Green's function obeys an integral equation with a kernel known at NLO~\cite{Fadin:1998py,Ciafaloni:1998gs,Fadin:1998jv,Fadin:2000kx,Fadin:2000hu,Fadin:2004zq,Fadin:2005zj}; higher-order corrections are actively studied~\cite{Caola:2021izf,Falcioni:2021dgr,DelDuca:2021vjq,Byrne:2022wzk,Fadin:2023roz,Byrne:2023nqx}.
The main limitation in BFKL-based predictions at NLL stems from the restricted set of NLO off-shell impact factors available. 
These include:
$i$) partonic ones (quarks/gluons)~\cite{Fadin:1999de,Fadin:1999df},
$ii$) forward jets~\cite{Bartels:2001ge,Bartels:2002yj,Caporale:2011cc,Ivanov:2012ms,Colferai:2015zfa}, and
$iii$) light hadrons~\cite{Ivanov:2012iv}.
Other NLO-computed impact factors address:
$iv$) virtual photon to vector meson transitions~\cite{Ivanov:2004pp},
$v$) light-by-light scattering~\cite{Bartels:2000gt,Bartels:2001mv,Bartels:2002uz,Bartels:2004bi,Fadin:2001ap,Balitsky:2012bs}, and
$vi$) Higgs production in the infinite top-mass limit~\cite{Hentschinski:2020tbi,Celiberto:2022fgx,Hentschinski:2022sko} and finite top-mass corrections~\cite{Celiberto:2024bbv,Celiberto:2025ece}, with NNLO extensions in~\cite{DelDuca:2025vux}.

Several LO channels have also been explored:
Drell--Yan~\cite{Hentschinski:2012poz,Motyka:2014lya},
heavy-quark pairs~\cite{Celiberto:2017nyx,Bolognino:2019ccd,Bolognino:2019yls}, and
forward $J/\psi$ at moderate transverse momenta~\cite{Boussarie:2017oae,Boussarie:2015jar,Boussarie:2016gaq,Boussarie:2017xdy}.
Gold-plated channels offer ideal probes for BFKL signatures at hadron colliders, including:
Mueller--Navelet dijets at NLO~\cite{Mueller:1986ey,Colferai:2010wu,Ducloue:2013hia,Caporale:2013uva,Colferai:2015zfa,Caporale:2015uva,Ducloue:2015jba,Celiberto:2015yba,Celiberto:2015mpa,Celiberto:2016ygs,Celiberto:2016vva,Caporale:2018qnm,deLeon:2020myv,deLeon:2021ecb,Celiberto:2022gji,Baldenegro:2024ndr},
correlated dihadron production~\cite{Celiberto:2016hae,Celiberto:2016zgb,Celiberto:2017ptm,Celiberto:2017uae,Celiberto:2017ydk},
multi-jet topologies~\cite{Caporale:2015vya,Caporale:2015int,Caporale:2016soq,Caporale:2016vxt,Caporale:2016xku,Celiberto:2016vhn,Caporale:2016djm,Caporale:2016pqe,Chachamis:2016qct,Chachamis:2016lyi,Caporale:2016lnh,Caporale:2016zkc,Caporale:2017jqj,Chachamis:2017vfa},
hadron-plus-jet~\cite{Bolognino:2018oth,Bolognino:2019cac,Bolognino:2019yqj,Celiberto:2020wpk,Celiberto:2020rxb,Celiberto:2022kxx},
Higgs-plus-jet~\cite{Celiberto:2020tmb,Celiberto:2021fjf,Celiberto:2021tky,Celiberto:2021txb,Celiberto:2021xpm},
heavy-light dijets~\cite{Bolognino:2021mrc,Bolognino:2021hxx}, and
heavy-flavor hadrons~\cite{Boussarie:2017oae,Celiberto:2017nyx,Bolognino:2019ouc,Bolognino:2019yls,Bolognino:2019ccd,Celiberto:2021dzy,Celiberto:2021fdp,Bolognino:2022wgl,Celiberto:2022dyf,Celiberto:2022grc,Bolognino:2022paj,Celiberto:2022qbh,Celiberto:2022keu,Celiberto:2022zdg,Celiberto:2022kza,Celiberto:2024omj}.

Clean tests of BFKL dynamics are provided by forward single-particle production, which probes the small-$x$ proton \ac{UGD} driven by BFKL evolution.
Examples include:
light vector meson leptoproduction~\cite{Anikin:2009bf,Anikin:2011sa,Besse:2013muy,Bolognino:2018rhb,Bolognino:2018mlw,Bolognino:2019bko,Bolognino:2019pba,Celiberto:2019slj,Luszczak:2022fkf,Bolognino:2021niq,Bolognino:2021gjm,Bolognino:2022uty,Celiberto:2022fam,Bolognino:2022ndh},
exclusive quarkonium photoproduction~\cite{Bautista:2016xnp,Garcia:2019tne,Hentschinski:2020yfm,Peredo:2023oym,Hentschinski:2025ovo},
inclusive Drell--Yan~\cite{Motyka:2014lya,Brzeminski:2016lwh,Motyka:2016lta,Celiberto:2018muu},
and $b$-tagged jet emission~\cite{Chachamis:2015ona,Chachamis:2013bwa,Chachamis:2009ks}.

Small-$x$ UGDs also inform collinear fits of \ac{PDFs} with small-$x$ resummation~\cite{Ball:2017otu,Abdolmaleki:2018jln,Bonvini:2019wxf,Silvetti:2022hyc,Silvetti:2023suu,Rinaudo:2024hdb,Celiberto:2025nnq}, and connect to twist-two gluon TMDs~\cite{Bacchetta:2020vty,Celiberto:2021zww,Bacchetta:2021oht,Bacchetta:2021lvw,Bacchetta:2021twk,Bacchetta:2022esb,Bacchetta:2022crh,Bacchetta:2022nyv,Celiberto:2022omz,Bacchetta:2023zir,Bacchetta:2024fci,Bacchetta:2024uxb}.
Refinements of this connection are found in~\cite{Hentschinski:2021lsh,Mukherjee:2023snp}, while the role of UGDs in dipole cross sections is explored in~\cite{Boroun:2023goy,Boroun:2023ldq}.

Analyses of small-$x$ resummed observables in Higgs and heavy-flavor production have also been developed via the {\Hell} method~\cite{Bonvini:2018ixe,Silvetti:2022hyc}, grounded in the \ac{ABF} framework~\cite{Ball:1995vc,Ball:1997vf,Altarelli:2001ji,Altarelli:2003hk,Altarelli:2005ni,Altarelli:2008aj,White:2006yh}.
This approach integrates collinear factorization with high-energy resummation, incorporating key theorems of high-energy factorization~\cite{Catani:1990xk,Catani:1990eg,Collins:1991ty,Catani:1993ww,Catani:1993rn,Catani:1994sq,Ball:2007ra,Caola:2010kv}, offering a complementary perspective on small-$x$ QCD.

A major challenge for BFKL predictions in Mueller--Navelet jet production lies in the destabilizing effect of NLL corrections.
These terms, although of the same formal order as LL ones, carry opposite sign, which leads to instabilities in the resummed series. This becomes critical when probing \ac{MHOUs} through variations of the energy scales.

As a result, predictions may become unphysical at large jet rapidity separations, and azimuthal correlations can show abnormal trends across rapidity ranges.
Several mitigation strategies have been proposed. Among them, the \ac{BLM} scale-setting scheme~\cite{Brodsky:1996sg,Brodsky:1997sd,Brodsky:1998kn,Brodsky:2002ka}, tailored for semi-hard observables~\cite{Caporale:2015uva}, has achieved partial stabilization, especially for azimuthal observables, with modest improvement in data agreement.

Still, BLM's effectiveness is limited in processes like dihadron or hadron-plus-jet emissions, where it suggests optimal renormalization scales well above the natural ones~\cite{Celiberto:2017ius,Bolognino:2018oth,Celiberto:2020wpk}.
This causes suppression of total cross sections and a deterioration in event statistics.

Encouraging signals of improved stability under higher-order effects and scale variations have recently emerged in Higgs-sensitive final states~\cite{Celiberto:2020tmb,Celiberto:2023rtu,Celiberto:2023uuk,Celiberto:2023eba,Celiberto:2023nym,Celiberto:2023dkr,Celiberto:2023rqp,Celiberto:2024mdt,Celiberto:2024bfu,Celiberto:2025edg}.
The trend was first observed in semi-inclusive production of $\Lambda_c$ hyperons~\cite{Celiberto:2021dzy} and bottom-flavored hadrons~\cite{Celiberto:2021fdp} at the LHC, and traced back to the behavior of ZM-VFNS collinear FFs, which drive heavy-flavor production at high $p_T$.

Follow-up studies on vector quarkonia~\cite{Celiberto:2022dyf,Celiberto:2023fzz}, charmed $B$ mesons~\cite{Celiberto:2022keu,Celiberto:2024omj}, exotic tetraquarks~\cite{Celiberto:2023rzw,Celiberto:2024mab,Celiberto:2024mrq,Celiberto:2024beg,Celiberto:2025dfe,Celiberto:2025ziy,Celiberto:2025vra}, and rare $\Omega$ baryons~\cite{Celiberto:2025ogy}, have confirmed that this so-called \emph{natural stability} of QCD high-energy resummation~\cite{Celiberto:2022grc} is a robust feature, inherently linked to final states involving heavy flavor.

\subsection{High-energy resummation at $\NLLp$}
\label{ssec:hybrid_factorization}

\begin{figure}[!t]
\centering

\includegraphics[width=0.575\textwidth]{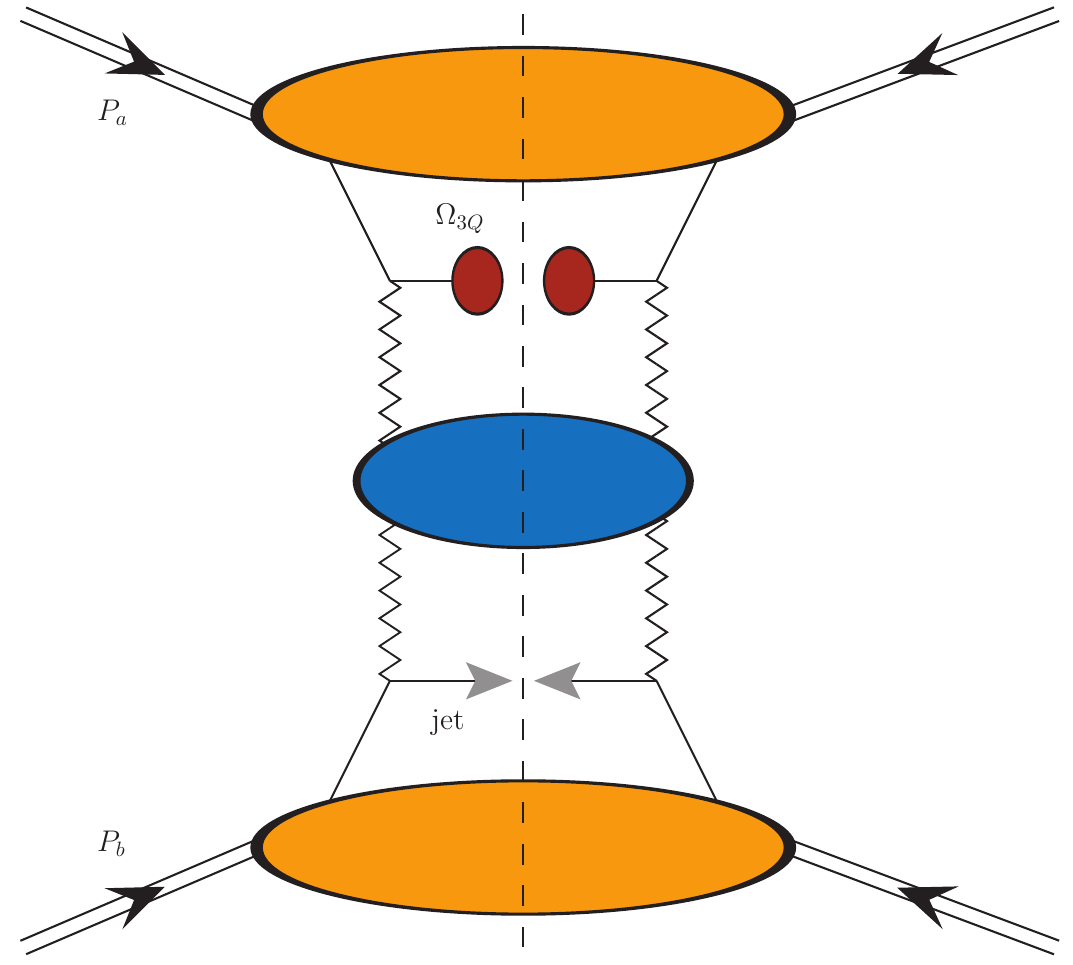}

\caption{Hybrid collinear factorization for semi-inclusive $\Omega_{3Q}$ plus jet production at hadron colliders.
Firebrick ovals represent the collinear FFs for rare baryons.
Gray arrows indicate the light-flavored jet.
The orange blob denotes the proton collinear PDFs.
The BFKL Green's function (blue oval) is linked to the two singly off-shell emission functions via Reggeized gluon lines.
Diagram created with {\tt JaxoDraw 2.0}~\cite{Binosi:2008ig}.}
\label{fig:process}
\end{figure}

We consider the process given in Fig.~\ref{fig:process}
\begin{equation}
\label{process}
\setlength{\jot}{10pt} 
\begin{split}
    {\rm p}(P_a) \;+\; {\rm p}(P_b) &\;\rightarrow\; \Omega_{3b}(\vec q_1, y_1) \;+\; {\cal X} \;+\; {\rm jet}(\vec q_2, y_2) \; ,
\end{split}
\end{equation}
where a fully bottomed, rare baryon $\Omega_{3b}$ is semi-inclusively detected together with a light jet.
The outgoing objects feature high transverse-momentum values, $|\vec q_{1,2}| \gg \Lambda_{\rm QCD}$, and large distance in rapidity is $ \DY = y_1 - y_2$. 

High transverse momenta and rapidity separations allow for with semi-hard final-state configurations.
Furthermore, the transverse-momentum ranges must be sufficiently high to ensure that ZM-VFNS collinear fragmentation dominates the production mechanism of heavy hadrons.

The incoming proton four-momenta are decomposed in terms of Sudakov vectors, which satisfy
$P_a^2= P_b^2=0$ and $2 (P_a\cdot P_b) = s$.
Therefore, one rewrites $q_1$ and $q_2$ as
\begin{equation}\label{sudakov}
q_{1,2} = x_{1,2} P_{a,b} - \frac{q_{1,2\perp}^{\,2}}{x_{1,2} s}P_{b,a} + q_{1,2\perp} \ , \qquad
\vec q_{1,2}^{\,2} \equiv -q_{1,2\perp}^2\;.
\end{equation}
Final-state object's longitudinal fractions, $x_{1,2}$, depend on rapidities as
$y_{1,2}=\pm\frac{1}{2}\ln\frac{x_{1,2}^2 s}
{\vec q_{1,2}^2}$.
Thus one have $\drv y_{1,2} = \pm \frac{\drv x_{1,2}}{x_{1,2}}$, and $\DY \equiv y_1 - y_2 = \ln\frac{x_1 x_2 s}{|\vec q_1||\vec q_2|}$.

At LO in pure collinear QCD, the cross section for the process in Eq.~\eqref{process} reduces to a one-dimensional convolution involving proton PDFs, hadron FFs, and partonic hard-scattering kernels. Explicitly, one obtains
\begin{equation}
\label{sigma_collinear}
\begin{split}
 \frac{\drv \sigma_{[p \;+\; p \;\;\to\;\; {\Omega_{3b}} \;+\; {\rm jet}]}^{\rm LO}}{\drv x_1 \drv x_2 \drv^2 \vec q_1 \drv^2 \vec q_2}
 &=\sum_{a,b} \int_0^1 \drv x_a \int_0^1 \drv x_b\ 
 f_a(x_a, \mu_F) f_b(x_b, \mu_F)
\int_{x_1}^1 \frac{\drv z}{z}D^{\Omega_{3b}}_{a}\left(\frac{x_1}{z}\right) 
\frac{\drv {\hat\sigma}_{a,b}}
{\drv x_1\drv x_2\drv z\,\drv ^2\vec q_1\drv ^2\vec q_2}\;,
\end{split}
\end{equation}
Here, the indices $(a,b)$ run over quarks, antiquarks, and gluons; $f_{a,b}$ denote the proton PDFs, $D^{\Omega_{3b}}_{a}$ are the FFs into the $\Omega_{3b}$ baryon, $x{a,b}$ are the longitudinal momentum fractions of the incoming partons, $z$ is the momentum fraction of the fragmenting parton carried by the $\Omega_{3b}$, and $\drv \hat\sigma_{a,b}$ represent the partonic cross sections for the corresponding subprocesses.

In contrast, deriving the high-energy resummed cross section within our hybrid factorization framework requires a two-step procedure.
The first step implements the high-energy factorization dictated by BFKL resummation, while the second introduces a collinear refinement through the inclusion of PDFs and FFs.
The resulting differential cross section naturally admits a Fourier expansion in the azimuthal-angle difference, with coefficients that encode the underlying dynamics of the process.
We get
\begin{equation}
 \label{dsigma_Fourier}
 \frac{\drv \sigma^\NLLp}{\drv y_1 \drv y_2 \drv \vec q_1 \drv \vec q_2 \drv \phi_1 \drv \phi_2} =
 \frac{1}{(2\pi)^2} \left[{\cal C}_0^\NLLp + 2 \sum_{n=1}^\infty \cos \left(n (\phi - \pi)\right) \,
 {\cal C}_n^\NLLp \right]\, ,
\end{equation}
with $\phi_{1,2}$ being the final-state azimuthal angles and $\phi = \phi_1 - \phi_2$.
Here, $\phi_{1,2}$ denote the azimuthal angles of the final-state particles, and $\phi = \phi_1 - \phi_2$ is their difference.
The azimuthal coefficients are computed within the BFKL framework and encode the resummation of LL and NLL energy logarithms.
Our calculation is performed in the $\MSb$ renormalization scheme~\cite{PhysRevD.18.3998}, which ensures a consistent treatment of perturbative corrections and ultraviolet divergences.
We obtain
\begin{equation}
\label{Cn_NLLp_MSb}
\begin{split}
 \CnNLLp &= \int_0^{2\pi} \drv \phi_1 \int_0^{2\pi} \drv \phi_2\,
 \cos \left(n (\phi - \pi)\right) \,
 \frac{\drv \sigma^\NLLp}{\drv y_1 \drv y_2\, \drv |\vec q_1| \, \drv |\vec q_2| \drv \phi_1 \drv \phi_2}\;
\\
 &= \; \frac{e^{\DY}}{s} 
 \int_{-\infty}^{+\infty} \drv \nu \, e^{{\DY} \bar \alpha_s(\mu_R)\chi^\NLO(n,\nu)}
\\
 &\times \; \alpha_s^2(\mu_R) \, 
 \biggl\{
 \F_1^\NLO(n,\nu,|\vec q_1|, x_1)[\F_2^\NLO(n,\nu,|\vec q_2|,x_2)]^*\,
\\ 
 &+ \,
 \left.
 \bar \alpha_s^2(\mu_R)
 \, \DY
 \frac{\beta_0}{4 N_c}\chi(n,\nu)f(\nu)
 \right\} \;.
\end{split}
\end{equation}
Here, we define $\bar{\alpha}_s(\mu_R) \equiv \alpha_s(\mu_R), N_c/\pi$, with $N_c$ the number of QCD colors, and $\beta_0 = (11 N_c - 2 n_f)/3$ the leading coefficient of the QCD $\beta$-function, where $n_f$ denotes the number of active quark flavors.
We adopt a two-loop running for the strong coupling, initialized at $\alpha_s(M_Z) = 0.118$, and consistently account for a variable flavor number $n_f$ along the evolution.

The resummation kernel entering the exponent of Eq.~\eqref{Cn_NLLp_MSb} reads
\begin{eqnarray}
 \label{chi}
 \chi^\NLO(n,\nu) = \chi(n,\nu) + \bar\alpha_s \hat \chi(n,\nu) \;,
\end{eqnarray}
where
\begin{eqnarray}
 \label{kernel_LO}
 \chi\left(n,\nu\right) = -2\gamma_{\rm E} - 2 \, {\rm Re} \left\{ \psi\left(\frac{1+n}{2} + i \nu \right) \right\} \, 
\end{eqnarray}
The functions $\chi(n,\nu)$ represent the LO eigenvalues of the BFKL kernel, $\gamma_{\rm E}$ is the Euler--Mascheroni constant, and $\psi(z) \equiv \Gamma^\prime(z)/\Gamma(z)$ denotes the digamma function, \emph{i.e.}, the logarithmic derivative of the Gamma function.
The term $\hat\chi(n,\nu)$ in Eq.~\eqref{chi} encodes the NLO correction to the kernel.
\begin{equation}
\begin{split}
\label{chi_NLO}
\hat \chi\left(n,\nu\right) &= \bar\chi(n,\nu)+\frac{\beta_0}{8 N_c}\chi(n,\nu)
\left(-\chi(n,\nu)+10/3+2\ln\frac{\mu_R^2}{\mu_C^2}\right) \;,
\end{split}
\end{equation}
The baryon mass is fixed to $m_{\Omega_{3b}} = 3 m_b$, with $m_b = 4.5$~GeV denoting the bottom mass.
We note that this choice, although widely adopted in phenomenological studies, does not account for the baryonic binding energy. For triply heavy systems like the $\Omega_{3b}$, this energy is non-negligible due to the weak repulsive interactions and the limited kinetic energy of bottom quarks, as discussed in quark model analyses~\cite{Yang:2011rp}. Recent lattice and potential-model studies~\cite{Zhou:2025fpp} estimate the $\Omega_{3b}$ mass to be slightly below $3m_b$, with corrections of a few hundred MeV. Since our predictions are only mildly sensitive to this variation, and the theoretical uncertainties on the initial FFs dominate, we opted for the fixed-mass prescription for simplicity.
On the light-flavor side, the jet does not carry a heavy mass scale, so its transverse mass reduces to the transverse momentum, $|\vec q_2|$.
The characteristic function $\bar\chi(n,\nu)$ appearing in the BFKL exponent was derived in Ref.~\cite{Kotikov:2000pm,Kotikov:2002ab}, and is given by
\begin{equation}
 \label{kernel_NLO}
 \bar \chi(n,\nu)\,=\, - \frac{1}{4}\left\{\frac{\pi^2 - 4}{3}\chi(n,\nu) - 6\zeta(3) - \frac{\drv^2 \chi}{\drv\nu^2} + \,2\,\phi(n,\nu) + \,2\,\phi(n,-\nu)
 \right.
\end{equation}
\[
 \left.
 +\; \frac{\pi^2\sinh(\pi\nu)}{2\,\nu\, \cosh^2(\pi\nu)}
 \left[
 \left(3+\left(1+\frac{n_f}{N_c^3}\right)\frac{11+12\nu^2}{16(1+\nu^2)}\right)
 \delta_{n0}
 -\left(1+\frac{n_f}{N_c^3}\right)\frac{1+4\nu^2}{32(1+\nu^2)}\delta_{n2}
\right]\right\} \, ,
\]
with
\begin{equation}
\label{kernel_NLO_phi}
 \phi(n,\nu)\,=\,-\int_0^1 \drv x\,\frac{x^{-1/2+i\nu+n/2}}{1+x}\left\{\frac{1}{2}\left(\psi^\prime\left(\frac{n+1}{2}\right)-\zeta(2)\right)+\mbox{Li}_2(x)+\mbox{Li}_2(-x)\right.
\end{equation}
\[
\left.
 +\; \ln x\left[\psi(n+1)-\psi(1)+\ln(1+x)+\sum_{k=1}^\infty\frac{(-x)^k}{k+n}\right]+\sum_{k=1}^\infty\frac{x^k}{(k+n)^2}\left[1-(-1)^k\right]\right\}
\]
\[
 =\; \sum_{k=0}^\infty\frac{(-1)^{k+1}}{k+(n+1)/2+i\nu}\left\{\psi^\prime(k+n+1)-\psi^\prime(k+1)\right.
\]
\[
 \left.
 +\; (-1)^{k+1}\left[\beta_{\psi}(k+n+1)+\beta_{\psi}(k+1)\right]-\frac{\psi(k+n+1)-\psi(k+1)}{k+(n+1)/2+i\nu}\right\} \; ,
\]
where
\begin{equation}
\label{kernel_NLO_phi_beta_psi}
 \beta_{\psi}(z)=\frac{1}{4}\left[\psi^\prime\left(\frac{z+1}{2}\right)
 -\psi^\prime\left(\frac{z}{2}\right)\right] \; ,
\end{equation}
and
\begin{equation}
\label{dilog}
\mbox{Li}_2(x) = \int^x_0 \drv \zeta \,\frac{\ln(1-\zeta)}\zeta \; .
\end{equation}
The singly off-shell emissions functions read
\begin{equation}
\label{IFs}
\F_{1,2}^\NLO(n,\nu,|\vec q_{1,2}|,x_{1,2}) =
\F_{1,2}(n,\nu,|\vec q_{1,2}|,x_{1,2}) +
\alpha_s(\mu_R) \, \hat \F_{1,2}(n,\nu,|\vec q_{1,2}|,x_{1,2}) \;.
\end{equation}
Here, the LO expressions portraying the production of a forward hadron and a forward jet are given by
\begin{equation}
\label{LOHEF}
\begin{split}
\F_h(n,\nu,|\vec q_h|,x_h) 
&= 2 \, \sqrt{\frac{C_F}{C_A}}
|\vec q_h|^{2i\nu-1}\,\int_{x_h}^1\frac{\drv z}{z}
\left(\frac{z}{x_h} \right)
^{2 i\nu-1} 
 \left[\frac{C_A}{C_F}f_g(z)D_g^h\left(\frac{x_h}{z}\right)
 +\sum_{a=q,\bar q}f_a(z)D_a^h\left(\frac{x_h}{z}\right)\right] 
\end{split}
\end{equation}
and
\begin{equation}
 \label{LOJEF}
 \F_J(n,\nu,|\vec q_J|,x_J) =  2 \sqrt{\frac{C_F}{C_A}}
 |\vec q_J|^{2i\nu-1}\,\left[\frac{C_A}{C_F}f_g(x_J)
 +\sum_{b=q,\bar q}f_b(x_J)\right] \;,
\end{equation}
with $C_F \equiv (N_c^2-1)/(2N_c)$ and $C_A \equiv N_c$ the Casimir factors in QCD.
The $f(\nu)$ term embodies the logarithmic derivative of LO functions
\begin{equation}
 f(\nu) = \frac{i}{2} \, \frac{\drv}{\drv \nu} \ln\left(\frac{\F_1}{\F_2^*}\right) + \ln\left(|\vec q_1| |\vec q_2|\right) \;.
\label{fnu}
\end{equation}

In Eq.~\eqref{Cn_NLLp_MSb}, the remaining ingredients are the NLO corrections to the emission functions, denoted by $\hat{\mathcal{F}}_{1,2}$.
The NLO term associated with the forward hadron was derived in Ref.~\cite{Ivanov:2012iv}, and its explicit expression is reported in Appendix~\hyperlink{app:NLOHEF}{A}.
For the forward jet, we follow the prescription outlined in Refs.~\cite{Ivanov:2012iv,Ivanov:2012ms}.

To enable practical implementation, we introduce a jet selection function,\footnote{Jet algorithms are typically categorized into two classes: \emph{cone-type} and \emph{sequential-clustering} algorithms. 
For an in-depth overview, see Refs.~\cite{Chekanov:2002rq,Salam:2010nqg}. A widely used example of the latter is the (anti-)$\kappa_\perp$ algorithm~\cite{Catani:1993hr,Cacciari:2008gp}.} evaluated within the \ac{SCA}~\cite{Furman:1981kf,Aversa:1988vb}, using its cone-type implementation~\cite{Colferai:2015zfa} with jet radius $R = 0.5$.
The analytical expression of the NLO jet emission function is given in Appendix~\hyperlink{app:NLOJEF}{B}.

Establishing a direct phenomenological comparison between our hybrid factorization approach and fixed-order QCD predictions would necessitate a dedicated computational framework able to evaluate NLO distributions for two-particle final states in hadronic collisions.
At present, such a tool does not appear to exist within the available literature or public codes.

As a practical alternative, we benchmark our results against fixed-order expectations by truncating the azimuthal-coefficient expansion in Eq.~\eqref{Cn_NLLp_MSb} at ${\cal O}(\alpha_s^3)$.
This defines an effective high-energy fixed-order approximation, denoted $\HENLOp$, which preserves the leading-power asymptotic behavior of full NLO calculations, while neglecting terms suppressed by inverse powers of the partonic center-of-mass energy.

The $\MSb$ expression for the azimuthal coefficients at $\HENLOp$ accuracy reads
\begin{equation}
\label{Cn_HENLO_MSb}
 \CnHENLOp =
 \frac{e^{\DY}}{s}
 \int_{-\infty}^{+\infty} \drv \nu \,
 \alpha_s^2(\mu_R) \,
 \left[ 1 + \bar \alpha_s(\mu_R) \DY \chi(n,\nu) \right] \,
 \F_1^\NLO(n,\nu,|\vec q_1|, x_1) \,[\F_2^\NLO(n,\nu,|\vec q_2|,x_2)]^*
 \;,
\end{equation}
with the resummation kernel being expanded and truncated at ${\cal O}(\alpha_s)$.
For the sake of comparison, we will also provide results within pure LL accuracy
\begin{equation}
\label{Cn_LL_MSb}
  \CnLL = \frac{e^{\DY}}{s} 
 \int_{-\infty}^{+\infty} \drv \nu \, e^{{\DY} \bar \alpha_s(\mu_R)\chi(n,\nu)} \, \alpha_s^2(\mu_R) \, \F_1(n,\nu,|\vec q_1|, x_1)[\F_2(n,\nu,|\vec q_2|,x_2)]^* \,.
\end{equation}

Equations~\eqref{Cn_NLLp_MSb} to~\eqref{Cn_LL_MSb} define the structure of our hybrid factorization formalism.
Following the BFKL approach, the hadronic cross section is written as a transverse-momentum convolution of the Green's function with two off-shell emission functions (impact factors), which embed collinear PDFs and FFs, thus bridging high-energy and collinear dynamics.

The $\NLLp$ label denotes full resummation of energy logarithms at next-to-leading-logarithmic accuracy, based on NLO-calculated perturbative inputs.
The `$+$' superscript in Eq.~\eqref{Cn_NLLp_MSb} highlights the inclusion of selected next-to-NLL terms, specifically the cross-product of the two NLO impact-factor corrections.

Renormalization and factorization scales are set according to the natural kinematic choice
\begin{equation}
\label{natural scales}
 \mu_R = \mu_F = \mu_N = m_{1 \perp} + m_{2 \perp} \;,
\end{equation}
with $m_{i \perp}$ standing transverse masses of final-state particles.

For collinear PDFs, we use the {\tt NNPDF4.0} NLO set~\cite{NNPDF:2021uiq,NNPDF:2021njg}, accessed via the {\tt LHAPDFv6.5.4} interface~\cite{Buckley:2014ana}.
These distributions are extracted from global fits based on the replica method~\cite{Forte:2002fg}, a strategy rooted in neural-network training and now broadly adopted in multi-dimensional analyses of proton structure~\cite{Bacchetta:2017gcc,Scimemi:2019cmh,Bacchetta:2019sam,Bacchetta:2022awv,Bury:2022czx,Moos:2023yfa}.
For a critical discussion of inter-set correlations and uncertainty propagation, see Ref.~\cite{Ball:2021dab}.

Unless otherwise stated, all results presented in this review are obtained within the $\MSb$ renormalization scheme~\cite{PhysRevD.18.3998}.

\section{Rare $\Omega$ baryons from HL-LHC to FCC with {\Jethad}}
\label{sec:phenomenology}

The numerical results shown in this review were produced using {\Jethad}, a flexible and modular framework built upon a hybrid architecture of \textsc{Python}- and \textsc{Fortran}-based components.
Originally designed to handle a wide range of physical distributions across diverse theoretical approaches~\cite{Celiberto:2020wpk,Celiberto:2022rfj,Celiberto:2023fzz,Celiberto:2024mrq,Celiberto:2024swu,Celiberto:2025_P5Q_review}, {\Jethad} provides a unified environment for the computation, control, and analysis of high-energy observables.
In our study, the generation of differential distributions relied on selected \textsc{Fortran 2008} modules, while data refinement and post-processing were performed using the embedded \textsc{Python 3.0} analyzer.

An overview of the current {\Jethad} version ({\tt v0.5.1}) is given in Section~\ref{ssec:jethad}; note that this version is not yet publicly released.
The uncertainty estimation strategy adopted throughout the analysis is discussed in Section~\ref{ssec:uncertainty}, and the applied selection criteria for final-state particles are described in Section~\ref{ssec:final_state}.
Phenomenological results for the $\Omega_{3b}$-plus-jet system, including rapidity-interval and transverse-momentum distributions, are presented in Sections~\ref{ssec:I_rates} and~\ref{ssec:pT_rates}, respectively.

This analysis builds upon and generalizes the findings of Section~4.2 in Ref.~\cite{Celiberto:2025ipt}, which was limited to $\Omega_{3c}$-plus-jet final states.

\subsection{The {\Jethad} {\tt v0.5.4} technology}
\label{ssec:jethad}

The {\Jethad} project was initiated in late 2017 to address the growing need for precise predictions of semi-hard hadron~\cite{Celiberto:2016hae,Celiberto:2017ptm} and jet~\cite{Celiberto:2015yba,Celiberto:2016ygs,Bolognino:2018oth} final states at the LHC, proposed as sensitive channels to probe high-energy resummation effects in QCD. 
From the outset, {\Jethad} has aimed to provide a dedicated numerical framework for the computation and analysis of high-energy observables.

The first named release, {\tt v0.2.7}, enabled a comparative study of BFKL and DGLAP dynamics in semi-inclusive hadron+jet production~\cite{Celiberto:2020wpk}. 
Subsequent versions introduced support for forward heavy-quark pair production ({\tt v0.3.0}~\cite{Bolognino:2019yls}), Higgs and $p_T$-spectrum observables ({\tt v0.4.2}\cite{Celiberto:2020tmb}), and a Python-based analyzer interfaced with the Fortran core ({\tt v0.4.3}~\cite{Bolognino:2021mrc}).

Advancements in heavy-flavor phenomenology followed with {\tt v0.4.4}\cite{Celiberto:2021dzy}, incorporating ZM-VFNS predictions at NLO. 
The forward Drell--Yan module \deffont{\rlap{D}Unamis} (\textsc{DYnamis})~\cite{Celiberto:2018muu} was added in {\tt v0.4.5}, while integration with the \textsc{LExA} code in {\tt v0.4.6}~\cite{Bolognino:2021niq} enabled low-$x$ studies using small-$x$ TMD densities.

With {\tt v0.4.7}~\cite{Celiberto:2022dyf}, {\Jethad} expanded into quarkonium production via NRQCD fragmentation. 
Versions {\tt v0.5.0}~\cite{Celiberto:2023fzz} and {\tt v0.5.1}~\cite{Celiberto:2024omj} brought improved MHOU treatment, a broader range of transverse-momentum observables~\cite{Celiberto:2022gji,Celiberto:2022kxx}, and support for matching to collinear factorization~\cite{Celiberto:2023rtu,Celiberto:2023uuk,Celiberto:2023eba,Celiberto:2023nym,Celiberto:2023rqp}.

The {\tt v0.5.2} release marked the debut of {\symJethad}, a \textsc{Mathematica}-based~\cite{Mathematica_V14-2} plugin for symbolic calculations in high-energy QCD. Successive versions added full support for exotic hadrons~\cite{Celiberto:2024beg} and rare-hadron production~\cite{Celiberto:2025ogy}, with {\tt v0.5.3} and {\tt v0.5.4} respectively.

From its core architecture to its specialized modules, {\Jethad} is designed for computational efficiency. Its multidimensional integrators are optimized for parallel execution and dynamically select the most effective algorithm based on the integration's features.

Process definitions are handled via a structure-based interface, where final-state particles are instantiated as \emph{objects} with properties like mass, charge, kinematic ranges, and rapidity labels. 
These are generated from a master database and injected into process modules through a dedicated controller.

Initial-state configurations are equally flexible. A \emph{particle-ascendancy} attribute identifies the production mechanism (hadro-, electro-, or photoproduction), activating only the relevant computational modules.

Built as an object-oriented system, {\Jethad} is agnostic to the specific physical process under study. 
While originally tailored to high-energy QCD and TMD factorization, its modular architecture allows the seamless integration of alternative theoretical frameworks through additional (super)modules. 
These are incorporated via a built-in point-to-routine mechanism, making {\Jethad} a versatile tool for phenomenological studies.

A public release of {\Jethad}, offering a standardized computational platform for diverse high-energy processes and formalisms, is planned in the near future.

\subsection{Uncertainty quantification}
\label{ssec:uncertainty}

A conventional method to assess the impact of MHOUs involves studying the sensitivity of observables to variations of the renormalization and factorization scales around their central value. 
MHOUs constitute a leading source of theoretical uncertainty~\cite{Celiberto:2022rfj}, and are estimated by jointly varying $\mu_R$ and $\mu_F$ within the interval $\mu_N/2$ to $2\mu_N$. 
The dimensionless parameter $C_\mu \equiv \mu_F/\mu_N = \mu_R/\mu_N$, shown in the plots of Sections~\ref{ssec:I_rates} and~\ref{ssec:pT_rates}, encodes this variation.

An additional source of uncertainty stems from proton PDFs. 
Previous studies on high-energy processes~\cite{Bolognino:2018oth,Celiberto:2020wpk,Celiberto:2021fdp,Celiberto:2022rfj} have shown that different PDF sets or replicas within the same set have only a marginal impact on predictions. 
We therefore adopt the central member of the {\tt NNPDF4.0} set throughout our analysis.

Further uncertainties may arise from the \emph{collinear improvement} of the NLO BFKL kernel, which incorporates renormalization-group (RG) terms to ensure a smooth interpolation with DGLAP in the collinear limit, as well as from the choice of renormalization scheme~\cite{Salam:1998tj,Ciafaloni:2003rd,Ciafaloni:2003ek,Ciafaloni:2000cb,Ciafaloni:1999yw,Ciafaloni:1998iv,SabioVera:2005tiv}. 
Nonetheless, their impact on semi-hard rapidity-sensitive observables remains within the MHOU bands~\cite{Celiberto:2022rfj}.

In particular, Ref.~\cite{Celiberto:2022rfj} explored the transition from the $\MSb$~\cite{PhysRevD.18.3998} to the MOM~\cite{Barbieri:1979be,PhysRevLett.42.1435} renormalization scheme, finding that MOM predictions for rapidity distributions are systematically enhanced, yet still compatible with $\MSb$ results within the MHOU envelope. 
A fully consistent MOM-based analysis would require PDFs and FFs evolved in that scheme, which are currently unavailable.

Finally, uncertainty bands in our predictions combine MHOUs with the statistical errors from multidimensional integration (see Section~\ref{ssec:final_state}). The latter are kept below the 1\% level, owing to the performance of the {\Jethad} integration routines.

\subsection{Observables and kinematics}
\label{ssec:final_state}

We focus on two primary observables: the rapidity-interval distribution, corresponding to the cross section differential in the rapidity separation $\Delta Y = y_1 - y_2$ between the two final-state particles, and the transverse-momentum distribution differential in $|\vec{q}_1|$.

The rapidity-interval observable is directly connected to the ${\cal C}_0$ angular coefficient, introduced in Section~\ref{ssec:hybrid_factorization}. It is obtained by integrating ${\cal C}_0$ over the transverse momenta and rapidities of the final-state particles, while keeping the rapidity gap $\Delta Y$ between the pentaquark and the light jet fixed. 
Its expression reads  
\begin{equation}
 \label{obs:I}
 \frac{\drv \sigma(\DY, s)}{\drv \DY} =
 \int_{|\vec q_2|^{\rm min}}^{|\vec q_2|^{\rm max}} 
 \!\!\drv |\vec q_1|
 \int_{|\vec q_2|^{\rm min}}^{|\vec q_2|^{\rm max}} 
 \!\!\drv |\vec q_2|
 \int_{\max (y_1^{\rm min}, \, \DY + y_2^{\rm min})}^{\min (y_1^{\rm max}, \, \DY + y_2^{\rm max})} \drv y_1
 \, \,
 {\cal C}_0^{\rm [res]}
\Bigm \lvert_{y_2 = y_1 - \DY}
 \;,
\end{equation}
Here, the label `${\rm [res]}$' identifies the resummation scheme adopted: $\NLLp$, $\HENLOp$, or $\LL$. 
To eliminate one of the rapidity integrals, specifically that over $y_2$, we insert the constraint $\delta(\Delta Y - (y_1 - y_2))$.

The transverse momentum of the forward $\Omega_{3Q}$ state is restricted to $60 < |\vec{q}_1|/{\rm GeV} < 120$, in accordance with the requirements of the ZM-VFNS approach, which assumes energy scales sufficiently above the heavy-quark evolution thresholds. 
The accompanying jet is selected in a slightly lower momentum window, $50 < |\vec{q}_2|/{\rm GeV} < 60$, in line with current LHC and prospective HL-LHC analyses~\cite{Khachatryan:2016udy}.

This \emph{asymmetric} $p_T$ configuration serves multiple purposes. 
It enhances sensitivity to genuine high-energy dynamics by reducing contamination from fixed-order effects~\cite{Celiberto:2015yba,Celiberto:2015mpa,Celiberto:2020wpk}, while suppressing large Sudakov logarithms arising from nearly back-to-back kinematics~\cite{Mueller:2013wwa,Marzani:2015oyb,Mueller:2015ael,Xiao:2018esv,Hatta:2020bgy,Hatta:2021jcd}, which would otherwise require additional resummation. 
Moreover, it reduces radiative instabilities~\cite{Andersen:2001kta,Fontannaz:2001nq} and helps preserve energy-momentum conservation at the partonic level~\cite{Ducloue:2014koa}.

As for rapidity coverage, we follow the current experimental setup. 
Rare baryons are assumed to be detected within the barrel calorimeter acceptance~\cite{Chatrchyan:2012xg}, $|y| < 2.5$, while jets are allowed in the extended range $|y| < 4.7$, which includes both barrel and endcap regions~\cite{Khachatryan:2016udy}.

The second observable under consideration is the transverse-momentum spectrum of the $\Omega_{3Q}$ state  
\begin{equation}
\label{obs:I-k1b}
 \frac{\drv \sigma(|\vec q_1|, s)}{\drv |\vec q_1|} =
 \int_{|\vec q_2|^{\rm min}}^{|\vec q_2|^{\rm max}} 
 \!\!\drv |\vec q_2|
 \, \,
 \int_{\DY^{\rm min}}^{\DY^{\rm max}} \drv \DY
 \int_{\max (y_1^{\rm min}, \, \DY + y_2^{\rm min})}^{\min (y_1^{\rm max}, \, \DY + y_2^{\rm max})} \drv y_1
 \, \,
 {\cal C}_{l=0}^{\rm [res]}
\Bigm \lvert_{y_2 = y_1 - \DY}
 \;.
\end{equation}
This distribution is differential in the transverse momentum of the $\Omega_{3b}$ baryon, $|\vec{q}_1|$, and integrated over the jet transverse-momentum range $40~\text{GeV} < |\vec{q}_2| < 120~\text{GeV}$ within a fixed $\Delta Y$ bin. 
The same rapidity cuts are applied to the $\Omega_{3b}$ and jet final states.

The $|\vec{q}_1|$ spectrum provides a valuable testbed for probing the interplay between the $\NLLp$ factorization framework and alternative resummation approaches. 
Varying $|\vec{q}_1|$ from 10 to 100~GeV allows us to explore a wide kinematic domain, where additional logarithmic enhancements may become relevant. 
Previous analyses on high-energy Higgs production~\cite{Celiberto:2020tmb} and heavy-jet tagging~\cite{Bolognino:2021mrc} have validated the robustness of our hybrid approach in the moderate-$p_T$ region, particularly when $|\vec{q}_1| \sim |\vec{q}_2|$.

At large $|\vec{q}_1|$, threshold logarithms are expected to become significant, marking the onset of the hard regime ($|\vec{q}_1| > |\vec{q}_2|^{\rm max}$). 
In contrast, the low-$p_T$ region ($|\vec{q}_1| \ll |\vec{q}_2|^{\rm min}$) is dominated by soft logarithms. 
Hence, studying the $|\vec{q}_1|$ spectrum offers both a nontrivial consistency check of our framework and a starting point for the inclusion of additional resummation techniques.

\subsection{Rapidity-differential distributions}
\label{ssec:I_rates}

\begin{figure}[!t]
\centering

   \hspace{0.00cm}
   \includegraphics[scale=0.375,clip]{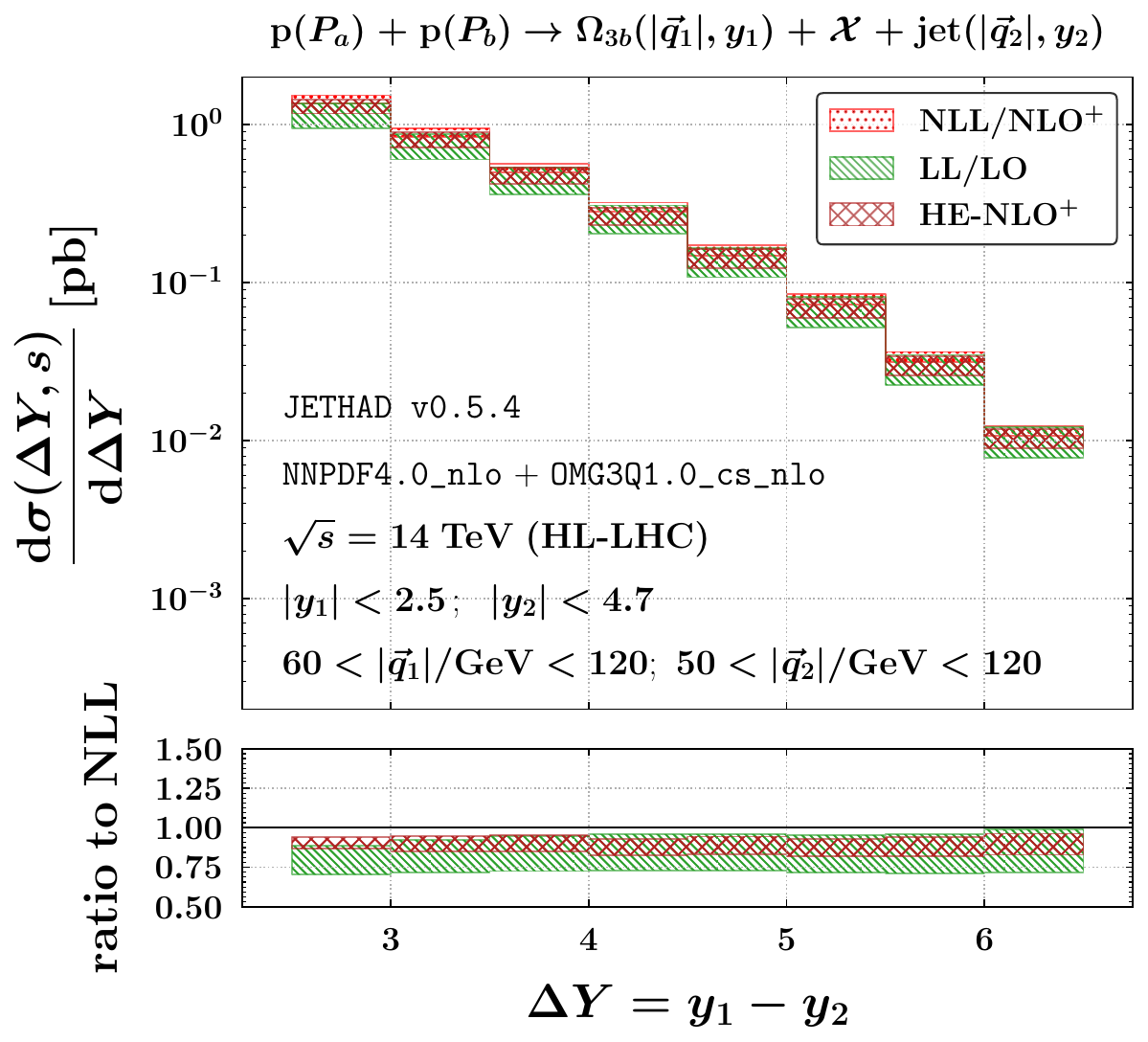}
   \hspace{0.30cm}
   \includegraphics[scale=0.375,clip]{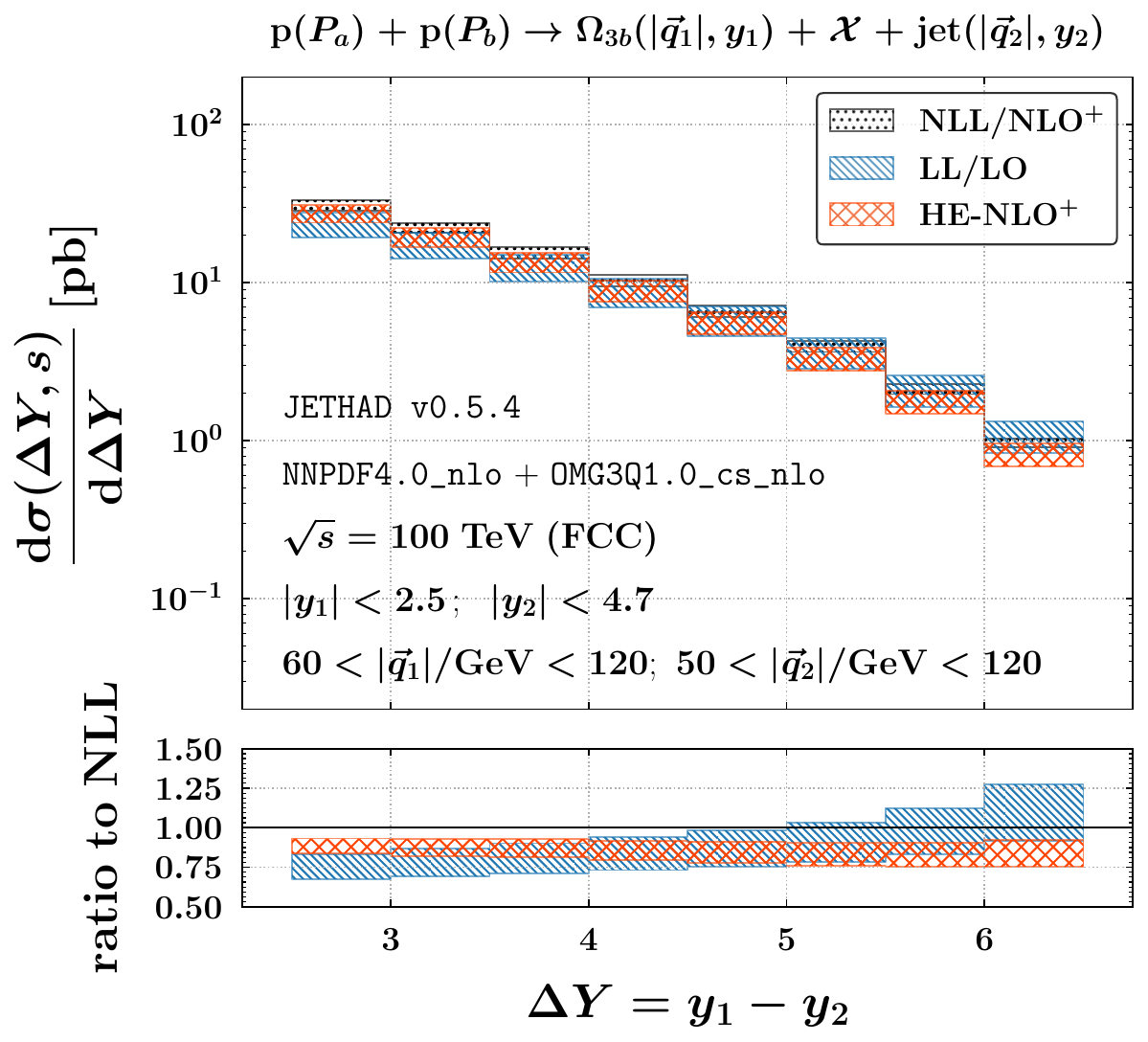}

\caption{Rapidity-differential distribution for the semi-inclusive detection of $\Omega_{3b}$ plus jet systems at $\sqrt{s} = 14$ TeV (HL-LHC, left) and $100$ TeV (nominal FCC, right).
Ancillary panels below the main plots show the ratio of $\LL$ or $\HENLOp$ predictions to the $\NLLp$ baseline.
The uncertainty bands account for the combined effects of MHOUs and multidimensional phase-space integration.
The analysis employs {\tt NNPDF4.0} NLO proton PDFs~\cite{NNPDF:2021uiq,NNPDF:2021njg} in combination with {\tt OMG3Q1.0} NLO FFs for heavy baryons~\cite{Celiberto:2025_OMG3Q10}.}
\label{fig:I_OQb}
\end{figure}

Figure~\ref{fig:I_OQb} displays the rapidity-differential distributions for the semi-inclusive production of $\Omega_{3b}$ plus jet systems at $\sqrt{s} = 14$~TeV (HL-LHC, left) and 100~TeV (nominal FCC, right). 
As mentioned, the analysis is carried out in the $\NLLp$ hybrid framework and uses the {\tt OMG3Q1.0} FFs in combination with {\tt NNPDF4.0} proton PDFs~\cite{NNPDF:2021uiq,NNPDF:2021njg}, as implemented in {\Jethad} {\tt v0.5.4}. 
Rapidity intervals are constructed using uniform $\Delta Y$ bins of fixed width $\Delta Y = 0.5$, covering the region $2.5 \leq \Delta Y \leq 6.5$.

Each main panel is accompanied by an ancillary subplot that shows the ratio of $\LL$ and $\HENLOp$ predictions to the NLL/NLO$^+$ baseline. 
This layout facilitates a transparent assessment of the impact of higher-order corrections and high-energy logarithmic enhancements across the full $\Delta Y$ range. Uncertainty bands combine multi-dimensional phase-space integration errors with MHOUs estimated through scale variations.

The predicted cross sections span a wide and phenomenologically relevant range, from approximately $10^{-2}$~pb in the smallest-$\Delta Y$ bin to above $30$~pb in the largest one. 
As expected, the overall yield is significantly enhanced at FCC energies, with an increase of roughly one order of magnitude compared to HL-LHC. 
This scaling reflects the amplified gluon luminosity at high energies and the corresponding boost in heavy-flavor fragmentation probabilities in the forward region. 
The large-$\Delta Y$ bins, although experimentally more challenging, are particularly promising for probing the high-energy dynamics encoded in the $\NLLp$ formalism.

\begin{figure}[!t]
\centering

   \hspace{0.00cm}
   \includegraphics[scale=0.375,clip]{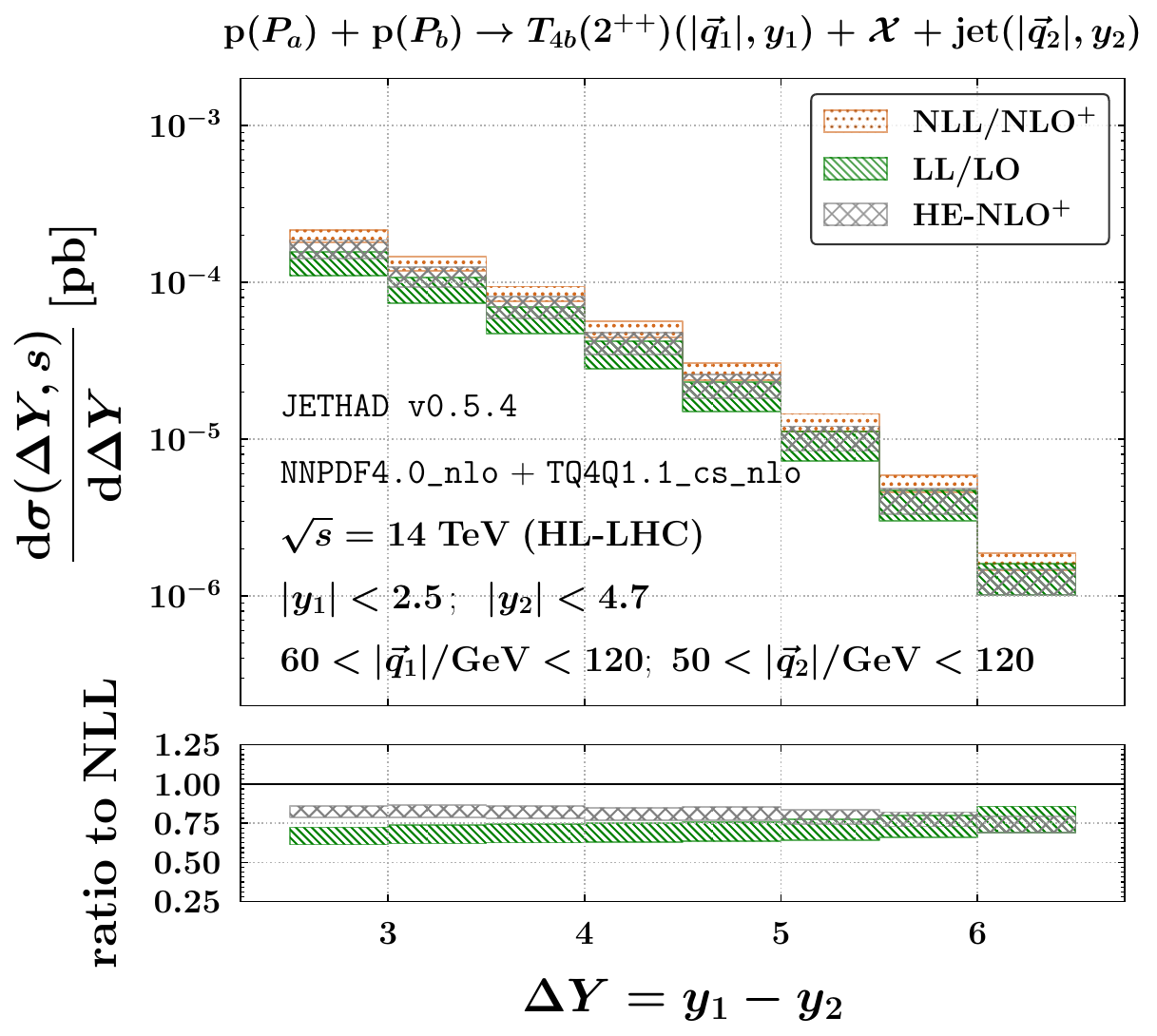}
   \hspace{0.30cm}
   \includegraphics[scale=0.375,clip]{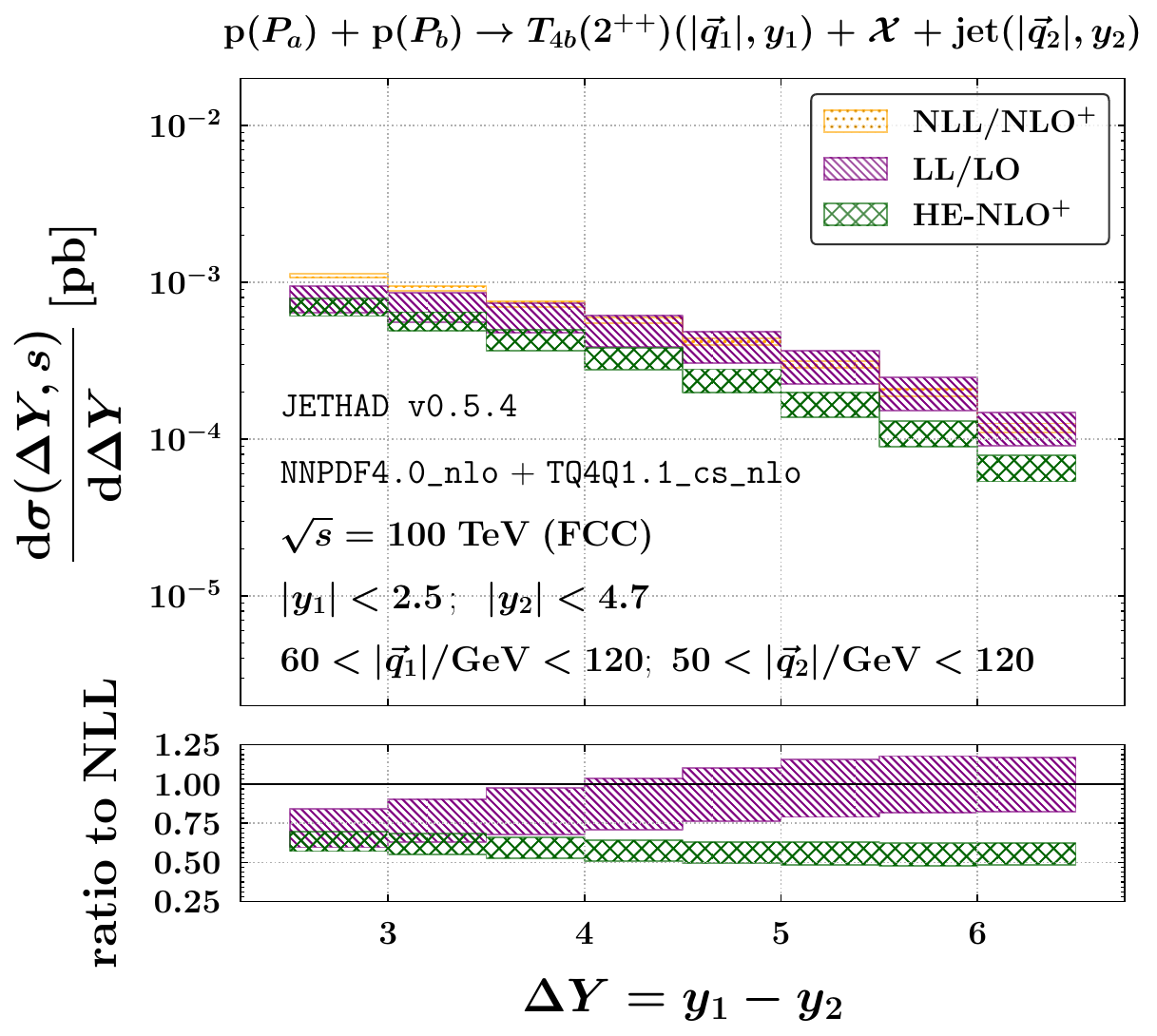}

\caption{Rapidity-differential distribution for the semi-inclusive detection of $\TQbTpp$ plus jet systems at $\sqrt{s} = 14$ TeV (HL-LHC, left) and $100$ TeV (nominal FCC, right).
Ancillary panels below the main plots show the ratio of $\LL$ or $\HENLOp$ predictions to the $\NLLp$ baseline.
The uncertainty bands account for the combined effects of MHOUs and multidimensional phase-space integration.
The analysis employs {\tt NNPDF4.0} NLO proton PDFs~\cite{NNPDF:2021uiq,NNPDF:2021njg} in combination with {\tt TQ4Q1.1} NLO FFs for tensor heavy tetraquarks~\cite{Celiberto:2025_OMG3Q10}.}
\label{fig:I_TQb}
\end{figure}

All $\Delta Y$ distributions exhibit a coherent trend: the differential cross section declines monotonically with increasing rapidity separation. This suppression originates from the competing influence of two mechanisms. On one hand, the partonic scattering amplitude grows with energy, and hence with $\Delta Y$, as dictated by the structure of high-energy logarithms and their resummation. 
On the other hand, this growth is mitigated by damping effects arising from the convolution with PDFs and FFs within the structure of the singly off-shell emission functions (see Eqs.~\eqref{LOHEF} and~\eqref{LOJEF}).

From the perspective of resummation, two important remarks are in order. 
First, the predictions demonstrate notable robustness with respect to MHOUs. 
The corresponding uncertainty bands remain confined within a factor of 1.5 for both HL-LHC and FCC kinematics (see first ancillary panels in Fig.~\ref{fig:I_OQb}). 
Although these bands might appear sizable compared to fixed-order expectations, they are remarkably narrow in the context of high-energy factorization. 
In fact, traditional BFKL-based analyses for light-hadron or jet production observables~\cite{Celiberto:2016hae,Celiberto:2017ptm,Bolognino:2018oth,Celiberto:2020wpk} frequently report higher-order corrections that span up to one or two orders of magnitude with respect to the leading baseline. 
In contrast, our results for $\Omega_{3b}$ production remain under tight control throughout the full $\Delta Y$ spectrum.

Second, the transition from $\LL$ to $\NLLp$ accuracy leads to a visible improvement in perturbative stability. 
The uncertainty bands at $\NLLp$ are consistently narrower than their LL counterparts and progressively converge toward them in the large-$\Delta Y$ region, eventually showing partial overlap. 
This convergence pattern is consistent with previous analyses of exotic hadron production in semi-hard regimes~\cite{Celiberto:2023rzw,Celiberto:2024mab,Celiberto:2025dfe,Celiberto:2025ziy,Celiberto:2025vra,Celiberto:2025ipt}, reinforcing the view that single-parton fragmentation into heavy-flavored rare baryons offers a particularly stable and reliable probe of high-energy QCD dynamics.

Figure~\ref{fig:I_TQb} presents the corresponding $\Delta Y$ distributions for the tensor $\TQbTpp$ resonance, often regarded as the bottomed analogue of the $X(6900)$ candidate~\cite{LHCb:2020bwg}.
These results enable a direct comparison between two distinct realizations of fully heavy-hadron production in the semihard regime of QCD: a baryonic configuration, represented by $\Omega_{3b}$, and a tetraquark one.
Despite sharing identical kinematic cuts and theoretical accuracies, the two channels exhibit markedly different behaviors, both in absolute magnitude and in their response to increasing collider energy.

The $\Omega_{3b}$ mode displays substantially larger cross sections than those obtained for $\TQbTpp$, with typical differences spanning two to three orders of magnitude across the explored $\Delta Y$ range.
Moreover, the growth in rate when moving from $\sqrt{s}=14$~TeV to $100$~TeV is more pronounced for the baryon, reflecting the stronger enhancement driven by the higher efficiency of heavy-flavor fragmentation in the triple-heavy system.
This behavior underscores the dual phenomenological role of $\Omega_{3b}$ production: it is simultaneously sensitive to the nonperturbative structure of the fragmentation mechanism and to the resummed logarithmic growth typical of high-energy dynamics.

From the viewpoint of theoretical stability, the LL uncertainty bands for $\Omega_{3b}$ remain more widely separated from the NLL baseline than in the tetraquark case, approaching it only gradually as $\sqrt{s}$ increases.
This indicates that predictive convergence under MHOUs and higher-order corrections emerges more clearly at FCC energies, where the resummation effects become fully manifest.
For the tetraquark, by contrast, LL and NLL bands already exhibit a moderate overlap at 14~TeV, pointing to a weaker sensitivity to the high-energy logarithmic structure of the kernel.

Finally, the $[\Omega_{3b} + \text{jet}]$ channel proves to be significantly more efficient than the tetraquark one in amplifying the genuine resummed signal relative to its fixed-order limit.
This is a nontrivial and noteworthy outcome, as rapidity-interval distributions are generally regarded as less responsive to high-energy dynamics than transverse-momentum observables.
In this case, however, the semihard production of a rare, fully heavy baryon exhibits a clear and robust imprint of high-energy resummation, establishing $\Omega_{3b}$ as a benchmark system for probing QCD in the forward regime and as a prime target for future experimental investigations at the HL-LHC and FCC.

\subsection{Transverse-momentum-differential distributions}
\label{ssec:pT_rates}

\begin{figure}[!t]
\centering

   \hspace{0.00cm}
   \includegraphics[scale=0.375,clip]{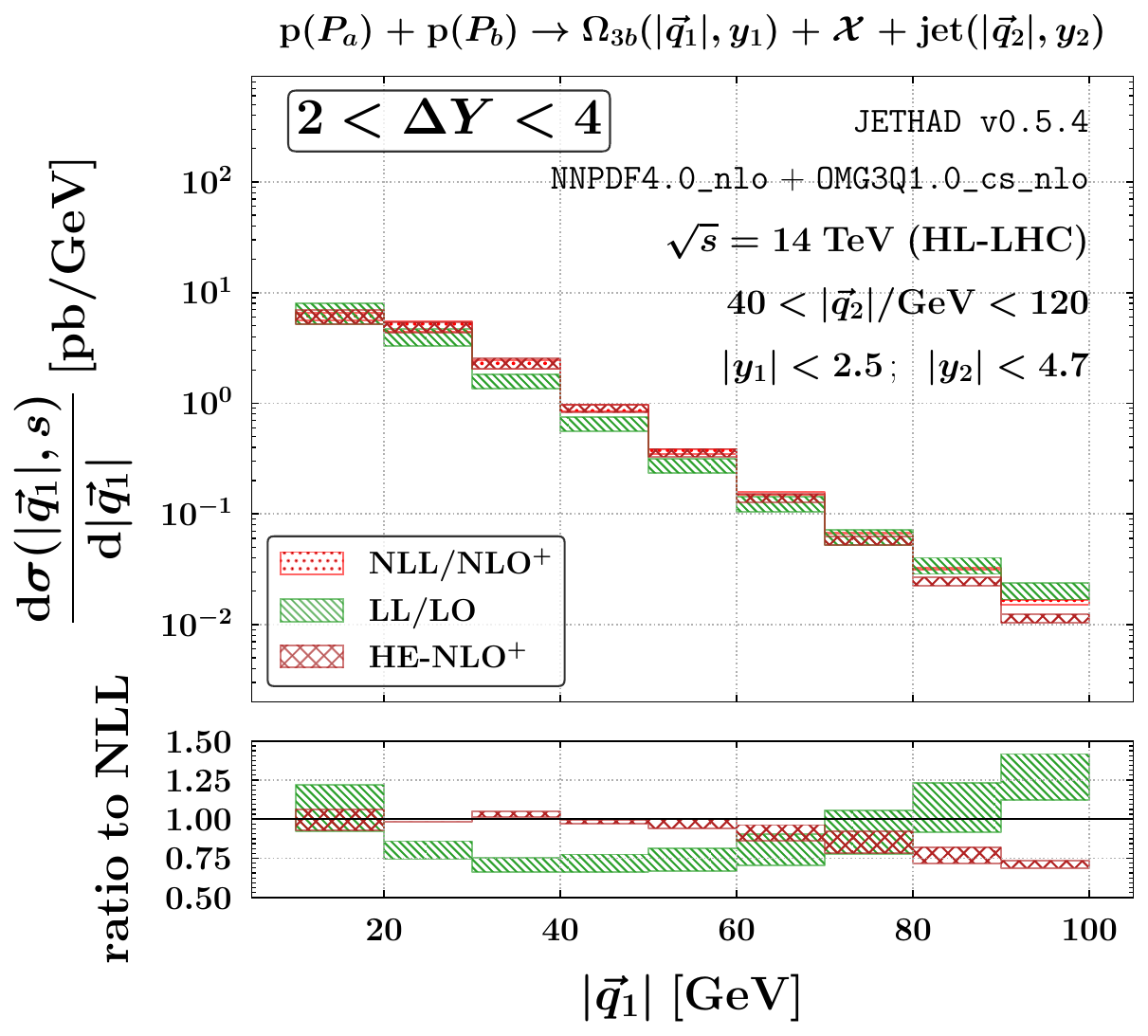}
   \hspace{0.30cm}
   \includegraphics[scale=0.375,clip]{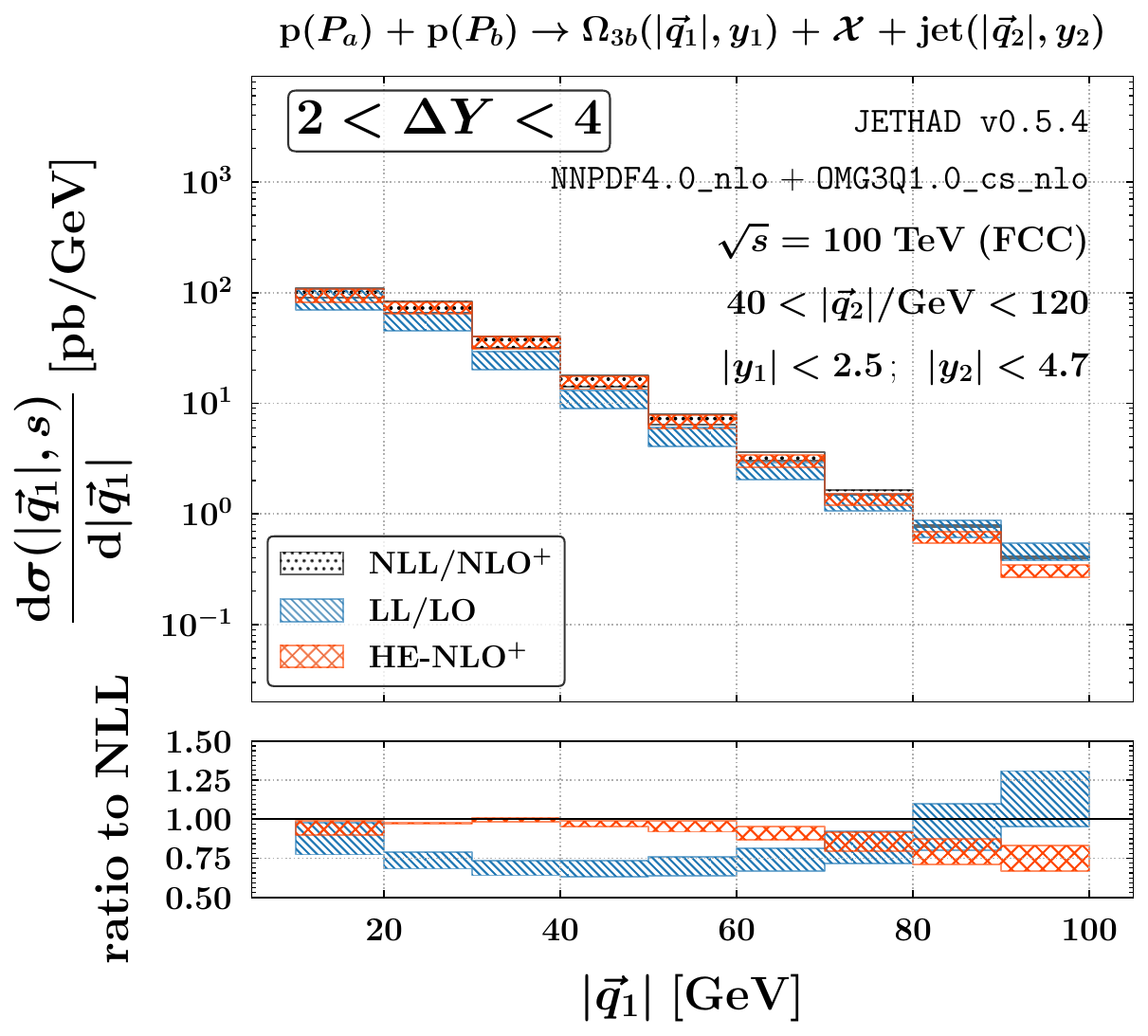}

   \vspace{0.45cm}  

   \hspace{0.00cm}
   \includegraphics[scale=0.375,clip]{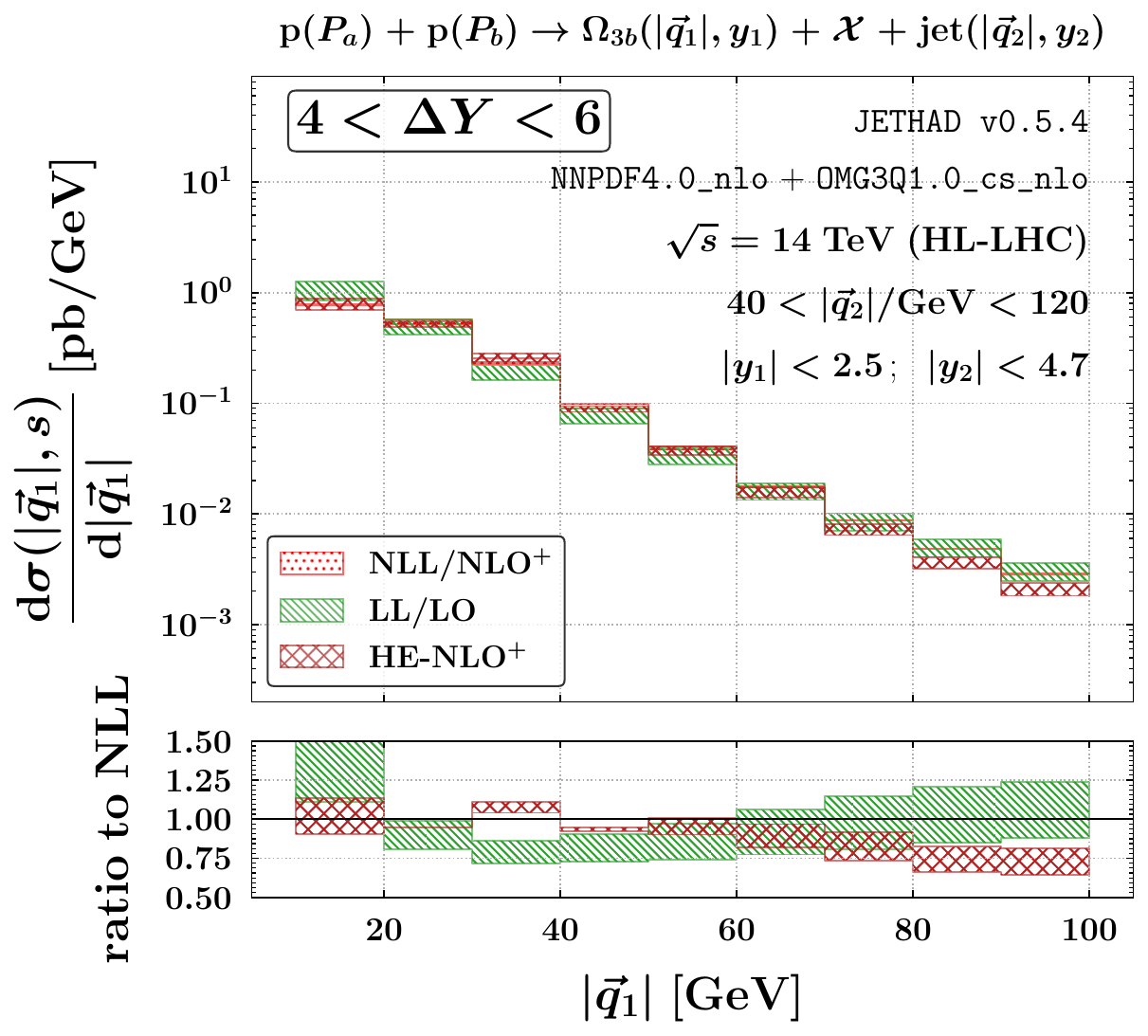}
   \hspace{0.30cm}
   \includegraphics[scale=0.375,clip]{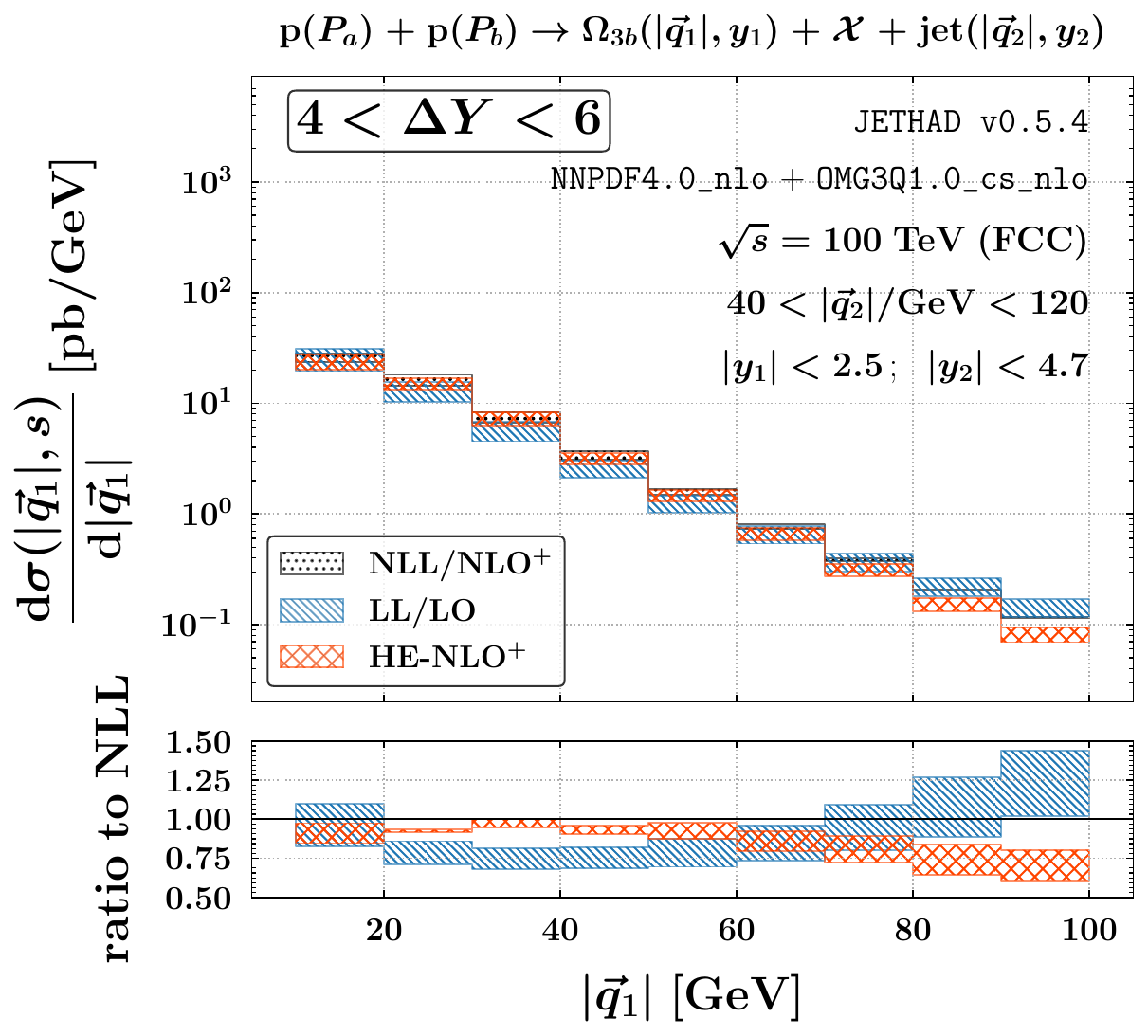}

\caption{Transverse-momentum-differential distributions for the semi-inclusive detection of $\Omega_{3b}$ plus jet systems at $\sqrt{s} = 14$ TeV (HL-LHC, left) and $100$ TeV (nominal FCC, right), and for $2 < \DY < 4$ (lower) or $4 < \DY < 6$ (lower).
Ancillary panels below the main plots show the ratio of $\LL$ or $\HENLOp$ predictions to the $\NLLp$ baseline.
The uncertainty bands account for the combined effects of MHOUs and multidimensional phase-space integration.
The analysis employs {\tt NNPDF4.0} NLO proton PDFs~\cite{NNPDF:2021uiq,NNPDF:2021njg} in combination with {\tt OMG3Q1.0} NLO FFs for heavy baryons~\cite{Celiberto:2025_OMG3Q10}.}
\label{fig:I-k1b}
\end{figure}

Figure~\ref{fig:I-k1b} presents the transverse-momentum-differential spectra for the $\Omega_{3Q}$ baryon, reconstructed in semi-inclusive $\Omega_{3Q}$ plus jet final states at $\sqrt{s} = 14$~TeV (HL-LHC, left panels) and $100$~TeV (FCC, right panels). 
Results are shown for two $\Delta Y$ slices, $2 < \Delta Y < 4$ (top) and $4 < \Delta Y < 6$ (bottom), with $|\vec{q}_1|$ sampled in uniform bins of $10$~GeV width.

The overall magnitude of the cross section is enhanced at higher collider energies, with the FCC yielding rates over two orders of magnitude larger than the HL-LHC, particularly in the low-$|\vec{q}_1|$ region. 
For fixed $\sqrt{s}$, increasing the $\Delta Y$ interval leads to a moderate suppression of the spectrum, driven by phase-space limitations and steep falloffs in parton luminosities.

All distributions exhibit a smooth and monotonic decline with growing $|\vec{q}_1|$, consistently across all energy and rapidity configurations. 
No local maxima or structures are present, as expected in collinear QCD where soft and semihard emissions dominate the forward phase space. The hybrid resummation captures these features by balancing energy logarithms with collinear constraints.

The $\NLLp$ curves at 14~TeV display a sharper decrease at high $|\vec{q}_1|$, while the corresponding FCC predictions are flatter and more extended, benefiting from enhanced fluxes at large partonic center-of-mass energies. 
In all setups, the separation between the $\NLLp$ baseline and its LL or HE-NLO$^+$ approximations widens with increasing $|\vec{q}_1|$, most notably at the FCC. 
This trend reinforces the role of transverse-momentum observables as powerful diagnostics of high-energy resummation effects in rare baryon production.

The increasing gap between $\NLLp$ and $\HENLOp$ at large $|\vec{q}_1|$ is physically motivated. While the BFKL formalism is optimal for symmetric configurations ($|\vec{q}_1| \simeq |\vec{q}_2|$), the region where $|\vec{q}_1| \gg |\vec{q}_2|$ lies outside its natural domain. 
In such asymmetric configurations, DGLAP-like collinear logarithms and possible threshold enhancements become relevant, pointing to the need for extended frameworks capable of interpolating between high-energy and threshold resummation~\cite{Bartels:2001ge,Ivanov:2012ms,Colferai:2015zfa}.

Moreover, the pattern of the $\LL$ over $\NLLp$ ratio reflects a subtle interplay among different NLO effects. 
Jet emission channels tend to receive negative corrections at NLO~\cite{Bartels:2001ge,Ivanov:2012ms,Colferai:2015zfa}, while the fragmentation side can behave differently: in particular, the gluon coefficient $C_{gg}$ is positively shifted at NLO~\cite{Ivanov:2012iv}, while other components experience negative contributions. 
Depending on the kinematic region and hadronic process, these effects may partially cancel or amplify each other. 

In summary, the transverse-momentum distributions of the $\Omega_{3Q}$ baryon remain stable under MHOU variations across the full range of $|\vec{q}_1|$, as evident from the moderate size of uncertainty bands in Fig.~\ref{fig:I-k1b}. 
These findings strengthen the case for using rare heavy baryons as clean probes of semihard QCD, both at current and future colliders, within a VFNS-based fragmentation picture.

\section{Future perspectives}
\label{sec:conclusions}

We presented a comprehensive study of the leading-power fragmentation of fully heavy $\Omega_{3Q}$ baryons, with particular attention to the triply bottom case, $\Omega_{3b}$, as a novel and promising probe of hadronic structure and high-energy QCD. Our investigation builds upon and extends previous analyses of the $\Omega_{3c}$ sector, for which we released the {\tt OMG3Q1.0} collinear FFs within a VFNS. 
These FFs were constructed from a diquark-inspired proxy model, widely used in the modeling of singly heavy baryons and heavy quarkonia, where the fragmentation process is described as a convolution between perturbative SDCs and a nonperturbative hadronization component.

A distinctive feature of the {\tt OMG3Q1.0} formulation was the simultaneous inclusion of both constituent-heavy-quark and gluon channels at the initial scale, which made $\Omega_{3c}$ baryons an ideal benchmark for testing the {\HFNRevo} evolution scheme~\cite{Celiberto:2025euy,Celiberto:2024mex,Celiberto:2024bxu,Celiberto:2024rxa,Celiberto:2025xvy}. 
Through a two-step analytic/numeric DGLAP evolution strategy, the effects of heavy thresholds and channel-specific matching were accurately implemented. 
These ingredients provided the basis for a robust hybrid NLL resummation analysis of $\Omega_{3c}$ plus jet production at HL-LHC and FCC energies.

Building on these results, the present work broadens the scope to triply bottom baryons and rare $\Omega_{3Q}$ states in general, aiming at a systematic exploration of their fragmentation properties and their phenomenological relevance in collider experiments. 
To this purpose, we introduced an updated class of hadron-structure-oriented collinear FFs that incorporate improved modeling of the nonperturbative component for heavy-quark fragmentation, capturing both compact and diquarklike configurations. 
The semi-inclusive production of $\Omega_{3b}$ plus jet systems was then analyzed within the {\Jethad} framework and its symbolic module {\symJethad}~\cite{Celiberto:2020wpk,Celiberto:2022rfj,Celiberto:2023fzz,Celiberto:2024mrq,Celiberto:2024swu,Celiberto:2025_P5Q_review}, combining NLL high-energy resummation with NLO$^+$ collinear dynamics.

Our results confirm and generalize the concept of \emph{natural stability}~\cite{Celiberto:2022grc} previously observed for $\Omega_{3c}$~\cite{Celiberto:2025ogy}, now seen to emerge also for $\Omega_{3b}$ observables across a broad energy spectrum. 
This stability arises from the VFNS collinear fragmentation at high transverse momentum and plays a crucial role in suppressing resummation-related instabilities, including unresummed threshold logarithms and NLL artifacts. 
Furthermore, transverse-momentum and rapidity-interval distributions for $\Omega_{3b}$ states display an excellent discriminating power between resummed and fixed-order predictions, further enhancing their value as precision tools.

The relevance of triply heavy baryons extends beyond perturbative QCD. 
Their valence structure, dominated by three heavy quarks, makes them sensitive to the partonic content of the colliding hadrons in nontrivial ways. 
In particular, rare $\Omega$ baryons can serve as indirect probes of intrinsic heavy flavor, especially intrinsic charm and bottom~\cite{Brodsky:1980pb,Brodsky:2015fna,Ball:2022qks,Guzzi:2022rca,NNPDF:2023tyk}, through the enhancement of specific fragmentation channels at large momentum fractions or forward rapidities. 
The simultaneous inclusion of heavy-quark and gluon inputs in our FFs ensures that such effects can be properly captured and tested.

Another compelling direction concerns the role of triply heavy baryons as multipurpose tools for probing soft-QCD phenomena. 
Their fragmentation patterns may encode spin and momentum correlations among constituents, offering a window into the nonperturbative dynamics of hadron formation. 
As such, $\Omega_{3Q}$ production provides a novel testbed for the interplay between hadronic structure and high-precision resummation techniques. 
Future progress may involve benchmarking our FFs against determinations from global fits, leveraging machine-learning methods already applied to the fragmentation of lighter hadrons~\cite{Nocera:2017qgb,Bertone:2017xsf,Bertone:2017tyb,Bertone:2018ecm,Khalek:2021gxf,Khalek:2022vgy,Soleymaninia:2022qjf,Soleymaninia:2022alt}.

In this context, future theoretical updates will aim to incorporate additional resummation strategies, such as soft-gluon~\cite{Hatta:2020bgy,Hatta:2021jcd,Caucal:2022ulg,Taels:2022tza} and jet-radius logarithmic resummations~\cite{Dasgupta:2014yra,Dasgupta:2016bnd,Banfi:2012jm,Banfi:2015pju,Liu:2017pbb}, as well as jet-angularity techniques~\cite{Luisoni:2015xha,Caletti:2021oor,Reichelt:2021svh}. 
Moreover, potential synergies with the saturation formalism at small-$x$~\cite{Gelis:2010nm,Kovchegov:2012mbw,Chirilli:2012jd,Boussarie:2014lxa,Benic:2016uku,Benic:2018hvb,Roy:2019hwr,Roy:2019cux,Beuf:2020dxl,Iancu:2021rup,Iancu:2023lel,vanHameren:2023oiq,Wallon:2023asa,Agostini:2024xqs,Altinoluk:2024zom,Altinoluk:2025dwd,Altinoluk:2025tms}, and with studies of angular asymmetries in dijet and heavy-hadron production, could further expand the reach of our framework~\cite{Caucal:2021ent,Caucal:2022ulg,Taels:2022tza,Kotko:2015ura,vanHameren:2016ftb,Altinoluk:2020qet,Altinoluk:2021ygv,Boussarie:2021ybe,Caucal:2023nci,Cheung:2024qvw,Caucal:2025mth,Kang:2013hta,Ma:2014mri,Ma:2015sia,Ma:2018qvc,Stebel:2021bbn,Mantysaari:2021ryb,Mantysaari:2022kdm}.

Altogether, our study lays the foundation for a detailed phenomenological program dedicated to triply heavy baryons, supporting their potential role in upcoming collider experiments. 
In this regard, precision measurements of rare $\Omega_{3Q}$ production may complement other channels, such as $J/\psi$ plus a heavy-flavor-tagged jet~\cite{Flore:2020jau}, in mapping the heavy-flavor content of the proton and exploring its link with exotic hadron structure.

Looking ahead, data from the future FCC~\cite{FCC:2018byv,FCC:2018evy,FCC:2018vvp,FCC:2018bvk,FCC:2025lpp,FCC:2025uan,FCC:2025jtd}, and other next-generation colliders~\cite{Chapon:2020heu,LHCspin:2025lvj,Anchordoqui:2021ghd,Feng:2022inv,Hentschinski:2022xnd,Accardi:2012qut,AbdulKhalek:2021gbh,Khalek:2022bzd,Acosta:2022ejc,AlexanderAryshev:2022pkx,LinearCollider:2025lya,LinearColliderVision:2025hlt,Brunner:2022usy,Arbuzov:2020cqg,Abazov:2021hku,Bernardi:2022hny,Amoroso:2022eow,Celiberto:2018hdy,Klein:2020nvu,2064676,MuonCollider:2022xlm,Aime:2022flm,MuonCollider:2022ded,Black:2022cth,Accettura:2023ked,InternationalMuonCollider:2024jyv,MuCoL:2024oxj,MuCoL:2025quu,InternationalMuonCollider:2025sys,Vignaroli:2023rxr,Dawson:2022zbb,Bose:2022obr,Begel:2022kwp,Abir:2023fpo,Accardi:2023chb,Gessner:2025acq,Altmann:2025feg}, will offer unique opportunities to test the predictions presented here. 
These efforts will deepen our understanding of hadronization mechanisms, probe potential manifestations of intrinsic charm or bottom, and uncover new connections between perturbative QCD and the nonperturbative landscape of exotic hadron formation.

\section*{Funding}
\label{sec:funding}
\addcontentsline{toc}{section}{\nameref{sec:funding}}

This research was funded by Comunidad de Madrid, grant number 2022-T1/TIC-24176.

\section*{Conflict of Interests}
\label{sec:coi}
\addcontentsline{toc}{section}{\nameref{sec:coi}}

The author declares no conflict of interest.

\section*{Data availability}
\label{sec:data_availability}
\addcontentsline{toc}{section}{\nameref{sec:data_availability}}

The two grids for the {\tt OMG3Q1.0} FF family
\begin{itemize}
    \item NLO, $\Omega_{3c}$\,: \,{\tt OMG3Q10\_cs\_O3c\_nlo};
    \item NLO, $\Omega_{3b}$\,: \,{\tt OMG3Q10\_cs\_O3b\_nlo},
\end{itemize}
can be publicly accessed from the following url: \url{https://github.com/FGCeliberto/Collinear_FFs/}.

Data underlying figures presented in this review can be made available upon a reasonable request.

\begin{appendices}

\printacronyms

\setcounter{appcnt}{0}
\hypertarget{app:NLOHEF}{
\section*{Appendix~A: NLO correction for the heavy-hadron singly off-shell emission function}}
\label{app:NLOHEF}

The next-to-leading order correction to the forward heavy-hadron singly off-shell emission function has been derived analytically in Ref.~\cite{Ivanov:2012iv}.
One has

\begin{equation}
  \label{NLOHEF}
  \hat \F_h(n,\nu,|\vec q_h|,x_h)=
  \frac{1}{\pi}\sqrt{\frac{C_F}{C_A}}
  \left(|\vec q_h|^2\right)^{i\nu-\frac{1}{2}}
  \int_{x_h}^1\frac{\drv x}{x}
  \int_{\frac{x_h}{x}}^1\frac{\drv \delta}{\delta}
  \left(\frac{x\delta}{x_h}\right)^{2i\nu-1}
\end{equation}
  \[ \times \,
  \left[
  \frac{C_A}{C_F}f_g(x)D_g^h\left(\frac{x_h}{x\delta}\right){\cal C}_{gg}
  \left(x,\delta\right)+\sum_{i=q\bar q}f_i(x)D_i^h
  \left(\frac{x_h}{x\delta}
  \right){\cal C}_{qq}\left(x,\delta\right)
  \right.
  \]
  \[ + \,
  \left.D_g^h\left(\frac{x_h}{x\delta}\right)
  \sum_{i=q\bar q}f_i(x){\cal C}_{qg}
  \left(x,\delta\right)+\frac{C_A}{C_F}f_g(x)\sum_{i=q\bar q}D_i^h
  \left(\frac{x_h}{x\delta}\right){\cal C}_{gq}\left(x,\delta\right)
  \right]\, ,
  \]
where
\begin{equation}
\stepcounter{appcnt}
\label{Cgg_hadron}
 {\cal C}_{gg}\left(x,\delta\right) =  P_{gg}(\delta)\left(1+\delta^{-2\gamma}\right)
 \ln \left( \frac {|\vec q_h|^2 x^2 \delta^2 }{\mu_F^2 x_h^2}\right)
 -\frac{\beta_0}{2}\ln \left( \frac {|\vec q_h|^2 x^2 \delta^2 }
 {\mu^2_R x_h^2}\right)
\end{equation}
\[
 + \, \delta(1-\delta)\left[C_A \ln\left(\frac{s_0 \, x^2}{|\vec q_h|^2 \,
 x_h^2 }\right) \chi(n,\gamma)
 - C_A\left(\frac{67}{18}-\frac{\pi^2}{2}\right)+\frac{5}{9}n_f
 \right.
\]
\[
 \left.
 +\frac{C_A}{2}\left(\psi^\prime\left(1+\gamma+\frac{n}{2}\right)
 -\psi^\prime\left(\frac{n}{2}-\gamma\right)
 -\chi^2(n,\gamma)\right) \right]
 + \, C_A \left(\frac{1}{\delta}+\frac{1}{(1-\delta)_+}-2+\delta\bar\delta\right)
\]
\[
 \times \, \left(\chi(n,\gamma)(1+\delta^{-2\gamma})-2(1+2\delta^{-2\gamma})\ln\delta
 +\frac{\bar \delta^2}{\delta^2}{\cal I}_2\right)
\]
\[
 + \, 2 \, C_A (1+\delta^{-2\gamma})
 \left(\left(\frac{1}{\delta}-2+\delta\bar\delta\right) \ln\bar\delta
 +\left(\frac{\ln(1-\delta)}{1-\delta}\right)_+\right) \ ,
\]

\begin{equation}
\stepcounter{appcnt}
\label{Cgq_hadron}
 {\cal C}_{gq}\left(x,\delta\right)=P_{qg}(\delta)\left(\frac{C_F}{C_A}+\delta^{-2\gamma}\right)\ln \left( \frac {|\vec q_h|^2 x^2 \delta^2 }{\mu_F^2 x_h^2}\right)
\end{equation}
\[
 + \, 2 \, \delta \bar\delta \, T_R \, \left(\frac{C_F}{C_A}+\delta^{-2\gamma}\right)+\, P_{qg}(\delta)\, \left(\frac{C_F}{C_A}\, \chi(n,\gamma)+2 \delta^{-2\gamma}\,\ln\frac{\bar\delta}{\delta} + \frac{\bar \delta}{\delta}{\cal I}_3\right) \ ,
\]

\begin{equation}
\stepcounter{appcnt}
\label{qg}
 {\cal C}_{qg}\left(x,\delta\right) =  P_{gq}(\delta)\left(\frac{C_A}{C_F}+\delta^{-2\gamma}\right)\ln \left( \frac {|\vec q_h|^2 x^2 \delta^2 }{\mu_F^2 x_h^2}\right)
\end{equation}
\[
 + \delta\left(C_F\delta^{-2\gamma}+C_A\right) + \, \frac{1+\bar \delta^2}{\delta}\left[C_F\delta^{-2\gamma}(\chi(n,\gamma)-2\ln\delta)+2C_A\ln\frac{\bar \delta}{\delta} + \frac{\bar \delta}{\delta}{\cal I}_1\right] \ ,
\]
and
\begin{equation}
\stepcounter{appcnt}
\label{Cqq_hadron}
 {\cal C}_{qq}\left(x,\delta\right)=P_{qq}(\delta)\left(1+\delta^{-2\gamma}\right)\ln \left( \frac {|\vec q_h|^2 x^2 \delta^2 }{\mu_F^2 x_h^2}\right)-\frac{\beta_0}{2}\ln \left( \frac {|\vec q_h|^2 x^2 \delta^2 }{\mu^2_R x_h^2}\right)
\end{equation}
\[
 + \, \delta(1-\delta)\left[C_A \ln\left(\frac{s_0 \, x_h^2}{|\vec q_h|^2 \, x^2 }\right) \chi(n,\gamma)+ C_A\left(\frac{85}{18}+\frac{\pi^2}{2}\right)-\frac{5}{9}n_f - 8\, C_F \right.
\]
\[
 \left. +\frac{C_A}{2}\left(\psi^\prime\left(1+\gamma+\frac{n}{2}\right)-\psi^\prime\left(\frac{n}{2}-\gamma\right)-\chi^2(n,\gamma)\right) \right] + \, C_F \,\bar \delta\,(1+\delta^{-2\gamma})
\]
\[
 +\left(1+\delta^2\right)\left[C_A (1+\delta^{-2\gamma})\frac{\chi(n,\gamma)}{2(1-\delta )_+}+\left(C_A-2\, C_F(1+\delta^{-2\gamma})\right)\frac{\ln \delta}{1-\delta}\right]
\]
\[
 +\, \left(C_F-\frac{C_A}{2}\right)\left(1+\delta^2\right)\left[2(1+\delta^{-2\gamma})\left(\frac{\ln (1-\delta)}{1-\delta}\right)_+ + \frac{\bar \delta}{\delta^2}{\cal I}_2\right] \; ,
\]

The scale $s_0$ is a typical BFKL energy-normalization parameter, commonly chosen as $s_0 = \mu_C$.
In addition, we define $\bar \delta \equiv 1 - \delta$ and $\gamma \equiv -\frac{1}{2} + i \nu$.
The LO DGLAP splitting functions $P_{ij}(\delta)$ are given by
\begin{eqnarray}
\stepcounter{appcnt}
\label{DGLAP_kernels}
 P_{gq}(z)&=&C_F\frac{1+(1-z)^2}{z} \; , \\ \nonumber
 P_{qg}(z)&=&T_R\left[z^2+(1-z)^2\right]\; , \\ \nonumber
 P_{qq}(z)&=&C_F\left( \frac{1+z^2}{1-z} \right)_+= C_F\left[ \frac{1+z^2}{(1-z)_+} +{3\over 2}\delta(1-z)\right]\; , \\ \nonumber
 P_{gg}(z)&=&2C_A\left[\frac{1}{(1-z)_+} +\frac{1}{z} -2+z(1-z)\right]+\left({11\over 6}C_A-\frac{n_f}{3}\right)\delta(1-z) \; ,
\end{eqnarray}
whereas the ${\cal I}_{2,1,3}$ functions are given by
\begin{equation}
\stepcounter{appcnt}
\label{I2}
{\cal I}_2=
\frac{\delta^2}{\bar \delta^2}\left[
\delta\left(\frac{{}_2F_1(1,1+\gamma-\frac{n}{2},2+\gamma-\frac{n}{2},\delta)}
{\frac{n}{2}-\gamma-1}-
\frac{{}_2F_1(1,1+\gamma+\frac{n}{2},2+\gamma+\frac{n}{2},\delta)}{\frac{n}{2}+
\gamma+1}\right)\right.
\end{equation}
\[
 \stepcounter{appcnt}
 \left.
 +\delta^{-2\gamma}\left(\frac{{}_2F_1(1,-\gamma-\frac{n}{2},1-\gamma-\frac{n}{2},\delta)}{\frac{n}{2}+\gamma}-\frac{{}_2F_1(1,-\gamma+\frac{n}{2},1-\gamma+\frac{n}{2},\delta)}{\frac{n}{2} -\gamma}\right)
\right.
\]
\[
 \left.
 +\left(1+\delta^{-2\gamma}\right)\left(\chi(n,\gamma)-2\ln \bar \delta \right)+2\ln{\delta}\right] \; ,
\]
\begin{equation}
\stepcounter{appcnt}
\label{I1}
 {\cal I}_1=\frac{\bar \delta}{2\delta}{\cal I}_2+\frac{\delta}{\bar \delta}\left[\ln \delta+\frac{1-\delta^{-2\gamma}}{2}\left(\chi(n,\gamma)-2\ln \bar \delta\right)\right] \; ,
\end{equation}
and
\begin{equation}
\stepcounter{appcnt}
\label{I3}
 {\cal I}_3=\frac{\bar \delta}{2\delta}{\cal I}_2-\frac{\delta}{\bar \delta}\left[\ln \delta+\frac{1-\delta^{-2\gamma}}{2}\left(\chi(n,\gamma)-2\ln \bar \delta\right)\right] \; ,
\end{equation}
with ${}_2F_1$ representing the Gauss hypergeometric function.

The \emph{plus~prescription} in Eqs.~\eqref{Cgg_hadron} and~\eqref{Cqq_hadron} stands as follows:
\begin{equation}
\label{plus-prescription}
\stepcounter{appcnt}
\int^1_\zeta \drv x \frac{f(x)}{(1-x)_+}
=\int^1_\zeta \drv x \frac{f(x)-f(1)}{(1-x)}
-\int^\zeta_0 \drv x \frac{f(1)}{(1-x)}\; ,
\end{equation}
with $f(x)$ denoting any regular-behaved generic function at $x=1$.

\setcounter{appcnt}{0}
\hypertarget{app:NLOJEF}{
\section*{Appendix~B: NLO correction for the light-jet singly off-shell emission function}}
\label{app:NLOJEF}

The NLO correction to the forward light-jet singly off-shell emission function has been derived analytically in the small-cone approximation~\cite{Colferai:2015zfa}.
One writes
\begin{equation}
\stepcounter{appcnt}
\label{NLOJEF}
 \hat \F_{J}(n,\nu,|\vec q_J|,x_J)=
 \frac{1}{\pi}\sqrt{\frac{C_F}{C_A}}
 \left(|\vec q_J|^2 \right)^{i\nu-1/2}
 \int^1_{x_J}\frac{\drv \delta}{\delta}
 \delta^{-\bar\alpha_s(\mu_R)\chi(n,\nu)}
\end{equation}
\[
\times\;
\left\{\sum_{i=q,\bar q} f_i \left(\frac{x_J}{ \delta}\right)\left[\left(P_{qq}(\delta)+\frac{C_A}{C_F}P_{gq}(\delta)\right)
\ln\frac{|\vec q_J|^2}{\mu_F^2}\right.\right.
\]
\[
-\;2\delta^{-2\gamma} \ln \frac{{\cal R}}{\max(\delta, \bar \delta)} \,
\left\{P_{qq}(\delta)+P_{gq}(\delta)\right\}-\frac{\beta_0}{2}
\ln\frac{|\vec q_J|^2}{\mu_R^2}\delta(1-\delta)
\]
\[
+\;C_A\delta(1-\delta)\left[\chi(n,\gamma)\ln\frac{s_0}{|\vec q_J|^2}
+\frac{85}{18}
\right.
\]
\[
\left.
+\;\frac{\pi^2}{2}+\frac{1}{2}\left(\psi^\prime
\left(1+\gamma+\frac{n}{2}\right)
-\psi^\prime\left(\frac{n}{2}-\gamma\right)-\chi^2(n,\gamma)\right)
\right]
\]
\[
+\;(1+\delta^2)\left\{C_A\left[\frac{(1+\delta^{-2\gamma})\,\chi(n,\gamma)}
{2(1-\delta)_+}-\delta^{-2\gamma}\left(\frac{\ln(1-\delta)}{1-\delta}\right)_+
\right]
\right.
\]
\[
\left.
+\;\left(C_F-\frac{C_A}{2}\right)\left[ \frac{\bar \delta}{\delta^2}{\cal I}_2
-\frac{2\ln\delta}{1-\delta}
+2\left(\frac{\ln(1-\delta)}{1-\delta}\right)_+ \right]\right\}
\]
\[
+\;\delta(1-\delta)\left(C_F\left(3\ln 2-\frac{\pi^2}{3}-\frac{9}{2}\right)
-\frac{5n_f}{9}\right)
+C_A\delta+C_F\bar \delta
\]
\[
\left.
+\;\frac{1+\bar \delta^2}{\delta}
\left(C_A\frac{\bar \delta}{\delta}{\cal I}_1+2C_A\ln\frac{\bar\delta}{\delta}
+C_F\delta^{-2\gamma}(\chi(n,\gamma)-2\ln \bar \delta)\right)\right]
\]
\[
+\;f_{g}\left(\frac{x_J}{\delta}\right)\frac{C_A}{C_F}
\left[
\left(P_{gg}(\delta)+2 \,n_f \frac{C_F}{C_A}P_{qg}(\delta)\right)
\ln\frac{|\vec q_J|^2}{\mu_F^2}
\right.
\]
\[
\left.
-\;2\delta^{-2\gamma} \ln \frac{{\cal R}}{\max(\delta, \bar \delta)} \left(P_{gg}(\delta)+2 \,n_f P_{qg}(\delta)\right)
-\frac{\beta_0}{2}\ln\frac{|\vec q_J|^2}{4\mu_R^2}\delta(1-\delta)
\right.
\]
\[
\left.
+\; C_A\delta(1-\delta)
\left(
\chi(n,\gamma)\ln\frac{s_0}{|\vec q_J|^2}+\frac{1}{12}+\frac{\pi^2}{6}
\right.\right.
\]
\[
\left.
+\;\frac{1}{2}\left[\psi^\prime\left(1+\gamma+\frac{n}{2}\right)
-\psi^\prime\left(\frac{n}{2}-\gamma\right)-\chi^2(n,\gamma)\right]
\right)
\]
\[
+\,2C_A (1-\delta^{-2\gamma})\left(\left(\frac{1}{\delta}-2
+\delta\bar\delta\right)\ln \bar \delta + \frac{\ln (1-\delta)}{1-\delta}\right)
\]
\[
+\,C_A\, \left[\frac{1}{\delta}+\frac{1}{(1- \delta)_+}-2+\delta\bar\delta\right]
\left((1+\delta^{-2\gamma})\chi(n,\gamma)-2\ln\delta+\frac{\bar \delta^2}
{\delta^2}{\cal I}_2\right)
\]
\[
\left.\left.
+\,n_f\left[\, 2\delta\bar \delta \, \frac{C_F}{C_A} +(\delta^2+\bar \delta^2)
\left(\frac{C_F}{C_A}\chi(n,\gamma)+\frac{\bar \delta}{\delta}{\cal I}_3\right)
-\frac{1}{12}\delta(1-\delta)\right]\right]\right\} \; ,
\]
where ${\cal R}$ depicts the radius the jet cone.

\end{appendices}

\bibliographystyle{elsarticle-num}

\bibliography{references}

\begin{thebibliography}{100}
\expandafter\ifx\csname url\endcsname\relax
  \def\url#1{\texttt{#1}}\fi
\expandafter\ifx\csname urlprefix\endcsname\relax\def\urlprefix{URL }\fi
\expandafter\ifx\csname href\endcsname\relax
  \def\href#1#2{#2} \def\path#1{#1}\fi

\bibitem{Apollinari:2015wtw}
G.~Apollinari, O.~Br\"uning, T.~Nakamoto, L.~Rossi, {High Luminosity Large Hadron Collider HL-LHC}, CERN Yellow Rep.~(5) (2015) 1--19.
\newblock \href {http://arxiv.org/abs/1705.08830} {\path{arXiv:1705.08830}}, \href {http://dx.doi.org/10.5170/CERN-2015-005.1} {\path{doi:10.5170/CERN-2015-005.1}}.

\bibitem{Apollinari:2015bam}
{High-Luminosity Large Hadron Collider (HL-LHC) : Preliminary Design Report}\href {http://dx.doi.org/10.5170/CERN-2015-005} {\path{doi:10.5170/CERN-2015-005}}.

\bibitem{Apollinari:2017lan}
{High-Luminosity Large Hadron Collider (HL-LHC)}: {Technical Design Report V. 0.1} 4/2017.
\newblock \href {http://dx.doi.org/10.23731/CYRM-2017-004} {\path{doi:10.23731/CYRM-2017-004}}.

\bibitem{Chapon:2020heu}
E.~Chapon, et~al., {Prospects for quarkonium studies at the high-luminosity LHC}, Prog. Part. Nucl. Phys. 122 (2022) 103906.
\newblock \href {http://arxiv.org/abs/2012.14161} {\path{arXiv:2012.14161}}, \href {http://dx.doi.org/10.1016/j.ppnp.2021.103906} {\path{doi:10.1016/j.ppnp.2021.103906}}.

\bibitem{AbdulKhalek:2021gbh}
R.~Abdul~Khalek, et~al., {Science Requirements and Detector Concepts for the Electron-Ion Collider}: {EIC Yellow Report}, Nucl. Phys. A 1026 (2022) 122447.
\newblock \href {http://arxiv.org/abs/2103.05419} {\path{arXiv:2103.05419}}, \href {http://dx.doi.org/10.1016/j.nuclphysa.2022.122447} {\path{doi:10.1016/j.nuclphysa.2022.122447}}.

\bibitem{Khalek:2022bzd}
R.~Abdul~Khalek, et~al., {Snowmass 2021 White Paper: Electron Ion Collider for High Energy Physics}, in: {2022 Snowmass Summer Study}, 2022.
\newblock \href {http://arxiv.org/abs/2203.13199} {\path{arXiv:2203.13199}}.

\bibitem{Hentschinski:2022xnd}
M.~Hentschinski, et~al., {White Paper on Forward Physics, BFKL, Saturation Physics and Diffraction}, Acta Phys. Polon. B 54~(3) (2023) 2.
\newblock \href {http://arxiv.org/abs/2203.08129} {\path{arXiv:2203.08129}}, \href {http://dx.doi.org/10.5506/APhysPolB.54.3-A2} {\path{doi:10.5506/APhysPolB.54.3-A2}}.

\bibitem{Amoroso:2022eow}
S.~Amoroso, et~al., {Snowmass 2021 whitepaper: Proton structure at the precision frontier}, Acta Phys. Polon. B 53~(12) (2022) A1.
\newblock \href {http://arxiv.org/abs/2203.13923} {\path{arXiv:2203.13923}}, \href {http://dx.doi.org/10.5506/APhysPolB.53.12-A1} {\path{doi:10.5506/APhysPolB.53.12-A1}}.

\bibitem{Abir:2023fpo}
R.~Abir, et~al., {The case for an EIC Theory Alliance: Theoretical Challenges of the EIC}\href {http://arxiv.org/abs/2305.14572} {\path{arXiv:2305.14572}}.

\bibitem{Allaire:2023fgp}
C.~Allaire, et~al., {Artificial Intelligence for the Electron Ion Collider (AI4EIC)}, Comput. Softw. Big Sci. 8 (2024) 5.
\newblock \href {http://arxiv.org/abs/2307.08593} {\path{arXiv:2307.08593}}, \href {http://dx.doi.org/10.1007/s41781-024-00113-4} {\path{doi:10.1007/s41781-024-00113-4}}.

\bibitem{FCC:2018byv}
A.~Abada, et~al., {FCC Physics Opportunities}: {Future Circular Collider Conceptual Design Report Volume 1}, Eur. Phys. J. C 79~(6) (2019) 474.
\newblock \href {http://dx.doi.org/10.1140/epjc/s10052-019-6904-3} {\path{doi:10.1140/epjc/s10052-019-6904-3}}.

\bibitem{FCC:2018evy}
A.~Abada, et~al., {FCC-ee: The Lepton Collider}: {Future Circular Collider Conceptual Design Report Volume 2}, Eur. Phys. J. ST 228~(2) (2019) 261--623.
\newblock \href {http://dx.doi.org/10.1140/epjst/e2019-900045-4} {\path{doi:10.1140/epjst/e2019-900045-4}}.

\bibitem{FCC:2018vvp}
A.~Abada, et~al., {FCC-hh: The Hadron Collider}: {Future Circular Collider Conceptual Design Report Volume 3}, Eur. Phys. J. ST 228~(4) (2019) 755--1107.
\newblock \href {http://dx.doi.org/10.1140/epjst/e2019-900087-0} {\path{doi:10.1140/epjst/e2019-900087-0}}.

\bibitem{FCC:2018bvk}
A.~Abada, et~al., {HE-LHC: The High-Energy Large Hadron Collider}: {Future Circular Collider Conceptual Design Report Volume 4}, Eur. Phys. J. ST 228~(5) (2019) 1109--1382.
\newblock \href {http://dx.doi.org/10.1140/epjst/e2019-900088-6} {\path{doi:10.1140/epjst/e2019-900088-6}}.

\bibitem{FCC:2025lpp}
M.~Benedikt, et~al., {Future Circular Collider Feasibility Study Report: Volume 1, Physics, Experiments, Detectors}, Eur. Phys. J. C 85~(12) (2025) 1468.
\newblock \href {http://arxiv.org/abs/2505.00272} {\path{arXiv:2505.00272}}, \href {http://dx.doi.org/10.1140/epjc/s10052-025-15077-x} {\path{doi:10.1140/epjc/s10052-025-15077-x}}.

\bibitem{FCC:2025uan}
M.~Benedikt, et~al., {Future Circular Collider Feasibility Study Report: Volume 2, Accelerators, Technical Infrastructure and Safety}, Eur. Phys. J. ST 234~(19) (2025) 5713--6197.
\newblock \href {http://arxiv.org/abs/2505.00274} {\path{arXiv:2505.00274}}, \href {http://dx.doi.org/10.1140/epjs/s11734-025-01967-4} {\path{doi:10.1140/epjs/s11734-025-01967-4}}.

\bibitem{FCC:2025jtd}
M.~Benedikt, et~al., {Future Circular Collider Feasibility Study Report: Volume 3, Civil Engineering, Implementation and Sustainability}, Eur. Phys. J. ST 234~(17) (2025) 5113--5383, [Erratum: Eur.Phys.J.ST None, (2025)].
\newblock \href {http://arxiv.org/abs/2505.00273} {\path{arXiv:2505.00273}}, \href {http://dx.doi.org/10.1140/epjs/s11734-025-01958-5} {\path{doi:10.1140/epjs/s11734-025-01958-5}}.

\bibitem{Gell-Mann:1962yej}
M.~Gell-Mann, {Symmetries of baryons and mesons}, Phys. Rev. 125 (1962) 1067--1084.
\newblock \href {http://dx.doi.org/10.1103/PhysRev.125.1067} {\path{doi:10.1103/PhysRev.125.1067}}.

\bibitem{Gell-Mann:1964ewy}
M.~Gell-Mann, {A Schematic Model of Baryons and Mesons}, Phys. Lett. 8 (1964) 214--215.
\newblock \href {http://dx.doi.org/10.1016/S0031-9163(64)92001-3} {\path{doi:10.1016/S0031-9163(64)92001-3}}.

\bibitem{Zweig:1964jf}
G.~Zweig, {An SU(3) model for strong interaction symmetry and its breaking. Version 2}, 1964, pp. 22--101.

\bibitem{Fritzsch:1973pi}
H.~Fritzsch, M.~Gell-Mann, H.~Leutwyler, {Advantages of the Color Octet Gluon Picture}, Phys. Lett. B 47 (1973) 365--368.
\newblock \href {http://dx.doi.org/10.1016/0370-2693(73)90625-4} {\path{doi:10.1016/0370-2693(73)90625-4}}.

\bibitem{Peccei:1977hh}
R.~D. Peccei, H.~R. Quinn, {CP Conservation in the Presence of Instantons}, Phys. Rev. Lett. 38 (1977) 1440--1443.
\newblock \href {http://dx.doi.org/10.1103/PhysRevLett.38.1440} {\path{doi:10.1103/PhysRevLett.38.1440}}.

\bibitem{Peccei:1977ur}
R.~D. Peccei, H.~R. Quinn, {Constraints Imposed by CP Conservation in the Presence of Instantons}, Phys. Rev. D 16 (1977) 1791--1797.
\newblock \href {http://dx.doi.org/10.1103/PhysRevD.16.1791} {\path{doi:10.1103/PhysRevD.16.1791}}.

\bibitem{Peccei:2006as}
R.~D. Peccei, {The Strong CP problem and axions}, Lect. Notes Phys. 741 (2008) 3--17.
\newblock \href {http://arxiv.org/abs/hep-ph/0607268} {\path{arXiv:hep-ph/0607268}}, \href {http://dx.doi.org/10.1007/978-3-540-73518-2_1} {\path{doi:10.1007/978-3-540-73518-2_1}}.

\bibitem{Duffy:2009ig}
L.~D. Duffy, K.~van Bibber, {Axions as Dark Matter Particles}, New J. Phys. 11 (2009) 105008.
\newblock \href {http://arxiv.org/abs/0904.3346} {\path{arXiv:0904.3346}}, \href {http://dx.doi.org/10.1088/1367-2630/11/10/105008} {\path{doi:10.1088/1367-2630/11/10/105008}}.

\bibitem{Forestell:2017wov}
L.~Forestell, D.~E. Morrissey, K.~Sigurdson, {Cosmological Bounds on Non-Abelian Dark Forces}, Phys. Rev. D 97~(7) (2018) 075029.
\newblock \href {http://arxiv.org/abs/1710.06447} {\path{arXiv:1710.06447}}, \href {http://dx.doi.org/10.1103/PhysRevD.97.075029} {\path{doi:10.1103/PhysRevD.97.075029}}.

\bibitem{Huang:2020crf}
W.-C. Huang, M.~Reichert, F.~Sannino, Z.-W. Wang, {Testing the dark SU(N) Yang-Mills theory confined landscape: From the lattice to gravitational waves}, Phys. Rev. D 104~(3) (2021) 035005.
\newblock \href {http://arxiv.org/abs/2012.11614} {\path{arXiv:2012.11614}}, \href {http://dx.doi.org/10.1103/PhysRevD.104.035005} {\path{doi:10.1103/PhysRevD.104.035005}}.

\bibitem{McLerran:2007qj}
L.~McLerran, R.~D. Pisarski, {Phases of cold, dense quarks at large N(c)}, Nucl. Phys. A 796 (2007) 83--100.
\newblock \href {http://arxiv.org/abs/0706.2191} {\path{arXiv:0706.2191}}, \href {http://dx.doi.org/10.1016/j.nuclphysa.2007.08.013} {\path{doi:10.1016/j.nuclphysa.2007.08.013}}.

\bibitem{Hidaka:2008yy}
Y.~Hidaka, L.~D. McLerran, R.~D. Pisarski, {Baryons and the phase diagram for a large number of colors and flavors}, Nucl. Phys. A 808 (2008) 117--123.
\newblock \href {http://arxiv.org/abs/0803.0279} {\path{arXiv:0803.0279}}, \href {http://dx.doi.org/10.1016/j.nuclphysa.2008.05.009} {\path{doi:10.1016/j.nuclphysa.2008.05.009}}.

\bibitem{McLerran:2018hbz}
L.~McLerran, S.~Reddy, {Quarkyonic Matter and Neutron Stars}, Phys. Rev. Lett. 122~(12) (2019) 122701.
\newblock \href {http://arxiv.org/abs/1811.12503} {\path{arXiv:1811.12503}}, \href {http://dx.doi.org/10.1103/PhysRevLett.122.122701} {\path{doi:10.1103/PhysRevLett.122.122701}}.

\bibitem{Buchmuller:1985jz}
W.~Buchmuller, D.~Wyler, {Effective Lagrangian Analysis of New Interactions and Flavor Conservation}, Nucl. Phys. B 268 (1986) 621--653.
\newblock \href {http://dx.doi.org/10.1016/0550-3213(86)90262-2} {\path{doi:10.1016/0550-3213(86)90262-2}}.

\bibitem{Witten:1979kh}
E.~Witten, {Baryons in the 1/n Expansion}, Nucl. Phys. B 160 (1979) 57--115.
\newblock \href {http://dx.doi.org/10.1016/0550-3213(79)90232-3} {\path{doi:10.1016/0550-3213(79)90232-3}}.

\bibitem{Dudek:2010wm}
J.~J. Dudek, R.~G. Edwards, M.~J. Peardon, D.~G. Richards, C.~E. Thomas, {Toward the excited meson spectrum of dynamical QCD}, Phys. Rev. D 82 (2010) 034508.
\newblock \href {http://arxiv.org/abs/1004.4930} {\path{arXiv:1004.4930}}, \href {http://dx.doi.org/10.1103/PhysRevD.82.034508} {\path{doi:10.1103/PhysRevD.82.034508}}.

\bibitem{Afonin:2019unu}
S.~S. Afonin, {The effect of higher dimensional QCD operators on the spectroscopy of bottom-up holographic models}, Universe 7~(4) (2021) 102.
\newblock \href {http://arxiv.org/abs/1905.13086} {\path{arXiv:1905.13086}}, \href {http://dx.doi.org/10.3390/universe7040102} {\path{doi:10.3390/universe7040102}}.

\bibitem{SLAC-SP-017:1974ind}
J.~E. Augustin, et~al., {Discovery of a Narrow Resonance in $e^+ e^-$ Annihilation}, Phys. Rev. Lett. 33 (1974) 1406--1408.
\newblock \href {http://dx.doi.org/10.1103/PhysRevLett.33.1406} {\path{doi:10.1103/PhysRevLett.33.1406}}.

\bibitem{E598:1974sol}
J.~J. Aubert, et~al., {Experimental Observation of a Heavy Particle $J$}, Phys. Rev. Lett. 33 (1974) 1404--1406.
\newblock \href {http://dx.doi.org/10.1103/PhysRevLett.33.1404} {\path{doi:10.1103/PhysRevLett.33.1404}}.

\bibitem{Bacci:1974za}
C.~Bacci, et~al., {Preliminary Result of Frascati (ADONE) on the Nature of a New 3.1-GeV Particle Produced in e+ e- Annihilation}, Phys. Rev. Lett. 33 (1974) 1408, [Erratum: Phys.Rev.Lett. 33, 1649 (1974)].
\newblock \href {http://dx.doi.org/10.1103/PhysRevLett.33.1408} {\path{doi:10.1103/PhysRevLett.33.1408}}.

\bibitem{Close:1991pf}
F.~Close, {Glueballs and exotic matter}, Nature 349 (1991) 368--369.
\newblock \href {http://dx.doi.org/10.1038/349368a0} {\path{doi:10.1038/349368a0}}.

\bibitem{Close:1997qda}
F.~E. Close, {Glueballs and hybrids: New states of matter}, Contemp. Phys. 38 (1997) 1--12.
\newblock \href {http://dx.doi.org/10.1080/001075197182522} {\path{doi:10.1080/001075197182522}}.

\bibitem{Close:1998zz}
F.~E. Close, {Glueballs and the pomeron: A central mystery}, in: {33rd Rencontres de Moriond: QCD and High-Energy Hadronic Interactions}, 1998, pp. 589--602.

\bibitem{Minkowski:1998mf}
P.~Minkowski, W.~Ochs, {Identification of the glueballs and the scalar meson nonet of lowest mass}, Eur. Phys. J. C 9 (1999) 283--312.
\newblock \href {http://arxiv.org/abs/hep-ph/9811518} {\path{arXiv:hep-ph/9811518}}, \href {http://dx.doi.org/10.1007/s100520050533} {\path{doi:10.1007/s100520050533}}.

\bibitem{Close:2000yg}
F.~E. Close, {Glueballs: A Central mystery}, Acta Phys. Polon. B 31 (2000) 2557--2565.
\newblock \href {http://arxiv.org/abs/hep-ph/0006288} {\path{arXiv:hep-ph/0006288}}.

\bibitem{Mathieu:2008me}
V.~Mathieu, N.~Kochelev, V.~Vento, {The Physics of Glueballs}, Int. J. Mod. Phys. E 18 (2009) 1--49.
\newblock \href {http://arxiv.org/abs/0810.4453} {\path{arXiv:0810.4453}}, \href {http://dx.doi.org/10.1142/S0218301309012124} {\path{doi:10.1142/S0218301309012124}}.

\bibitem{Hsiao:2013dta}
Y.~K. Hsiao, C.~Q. Geng, {Identifying Glueball at 3.02 GeV in Baryonic $B$ Decays}, Phys. Lett. B 727 (2013) 168--171.
\newblock \href {http://arxiv.org/abs/1302.3331} {\path{arXiv:1302.3331}}, \href {http://dx.doi.org/10.1016/j.physletb.2013.10.008} {\path{doi:10.1016/j.physletb.2013.10.008}}.

\bibitem{D0:2020tig}
V.~M. Abazov, et~al., {Odderon Exchange from Elastic Scattering Differences between $pp$ and $p \bar{p}$ Data at 1.96~TeV and from pp Forward Scattering Measurements}, Phys. Rev. Lett. 127~(6) (2021) 062003.
\newblock \href {http://arxiv.org/abs/2012.03981} {\path{arXiv:2012.03981}}, \href {http://dx.doi.org/10.1103/PhysRevLett.127.062003} {\path{doi:10.1103/PhysRevLett.127.062003}}.

\bibitem{Csorgo:2019ewn}
T.~Cs\"org\H{o}, T.~Novak, R.~Pasechnik, A.~Ster, I.~Szanyi, {Evidence of Odderon-exchange from scaling properties of elastic scattering at TeV energies}, Eur. Phys. J. C 81~(2) (2021) 180.
\newblock \href {http://arxiv.org/abs/1912.11968} {\path{arXiv:1912.11968}}, \href {http://dx.doi.org/10.1140/epjc/s10052-021-08867-6} {\path{doi:10.1140/epjc/s10052-021-08867-6}}.

\bibitem{Jaffe:1976ig}
R.~L. Jaffe, {Multi-Quark Hadrons. 1. The Phenomenology of (2 Quark 2 anti-Quark) Mesons}, Phys. Rev. D 15 (1977) 267.
\newblock \href {http://dx.doi.org/10.1103/PhysRevD.15.267} {\path{doi:10.1103/PhysRevD.15.267}}.

\bibitem{Jaffe:1976ih}
R.~L. Jaffe, {Multi-Quark Hadrons. 2. Methods}, Phys. Rev. D 15 (1977) 281.
\newblock \href {http://dx.doi.org/10.1103/PhysRevD.15.281} {\path{doi:10.1103/PhysRevD.15.281}}.

\bibitem{Ader:1981db}
J.~P. Ader, J.~M. Richard, P.~Taxil, {DO NARROW HEAVY MULTI - QUARK STATES EXIST?}, Phys. Rev. D 25 (1982) 2370.
\newblock \href {http://dx.doi.org/10.1103/PhysRevD.25.2370} {\path{doi:10.1103/PhysRevD.25.2370}}.

\bibitem{Belle:2003nnu}
S.~K. Choi, et~al., {Observation of a narrow charmonium-like state in exclusive $B^\pm \to K^\pm \pi^+ \pi^- J/\psi$ decays}, Phys. Rev. Lett. 91 (2003) 262001.
\newblock \href {http://arxiv.org/abs/hep-ex/0309032} {\path{arXiv:hep-ex/0309032}}, \href {http://dx.doi.org/10.1103/PhysRevLett.91.262001} {\path{doi:10.1103/PhysRevLett.91.262001}}.

\bibitem{LHCb:2020bls}
R.~Aaij, et~al., {A model-independent study of resonant structure in $B^+\to D^+D^-K^+$ decays}, Phys. Rev. Lett. 125 (2020) 242001.
\newblock \href {http://arxiv.org/abs/2009.00025} {\path{arXiv:2009.00025}}, \href {http://dx.doi.org/10.1103/PhysRevLett.125.242001} {\path{doi:10.1103/PhysRevLett.125.242001}}.

\bibitem{Bjorken:1985ei}
J.~D. Bjorken, {Is the ccc a new deal for baryon spectroscopy?}, AIP Conf. Proc. 132 (1985) 390--403.
\newblock \href {http://dx.doi.org/10.1063/1.35379} {\path{doi:10.1063/1.35379}}.

\bibitem{Fleck:1989mb}
S.~Fleck, J.~M. Richard, {Baryons with double charm}, Prog. Theor. Phys. 82 (1989) 760--774.
\newblock \href {http://dx.doi.org/10.1143/PTP.82.760} {\path{doi:10.1143/PTP.82.760}}.

\bibitem{Martynenko:2007je}
A.~P. Martynenko, {Ground-state triply and doubly heavy baryons in a relativistic three-quark model}, Phys. Lett. B 663 (2008) 317--321.
\newblock \href {http://arxiv.org/abs/0708.2033} {\path{arXiv:0708.2033}}, \href {http://dx.doi.org/10.1016/j.physletb.2008.04.030} {\path{doi:10.1016/j.physletb.2008.04.030}}.

\bibitem{Martynenko:2013eoa}
A.~P. Martynenko, A.~M. Trunin, {Relativistic corrections to the pair double heavy diquark production in $e^+e^-$ annihilation}, Phys. Rev. D 89~(1) (2014) 014004.
\newblock \href {http://arxiv.org/abs/1308.3998} {\path{arXiv:1308.3998}}, \href {http://dx.doi.org/10.1103/PhysRevD.89.014004} {\path{doi:10.1103/PhysRevD.89.014004}}.

\bibitem{Karliner:2014gca}
M.~Karliner, J.~L. Rosner, {Baryons with two heavy quarks: Masses, production, decays, and detection}, Phys. Rev. D 90~(9) (2014) 094007.
\newblock \href {http://arxiv.org/abs/1408.5877} {\path{arXiv:1408.5877}}, \href {http://dx.doi.org/10.1103/PhysRevD.90.094007} {\path{doi:10.1103/PhysRevD.90.094007}}.

\bibitem{Yoshida:2015tia}
T.~Yoshida, E.~Hiyama, A.~Hosaka, M.~Oka, K.~Sadato, {Spectrum of heavy baryons in the quark model}, Phys. Rev. D 92~(11) (2015) 114029.
\newblock \href {http://arxiv.org/abs/1510.01067} {\path{arXiv:1510.01067}}, \href {http://dx.doi.org/10.1103/PhysRevD.92.114029} {\path{doi:10.1103/PhysRevD.92.114029}}.

\bibitem{Ebert:2002ig}
D.~Ebert, R.~N. Faustov, V.~O. Galkin, A.~P. Martynenko, {Mass spectra of doubly heavy baryons in the relativistic quark model}, Phys. Rev. D 66 (2002) 014008.
\newblock \href {http://arxiv.org/abs/hep-ph/0201217} {\path{arXiv:hep-ph/0201217}}, \href {http://dx.doi.org/10.1103/PhysRevD.66.014008} {\path{doi:10.1103/PhysRevD.66.014008}}.

\bibitem{Roberts:2007ni}
W.~Roberts, M.~Pervin, {Heavy baryons in a quark model}, Int. J. Mod. Phys. A 23 (2008) 2817--2860.
\newblock \href {http://arxiv.org/abs/0711.2492} {\path{arXiv:0711.2492}}, \href {http://dx.doi.org/10.1142/S0217751X08041219} {\path{doi:10.1142/S0217751X08041219}}.

\bibitem{Chen:2011mb}
Y.-Q. Chen, S.-Z. Wu, {Production of Triply Heavy Baryons at LHC}, JHEP 08 (2011) 144, [Erratum: JHEP 09, 089 (2011)].
\newblock \href {http://arxiv.org/abs/1106.0193} {\path{arXiv:1106.0193}}, \href {http://dx.doi.org/10.1007/JHEP08(2011)144} {\path{doi:10.1007/JHEP08(2011)144}}.

\bibitem{Padmanath:2013zfa}
M.~Padmanath, R.~G. Edwards, N.~Mathur, M.~Peardon, {Spectroscopy of triply-charmed baryons from lattice QCD}, Phys. Rev. D 90~(7) (2014) 074504.
\newblock \href {http://arxiv.org/abs/1307.7022} {\path{arXiv:1307.7022}}, \href {http://dx.doi.org/10.1103/PhysRevD.90.074504} {\path{doi:10.1103/PhysRevD.90.074504}}.

\bibitem{Brown:2014ena}
Z.~S. Brown, W.~Detmold, S.~Meinel, K.~Orginos, {Charmed bottom baryon spectroscopy from lattice QCD}, Phys. Rev. D 90~(9) (2014) 094507.
\newblock \href {http://arxiv.org/abs/1409.0497} {\path{arXiv:1409.0497}}, \href {http://dx.doi.org/10.1103/PhysRevD.90.094507} {\path{doi:10.1103/PhysRevD.90.094507}}.

\bibitem{Meinel:2012qz}
S.~Meinel, {Excited-state spectroscopy of triply-bottom baryons from lattice QCD}, Phys. Rev. D 85 (2012) 114510.
\newblock \href {http://arxiv.org/abs/1202.1312} {\path{arXiv:1202.1312}}, \href {http://dx.doi.org/10.1103/PhysRevD.85.114510} {\path{doi:10.1103/PhysRevD.85.114510}}.

\bibitem{Flynn:2003vz}
J.~M. Flynn, F.~Mescia, A.~S.~B. Tariq, {Spectroscopy of doubly charmed baryons in lattice QCD}, JHEP 07 (2003) 066.
\newblock \href {http://arxiv.org/abs/hep-lat/0307025} {\path{arXiv:hep-lat/0307025}}, \href {http://dx.doi.org/10.1088/1126-6708/2003/07/066} {\path{doi:10.1088/1126-6708/2003/07/066}}.

\bibitem{Shah:2016vmd}
Z.~Shah, K.~Thakkar, A.~K. Rai, {Excited State Mass spectra of doubly heavy baryons $\Omega_{cc}$, $\Omega_{bb}$ and $\Omega_{bc}$}, Eur. Phys. J. C 76~(10) (2016) 530.
\newblock \href {http://arxiv.org/abs/1609.03030} {\path{arXiv:1609.03030}}, \href {http://dx.doi.org/10.1140/epjc/s10052-016-4379-z} {\path{doi:10.1140/epjc/s10052-016-4379-z}}.

\bibitem{Shah:2017jkr}
Z.~Shah, A.~K. Rai, {Masses and Regge trajectories of triply heavy $\Omega_{ccc}$ and $\Omega_{bbb}$ baryons}, Eur. Phys. J. A 53~(10) (2017) 195.
\newblock \href {http://dx.doi.org/10.1140/epja/i2017-12386-2} {\path{doi:10.1140/epja/i2017-12386-2}}.

\bibitem{Bagan:1994dy}
E.~Bagan, H.~G. Dosch, P.~Gosdzinsky, S.~Narison, J.~M. Richard, {Hadrons with charm and beauty}, Z. Phys. C 64 (1994) 57--72.
\newblock \href {http://arxiv.org/abs/hep-ph/9403208} {\path{arXiv:hep-ph/9403208}}, \href {http://dx.doi.org/10.1007/BF01557235} {\path{doi:10.1007/BF01557235}}.

\bibitem{Chang:2006eu}
C.-H. Chang, C.-F. Qiao, J.-X. Wang, X.-G. Wu, {Estimate of the hadronic production of the doubly charmed baryon Xi(cc) under GM-VFN scheme}, Phys. Rev. D 73 (2006) 094022.
\newblock \href {http://arxiv.org/abs/hep-ph/0601032} {\path{arXiv:hep-ph/0601032}}, \href {http://dx.doi.org/10.1103/PhysRevD.73.094022} {\path{doi:10.1103/PhysRevD.73.094022}}.

\bibitem{GomshiNobary:2005ur}
M.~A. Gomshi~Nobary, R.~Sepahvand, {An Ivestigation of triply heavy baryon production at hadron colliders}, Nucl. Phys. B 741 (2006) 34--41.
\newblock \href {http://arxiv.org/abs/hep-ph/0508115} {\path{arXiv:hep-ph/0508115}}, \href {http://dx.doi.org/10.1016/j.nuclphysb.2006.01.043} {\path{doi:10.1016/j.nuclphysb.2006.01.043}}.

\bibitem{Gershtein:2000nx}
S.~S. Gershtein, V.~V. Kiselev, A.~K. Likhoded, A.~I. Onishchenko, {Spectroscopy of doubly heavy baryons}, Phys. Rev. D 62 (2000) 054021.
\newblock \href {http://dx.doi.org/10.1103/PhysRevD.62.054021} {\path{doi:10.1103/PhysRevD.62.054021}}.

\bibitem{LHCb:2015yax}
R.~Aaij, et~al., {Observation of $J/\psi p$ Resonances Consistent with Pentaquark States in $\Lambda_b^0 \to J/\psi K^- p$ Decays}, Phys. Rev. Lett. 115 (2015) 072001.
\newblock \href {http://arxiv.org/abs/1507.03414} {\path{arXiv:1507.03414}}, \href {http://dx.doi.org/10.1103/PhysRevLett.115.072001} {\path{doi:10.1103/PhysRevLett.115.072001}}.

\bibitem{LHCb:2019kea}
R.~Aaij, et~al., {Observation of a narrow pentaquark state, $P_c(4312)^+$, and of two-peak structure of the $P_c(4450)^+$}, Phys. Rev. Lett. 122~(22) (2019) 222001.
\newblock \href {http://arxiv.org/abs/1904.03947} {\path{arXiv:1904.03947}}, \href {http://dx.doi.org/10.1103/PhysRevLett.122.222001} {\path{doi:10.1103/PhysRevLett.122.222001}}.

\bibitem{LHCb:2020jpq}
R.~Aaij, et~al., {Evidence of a $J/\psi\Lambda$ structure and observation of excited $\Xi^-$ states in the $\Xi^-_b \to J/\psi\Lambda K^-$ decay}, Sci. Bull. 66 (2021) 1278--1287.
\newblock \href {http://arxiv.org/abs/2012.10380} {\path{arXiv:2012.10380}}, \href {http://dx.doi.org/10.1016/j.scib.2021.02.030} {\path{doi:10.1016/j.scib.2021.02.030}}.

\bibitem{Brambilla:2019esw}
N.~Brambilla, S.~Eidelman, C.~Hanhart, A.~Nefediev, C.-P. Shen, C.~E. Thomas, A.~Vairo, C.-Z. Yuan, {The $XYZ$ states: experimental and theoretical status and perspectives}, Phys. Rept. 873 (2020) 1--154.
\newblock \href {http://arxiv.org/abs/1907.07583} {\path{arXiv:1907.07583}}, \href {http://dx.doi.org/10.1016/j.physrep.2020.05.001} {\path{doi:10.1016/j.physrep.2020.05.001}}.

\bibitem{Esposito:2016noz}
A.~Esposito, A.~Pilloni, A.~D. Polosa, {Multiquark Resonances}, Phys. Rept. 668 (2017) 1--97.
\newblock \href {http://arxiv.org/abs/1611.07920} {\path{arXiv:1611.07920}}, \href {http://dx.doi.org/10.1016/j.physrep.2016.11.002} {\path{doi:10.1016/j.physrep.2016.11.002}}.

\bibitem{Lebed:2016hpi}
R.~F. Lebed, R.~E. Mitchell, E.~S. Swanson, {Heavy-Quark QCD Exotica}, Prog. Part. Nucl. Phys. 93 (2017) 143--194.
\newblock \href {http://arxiv.org/abs/1610.04528} {\path{arXiv:1610.04528}}, \href {http://dx.doi.org/10.1016/j.ppnp.2016.11.003} {\path{doi:10.1016/j.ppnp.2016.11.003}}.

\bibitem{Faustov:2021qqf}
R.~N. Faustov, V.~O. Galkin, {Triply heavy baryon spectroscopy in the relativistic quark model}, Phys. Rev. D 105~(1) (2022) 014013.
\newblock \href {http://arxiv.org/abs/2111.07702} {\path{arXiv:2111.07702}}, \href {http://dx.doi.org/10.1103/PhysRevD.105.014013} {\path{doi:10.1103/PhysRevD.105.014013}}.

\bibitem{Mathur:2018rwu}
N.~Mathur, M.~Padmanath, {Lattice QCD study of doubly-charmed strange baryons}, Phys. Rev. D 99~(3) (2019) 031501.
\newblock \href {http://arxiv.org/abs/1807.00174} {\path{arXiv:1807.00174}}, \href {http://dx.doi.org/10.1103/PhysRevD.99.031501} {\path{doi:10.1103/PhysRevD.99.031501}}.

\bibitem{Francis:2018jyb}
A.~Francis, R.~J. Hudspith, R.~Lewis, K.~Maltman, {Evidence for charm-bottom tetraquarks and the mass dependence of heavy-light tetraquark states from lattice QCD} (2019).
\newblock \href {http://arxiv.org/abs/1810.10550} {\path{arXiv:1810.10550}}, \href {http://dx.doi.org/10.1103/PhysRevD.99.054505} {\path{doi:10.1103/PhysRevD.99.054505}}.

\bibitem{Eichten:1994gt}
E.~J. Eichten, C.~Quigg, {Mesons with beauty and charm: Spectroscopy}, Phys. Rev. D 49 (1994) 5845--5856.
\newblock \href {http://arxiv.org/abs/hep-ph/9402210} {\path{arXiv:hep-ph/9402210}}, \href {http://dx.doi.org/10.1103/PhysRevD.49.5845} {\path{doi:10.1103/PhysRevD.49.5845}}.

\bibitem{Godfrey:1985xj}
S.~Godfrey, N.~Isgur, {Mesons in a Relativized Quark Model with Chromodynamics}, Phys. Rev. D 32 (1985) 189--231.
\newblock \href {http://dx.doi.org/10.1103/PhysRevD.32.189} {\path{doi:10.1103/PhysRevD.32.189}}.

\bibitem{Ali:2017jda}
A.~Ali, J.~S. Lange, S.~Stone, {Exotics: Heavy Pentaquarks and Tetraquarks}, Prog. Part. Nucl. Phys. 97 (2017) 123--198.
\newblock \href {http://arxiv.org/abs/1706.00610} {\path{arXiv:1706.00610}}, \href {http://dx.doi.org/10.1016/j.ppnp.2017.08.003} {\path{doi:10.1016/j.ppnp.2017.08.003}}.

\bibitem{LHCb:2020bwg}
R.~Aaij, et~al., {Observation of structure in the $J /\psi$ -pair mass spectrum}, Sci. Bull. 65~(23) (2020) 1983--1993.
\newblock \href {http://arxiv.org/abs/2006.16957} {\path{arXiv:2006.16957}}, \href {http://dx.doi.org/10.1016/j.scib.2020.08.032} {\path{doi:10.1016/j.scib.2020.08.032}}.

\bibitem{Karliner:2020vsi}
M.~Karliner, J.~L. Rosner, {First exotic hadron with open heavy flavor: $cs\bar u\bar d$ tetraquark}, Phys. Rev. D 102~(9) (2020) 094016.
\newblock \href {http://arxiv.org/abs/2008.05993} {\path{arXiv:2008.05993}}, \href {http://dx.doi.org/10.1103/PhysRevD.102.094016} {\path{doi:10.1103/PhysRevD.102.094016}}.

\bibitem{Pineda:2011dg}
A.~Pineda, {Review of Heavy Quarkonium at weak coupling}, Prog. Part. Nucl. Phys. 67 (2012) 735--785.
\newblock \href {http://arxiv.org/abs/1111.0165} {\path{arXiv:1111.0165}}, \href {http://dx.doi.org/10.1016/j.ppnp.2012.01.038} {\path{doi:10.1016/j.ppnp.2012.01.038}}.

\bibitem{Celiberto:2024mab}
F.~G. Celiberto, G.~Gatto, A.~Papa, {Fully charmed tetraquarks from LHC to FCC: natural stability from fragmentation}, Eur. Phys. J. C 84~(10) (2024) 1071.
\newblock \href {http://arxiv.org/abs/2405.14773} {\path{arXiv:2405.14773}}, \href {http://dx.doi.org/10.1140/epjc/s10052-024-13345-w} {\path{doi:10.1140/epjc/s10052-024-13345-w}}.

\bibitem{Celiberto:2024beg}
F.~G. Celiberto, G.~Gatto, {Bottomoniumlike states in proton collisions: Fragmentation and resummation}, Phys. Rev. D 111~(3) (2025) 034037.
\newblock \href {http://arxiv.org/abs/2412.10549} {\path{arXiv:2412.10549}}, \href {http://dx.doi.org/10.1103/PhysRevD.111.034037} {\path{doi:10.1103/PhysRevD.111.034037}}.

\bibitem{Celiberto:2025dfe}
F.~G. Celiberto, {Fragmentation functions for axial-vector heavy tetraquarks: A TQ4Q1.1 update}, Phys. Rev. D 111~(11) (2025) L111501.
\newblock \href {http://arxiv.org/abs/2504.03949} {\path{arXiv:2504.03949}}, \href {http://dx.doi.org/10.1103/ympl-ly2l} {\path{doi:10.1103/ympl-ly2l}}.

\bibitem{Celiberto:2025ziy}
F.~G. Celiberto, {Fragmentation of fully heavy tetraquarks: The TQ4Q1.1 functions as a case study}, Phys. Rev. D 112~(7) (2025) 074041.
\newblock \href {http://arxiv.org/abs/2507.09744} {\path{arXiv:2507.09744}}, \href {http://dx.doi.org/10.1103/375n-fw5h} {\path{doi:10.1103/375n-fw5h}}.

\bibitem{Celiberto:2025vra}
F.~G. Celiberto, A.~V. Giannini, V.~P. Gon{\c{c}}alves, Y.~N. Lima, {Fully charmed tetraquark production in forward rapidity $pp$ collisions at LHC and FCC energies}, Phys. Rev. D 113~(5) (2026) 054014.
\newblock \href {http://arxiv.org/abs/2511.18984} {\path{arXiv:2511.18984}}, \href {http://dx.doi.org/10.1103/tq47-w7jn} {\path{doi:10.1103/tq47-w7jn}}.

\bibitem{Llanes-Estrada:2011gwu}
F.~J. Llanes-Estrada, O.~I. Pavlova, R.~Williams, {A First Estimate of Triply Heavy Baryon Masses from the pNRQCD Perturbative Static Potential}, Eur. Phys. J. C 72 (2012) 2019.
\newblock \href {http://arxiv.org/abs/1111.7087} {\path{arXiv:1111.7087}}, \href {http://dx.doi.org/10.1140/epjc/s10052-012-2019-9} {\path{doi:10.1140/epjc/s10052-012-2019-9}}.

\bibitem{Wei:2016jyk}
K.-W. Wei, B.~Chen, N.~Liu, Q.-Q. Wang, X.-H. Guo, {Spectroscopy of singly, doubly, and triply bottom baryons}, Phys. Rev. D 95~(11) (2017) 116005.
\newblock \href {http://arxiv.org/abs/1609.02512} {\path{arXiv:1609.02512}}, \href {http://dx.doi.org/10.1103/PhysRevD.95.116005} {\path{doi:10.1103/PhysRevD.95.116005}}.

\bibitem{Yang:2019lsg}
G.~Yang, J.~Ping, P.~G. Ortega, J.~Segovia, {Triply heavy baryons in the constituent quark model}, Chin. Phys. C 44~(2) (2020) 023102.
\newblock \href {http://arxiv.org/abs/1904.10166} {\path{arXiv:1904.10166}}, \href {http://dx.doi.org/10.1088/1674-1137/44/2/023102} {\path{doi:10.1088/1674-1137/44/2/023102}}.

\bibitem{Gomez-Rocha:2023jfr}
M.~G\'omez-Rocha, J.~More, K.~Serafin, {Baryon Masses Estimate in Heavy Flavor QCD}: {An Effective Particle Approach to Hadron Spectra}, Few Body Syst. 64~(3) (2023) 44.
\newblock \href {http://arxiv.org/abs/2305.06728} {\path{arXiv:2305.06728}}, \href {http://dx.doi.org/10.1007/s00601-023-01818-9} {\path{doi:10.1007/s00601-023-01818-9}}.

\bibitem{Najjar:2024deh}
Z.~R. Najjar, K.~Azizi, H.~R. Moshfegh, {Properties of the ground and excited states of triply heavy spin-1/2 baryons}, Eur. Phys. J. C 84~(6) (2024) 612.
\newblock \href {http://arxiv.org/abs/2402.14348} {\path{arXiv:2402.14348}}, \href {http://dx.doi.org/10.1140/epjc/s10052-024-12960-x} {\path{doi:10.1140/epjc/s10052-024-12960-x}}.

\bibitem{deArenaza:2024dhe}
N.~M. de~Arenaza, J.~J. G\'alvez-Viruet, F.~J. Llanes-Estrada, {Triply-heavy/strange baryons with Cornell potential on a quantum computer}\href {http://arxiv.org/abs/2407.07232} {\path{arXiv:2407.07232}}.

\bibitem{Belle-II:2010dht}
T.~Abe, et~al., {Belle II Technical Design Report}\href {http://arxiv.org/abs/1011.0352} {\path{arXiv:1011.0352}}.

\bibitem{ATLAS:2016bek}
M.~Aaboud, et~al., {Search for new phenomena in final states with an energetic jet and large missing transverse momentum in $pp$ collisions at $\sqrt{s}=13$ TeV using the ATLAS detector}, Phys. Rev. D 94~(3) (2016) 032005.
\newblock \href {http://arxiv.org/abs/1604.07773} {\path{arXiv:1604.07773}}, \href {http://dx.doi.org/10.1103/PhysRevD.94.032005} {\path{doi:10.1103/PhysRevD.94.032005}}.

\bibitem{CMS:2024aqx}
A.~Hayrapetyan, et~al., {Performance of the CMS high-level trigger during LHC Run 2}, JINST 19~(11) (2024) P11021.
\newblock \href {http://arxiv.org/abs/2410.17038} {\path{arXiv:2410.17038}}, \href {http://dx.doi.org/10.1088/1748-0221/19/11/P11021} {\path{doi:10.1088/1748-0221/19/11/P11021}}.

\bibitem{LHCb:2020frr}
R.~Aaij, et~al., {Measurement of differential $ b\overline{b} $- and $ c\overline{c} $-dijet cross-sections in the forward region of $pp$ collisions at $ \sqrt{s} $ = 13 TeV}, JHEP 02 (2021) 023.
\newblock \href {http://arxiv.org/abs/2010.09437} {\path{arXiv:2010.09437}}, \href {http://dx.doi.org/10.1007/JHEP02(2021)023} {\path{doi:10.1007/JHEP02(2021)023}}.

\bibitem{Braaten:1994bz}
E.~Braaten, K.-m. Cheung, S.~Fleming, T.~C. Yuan, {Perturbative QCD fragmentation functions as a model for heavy quark fragmentation}, Phys. Rev. D 51 (1995) 4819--4829.
\newblock \href {http://arxiv.org/abs/hep-ph/9409316} {\path{arXiv:hep-ph/9409316}}, \href {http://dx.doi.org/10.1103/PhysRevD.51.4819} {\path{doi:10.1103/PhysRevD.51.4819}}.

\bibitem{Braaten:1993rw}
E.~Braaten, T.~C. Yuan, {Gluon fragmentation into heavy quarkonium}, Phys. Rev. Lett. 71 (1993) 1673--1676.
\newblock \href {http://arxiv.org/abs/hep-ph/9303205} {\path{arXiv:hep-ph/9303205}}, \href {http://dx.doi.org/10.1103/PhysRevLett.71.1673} {\path{doi:10.1103/PhysRevLett.71.1673}}.

\bibitem{Braaten:1993mp}
E.~Braaten, K.-m. Cheung, T.~C. Yuan, {Z0 decay into charmonium via charm quark fragmentation}, Phys. Rev. D 48 (1993) 4230--4235.
\newblock \href {http://arxiv.org/abs/hep-ph/9302307} {\path{arXiv:hep-ph/9302307}}, \href {http://dx.doi.org/10.1103/PhysRevD.48.4230} {\path{doi:10.1103/PhysRevD.48.4230}}.

\bibitem{Braaten:1994xb}
E.~Braaten, M.~A. Doncheski, S.~Fleming, M.~L. Mangano, {Fragmentation production of $J/\psi$ and $\psi^\prime$ at the Tevatron}, Phys. Lett. B 333 (1994) 548--554.
\newblock \href {http://arxiv.org/abs/hep-ph/9405407} {\path{arXiv:hep-ph/9405407}}, \href {http://dx.doi.org/10.1016/0370-2693(94)90182-1} {\path{doi:10.1016/0370-2693(94)90182-1}}.

\bibitem{Braaten:1993jn}
E.~Braaten, K.-m. Cheung, T.~C. Yuan, {Perturbative QCD fragmentation functions for $B_c$ and $B_{c}$ * production}, Phys. Rev. D 48~(11) (1993) R5049.
\newblock \href {http://arxiv.org/abs/hep-ph/9305206} {\path{arXiv:hep-ph/9305206}}, \href {http://dx.doi.org/10.1103/PhysRevD.48.R5049} {\path{doi:10.1103/PhysRevD.48.R5049}}.

\bibitem{Kiselev:1994pu}
V.~V. Kiselev, A.~K. Likhoded, M.~V. Shevlyagin, {Double charmed baryon production at B factory}, Phys. Lett. B 332 (1994) 411--414.
\newblock \href {http://arxiv.org/abs/hep-ph/9408407} {\path{arXiv:hep-ph/9408407}}, \href {http://dx.doi.org/10.1016/0370-2693(94)91273-4} {\path{doi:10.1016/0370-2693(94)91273-4}}.

\bibitem{Anselmino:1992vg}
M.~Anselmino, E.~Predazzi, S.~Ekelin, S.~Fredriksson, D.~B. Lichtenberg, {Diquarks}, Rev. Mod. Phys. 65 (1993) 1199--1234.
\newblock \href {http://dx.doi.org/10.1103/RevModPhys.65.1199} {\path{doi:10.1103/RevModPhys.65.1199}}.

\bibitem{Ebert:1995fp}
D.~Ebert, T.~Feldmann, C.~Kettner, H.~Reinhardt, {A Diquark model for baryons containing one heavy quark}, Z. Phys. C 71 (1996) 329--336.
\newblock \href {http://arxiv.org/abs/hep-ph/9506298} {\path{arXiv:hep-ph/9506298}}, \href {http://dx.doi.org/10.1007/BF02906991} {\path{doi:10.1007/BF02906991}}.

\bibitem{MoosaviNejad:2016qdx}
S.~M. Moosavi~Nejad, P.~Sartipi~Yarahmadi, {Heavy quark fragmentation functions at next-to-leading perturbative QCD}, Eur. Phys. J. A 52~(10) (2016) 315.
\newblock \href {http://arxiv.org/abs/1609.07422} {\path{arXiv:1609.07422}}, \href {http://dx.doi.org/10.1140/epja/i2016-16315-7} {\path{doi:10.1140/epja/i2016-16315-7}}.

\bibitem{Caswell:1985ui}
W.~E. Caswell, G.~P. Lepage, {Effective Lagrangians for Bound State Problems in QED, QCD, and Other Field Theories}, Phys. Lett. B 167 (1986) 437--442.
\newblock \href {http://dx.doi.org/10.1016/0370-2693(86)91297-9} {\path{doi:10.1016/0370-2693(86)91297-9}}.

\bibitem{Bodwin:1994jh}
G.~T. Bodwin, E.~Braaten, G.~P. Lepage, {Rigorous QCD analysis of inclusive annihilation and production of heavy quarkonium}, Phys. Rev. D 51 (1995) 1125--1171, [Erratum: Phys.Rev.D 55, 5853 (1997)].
\newblock \href {http://arxiv.org/abs/hep-ph/9407339} {\path{arXiv:hep-ph/9407339}}, \href {http://dx.doi.org/10.1103/PhysRevD.55.5853} {\path{doi:10.1103/PhysRevD.55.5853}}.

\bibitem{Cho:1995vh}
P.~L. Cho, A.~K. Leibovich, {Color octet quarkonia production}, Phys. Rev. D 53 (1996) 150--162.
\newblock \href {http://arxiv.org/abs/hep-ph/9505329} {\path{arXiv:hep-ph/9505329}}, \href {http://dx.doi.org/10.1103/PhysRevD.53.150} {\path{doi:10.1103/PhysRevD.53.150}}.

\bibitem{Cho:1995ce}
P.~L. Cho, A.~K. Leibovich, {Color octet quarkonia production. 2.}, Phys. Rev. D 53 (1996) 6203--6217.
\newblock \href {http://arxiv.org/abs/hep-ph/9511315} {\path{arXiv:hep-ph/9511315}}, \href {http://dx.doi.org/10.1103/PhysRevD.53.6203} {\path{doi:10.1103/PhysRevD.53.6203}}.

\bibitem{Bodwin:2005hm}
G.~T. Bodwin, E.~Braaten, J.~Lee, {Comparison of the color-evaporation model and the NRQCD factorization approach in charmonium production}, Phys. Rev. D 72 (2005) 014004.
\newblock \href {http://arxiv.org/abs/hep-ph/0504014} {\path{arXiv:hep-ph/0504014}}, \href {http://dx.doi.org/10.1103/PhysRevD.72.014004} {\path{doi:10.1103/PhysRevD.72.014004}}.

\bibitem{Adamov:1997yk}
A.~D. Adamov, G.~R. Goldstein, {Fragmentation functions for baryons in a quark - diquark model}, Phys. Rev. D 56 (1997) 7381--7391.
\newblock \href {http://arxiv.org/abs/hep-ph/9706491} {\path{arXiv:hep-ph/9706491}}, \href {http://dx.doi.org/10.1103/PhysRevD.56.7381} {\path{doi:10.1103/PhysRevD.56.7381}}.

\bibitem{Yang:2002gh}
J.-J. Yang, {Flavor and spin structure of quark fragmentation functions in a diquark model for octet baryons}, Phys. Rev. D 65 (2002) 094035.
\newblock \href {http://dx.doi.org/10.1103/PhysRevD.65.094035} {\path{doi:10.1103/PhysRevD.65.094035}}.

\bibitem{GomshiNobary:2004mq}
M.~A. Gomshi~Nobary, R.~Sepahvand, {Fragmentation of triply heavy baryons}, Phys. Rev. D 71 (2005) 034024.
\newblock \href {http://arxiv.org/abs/hep-ph/0406148} {\path{arXiv:hep-ph/0406148}}, \href {http://dx.doi.org/10.1103/PhysRevD.71.034024} {\path{doi:10.1103/PhysRevD.71.034024}}.

\bibitem{MoosaviNejad:2017bda}
S.~M. Moosavi~Nejad, M.~Delpasand, {Polarized heavy baryon production in quark-diquark model considering two different scenarios}, Eur. Phys. J. A 53~(9) (2017) 174.
\newblock \href {http://dx.doi.org/10.1140/epja/i2017-12364-8} {\path{doi:10.1140/epja/i2017-12364-8}}.

\bibitem{MoosaviNejad:2017rvi}
S.~M. Moosavi~Nejad, {NLO QCD corrections to triply heavy baryon fragmentation function considering the effect of nonperturbative dynamics of baryon bound states}, Phys. Rev. D 96~(11) (2017) 114021.
\newblock \href {http://dx.doi.org/10.1103/PhysRevD.96.114021} {\path{doi:10.1103/PhysRevD.96.114021}}.

\bibitem{Delpasand:2019xpk}
M.~Delpasand, S.~M. Moosavi~Nejad, {Gluon fragmentation into triply heavy baryons considering two various scenarios}, Phys. Rev. D 99~(11) (2019) 114028.
\newblock \href {http://dx.doi.org/10.1103/PhysRevD.99.114028} {\path{doi:10.1103/PhysRevD.99.114028}}.

\bibitem{Cacciari:1993mq}
M.~Cacciari, M.~Greco, {Large $p_{T}$ hadroproduction of heavy quarks}, Nucl. Phys. B 421 (1994) 530--544.
\newblock \href {http://arxiv.org/abs/hep-ph/9311260} {\path{arXiv:hep-ph/9311260}}, \href {http://dx.doi.org/10.1016/0550-3213(94)90515-0} {\path{doi:10.1016/0550-3213(94)90515-0}}.

\bibitem{Buza:1996wv}
M.~Buza, Y.~Matiounine, J.~Smith, W.~L. van Neerven, {Charm electroproduction viewed in the variable flavor number scheme versus fixed order perturbation theory}, Eur. Phys. J. C 1 (1998) 301--320.
\newblock \href {http://arxiv.org/abs/hep-ph/9612398} {\path{arXiv:hep-ph/9612398}}, \href {http://dx.doi.org/10.1007/BF01245820} {\path{doi:10.1007/BF01245820}}.

\bibitem{Cacciari:2001cw}
M.~Cacciari, S.~Catani, {Soft gluon resummation for the fragmentation of light and heavy quarks at large x}, Nucl. Phys. B 617 (2001) 253--290.
\newblock \href {http://arxiv.org/abs/hep-ph/0107138} {\path{arXiv:hep-ph/0107138}}, \href {http://dx.doi.org/10.1016/S0550-3213(01)00469-2} {\path{doi:10.1016/S0550-3213(01)00469-2}}.

\bibitem{Mitov:2006wy}
A.~Mitov, S.-O. Moch, {QCD Corrections to Semi-Inclusive Hadron Production in Electron-Positron Annihilation at Two Loops}, Nucl. Phys. B 751 (2006) 18--52.
\newblock \href {http://arxiv.org/abs/hep-ph/0604160} {\path{arXiv:hep-ph/0604160}}, \href {http://dx.doi.org/10.1016/j.nuclphysb.2006.05.018} {\path{doi:10.1016/j.nuclphysb.2006.05.018}}.

\bibitem{An:2019idk}
H.-T. An, Q.-S. Zhou, Z.-W. Liu, Y.-R. Liu, X.~Liu, {Exotic pentaquark states with the $qqQQ\bar{Q}$ configuration}, Phys. Rev. D 100~(5) (2019) 056004.
\newblock \href {http://arxiv.org/abs/1905.07858} {\path{arXiv:1905.07858}}, \href {http://dx.doi.org/10.1103/PhysRevD.100.056004} {\path{doi:10.1103/PhysRevD.100.056004}}.

\bibitem{Ortiz-Pacheco:2023kjn}
E.~Ortiz-Pacheco, R.~Bijker, {Masses and radiative decay widths of S- and P-wave singly, doubly, and triply heavy charm and bottom baryons}, Phys. Rev. D 108~(5) (2023) 054014.
\newblock \href {http://arxiv.org/abs/2307.04939} {\path{arXiv:2307.04939}}, \href {http://dx.doi.org/10.1103/PhysRevD.108.054014} {\path{doi:10.1103/PhysRevD.108.054014}}.

\bibitem{Liu:2024mwn}
C.-L. Liu, W.-X. Zhang, D.~Jia, {Masses and decays of triply-heavy pentaquarks*}, Chin. Phys. C 48~(10) (2024) 103110.
\newblock \href {http://arxiv.org/abs/2403.13456} {\path{arXiv:2403.13456}}, \href {http://dx.doi.org/10.1088/1674-1137/ad595a} {\path{doi:10.1088/1674-1137/ad595a}}.

\bibitem{Chen:2016qju}
H.-X. Chen, W.~Chen, X.~Liu, S.-L. Zhu, {The hidden-charm pentaquark and tetraquark states}, Phys. Rept. 639 (2016) 1--121.
\newblock \href {http://arxiv.org/abs/1601.02092} {\path{arXiv:1601.02092}}, \href {http://dx.doi.org/10.1016/j.physrep.2016.05.004} {\path{doi:10.1016/j.physrep.2016.05.004}}.

\bibitem{Wang:2018utj}
W.~Wang, J.~Xu, {Weak Decays of Triply Heavy Baryons}, Phys. Rev. D 97~(9) (2018) 093007.
\newblock \href {http://arxiv.org/abs/1803.01476} {\path{arXiv:1803.01476}}, \href {http://dx.doi.org/10.1103/PhysRevD.97.093007} {\path{doi:10.1103/PhysRevD.97.093007}}.

\bibitem{Celiberto:2025ogy}
F.~G. Celiberto, {Unwinding the rare $\Omega$ sector: Fragmentation of fully charmed baryons from HL-LHC to FCC}, Phys. Rev. D 112~(7) (2025) 074023.
\newblock \href {http://arxiv.org/abs/2506.00776} {\path{arXiv:2506.00776}}, \href {http://dx.doi.org/10.1103/sb8n-9nt6} {\path{doi:10.1103/sb8n-9nt6}}.

\bibitem{Celiberto:2025_OMG3Q10}
F.~G. Celiberto, \href{{https://github.com/FGCeliberto/Collinear_FFs/}}{{OMG3Q1.0: OMeGa baryons with 3 heavy Quarks VFNS FFs}}, 2025.
\newline\urlprefix\url{{https://github.com/FGCeliberto/Collinear_FFs/}}

\bibitem{Celiberto:2025ipt}
F.~G. Celiberto, {Heavy-flavor multimodal fragmentation to S-wave pentacharms at next-generation hadron colliders}, Eur. Phys. J. C 85~(12) (2025) 1395.
\newblock \href {http://arxiv.org/abs/2502.11136} {\path{arXiv:2502.11136}}, \href {http://dx.doi.org/10.1140/epjc/s10052-025-15079-9} {\path{doi:10.1140/epjc/s10052-025-15079-9}}.

\bibitem{Mele:1990cw}
B.~Mele, P.~Nason, {The Fragmentation function for heavy quarks in QCD}, Nucl. Phys. B 361 (1991) 626--644, [Erratum: Nucl.Phys.B 921, 841--842 (2017)].
\newblock \href {http://dx.doi.org/10.1016/0550-3213(91)90597-Q} {\path{doi:10.1016/0550-3213(91)90597-Q}}.

\bibitem{Celiberto:2025euy}
F.~G. Celiberto, F.~Lonigro, {Pseudoscalar heavy-quarkonium hadroproduction from nonrelativistic fragmentation at NLL/NLO+}, Phys. Rev. D 112~(11) (2025) 114040.
\newblock \href {http://arxiv.org/abs/2510.10593} {\path{arXiv:2510.10593}}, \href {http://dx.doi.org/10.1103/rmsq-bq3b} {\path{doi:10.1103/rmsq-bq3b}}.

\bibitem{Celiberto:2024mex}
F.~G. Celiberto, {Towards Quarkonium Fragmentation from NRQCD in a Variable-Flavor Number Scheme}, in: {58th Rencontres de Moriond on QCD and High Energy Interactions}, 2024.
\newblock \href {http://arxiv.org/abs/2405.08221} {\path{arXiv:2405.08221}}.

\bibitem{Celiberto:2024bxu}
F.~G. Celiberto, {Quarkonium fragmentation in a variable-flavor number scheme: Towards NRFF1.0}, PoS DIS2024 (2025) 168.
\newblock \href {http://arxiv.org/abs/2406.10779} {\path{arXiv:2406.10779}}, \href {http://dx.doi.org/10.22323/1.469.0168} {\path{doi:10.22323/1.469.0168}}.

\bibitem{Celiberto:2024rxa}
F.~G. Celiberto, {On the Quarkonium-in-jet Collinear Fragmentation at Moderate-to-large Transverse Momentum}, Acta Phys. Polon. Supp. 18~(1) (2025) 1--A22.
\newblock \href {http://arxiv.org/abs/2412.05661} {\path{arXiv:2412.05661}}, \href {http://dx.doi.org/10.5506/APhysPolBSupp.18.1-A22} {\path{doi:10.5506/APhysPolBSupp.18.1-A22}}.

\bibitem{Celiberto:2025xvy}
F.~G. Celiberto, F.~Lonigro, {Heavy-Flavor Fragmentation and Jet Structure from HF-NRevo: Bridging to Heavy-Ion Collisions}, PoS EPS-HEP2025 (2026) 191.
\newblock \href {http://arxiv.org/abs/2510.22449} {\path{arXiv:2510.22449}}, \href {http://dx.doi.org/10.22323/1.485.0191} {\path{doi:10.22323/1.485.0191}}.

\bibitem{Buckley:2014ana}
A.~Buckley, J.~Ferrando, S.~Lloyd, K.~Nordstr\"om, B.~Page, M.~R\"ufenacht, M.~Sch\"onherr, G.~Watt, {LHAPDF6: parton density access in the LHC precision era}, Eur. Phys. J. C 75 (2015) 132.
\newblock \href {http://arxiv.org/abs/1412.7420} {\path{arXiv:1412.7420}}, \href {http://dx.doi.org/10.1140/epjc/s10052-015-3318-8} {\path{doi:10.1140/epjc/s10052-015-3318-8}}.

\bibitem{Celiberto:2022gji}
F.~G. Celiberto, A.~Papa, {Mueller-Navelet jets at the LHC: Hunting data with azimuthal distributions}, Phys. Rev. D 106~(11) (2022) 114004.
\newblock \href {http://arxiv.org/abs/2207.05015} {\path{arXiv:2207.05015}}, \href {http://dx.doi.org/10.1103/PhysRevD.106.114004} {\path{doi:10.1103/PhysRevD.106.114004}}.

\bibitem{Bolognino:2021mrc}
A.~D. Bolognino, F.~G. Celiberto, M.~Fucilla, D.~{\relax Yu}. Ivanov, A.~Papa, {Inclusive production of a heavy-light dijet system in hybrid high-energy and collinear factorization}, Phys. Rev. D 103~(9) (2021) 094004.
\newblock \href {http://arxiv.org/abs/2103.07396} {\path{arXiv:2103.07396}}, \href {http://dx.doi.org/10.1103/PhysRevD.103.094004} {\path{doi:10.1103/PhysRevD.103.094004}}.

\bibitem{Celiberto:2022dyf}
F.~G. Celiberto, M.~Fucilla, {Diffractive semi-hard production of a $J/\psi $ or a $\Upsilon $ from single-parton fragmentation plus a jet in hybrid factorization}, Eur. Phys. J. C 82~(10) (2022) 929.
\newblock \href {http://arxiv.org/abs/2202.12227} {\path{arXiv:2202.12227}}, \href {http://dx.doi.org/10.1140/epjc/s10052-022-10818-8} {\path{doi:10.1140/epjc/s10052-022-10818-8}}.

\bibitem{Mathematica_V14-2}
W.~R. Inc., \href{https://www.wolfram.com/mathematica}{Mathematica, {V}ersion 14.2}, champaign, IL, 2024.
\newline\urlprefix\url{https://www.wolfram.com/mathematica}

\bibitem{Celiberto:2020wpk}
F.~G. Celiberto, {Hunting BFKL in semi-hard reactions at the LHC}, Eur. Phys. J. C 81~(8) (2021) 691.
\newblock \href {http://arxiv.org/abs/2008.07378} {\path{arXiv:2008.07378}}, \href {http://dx.doi.org/10.1140/epjc/s10052-021-09384-2} {\path{doi:10.1140/epjc/s10052-021-09384-2}}.

\bibitem{Celiberto:2022rfj}
F.~G. Celiberto, {High-energy emissions of light mesons plus heavy flavor at the LHC and the Forward Physics Facility}, Phys. Rev. D 105~(11) (2022) 114008.
\newblock \href {http://arxiv.org/abs/2204.06497} {\path{arXiv:2204.06497}}, \href {http://dx.doi.org/10.1103/PhysRevD.105.114008} {\path{doi:10.1103/PhysRevD.105.114008}}.

\bibitem{Celiberto:2023fzz}
F.~G. Celiberto, {Vector Quarkonia at the LHC with JETHAD: A High-Energy Viewpoint}, Universe 9~(7) (2023) 324.
\newblock \href {http://arxiv.org/abs/2305.14295} {\path{arXiv:2305.14295}}, \href {http://dx.doi.org/10.3390/universe9070324} {\path{doi:10.3390/universe9070324}}.

\bibitem{Celiberto:2024mrq}
F.~G. Celiberto, {Exotic Tetraquarks at the HL-LHC with JETHAD: A High-Energy Viewpoint}, Symmetry 16~(5) (2024) 550.
\newblock \href {http://arxiv.org/abs/2403.15639} {\path{arXiv:2403.15639}}, \href {http://dx.doi.org/10.3390/sym16050550} {\path{doi:10.3390/sym16050550}}.

\bibitem{Celiberto:2024swu}
F.~G. Celiberto, {Forward \& Far-Forward Heavy Hadrons with JETHAD: A High-Energy Viewpoint}, Particles 7~(3) (2024) 502--542.
\newblock \href {http://arxiv.org/abs/2405.09526} {\path{arXiv:2405.09526}}, \href {http://dx.doi.org/10.3390/particles7030029} {\path{doi:10.3390/particles7030029}}.

\bibitem{Celiberto:2025_P5Q_review}
F.~G. Celiberto, {Fully Heavy Pentaquarks with JETHAD: A High-Energy Viewpoint}, Particles 9~(1) (2026) 23.
\newblock \href {http://dx.doi.org/10.3390/sym18010029} {\path{doi:10.3390/sym18010029}}.

\bibitem{Cacciari:1996wr}
M.~Cacciari, M.~Greco, S.~Rolli, A.~Tanzini, {Charmed mesons fragmentation functions}, Phys. Rev. D 55 (1997) 2736--2740.
\newblock \href {http://arxiv.org/abs/hep-ph/9608213} {\path{arXiv:hep-ph/9608213}}, \href {http://dx.doi.org/10.1103/PhysRevD.55.2736} {\path{doi:10.1103/PhysRevD.55.2736}}.

\bibitem{Jaffe:1993ie}
R.~L. Jaffe, L.~Randall, {Heavy quark fragmentation into heavy mesons}, Nucl. Phys. B 412 (1994) 79--105.
\newblock \href {http://arxiv.org/abs/hep-ph/9306201} {\path{arXiv:hep-ph/9306201}}, \href {http://dx.doi.org/10.1016/0550-3213(94)90495-2} {\path{doi:10.1016/0550-3213(94)90495-2}}.

\bibitem{Kniehl:2005mk}
B.~A. Kniehl, G.~Kramer, I.~Schienbein, H.~Spiesberger, {Collinear subtractions in hadroproduction of heavy quarks}, Eur. Phys. J. C 41 (2005) 199--212.
\newblock \href {http://arxiv.org/abs/hep-ph/0502194} {\path{arXiv:hep-ph/0502194}}, \href {http://dx.doi.org/10.1140/epjc/s2005-02200-7} {\path{doi:10.1140/epjc/s2005-02200-7}}.

\bibitem{Helenius:2018uul}
I.~Helenius, H.~Paukkunen, {Revisiting the D-meson hadroproduction in general-mass variable flavour number scheme}, JHEP 05 (2018) 196.
\newblock \href {http://arxiv.org/abs/1804.03557} {\path{arXiv:1804.03557}}, \href {http://dx.doi.org/10.1007/JHEP05(2018)196} {\path{doi:10.1007/JHEP05(2018)196}}.

\bibitem{Helenius:2023wkn}
I.~Helenius, H.~Paukkunen, {B-meson hadroproduction in the SACOT-m$_{T}$ scheme}, JHEP 07 (2023) 054.
\newblock \href {http://arxiv.org/abs/2303.17864} {\path{arXiv:2303.17864}}, \href {http://dx.doi.org/10.1007/JHEP07(2023)054} {\path{doi:10.1007/JHEP07(2023)054}}.

\bibitem{Generet:2023vte}
T.~Generet, {$B$-hadron production at higher orders in QCD}, Ph.D. thesis, RWTH Aachen University, RWTH Aachen U. (2023).

\bibitem{Mele:1990yq}
B.~Mele, P.~Nason, {Next-to-leading QCD calculation of the heavy quark fragmentation function}, Phys. Lett. B 245 (1990) 635--639.
\newblock \href {http://dx.doi.org/10.1016/0370-2693(90)90704-A} {\path{doi:10.1016/0370-2693(90)90704-A}}.

\bibitem{Rijken:1996vr}
P.~J. Rijken, W.~L. van Neerven, {O (alpha-s**2) contributions to the longitudinal fragmentation function in e+ e- annihilation}, Phys. Lett. B 386 (1996) 422--428.
\newblock \href {http://arxiv.org/abs/hep-ph/9604436} {\path{arXiv:hep-ph/9604436}}, \href {http://dx.doi.org/10.1016/0370-2693(96)00898-2} {\path{doi:10.1016/0370-2693(96)00898-2}}.

\bibitem{Blumlein:2006rr}
J.~Blumlein, V.~Ravindran, {O (alpha**2(s)) Timelike Wilson Coefficients for Parton-Fragmentation Functions in Mellin Space}, Nucl. Phys. B 749 (2006) 1--24.
\newblock \href {http://arxiv.org/abs/hep-ph/0604019} {\path{arXiv:hep-ph/0604019}}, \href {http://dx.doi.org/10.1016/j.nuclphysb.2006.04.032} {\path{doi:10.1016/j.nuclphysb.2006.04.032}}.

\bibitem{Melnikov:2004bm}
K.~Melnikov, A.~Mitov, {Perturbative heavy quark fragmentation function through $\mathcal{O}(\alpha^2_s)$}, Phys. Rev. D 70 (2004) 034027.
\newblock \href {http://arxiv.org/abs/hep-ph/0404143} {\path{arXiv:hep-ph/0404143}}, \href {http://dx.doi.org/10.1103/PhysRevD.70.034027} {\path{doi:10.1103/PhysRevD.70.034027}}.

\bibitem{Mitov:2004du}
A.~Mitov, {Perturbative heavy quark fragmentation function through $\mathcal{O}(\alpha^2_s)$: Gluon initiated contribution}, Phys. Rev. D 71 (2005) 054021.
\newblock \href {http://arxiv.org/abs/hep-ph/0410205} {\path{arXiv:hep-ph/0410205}}, \href {http://dx.doi.org/10.1103/PhysRevD.71.054021} {\path{doi:10.1103/PhysRevD.71.054021}}.

\bibitem{Biello:2024zti}
C.~Biello, L.~Bonino, {Time-Like heavy-flavour thresholds for fragmentation functions: the light-quark matching condition at NNLO}, Eur. Phys. J. C 84~(11) (2024) 1192.
\newblock \href {http://arxiv.org/abs/2407.07623} {\path{arXiv:2407.07623}}, \href {http://dx.doi.org/10.1140/epjc/s10052-024-13532-9} {\path{doi:10.1140/epjc/s10052-024-13532-9}}.

\bibitem{Kartvelishvili:1977pi}
V.~Kartvelishvili, A.~Likhoded, V.~Petrov, {On the Fragmentation Functions of Heavy Quarks Into Hadrons}, Phys. Lett. B 78 (1978) 615--617.
\newblock \href {http://dx.doi.org/10.1016/0370-2693(78)90653-6} {\path{doi:10.1016/0370-2693(78)90653-6}}.

\bibitem{Bowler:1981sb}
M.~G. Bowler, {e+ e- Production of Heavy Quarks in the String Model}, Z. Phys. C 11 (1981) 169.
\newblock \href {http://dx.doi.org/10.1007/BF01574001} {\path{doi:10.1007/BF01574001}}.

\bibitem{Peterson:1982ak}
C.~Peterson, D.~Schlatter, I.~Schmitt, P.~M. Zerwas, {Scaling Violations in Inclusive e+ e- Annihilation Spectra}, Phys. Rev. D 27 (1983) 105.
\newblock \href {http://dx.doi.org/10.1103/PhysRevD.27.105} {\path{doi:10.1103/PhysRevD.27.105}}.

\bibitem{Andersson:1983jt}
B.~Andersson, G.~Gustafson, B.~Soderberg, {A General Model for Jet Fragmentation}, Z. Phys. C 20 (1983) 317.
\newblock \href {http://dx.doi.org/10.1007/BF01407824} {\path{doi:10.1007/BF01407824}}.

\bibitem{Collins:1984ms}
P.~D.~B. Collins, T.~P. Spiller, {The Fragmentation of Heavy Quarks}, J. Phys. G 11 (1985) 1289.
\newblock \href {http://dx.doi.org/10.1088/0305-4616/11/12/006} {\path{doi:10.1088/0305-4616/11/12/006}}.

\bibitem{Colangelo:1992kh}
G.~Colangelo, P.~Nason, {A Theoretical study of the c and b fragmentation function from e+ e- annihilation}, Phys. Lett. B 285 (1992) 167--171.
\newblock \href {http://dx.doi.org/10.1016/0370-2693(92)91317-3} {\path{doi:10.1016/0370-2693(92)91317-3}}.

\bibitem{Georgi:1990um}
H.~Georgi, {An Effective Field Theory for Heavy Quarks at Low-energies}, Phys. Lett. B 240 (1990) 447--450.
\newblock \href {http://dx.doi.org/10.1016/0370-2693(90)91128-X} {\path{doi:10.1016/0370-2693(90)91128-X}}.

\bibitem{Eichten:1989zv}
E.~Eichten, B.~R. Hill, {An Effective Field Theory for the Calculation of Matrix Elements Involving Heavy Quarks}, Phys. Lett. B 234 (1990) 511--516.
\newblock \href {http://dx.doi.org/10.1016/0370-2693(90)92049-O} {\path{doi:10.1016/0370-2693(90)92049-O}}.

\bibitem{Grinstein:1992ss}
B.~Grinstein, {Light quark, heavy quark systems}, Ann. Rev. Nucl. Part. Sci. 42 (1992) 101--145.
\newblock \href {http://dx.doi.org/10.1146/annurev.ns.42.120192.000533} {\path{doi:10.1146/annurev.ns.42.120192.000533}}.

\bibitem{Neubert:1993mb}
M.~Neubert, {Heavy quark symmetry}, Phys. Rept. 245 (1994) 259--396.
\newblock \href {http://arxiv.org/abs/hep-ph/9306320} {\path{arXiv:hep-ph/9306320}}, \href {http://dx.doi.org/10.1016/0370-1573(94)90091-4} {\path{doi:10.1016/0370-1573(94)90091-4}}.

\bibitem{Cacciari:1997du}
M.~Cacciari, M.~Greco, {D* production from e+ e- to e p collisions in NLO QCD}, Phys. Rev. D 55 (1997) 7134--7143.
\newblock \href {http://arxiv.org/abs/hep-ph/9702389} {\path{arXiv:hep-ph/9702389}}, \href {http://dx.doi.org/10.1103/PhysRevD.55.7134} {\path{doi:10.1103/PhysRevD.55.7134}}.

\bibitem{Thacker:1990bm}
B.~A. Thacker, G.~P. Lepage, {Heavy quark bound states in lattice QCD}, Phys. Rev. D 43 (1991) 196--208.
\newblock \href {http://dx.doi.org/10.1103/PhysRevD.43.196} {\path{doi:10.1103/PhysRevD.43.196}}.

\bibitem{Leibovich:1996pa}
A.~K. Leibovich, {Psi-prime polarization due to color octet quarkonia production}, Phys. Rev. D 56 (1997) 4412--4415.
\newblock \href {http://arxiv.org/abs/hep-ph/9610381} {\path{arXiv:hep-ph/9610381}}, \href {http://dx.doi.org/10.1103/PhysRevD.56.4412} {\path{doi:10.1103/PhysRevD.56.4412}}.

\bibitem{Grinstein:1998xb}
B.~Grinstein, {A Modern introduction to quarkonium theory}, Int. J. Mod. Phys. A 15 (2000) 461--496.
\newblock \href {http://arxiv.org/abs/hep-ph/9811264} {\path{arXiv:hep-ph/9811264}}, \href {http://dx.doi.org/10.1142/S0217751X00000227} {\path{doi:10.1142/S0217751X00000227}}.

\bibitem{Kramer:2001hh}
M.~Kr\"amer, {Quarkonium production at high-energy colliders}, Prog. Part. Nucl. Phys. 47 (2001) 141--201.
\newblock \href {http://arxiv.org/abs/hep-ph/0106120} {\path{arXiv:hep-ph/0106120}}, \href {http://dx.doi.org/10.1016/S0146-6410(01)00154-5} {\path{doi:10.1016/S0146-6410(01)00154-5}}.

\bibitem{QuarkoniumWorkingGroup:2004kpm}
N.~Brambilla, et~al., {Heavy quarkonium physics}\href {http://arxiv.org/abs/hep-ph/0412158} {\path{arXiv:hep-ph/0412158}}, \href {http://dx.doi.org/10.5170/CERN-2005-005} {\path{doi:10.5170/CERN-2005-005}}.

\bibitem{Lansberg:2005aw}
J.-P. Lansberg, {Quarkonium production at high-energy hadron colliders: A Systematic gauge-invariant approach to relativistic effects of $J/\psi$, $\psi^\prime$ and $\upsilon$ production}, Phd thesis, Liege University (4 2005).
\newblock \href {http://arxiv.org/abs/hep-ph/0507175} {\path{arXiv:hep-ph/0507175}}.

\bibitem{Lansberg:2019adr}
J.-P. Lansberg, {New Observables in Inclusive Production of Quarkonia}, Phys. Rept. 889 (2020) 1--106.
\newblock \href {http://arxiv.org/abs/1903.09185} {\path{arXiv:1903.09185}}, \href {http://dx.doi.org/10.1016/j.physrep.2020.08.007} {\path{doi:10.1016/j.physrep.2020.08.007}}.

\bibitem{Alekhin:2009ni}
S.~Alekhin, J.~Blumlein, S.~Klein, S.~Moch, {The 3, 4, and 5-flavor NNLO Parton from Deep-Inelastic-Scattering Data and at Hadron Colliders}, Phys. Rev. D 81 (2010) 014032.
\newblock \href {http://arxiv.org/abs/0908.2766} {\path{arXiv:0908.2766}}, \href {http://dx.doi.org/10.1103/PhysRevD.81.014032} {\path{doi:10.1103/PhysRevD.81.014032}}.

\bibitem{Fleming:2012wy}
S.~Fleming, A.~K. Leibovich, T.~Mehen, I.~Z. Rothstein, {The Systematics of Quarkonium Production at the LHC and Double Parton Fragmentation}, Phys. Rev. D 86 (2012) 094012.
\newblock \href {http://arxiv.org/abs/1207.2578} {\path{arXiv:1207.2578}}, \href {http://dx.doi.org/10.1103/PhysRevD.86.094012} {\path{doi:10.1103/PhysRevD.86.094012}}.

\bibitem{Kang:2014tta}
Z.-B. Kang, Y.-Q. Ma, J.-W. Qiu, G.~Sterman, {Heavy Quarkonium Production at Collider Energies: Factorization and Evolution}, Phys. Rev. D 90~(3) (2014) 034006.
\newblock \href {http://arxiv.org/abs/1401.0923} {\path{arXiv:1401.0923}}, \href {http://dx.doi.org/10.1103/PhysRevD.90.034006} {\path{doi:10.1103/PhysRevD.90.034006}}.

\bibitem{Echevarria:2019ynx}
M.~G. Echevarria, {Proper TMD factorization for quarkonia production: $pp\to\eta_{c,b}$ as a study case}, JHEP 10 (2019) 144.
\newblock \href {http://arxiv.org/abs/1907.06494} {\path{arXiv:1907.06494}}, \href {http://dx.doi.org/10.1007/JHEP10(2019)144} {\path{doi:10.1007/JHEP10(2019)144}}.

\bibitem{Boer:2023zit}
D.~Boer, J.~Bor, L.~Maxia, C.~Pisano, F.~Yuan, {Transverse momentum dependent shape function for J/\ensuremath{\psi} production in SIDIS}, JHEP 08 (2023) 105.
\newblock \href {http://arxiv.org/abs/2304.09473} {\path{arXiv:2304.09473}}, \href {http://dx.doi.org/10.1007/JHEP08(2023)105} {\path{doi:10.1007/JHEP08(2023)105}}.

\bibitem{Zheng:2019gnb}
X.-C. Zheng, C.-H. Chang, T.-F. Feng, X.-G. Wu, {QCD NLO fragmentation functions for c or b\textasciimacron{} quark to Bc or Bc* meson and their application}, Phys. Rev. D 100~(3) (2019) 034004.
\newblock \href {http://arxiv.org/abs/1901.03477} {\path{arXiv:1901.03477}}, \href {http://dx.doi.org/10.1103/PhysRevD.100.034004} {\path{doi:10.1103/PhysRevD.100.034004}}.

\bibitem{Zheng:2021sdo}
X.-C. Zheng, C.-H. Chang, X.-G. Wu, {Fragmentation functions for gluon into $B_c$ or $ {B}_c^{\ast } $ meson}, JHEP 05 (2022) 036.
\newblock \href {http://arxiv.org/abs/2112.10520} {\path{arXiv:2112.10520}}, \href {http://dx.doi.org/10.1007/JHEP05(2022)036} {\path{doi:10.1007/JHEP05(2022)036}}.

\bibitem{Celiberto:2022keu}
F.~G. Celiberto, {The high-energy spectrum of QCD from inclusive emissions of charmed B-mesons}, Phys. Lett. B 835 (2022) 137554.
\newblock \href {http://arxiv.org/abs/2206.09413} {\path{arXiv:2206.09413}}, \href {http://dx.doi.org/10.1016/j.physletb.2022.137554} {\path{doi:10.1016/j.physletb.2022.137554}}.

\bibitem{Celiberto:2024omj}
F.~G. Celiberto, {High-energy QCD dynamics from bottom flavor fragmentation at the Hi-Lumi LHC}, Eur. Phys. J. C 84~(4) (2024) 384.
\newblock \href {http://arxiv.org/abs/2401.01410} {\path{arXiv:2401.01410}}, \href {http://dx.doi.org/10.1140/epjc/s10052-024-12704-x} {\path{doi:10.1140/epjc/s10052-024-12704-x}}.

\bibitem{LHCb:2014iah}
R.~Aaij, et~al., {Precision measurement of $CP$ violation in $B_s^0 \to J/\psi K^+K^-$ decays}, Phys. Rev. Lett. 114~(4) (2015) 041801.
\newblock \href {http://arxiv.org/abs/1411.3104} {\path{arXiv:1411.3104}}, \href {http://dx.doi.org/10.1103/PhysRevLett.114.041801} {\path{doi:10.1103/PhysRevLett.114.041801}}.

\bibitem{LHCb:2016qpe}
R.~Aaij, et~al., {Measurement of the $b$-quark production cross-section in 7 and 13 TeV $pp$ collisions}, Phys. Rev. Lett. 118~(5) (2017) 052002, [Erratum: Phys.Rev.Lett. 119, 169901 (2017)].
\newblock \href {http://arxiv.org/abs/1612.05140} {\path{arXiv:1612.05140}}, \href {http://dx.doi.org/10.1103/PhysRevLett.118.052002} {\path{doi:10.1103/PhysRevLett.118.052002}}.

\bibitem{ATLAS:2023bft}
G.~Aad, et~al., {Observation of an Excess of Dicharmonium Events in the Four-Muon Final State with the ATLAS Detector}, Phys. Rev. Lett. 131~(15) (2023) 151902.
\newblock \href {http://arxiv.org/abs/2304.08962} {\path{arXiv:2304.08962}}, \href {http://dx.doi.org/10.1103/PhysRevLett.131.151902} {\path{doi:10.1103/PhysRevLett.131.151902}}.

\bibitem{CMS:2023owd}
A.~Hayrapetyan, et~al., {New Structures in the J/\ensuremath{\psi}J/\ensuremath{\psi} Mass Spectrum in Proton-Proton Collisions at s=13\,\,TeV}, Phys. Rev. Lett. 132~(11) (2024) 111901.
\newblock \href {http://arxiv.org/abs/2306.07164} {\path{arXiv:2306.07164}}, \href {http://dx.doi.org/10.1103/PhysRevLett.132.111901} {\path{doi:10.1103/PhysRevLett.132.111901}}.

\bibitem{Zhang:2020hoh}
H.-F. Zhang, Y.-Q. Ma, W.-L. Sang, {Perturbative QCD evidence for spin-2 particles in the di-$J/\psi$ resonances}, Sci. Bull. 70 (2025) 1915--1917.
\newblock \href {http://arxiv.org/abs/2009.08376} {\path{arXiv:2009.08376}}, \href {http://dx.doi.org/10.1016/j.scib.2025.04.035} {\path{doi:10.1016/j.scib.2025.04.035}}.

\bibitem{Zhu:2020xni}
R.~Zhu, {Fully-heavy tetraquark spectra and production at hadron colliders}, Nucl. Phys. B 966 (2021) 115393.
\newblock \href {http://arxiv.org/abs/2010.09082} {\path{arXiv:2010.09082}}, \href {http://dx.doi.org/10.1016/j.nuclphysb.2021.115393} {\path{doi:10.1016/j.nuclphysb.2021.115393}}.

\bibitem{Feng:2020riv}
F.~Feng, Y.~Huang, Y.~Jia, W.-L. Sang, X.~Xiong, J.-Y. Zhang, {Fragmentation production of fully-charmed tetraquarks at the LHC}, Phys. Rev. D 106~(11) (2022) 114029.
\newblock \href {http://arxiv.org/abs/2009.08450} {\path{arXiv:2009.08450}}, \href {http://dx.doi.org/10.1103/PhysRevD.106.114029} {\path{doi:10.1103/PhysRevD.106.114029}}.

\bibitem{Suzuki:1977km}
M.~Suzuki, {Fragmentation of Hadrons from Heavy Quark Partons}, Phys. Lett. B 71 (1977) 139--141.
\newblock \href {http://dx.doi.org/10.1016/0370-2693(77)90761-4} {\path{doi:10.1016/0370-2693(77)90761-4}}.

\bibitem{Suzuki:1985up}
M.~Suzuki, {Spin Property of Heavy Hadron in Heavy Quark Fragmentation: A Simple Model}, Phys. Rev. D 33 (1986) 676.
\newblock \href {http://dx.doi.org/10.1103/PhysRevD.33.676} {\path{doi:10.1103/PhysRevD.33.676}}.

\bibitem{Amiri:1986zv}
F.~Amiri, C.-R. Ji, {Perturbative Quantum Chromodynamic Prediction for the Heavy Quark Fragmentation Function}, Phys. Lett. B 195 (1987) 593--598.
\newblock \href {http://dx.doi.org/10.1016/0370-2693(87)91579-6} {\path{doi:10.1016/0370-2693(87)91579-6}}.

\bibitem{Nejad:2021mmp}
S.~M. {\relax Moosavi}~Nejad, N.~Amiri, {Ground state heavy tetraquark production in heavy quark fragmentation}, Phys. Rev. D 105~(3) (2022) 034001.
\newblock \href {http://arxiv.org/abs/2110.15251} {\path{arXiv:2110.15251}}, \href {http://dx.doi.org/10.1103/PhysRevD.105.034001} {\path{doi:10.1103/PhysRevD.105.034001}}.

\bibitem{Celiberto:2023rzw}
F.~G. Celiberto, A.~Papa, {A high-energy QCD portal to exotic matter: Heavy-light tetraquarks at the HL-LHC}, Phys. Lett. B 848 (2024) 138406.
\newblock \href {http://arxiv.org/abs/2308.00809} {\path{arXiv:2308.00809}}, \href {http://dx.doi.org/10.1016/j.physletb.2023.138406} {\path{doi:10.1016/j.physletb.2023.138406}}.

\bibitem{Bai:2024ezn}
X.-W. Bai, F.~Feng, C.-M. Gan, Y.~Huang, W.-L. Sang, H.-F. Zhang, {Producing fully-charmed tetraquarks via charm quark fragmentation in colliders}, JHEP 09 (2024) 002.
\newblock \href {http://arxiv.org/abs/2404.13889} {\path{arXiv:2404.13889}}, \href {http://dx.doi.org/10.1007/JHEP09(2024)002} {\path{doi:10.1007/JHEP09(2024)002}}.

\bibitem{Ma:2025ryo}
H.-H. Ma, Z.-K. Tao, J.-J. Niu, {Application of fragmentation function to the indirect production of fully charmed tetraquark}, Eur. Phys. J. C 85~(4) (2025) 397.
\newblock \href {http://arxiv.org/abs/2502.20891} {\path{arXiv:2502.20891}}, \href {http://dx.doi.org/10.1140/epjc/s10052-025-14128-7} {\path{doi:10.1140/epjc/s10052-025-14128-7}}.

\bibitem{Nakhaei:2025zty}
H.~S. Nakhaei, G.~R. Boroun, {Analysis of the fragmentation function of gluon at next-to-leading order approximation}, Phys. Rev. D 112~(5) (2025) 054039.
\newblock \href {http://arxiv.org/abs/2508.05256} {\path{arXiv:2508.05256}}, \href {http://dx.doi.org/10.1103/x9vl-d6pm} {\path{doi:10.1103/x9vl-d6pm}}.

\bibitem{Farashaeian:2024son}
R.~Farashaeian, S.~M. Moosavi~Nejad, {Ground state fully heavy pentaquark production in the pair annihilation process}, Eur. Phys. J. A 60~(3) (2024) 65.
\newblock \href {http://dx.doi.org/10.1140/epja/s10050-024-01294-7} {\path{doi:10.1140/epja/s10050-024-01294-7}}.

\bibitem{Farashaeian:2024cpd}
R.~Farashaeian, S.~M. Moosavi~Nejad, {Fragmentation production of S-wave heavy pentaquark in diquark model}, Eur. Phys. J. A 60~(7) (2024) 143.
\newblock \href {http://dx.doi.org/10.1140/epja/s10050-024-01360-0} {\path{doi:10.1140/epja/s10050-024-01360-0}}.

\bibitem{Maiani:2004vq}
L.~Maiani, F.~Piccinini, A.~D. Polosa, V.~Riquer, {Diquark-antidiquarks with hidden or open charm and the nature of X(3872)}, Phys. Rev. D 71 (2005) 014028.
\newblock \href {http://arxiv.org/abs/hep-ph/0412098} {\path{arXiv:hep-ph/0412098}}, \href {http://dx.doi.org/10.1103/PhysRevD.71.014028} {\path{doi:10.1103/PhysRevD.71.014028}}.

\bibitem{Jaffe:2003sg}
R.~L. Jaffe, F.~Wilczek, {Diquarks and exotic spectroscopy}, Phys. Rev. Lett. 91 (2003) 232003.
\newblock \href {http://arxiv.org/abs/hep-ph/0307341} {\path{arXiv:hep-ph/0307341}}, \href {http://dx.doi.org/10.1103/PhysRevLett.91.232003} {\path{doi:10.1103/PhysRevLett.91.232003}}.

\bibitem{Guo:2013xga}
F.-K. Guo, C.~Hidalgo-Duque, J.~Nieves, M.~P. Valderrama, {Heavy-antiquark\textendash{}diquark symmetry and heavy hadron molecules: Are there triply heavy pentaquarks?}, Phys. Rev. D 88~(5) (2013) 054014.
\newblock \href {http://arxiv.org/abs/1305.4052} {\path{arXiv:1305.4052}}, \href {http://dx.doi.org/10.1103/PhysRevD.88.054014} {\path{doi:10.1103/PhysRevD.88.054014}}.

\bibitem{DeSanctis:2016zph}
M.~De~Sanctis, J.~Ferretti, R.~Maga\~na Vsevolodovna, P.~Saracco, E.~Santopinto, {An interacting quark-diquark model. Strange and nonstrange baryon spectroscopy and other observables}, Few Body Syst. 57~(12) (2016) 1177--1184.
\newblock \href {http://arxiv.org/abs/1608.00387} {\path{arXiv:1608.00387}}, \href {http://dx.doi.org/10.1007/s00601-016-1139-4} {\path{doi:10.1007/s00601-016-1139-4}}.

\bibitem{Bacchetta:2008af}
A.~Bacchetta, F.~Conti, M.~Radici, {Transverse-momentum distributions in a diquark spectator model}, Phys. Rev. D78 (2008) 074010.
\newblock \href {http://arxiv.org/abs/0807.0323} {\path{arXiv:0807.0323}}, \href {http://dx.doi.org/10.1103/PhysRevD.78.074010} {\path{doi:10.1103/PhysRevD.78.074010}}.

\bibitem{Bacchetta:2010si}
A.~Bacchetta, M.~Radici, F.~Conti, M.~Guagnelli, {Weighted azimuthal asymmetries in a diquark spectator model}, Eur. Phys. J. A45 (2010) 373--388.
\newblock \href {http://arxiv.org/abs/1003.1328} {\path{arXiv:1003.1328}}, \href {http://dx.doi.org/10.1140/epja/i2010-11016-y} {\path{doi:10.1140/epja/i2010-11016-y}}.

\bibitem{Bacchetta:2020vty}
A.~Bacchetta, F.~G. Celiberto, M.~Radici, P.~Taels, {Transverse-momentum-dependent gluon distribution functions in a spectator model}, Eur. Phys. J. C 80~(8) (2020) 733.
\newblock \href {http://arxiv.org/abs/2005.02288} {\path{arXiv:2005.02288}}, \href {http://dx.doi.org/10.1140/epjc/s10052-020-8327-6} {\path{doi:10.1140/epjc/s10052-020-8327-6}}.

\bibitem{Bacchetta:2024fci}
A.~Bacchetta, F.~G. Celiberto, M.~Radici, {T-odd gluon distribution functions in a spectator model}, Eur. Phys. J. C 84~(6) (2024) 576.
\newblock \href {http://arxiv.org/abs/2402.17556} {\path{arXiv:2402.17556}}, \href {http://dx.doi.org/10.1140/epjc/s10052-024-12927-y} {\path{doi:10.1140/epjc/s10052-024-12927-y}}.

\bibitem{Chakrabarti:2023djs}
D.~Chakrabarti, P.~Choudhary, B.~Gurjar, R.~Kishore, T.~Maji, C.~Mondal, A.~Mukherjee, {Gluon distributions in the proton in a light-front spectator model}, Phys. Rev. D 108~(1) (2023) 014009.
\newblock \href {http://arxiv.org/abs/2304.09908} {\path{arXiv:2304.09908}}, \href {http://dx.doi.org/10.1103/PhysRevD.108.014009} {\path{doi:10.1103/PhysRevD.108.014009}}.

\bibitem{Banu:2024ywv}
K.~Banu, A.~Mukherjee, A.~Pawar, S.~Rajesh, {Unraveling gluon TMDs in J/\ensuremath{\psi} and pion production at the EIC}, Phys. Rev. D 110~(5) (2024) 054009.
\newblock \href {http://arxiv.org/abs/2406.00271} {\path{arXiv:2406.00271}}, \href {http://dx.doi.org/10.1103/PhysRevD.110.054009} {\path{doi:10.1103/PhysRevD.110.054009}}.

\bibitem{Nzar:1995wb}
M.~Nzar, P.~Hoodbhoy, {Quark fragmentation functions in a diquark model for proton and Lambda hyperon production}, Phys. Rev. D 51 (1995) 32--36.
\newblock \href {http://arxiv.org/abs/hep-ph/9502349} {\path{arXiv:hep-ph/9502349}}, \href {http://dx.doi.org/10.1103/PhysRevD.51.32} {\path{doi:10.1103/PhysRevD.51.32}}.

\bibitem{Ma:2001ri}
B.-Q. Ma, I.~Schmidt, J.~Soffer, J.-J. Yang, {Quark distributions of octet baryons from SU(3) symmetry}, Phys. Rev. D 65 (2002) 034004.
\newblock \href {http://arxiv.org/abs/hep-ph/0110029} {\path{arXiv:hep-ph/0110029}}, \href {http://dx.doi.org/10.1103/PhysRevD.65.034004} {\path{doi:10.1103/PhysRevD.65.034004}}.

\bibitem{Falk:1993gb}
A.~F. Falk, M.~E. Luke, M.~J. Savage, M.~B. Wise, {Heavy quark fragmentation to baryons containing two heavy quarks}, Phys. Rev. D 49 (1994) 555--558.
\newblock \href {http://arxiv.org/abs/hep-ph/9305315} {\path{arXiv:hep-ph/9305315}}, \href {http://dx.doi.org/10.1103/PhysRevD.49.555} {\path{doi:10.1103/PhysRevD.49.555}}.

\bibitem{Maiani:2015vwa}
L.~Maiani, A.~D. Polosa, V.~Riquer, {The New Pentaquarks in the Diquark Model}, Phys. Lett. B 749 (2015) 289--291.
\newblock \href {http://dx.doi.org/10.1016/j.physletb.2015.08.008} {\path{doi:10.1016/j.physletb.2015.08.008}}.

\bibitem{Faustov:2020qfm}
R.~N. Faustov, V.~O. Galkin, E.~M. Savchenko, {Masses of the $QQ\bar Q\bar Q$ tetraquarks in the relativistic diquark--antidiquark picture}, Phys. Rev. D 102~(11) (2020) 114030.
\newblock \href {http://arxiv.org/abs/2009.13237} {\path{arXiv:2009.13237}}, \href {http://dx.doi.org/10.1103/PhysRevD.102.114030} {\path{doi:10.1103/PhysRevD.102.114030}}.

\bibitem{Faustov:2021hjs}
R.~N. Faustov, V.~O. Galkin, E.~M. Savchenko, {Heavy tetraquarks in the relativistic quark model}, Universe 7~(4) (2021) 94.
\newblock \href {http://arxiv.org/abs/2103.01763} {\path{arXiv:2103.01763}}, \href {http://dx.doi.org/10.3390/universe7040094} {\path{doi:10.3390/universe7040094}}.

\bibitem{Faustov:2022mvs}
R.~N. Faustov, V.~O. Galkin, E.~M. Savchenko, {Fully Heavy Tetraquark Spectroscopy in the Relativistic Quark Model}, Symmetry 14~(12) (2022) 2504.
\newblock \href {http://arxiv.org/abs/2210.16015} {\path{arXiv:2210.16015}}, \href {http://dx.doi.org/10.3390/sym14122504} {\path{doi:10.3390/sym14122504}}.

\bibitem{Martynenko:1996bt}
A.~P. Martynenko, V.~A. Saleev, {Heavy quark fragmentation into baryons in a quark - diquark model}, Phys. Lett. B 385 (1996) 297--303.
\newblock \href {http://arxiv.org/abs/hep-ph/9604259} {\path{arXiv:hep-ph/9604259}}, \href {http://dx.doi.org/10.1016/0370-2693(96)00848-9} {\path{doi:10.1016/0370-2693(96)00848-9}}.

\bibitem{GomshiNobary:2007xk}
M.~A. Gomshi~Nobary, R.~Sepahvand, {Ground state heavy baryon production in a relativistic quark-diquark model}, Phys. Rev. D 76 (2007) 114006.
\newblock \href {http://arxiv.org/abs/0711.0187} {\path{arXiv:0711.0187}}, \href {http://dx.doi.org/10.1103/PhysRevD.76.114006} {\path{doi:10.1103/PhysRevD.76.114006}}.

\bibitem{Delpasand:2020mqv}
M.~Delpasand, S.~M. Moosavi~Nejad, {$\Omega _{ccc}$ baryon production from gluon in vector diquark fragmentation}, Eur. Phys. J. A 56~(2) (2020) 56.
\newblock \href {http://dx.doi.org/10.1140/epja/s10050-020-00069-0} {\path{doi:10.1140/epja/s10050-020-00069-0}}.

\bibitem{Binosi:2008ig}
D.~Binosi, J.~Collins, C.~Kaufhold, L.~Theussl, {JaxoDraw: A Graphical user interface for drawing Feynman diagrams. Version 2.0 release notes}, Comput. Phys. Commun. 180 (2009) 1709--1715.
\newblock \href {http://arxiv.org/abs/0811.4113} {\path{arXiv:0811.4113}}, \href {http://dx.doi.org/10.1016/j.cpc.2009.02.020} {\path{doi:10.1016/j.cpc.2009.02.020}}.

\bibitem{Lepage:1980fj}
G.~P. Lepage, S.~J. Brodsky, {Exclusive Processes in Perturbative Quantum Chromodynamics}, Phys. Rev. D 22 (1980) 2157.
\newblock \href {http://dx.doi.org/10.1103/PhysRevD.22.2157} {\path{doi:10.1103/PhysRevD.22.2157}}.

\bibitem{Bjorken:1977md}
J.~D. Bjorken, {Properties of Hadron Distributions in Reactions Containing Very Heavy Quarks}, Phys. Rev. D 17 (1978) 171--173.
\newblock \href {http://dx.doi.org/10.1103/PhysRevD.17.171} {\path{doi:10.1103/PhysRevD.17.171}}.

\bibitem{Kinoshita:1985mh}
K.~Kinoshita, {A COVARIANT PARTON MODEL FOR HEAVY QUARK FRAGMENTATION}, Prog. Theor. Phys. 75 (1986) 84.
\newblock \href {http://dx.doi.org/10.1143/PTP.75.84} {\path{doi:10.1143/PTP.75.84}}.

\bibitem{Mertig:1990an}
R.~Mertig, M.~Bohm, A.~Denner, {FEYN CALC: Computer algebraic calculation of Feynman amplitudes}, Comput. Phys. Commun. 64 (1991) 345--359.
\newblock \href {http://dx.doi.org/10.1016/0010-4655(91)90130-D} {\path{doi:10.1016/0010-4655(91)90130-D}}.

\bibitem{Shtabovenko:2016sxi}
V.~Shtabovenko, R.~Mertig, F.~Orellana, {New Developments in FeynCalc 9.0}, Comput. Phys. Commun. 207 (2016) 432--444.
\newblock \href {http://arxiv.org/abs/1601.01167} {\path{arXiv:1601.01167}}, \href {http://dx.doi.org/10.1016/j.cpc.2016.06.008} {\path{doi:10.1016/j.cpc.2016.06.008}}.

\bibitem{Shtabovenko:2020gxv}
V.~Shtabovenko, R.~Mertig, F.~Orellana, {FeynCalc 9.3: New features and improvements}, Comput. Phys. Commun. 256 (2020) 107478.
\newblock \href {http://arxiv.org/abs/2001.04407} {\path{arXiv:2001.04407}}, \href {http://dx.doi.org/10.1016/j.cpc.2020.107478} {\path{doi:10.1016/j.cpc.2020.107478}}.

\bibitem{ParticleDataGroup:2020ssz}
P.~A. Zyla, et~al., {Review of Particle Physics}, PTEP 2020~(8) (2020) 083C01.
\newblock \href {http://dx.doi.org/10.1093/ptep/ptaa104} {\path{doi:10.1093/ptep/ptaa104}}.

\bibitem{GomshiNobary:1994eq}
M.~A. Gomshi~Nobary, {Heavy quark fragmentation functions}, J. Phys. G 20 (1994) 65--72.
\newblock \href {http://dx.doi.org/10.1088/0954-3899/20/1/008} {\path{doi:10.1088/0954-3899/20/1/008}}.

\bibitem{Celiberto:2016hae}
F.~G. Celiberto, D.~{\relax Yu}. Ivanov, B.~Murdaca, A.~Papa, {High energy resummation in dihadron production at the LHC}, Phys. Rev. D 94~(3) (2016) 034013.
\newblock \href {http://arxiv.org/abs/1604.08013} {\path{arXiv:1604.08013}}, \href {http://dx.doi.org/10.1103/PhysRevD.94.034013} {\path{doi:10.1103/PhysRevD.94.034013}}.

\bibitem{Celiberto:2017ptm}
F.~G. Celiberto, D.~{\relax Yu}. Ivanov, B.~Murdaca, A.~Papa, {Dihadron production at the LHC: full next-to-leading BFKL calculation}, Eur. Phys. J. C 77~(6) (2017) 382.
\newblock \href {http://arxiv.org/abs/1701.05077} {\path{arXiv:1701.05077}}, \href {http://dx.doi.org/10.1140/epjc/s10052-017-4949-8} {\path{doi:10.1140/epjc/s10052-017-4949-8}}.

\bibitem{Bolognino:2018oth}
A.~D. Bolognino, F.~G. Celiberto, D.~{\relax Yu}. Ivanov, M.~M.~A. Mohammed, A.~Papa, {Hadron-jet correlations in high-energy hadronic collisions at the LHC}, Eur. Phys. J. C 78~(9) (2018) 772.
\newblock \href {http://arxiv.org/abs/1808.05483} {\path{arXiv:1808.05483}}, \href {http://dx.doi.org/10.1140/epjc/s10052-018-6253-7} {\path{doi:10.1140/epjc/s10052-018-6253-7}}.

\bibitem{Celiberto:2021dzy}
F.~G. Celiberto, M.~Fucilla, D.~{\relax Yu}. Ivanov, A.~Papa, {High-energy resummation in $\Lambda_c$ baryon production}, Eur. Phys. J. C 81~(8) (2021) 780.
\newblock \href {http://arxiv.org/abs/2105.06432} {\path{arXiv:2105.06432}}, \href {http://dx.doi.org/10.1140/epjc/s10052-021-09448-3} {\path{doi:10.1140/epjc/s10052-021-09448-3}}.

\bibitem{Celiberto:2021fdp}
F.~G. Celiberto, M.~Fucilla, D.~{\relax Yu}. Ivanov, M.~M.~A. Mohammed, A.~Papa, {Bottom-flavored inclusive emissions in the variable-flavor number scheme: A high-energy analysis}, Phys. Rev. D 104~(11) (2021) 114007.
\newblock \href {http://arxiv.org/abs/2109.11875} {\path{arXiv:2109.11875}}, \href {http://dx.doi.org/10.1103/PhysRevD.104.114007} {\path{doi:10.1103/PhysRevD.104.114007}}.

\bibitem{Bertone:2013vaa}
V.~Bertone, S.~Carrazza, J.~Rojo, {APFEL: A PDF Evolution Library with QED corrections}, Comput. Phys. Commun. 185 (2014) 1647--1668.
\newblock \href {http://arxiv.org/abs/1310.1394} {\path{arXiv:1310.1394}}, \href {http://dx.doi.org/10.1016/j.cpc.2014.03.007} {\path{doi:10.1016/j.cpc.2014.03.007}}.

\bibitem{Carrazza:2014gfa}
S.~Carrazza, A.~Ferrara, D.~Palazzo, J.~Rojo, {APFEL Web}: {a web-based application for the graphical visualization of parton distribution functions}, J. Phys. G 42~(5) (2015) 057001.
\newblock \href {http://arxiv.org/abs/1410.5456} {\path{arXiv:1410.5456}}, \href {http://dx.doi.org/10.1088/0954-3899/42/5/057001} {\path{doi:10.1088/0954-3899/42/5/057001}}.

\bibitem{Bertone:2017gds}
V.~Bertone, {APFEL++: A new PDF evolution library in C++}, PoS DIS2017 (2018) 201.
\newblock \href {http://arxiv.org/abs/1708.00911} {\path{arXiv:1708.00911}}, \href {http://dx.doi.org/10.22323/1.297.0201} {\path{doi:10.22323/1.297.0201}}.

\bibitem{Candido:2022tld}
A.~Candido, F.~Hekhorn, G.~Magni, {EKO: evolution kernel operators}, Eur. Phys. J. C 82~(10) (2022) 976.
\newblock \href {http://arxiv.org/abs/2202.02338} {\path{arXiv:2202.02338}}, \href {http://dx.doi.org/10.1140/epjc/s10052-022-10878-w} {\path{doi:10.1140/epjc/s10052-022-10878-w}}.

\bibitem{Hekhorn:2023gul}
F.~Hekhorn, G.~Magni, {DGLAP evolution of parton distributions at approximate N$^3$LO}\href {http://arxiv.org/abs/2306.15294} {\path{arXiv:2306.15294}}.

\bibitem{Artoisenet:2014lpa}
P.~Artoisenet, E.~Braaten, {Gluon fragmentation into quarkonium at next-to-leading order}, JHEP 04 (2015) 121.
\newblock \href {http://arxiv.org/abs/1412.3834} {\path{arXiv:1412.3834}}, \href {http://dx.doi.org/10.1007/JHEP04(2015)121} {\path{doi:10.1007/JHEP04(2015)121}}.

\bibitem{Zhang:2018mlo}
P.~Zhang, C.-Y. Wang, X.~Liu, Y.-Q. Ma, C.~Meng, K.-T. Chao, {Semi-analytical calculation of gluon fragmentation into$^{1}$S$_{0}^{[1,8]}$ quarkonia at next-to-leading order}, JHEP 04 (2019) 116.
\newblock \href {http://arxiv.org/abs/1810.07656} {\path{arXiv:1810.07656}}, \href {http://dx.doi.org/10.1007/JHEP04(2019)116} {\path{doi:10.1007/JHEP04(2019)116}}.

\bibitem{Zheng:2021mqr}
X.-C. Zheng, Z.-Y. Zhang, X.-G. Wu, {Fragmentation functions for a quark into a spin-singlet quarkonium: Different flavor case}, Phys. Rev. D 103~(7) (2021) 074004.
\newblock \href {http://arxiv.org/abs/2101.01527} {\path{arXiv:2101.01527}}, \href {http://dx.doi.org/10.1103/PhysRevD.103.074004} {\path{doi:10.1103/PhysRevD.103.074004}}.

\bibitem{Zheng:2021ylc}
X.-C. Zheng, X.-G. Wu, X.-D. Huang, {NLO fragmentation functions for a quark into a spin-singlet quarkonium: same flavor case}, JHEP 07 (2021) 014.
\newblock \href {http://arxiv.org/abs/2105.14580} {\path{arXiv:2105.14580}}, \href {http://dx.doi.org/10.1007/JHEP07(2021)014} {\path{doi:10.1007/JHEP07(2021)014}}.

\bibitem{Zheng:2019dfk}
X.-C. Zheng, C.-H. Chang, X.-G. Wu, {NLO fragmentation functions of heavy quarks into heavy quarkonia}, Phys. Rev. D 100~(1) (2019) 014005.
\newblock \href {http://arxiv.org/abs/1905.09171} {\path{arXiv:1905.09171}}, \href {http://dx.doi.org/10.1103/PhysRevD.100.014005} {\path{doi:10.1103/PhysRevD.100.014005}}.

\bibitem{Celiberto:2022kxx}
F.~G. Celiberto, {Emergence of high-energy dynamics from cascade-baryon detections at the LHC}, Eur. Phys. J. C 83~(4) (2023) 332.
\newblock \href {http://arxiv.org/abs/2208.14577} {\path{arXiv:2208.14577}}, \href {http://dx.doi.org/10.1140/epjc/s10052-023-11417-x} {\path{doi:10.1140/epjc/s10052-023-11417-x}}.

\bibitem{Celiberto:2022grc}
F.~G. Celiberto, {Stabilizing BFKL via Heavy-flavor and NRQCD Fragmentation}, Acta Phys. Polon. Supp. 16~(5) (2023) 41.
\newblock \href {http://arxiv.org/abs/2211.11780} {\path{arXiv:2211.11780}}, \href {http://dx.doi.org/10.5506/APhysPolBSupp.16.5-A41} {\path{doi:10.5506/APhysPolBSupp.16.5-A41}}.

\bibitem{Collins:1989gx}
J.~C. Collins, D.~E. Soper, G.~F. Sterman, {Factorization of Hard Processes in QCD}, Adv. Ser. Direct. High Energy Phys. 5 (1989) 1--91.
\newblock \href {http://arxiv.org/abs/hep-ph/0409313} {\path{arXiv:hep-ph/0409313}}, \href {http://dx.doi.org/10.1142/9789814503266_0001} {\path{doi:10.1142/9789814503266_0001}}.

\bibitem{Sterman:1995fz}
G.~F. Sterman, {Partons, factorization and resummation, TASI 95}, in: {Theoretical Advanced Study Institute in Elementary Particle Physics (TASI 95): QCD and Beyond}, 1995, pp. 327--408.
\newblock \href {http://arxiv.org/abs/hep-ph/9606312} {\path{arXiv:hep-ph/9606312}}.

\bibitem{Gribov:1983ivg}
L.~V. Gribov, E.~M. Levin, M.~G. Ryskin, {Semihard Processes in QCD}, Phys. Rept. 100 (1983) 1--150.
\newblock \href {http://dx.doi.org/10.1016/0370-1573(83)90022-4} {\path{doi:10.1016/0370-1573(83)90022-4}}.

\bibitem{Celiberto:2017ius}
F.~G. Celiberto, {High-energy resummation in semi-hard processes at the LHC}, Phd thesis, Universit\`a della Calabria and INFN-Cosenza (2017).
\newblock \href {http://arxiv.org/abs/1707.04315} {\path{arXiv:1707.04315}}.

\bibitem{Bolognino:2021bjd}
A.~D. Bolognino, {From semi-hard processes to the unintegrated gluon distribution: a phenomenological path in the high-energy framework}, Phd thesis, Calabria U. (2021).
\newblock \href {http://arxiv.org/abs/2109.03033} {\path{arXiv:2109.03033}}.

\bibitem{Mohammed:2022gbk}
M.~M.~A. Mohammed, {Hunting stabilization effects of the high-energy resummation at the LHC}, Phd thesis, Universit\`a della Calabria and INFN-Cosenza (4 2022).
\newblock \href {http://arxiv.org/abs/2204.11606} {\path{arXiv:2204.11606}}.

\bibitem{Gatto:2025kfl}
G.~Gatto, {High-energy dynamics of QCD: Theoretical and phenomenological results}, Phd thesis, Universit\`a della Calabria and INFN-Cosenza (6 2025).
\newblock \href {http://arxiv.org/abs/2506.03222} {\path{arXiv:2506.03222}}.

\bibitem{Fadin:1975cb}
V.~S. Fadin, E.~Kuraev, L.~Lipatov, {On the Pomeranchuk Singularity in Asymptotically Free Theories}, Phys. Lett. B 60 (1975) 50--52.
\newblock \href {http://dx.doi.org/10.1016/0370-2693(75)90524-9} {\path{doi:10.1016/0370-2693(75)90524-9}}.

\bibitem{Kuraev:1976ge}
E.~A. Kuraev, L.~N. Lipatov, V.~S. Fadin, {Multi - Reggeon Processes in the Yang-Mills Theory}, Sov. Phys. JETP 44 (1976) 443--450.

\bibitem{Kuraev:1977fs}
E.~Kuraev, L.~Lipatov, V.~S. Fadin, {The Pomeranchuk Singularity in Nonabelian Gauge Theories}, Sov.\ Phys.\ JETP 45 (1977) 199--204.

\bibitem{Balitsky:1978ic}
I.~Balitsky, L.~Lipatov, {The Pomeranchuk Singularity in Quantum Chromodynamics}, Sov.\ J.\ Nucl.\ Phys. 28 (1978) 822--829.

\bibitem{Fadin:1998py}
V.~S. Fadin, L.~N. Lipatov, {BFKL pomeron in the next-to-leading approximation}, Phys. Lett. B 429 (1998) 127--134.
\newblock \href {http://arxiv.org/abs/hep-ph/9802290} {\path{arXiv:hep-ph/9802290}}, \href {http://dx.doi.org/10.1016/S0370-2693(98)00473-0} {\path{doi:10.1016/S0370-2693(98)00473-0}}.

\bibitem{Ciafaloni:1998gs}
M.~Ciafaloni, G.~Camici, {Energy scale(s) and next-to-leading BFKL equation}, Phys. Lett. B 430 (1998) 349--354.
\newblock \href {http://arxiv.org/abs/hep-ph/9803389} {\path{arXiv:hep-ph/9803389}}, \href {http://dx.doi.org/10.1016/S0370-2693(98)00551-6} {\path{doi:10.1016/S0370-2693(98)00551-6}}.

\bibitem{Fadin:1998jv}
V.~S. Fadin, R.~Fiore, A.~Papa, {The Quark part of the nonforward BFKL kernel and the 'bootstrap' for the gluon Reggeization}, Phys. Rev. D 60 (1999) 074025.
\newblock \href {http://arxiv.org/abs/hep-ph/9812456} {\path{arXiv:hep-ph/9812456}}, \href {http://dx.doi.org/10.1103/PhysRevD.60.074025} {\path{doi:10.1103/PhysRevD.60.074025}}.

\bibitem{Fadin:2000kx}
V.~S. Fadin, D.~A. Gorbachev, {Nonforward color octet BFKL kernel}, JETP Lett. 71 (2000) 222--226.
\newblock \href {http://dx.doi.org/10.1134/1.568320} {\path{doi:10.1134/1.568320}}.

\bibitem{Fadin:2000hu}
V.~S. Fadin, D.~A. Gorbachev, {Nonforward color-octet kernel of the Balitsky-Fadin-Kuraev-Lipatov equation}, Phys. Atom. Nucl. 63 (2000) 2157--2172.
\newblock \href {http://dx.doi.org/10.1134/1.1333885} {\path{doi:10.1134/1.1333885}}.

\bibitem{Fadin:2004zq}
V.~S. Fadin, R.~Fiore, {Non-forward BFKL pomeron at next-to-leading order}, Phys. Lett. B 610 (2005) 61--66, [Erratum: Phys.Lett.B 621, 320 (2005)].
\newblock \href {http://arxiv.org/abs/hep-ph/0412386} {\path{arXiv:hep-ph/0412386}}, \href {http://dx.doi.org/10.1016/j.physletb.2005.06.074} {\path{doi:10.1016/j.physletb.2005.06.074}}.

\bibitem{Fadin:2005zj}
V.~S. Fadin, R.~Fiore, {Non-forward NLO BFKL kernel}, Phys. Rev. D 72 (2005) 014018.
\newblock \href {http://arxiv.org/abs/hep-ph/0502045} {\path{arXiv:hep-ph/0502045}}, \href {http://dx.doi.org/10.1103/PhysRevD.72.014018} {\path{doi:10.1103/PhysRevD.72.014018}}.

\bibitem{Caola:2021izf}
F.~Caola, A.~Chakraborty, G.~Gambuti, A.~von Manteuffel, L.~Tancredi, {Three-Loop Gluon Scattering in QCD and the Gluon Regge Trajectory}, Phys. Rev. Lett. 128~(21) (2022) 212001.
\newblock \href {http://arxiv.org/abs/2112.11097} {\path{arXiv:2112.11097}}, \href {http://dx.doi.org/10.1103/PhysRevLett.128.212001} {\path{doi:10.1103/PhysRevLett.128.212001}}.

\bibitem{Falcioni:2021dgr}
G.~Falcioni, E.~Gardi, N.~Maher, C.~Milloy, L.~Vernazza, {Disentangling the Regge Cut and Regge Pole in Perturbative QCD}, Phys. Rev. Lett. 128~(13) (2022) 132001.
\newblock \href {http://arxiv.org/abs/2112.11098} {\path{arXiv:2112.11098}}, \href {http://dx.doi.org/10.1103/PhysRevLett.128.132001} {\path{doi:10.1103/PhysRevLett.128.132001}}.

\bibitem{DelDuca:2021vjq}
V.~Del~Duca, R.~Marzucca, B.~Verbeek, {The gluon Regge trajectory at three loops from planar Yang-Mills theory}, JHEP 01 (2022) 149.
\newblock \href {http://arxiv.org/abs/2111.14265} {\path{arXiv:2111.14265}}, \href {http://dx.doi.org/10.1007/JHEP01(2022)149} {\path{doi:10.1007/JHEP01(2022)149}}.

\bibitem{Byrne:2022wzk}
E.~P. Byrne, V.~Del~Duca, L.~J. Dixon, E.~Gardi, J.~M. Smillie, {One-loop central-emission vertex for two gluons in $ \mathcal{N} $ = 4 super Yang-Mills theory}, JHEP 08 (2022) 271.
\newblock \href {http://arxiv.org/abs/2204.12459} {\path{arXiv:2204.12459}}, \href {http://dx.doi.org/10.1007/JHEP08(2022)271} {\path{doi:10.1007/JHEP08(2022)271}}.

\bibitem{Fadin:2023roz}
V.~S. Fadin, M.~Fucilla, A.~Papa, {One-loop Lipatov vertex in QCD with higher \ensuremath{\epsilon}-accuracy}, JHEP 04 (2023) 137.
\newblock \href {http://arxiv.org/abs/2302.09868} {\path{arXiv:2302.09868}}, \href {http://dx.doi.org/10.1007/JHEP04(2023)137} {\path{doi:10.1007/JHEP04(2023)137}}.

\bibitem{Byrne:2023nqx}
E.~P. Byrne, {One-loop five-parton amplitudes in the NMRK limit}, JHEP 07 (2024) 284.
\newblock \href {http://arxiv.org/abs/2312.15051} {\path{arXiv:2312.15051}}, \href {http://dx.doi.org/10.1007/JHEP07(2024)284} {\path{doi:10.1007/JHEP07(2024)284}}.

\bibitem{Fadin:1999de}
V.~S. Fadin, R.~Fiore, M.~I. Kotsky, A.~Papa, {The Gluon impact factors}, Phys. Rev. D 61 (2000) 094005.
\newblock \href {http://arxiv.org/abs/hep-ph/9908264} {\path{arXiv:hep-ph/9908264}}, \href {http://dx.doi.org/10.1103/PhysRevD.61.094005} {\path{doi:10.1103/PhysRevD.61.094005}}.

\bibitem{Fadin:1999df}
V.~S. Fadin, R.~Fiore, M.~I. Kotsky, A.~Papa, {The Quark impact factors}, Phys. Rev. D 61 (2000) 094006.
\newblock \href {http://arxiv.org/abs/hep-ph/9908265} {\path{arXiv:hep-ph/9908265}}, \href {http://dx.doi.org/10.1103/PhysRevD.61.094006} {\path{doi:10.1103/PhysRevD.61.094006}}.

\bibitem{Bartels:2001ge}
J.~Bartels, D.~Colferai, G.~P. Vacca, {The NLO jet vertex for Mueller-Navelet and forward jets: The Quark part}, Eur. Phys. J. C 24 (2002) 83--99.
\newblock \href {http://arxiv.org/abs/hep-ph/0112283} {\path{arXiv:hep-ph/0112283}}, \href {http://dx.doi.org/10.1007/s100520200919} {\path{doi:10.1007/s100520200919}}.

\bibitem{Bartels:2002yj}
J.~Bartels, D.~Colferai, G.~P. Vacca, {The NLO jet vertex for Mueller-Navelet and forward jets: The Gluon part}, Eur. Phys. J. C 29 (2003) 235--249.
\newblock \href {http://arxiv.org/abs/hep-ph/0206290} {\path{arXiv:hep-ph/0206290}}, \href {http://dx.doi.org/10.1140/epjc/s2003-01169-5} {\path{doi:10.1140/epjc/s2003-01169-5}}.

\bibitem{Caporale:2011cc}
F.~Caporale, D.~{\relax Yu}. Ivanov, B.~Murdaca, A.~Papa, A.~Perri, {The next-to-leading order jet vertex for Mueller-Navelet and forward jets revisited}, JHEP 02 (2012) 101.
\newblock \href {http://arxiv.org/abs/1112.3752} {\path{arXiv:1112.3752}}, \href {http://dx.doi.org/10.1007/JHEP02(2012)101} {\path{doi:10.1007/JHEP02(2012)101}}.

\bibitem{Ivanov:2012ms}
D.~{\relax Yu}. Ivanov, A.~Papa, {The next-to-leading order forward jet vertex in the small-cone approximation}, JHEP 05 (2012) 086.
\newblock \href {http://arxiv.org/abs/1202.1082} {\path{arXiv:1202.1082}}, \href {http://dx.doi.org/10.1007/JHEP05(2012)086} {\path{doi:10.1007/JHEP05(2012)086}}.

\bibitem{Colferai:2015zfa}
D.~Colferai, A.~Niccoli, {The NLO jet vertex in the small-cone approximation for kt and cone algorithms}, JHEP 04 (2015) 071.
\newblock \href {http://arxiv.org/abs/1501.07442} {\path{arXiv:1501.07442}}, \href {http://dx.doi.org/10.1007/JHEP04(2015)071} {\path{doi:10.1007/JHEP04(2015)071}}.

\bibitem{Ivanov:2012iv}
D.~{\relax Yu}. Ivanov, A.~Papa, {Inclusive production of a pair of hadrons separated by a large interval of rapidity in proton collisions}, JHEP 07 (2012) 045.
\newblock \href {http://arxiv.org/abs/1205.6068} {\path{arXiv:1205.6068}}, \href {http://dx.doi.org/10.1007/JHEP07(2012)045} {\path{doi:10.1007/JHEP07(2012)045}}.

\bibitem{Ivanov:2004pp}
D.~{\relax Yu}. Ivanov, M.~I. Kotsky, A.~Papa, {The Impact factor for the virtual photon to light vector meson transition}, Eur. Phys. J. C 38 (2004) 195--213.
\newblock \href {http://arxiv.org/abs/hep-ph/0405297} {\path{arXiv:hep-ph/0405297}}, \href {http://dx.doi.org/10.1140/epjc/s2004-02039-4} {\path{doi:10.1140/epjc/s2004-02039-4}}.

\bibitem{Bartels:2000gt}
J.~Bartels, S.~Gieseke, C.~F. Qiao, {The (gamma* ---\ensuremath{>} q anti-q) Reggeon vertex in next-to-leading order QCD}, Phys. Rev. D 63 (2001) 056014, [Erratum: Phys.Rev.D 65, 079902 (2002)].
\newblock \href {http://arxiv.org/abs/hep-ph/0009102} {\path{arXiv:hep-ph/0009102}}, \href {http://dx.doi.org/10.1103/PhysRevD.63.056014} {\path{doi:10.1103/PhysRevD.63.056014}}.

\bibitem{Bartels:2001mv}
J.~Bartels, S.~Gieseke, A.~Kyrieleis, {The Process gamma*(L) + q ---\ensuremath{>} (q anti-q g) + q: Real corrections to the virtual photon impact factor}, Phys. Rev. D 65 (2002) 014006.
\newblock \href {http://arxiv.org/abs/hep-ph/0107152} {\path{arXiv:hep-ph/0107152}}, \href {http://dx.doi.org/10.1103/PhysRevD.65.014006} {\path{doi:10.1103/PhysRevD.65.014006}}.

\bibitem{Bartels:2002uz}
J.~Bartels, D.~Colferai, S.~Gieseke, A.~Kyrieleis, {NLO corrections to the photon impact factor: Combining real and virtual corrections}, Phys. Rev. D 66 (2002) 094017.
\newblock \href {http://arxiv.org/abs/hep-ph/0208130} {\path{arXiv:hep-ph/0208130}}, \href {http://dx.doi.org/10.1103/PhysRevD.66.094017} {\path{doi:10.1103/PhysRevD.66.094017}}.

\bibitem{Bartels:2004bi}
J.~Bartels, A.~Kyrieleis, {NLO corrections to the gamma* impact factor: First numerical results for the real corrections to gamma*(L)}, Phys. Rev. D 70 (2004) 114003.
\newblock \href {http://arxiv.org/abs/hep-ph/0407051} {\path{arXiv:hep-ph/0407051}}, \href {http://dx.doi.org/10.1103/PhysRevD.70.114003} {\path{doi:10.1103/PhysRevD.70.114003}}.

\bibitem{Fadin:2001ap}
V.~S. Fadin, D.~{\relax Yu}. Ivanov, M.~I. Kotsky, {Photon Reggeon interaction vertices in the NLA}, Phys. Atom. Nucl. 65 (2002) 1513--1527.
\newblock \href {http://arxiv.org/abs/hep-ph/0106099} {\path{arXiv:hep-ph/0106099}}, \href {http://dx.doi.org/10.1134/1.1501664} {\path{doi:10.1134/1.1501664}}.

\bibitem{Balitsky:2012bs}
I.~Balitsky, G.~A. Chirilli, {Photon impact factor and $k_T$-factorization for DIS in the next-to-leading order}, Phys. Rev. D 87~(1) (2013) 014013.
\newblock \href {http://arxiv.org/abs/1207.3844} {\path{arXiv:1207.3844}}, \href {http://dx.doi.org/10.1103/PhysRevD.87.014013} {\path{doi:10.1103/PhysRevD.87.014013}}.

\bibitem{Hentschinski:2020tbi}
M.~Hentschinski, K.~Kutak, A.~van Hameren, {Forward Higgs production within high energy factorization in the heavy quark limit at next-to-leading order accuracy}, Eur. Phys. J. C 81~(2) (2021) 112, [Erratum: Eur. Phys. J. C 81, 262 (2021)].
\newblock \href {http://arxiv.org/abs/2011.03193} {\path{arXiv:2011.03193}}, \href {http://dx.doi.org/10.1140/epjc/s10052-021-08902-6} {\path{doi:10.1140/epjc/s10052-021-08902-6}}.

\bibitem{Celiberto:2022fgx}
F.~G. Celiberto, M.~Fucilla, D.~{\relax Yu}. Ivanov, M.~M.~A. Mohammed, A.~Papa, {The next-to-leading order Higgs impact factor in the infinite top-mass limit}, JHEP 08 (2022) 092.
\newblock \href {http://arxiv.org/abs/2205.02681} {\path{arXiv:2205.02681}}, \href {http://dx.doi.org/10.1007/JHEP08(2022)092} {\path{doi:10.1007/JHEP08(2022)092}}.

\bibitem{Hentschinski:2022sko}
M.~Hentschinski, {Forward Higgs production at NLO using Lipatov's high energy effective action}, SciPost Phys. Proc. 8 (2022) 136.
\newblock \href {http://dx.doi.org/10.21468/SciPostPhysProc.8.136} {\path{doi:10.21468/SciPostPhysProc.8.136}}.

\bibitem{Celiberto:2024bbv}
F.~G. Celiberto, L.~Delle~Rose, M.~Fucilla, G.~Gatto, A.~Papa, {The next-to-leading order Higgs impact factor at physical top mass: the real corrections}, JHEP 12 (2024) 061.
\newblock \href {http://arxiv.org/abs/2409.20354} {\path{arXiv:2409.20354}}, \href {http://dx.doi.org/10.1007/JHEP12(2024)061} {\path{doi:10.1007/JHEP12(2024)061}}.

\bibitem{Celiberto:2025ece}
F.~G. Celiberto, L.~Delle~Rose, G.~Gatto, A.~Papa, M.~Fucilla, {The Real Corrections to the Higgs Impact Factor at Next-to-leading Order with Finite Top Mass}, Acta Phys. Polon. Supp. 18~(1) (2025) 1--A37.
\newblock \href {http://arxiv.org/abs/2502.02228} {\path{arXiv:2502.02228}}, \href {http://dx.doi.org/10.5506/APhysPolBSupp.18.1-A37} {\path{doi:10.5506/APhysPolBSupp.18.1-A37}}.

\bibitem{DelDuca:2025vux}
V.~Del~Duca, G.~Falcioni, {The two-loop Higgs impact factor}, JHEP 07 (2025) 018.
\newblock \href {http://arxiv.org/abs/2504.06184} {\path{arXiv:2504.06184}}, \href {http://dx.doi.org/10.1007/JHEP07(2025)018} {\path{doi:10.1007/JHEP07(2025)018}}.

\bibitem{Hentschinski:2012poz}
M.~Hentschinski, C.~Salas, {Forward Drell-Yan plus backward jet as a test of BFKL evolution}, in: {20th International Workshop on Deep-Inelastic Scattering and Related Subjects}, 2012.
\newblock \href {http://arxiv.org/abs/1301.1227} {\path{arXiv:1301.1227}}, \href {http://dx.doi.org/10.3204/DESY-PROC-2012-02/115} {\path{doi:10.3204/DESY-PROC-2012-02/115}}.

\bibitem{Motyka:2014lya}
L.~Motyka, M.~Sadzikowski, T.~Stebel, {Twist expansion of Drell-Yan structure functions in color dipole approach}, JHEP 05 (2015) 087.
\newblock \href {http://arxiv.org/abs/1412.4675} {\path{arXiv:1412.4675}}, \href {http://dx.doi.org/10.1007/JHEP05(2015)087} {\path{doi:10.1007/JHEP05(2015)087}}.

\bibitem{Celiberto:2017nyx}
F.~G. Celiberto, D.~{\relax Yu}. Ivanov, B.~Murdaca, A.~Papa, {High-energy resummation in heavy-quark pair photoproduction}, Phys. Lett. B 777 (2018) 141--150.
\newblock \href {http://arxiv.org/abs/1709.10032} {\path{arXiv:1709.10032}}, \href {http://dx.doi.org/10.1016/j.physletb.2017.12.020} {\path{doi:10.1016/j.physletb.2017.12.020}}.

\bibitem{Bolognino:2019ccd}
A.~D. Bolognino, F.~G. Celiberto, M.~Fucilla, D.~{\relax Yu}. Ivanov, B.~Murdaca, A.~Papa, {Inclusive production of two rapidity-separated heavy quarks as a probe of BFKL dynamics}, PoS DIS2019 (2019) 067.
\newblock \href {http://arxiv.org/abs/1906.05940} {\path{arXiv:1906.05940}}, \href {http://dx.doi.org/10.22323/1.352.0067} {\path{doi:10.22323/1.352.0067}}.

\bibitem{Bolognino:2019yls}
A.~D. Bolognino, F.~G. Celiberto, M.~Fucilla, D.~{\relax Yu}. Ivanov, A.~Papa, {High-energy resummation in heavy-quark pair hadroproduction}, Eur. Phys. J. C 79~(11) (2019) 939.
\newblock \href {http://arxiv.org/abs/1909.03068} {\path{arXiv:1909.03068}}, \href {http://dx.doi.org/10.1140/epjc/s10052-019-7392-1} {\path{doi:10.1140/epjc/s10052-019-7392-1}}.

\bibitem{Boussarie:2017oae}
R.~Boussarie, B.~Duclou\'e, L.~Szymanowski, S.~Wallon, {Forward $J/\psi$ and very backward jet inclusive production at the LHC}, Phys. Rev. D 97~(1) (2018) 014008.
\newblock \href {http://arxiv.org/abs/1709.01380} {\path{arXiv:1709.01380}}, \href {http://dx.doi.org/10.1103/PhysRevD.97.014008} {\path{doi:10.1103/PhysRevD.97.014008}}.

\bibitem{Boussarie:2015jar}
R.~Boussarie, B.~Duclou\'e, L.~Szymanowski, S.~Wallon, {Production of a forward J/psi and a backward jet at the LHC}, in: {International Conference on the Structure and Interactions of the Photon and 21st International Workshop on Photon-Photon Collisions and International Workshop on High Energy Photon Linear Colliders}, 2015.
\newblock \href {http://arxiv.org/abs/1511.02181} {\path{arXiv:1511.02181}}.

\bibitem{Boussarie:2016gaq}
R.~Boussarie, B.~Ducloue, L.~Szymanowski, S.~Wallon, {Production of a forward $J/\psi$ and a backward jet at the LHC}, PoS DIS2016 (2016) 204.
\newblock \href {http://dx.doi.org/10.22323/1.265.0204} {\path{doi:10.22323/1.265.0204}}.

\bibitem{Boussarie:2017xdy}
R.~Boussarie, B.~Duclou\'e, L.~Szymanowski, S.~Wallon, {QCD resummation effects in forward $J/\psi$ and very backward jet inclusive production at the LHC}, PoS DIS2017 (2018) 063.
\newblock \href {http://arxiv.org/abs/1709.02671} {\path{arXiv:1709.02671}}, \href {http://dx.doi.org/10.22323/1.297.0063} {\path{doi:10.22323/1.297.0063}}.

\bibitem{Mueller:1986ey}
A.~H. Mueller, H.~Navelet, {An Inclusive Minijet Cross-Section and the Bare Pomeron in QCD}, Nucl. Phys. B 282 (1987) 727--744.
\newblock \href {http://dx.doi.org/10.1016/0550-3213(87)90705-X} {\path{doi:10.1016/0550-3213(87)90705-X}}.

\bibitem{Colferai:2010wu}
D.~Colferai, F.~Schwennsen, L.~Szymanowski, S.~Wallon, {Mueller Navelet jets at LHC - complete NLL BFKL calculation}, JHEP 12 (2010) 026.
\newblock \href {http://arxiv.org/abs/1002.1365} {\path{arXiv:1002.1365}}, \href {http://dx.doi.org/10.1007/JHEP12(2010)026} {\path{doi:10.1007/JHEP12(2010)026}}.

\bibitem{Ducloue:2013hia}
B.~Duclou\'e, L.~Szymanowski, S.~Wallon, {Confronting Mueller-Navelet jets in NLL BFKL with LHC experiments at 7 TeV}, JHEP 05 (2013) 096.
\newblock \href {http://arxiv.org/abs/1302.7012} {\path{arXiv:1302.7012}}, \href {http://dx.doi.org/10.1007/JHEP05(2013)096} {\path{doi:10.1007/JHEP05(2013)096}}.

\bibitem{Caporale:2013uva}
F.~Caporale, B.~Murdaca, A.~Sabio~Vera, C.~Salas, {Scale choice and collinear contributions to Mueller-Navelet jets at LHC energies}, Nucl. Phys. B 875 (2013) 134--151.
\newblock \href {http://arxiv.org/abs/1305.4620} {\path{arXiv:1305.4620}}, \href {http://dx.doi.org/10.1016/j.nuclphysb.2013.07.005} {\path{doi:10.1016/j.nuclphysb.2013.07.005}}.

\bibitem{Caporale:2015uva}
F.~Caporale, D.~{\relax Yu}. Ivanov, B.~Murdaca, A.~Papa, {Brodsky-Lepage-Mackenzie optimal renormalization scale setting for semihard processes}, Phys. Rev. D 91~(11) (2015) 114009.
\newblock \href {http://arxiv.org/abs/1504.06471} {\path{arXiv:1504.06471}}, \href {http://dx.doi.org/10.1103/PhysRevD.91.114009} {\path{doi:10.1103/PhysRevD.91.114009}}.

\bibitem{Ducloue:2015jba}
B.~Duclou\'e, L.~Szymanowski, S.~Wallon, {Evaluating the double parton scattering contribution to Mueller-Navelet jets production at the LHC}, Phys. Rev. D 92~(7) (2015) 076002.
\newblock \href {http://arxiv.org/abs/1507.04735} {\path{arXiv:1507.04735}}, \href {http://dx.doi.org/10.1103/PhysRevD.92.076002} {\path{doi:10.1103/PhysRevD.92.076002}}.

\bibitem{Celiberto:2015yba}
F.~G. Celiberto, D.~{\relax Yu}. Ivanov, B.~Murdaca, A.~Papa, {Mueller--Navelet Jets at LHC: BFKL Versus High-Energy DGLAP}, Eur. Phys. J. C 75~(6) (2015) 292.
\newblock \href {http://arxiv.org/abs/1504.08233} {\path{arXiv:1504.08233}}, \href {http://dx.doi.org/10.1140/epjc/s10052-015-3522-6} {\path{doi:10.1140/epjc/s10052-015-3522-6}}.

\bibitem{Celiberto:2015mpa}
F.~G. Celiberto, D.~{\relax Yu}. Ivanov, B.~Murdaca, A.~Papa, {Mueller--Navelet Jets at the LHC: Discriminating BFKL from DGLAP by Asymmetric Cuts}, Acta Phys. Polon. Supp. 8 (2015) 935.
\newblock \href {http://arxiv.org/abs/1510.01626} {\path{arXiv:1510.01626}}, \href {http://dx.doi.org/10.5506/APhysPolBSupp.8.935} {\path{doi:10.5506/APhysPolBSupp.8.935}}.

\bibitem{Celiberto:2016ygs}
F.~G. Celiberto, D.~{\relax Yu}. Ivanov, B.~Murdaca, A.~Papa, {Mueller--Navelet jets at 13 TeV LHC: dependence on dynamic constraints in the central rapidity region}, Eur. Phys. J. C 76~(4) (2016) 224.
\newblock \href {http://arxiv.org/abs/1601.07847} {\path{arXiv:1601.07847}}, \href {http://dx.doi.org/10.1140/epjc/s10052-016-4053-5} {\path{doi:10.1140/epjc/s10052-016-4053-5}}.

\bibitem{Celiberto:2016vva}
F.~G. Celiberto, D.~{\relax Yu}. Ivanov, B.~Murdaca, A.~Papa, {BFKL effects and central rapidity dependence in Mueller-Navelet jet production at 13 TeV LHC}, PoS DIS2016 (2016) 176.
\newblock \href {http://arxiv.org/abs/1606.08892} {\path{arXiv:1606.08892}}, \href {http://dx.doi.org/10.22323/1.265.0176} {\path{doi:10.22323/1.265.0176}}.

\bibitem{Caporale:2018qnm}
F.~Caporale, F.~G. Celiberto, G.~Chachamis, D.~Gordo~G\'omez, A.~Sabio~Vera, {Inclusive dijet hadroproduction with a rapidity veto constraint}, Nucl. Phys. B 935 (2018) 412--434.
\newblock \href {http://arxiv.org/abs/1806.06309} {\path{arXiv:1806.06309}}, \href {http://dx.doi.org/10.1016/j.nuclphysb.2018.09.002} {\path{doi:10.1016/j.nuclphysb.2018.09.002}}.

\bibitem{deLeon:2020myv}
N.~B. de~Le{\'o}n, G.~Chachamis, A.~Sabio~Vera, {Multiperipheral final states in crowded twin-jet events at the LHC}, Nucl. Phys. B 971 (2021) 115518.
\newblock \href {http://arxiv.org/abs/2012.09664} {\path{arXiv:2012.09664}}, \href {http://dx.doi.org/10.1016/j.nuclphysb.2021.115518} {\path{doi:10.1016/j.nuclphysb.2021.115518}}.

\bibitem{deLeon:2021ecb}
N.~B. de~Le\'on, G.~Chachamis, A.~Sabio~Vera, {Average minijet rapidity ratios in Mueller\textendash{}Navelet jets}, Eur. Phys. J. C 81~(11) (2021) 1019.
\newblock \href {http://arxiv.org/abs/2106.11255} {\path{arXiv:2106.11255}}, \href {http://dx.doi.org/10.1140/epjc/s10052-021-09811-4} {\path{doi:10.1140/epjc/s10052-021-09811-4}}.

\bibitem{Baldenegro:2024ndr}
C.~Baldenegro, G.~Chachamis, M.~Kampshoff, M.~Klasen, G.~J. Milhano, C.~Royon, A.~Sabio~Vera, {Multijet event shape variables for Mueller-Navelet jet topologies}, Phys. Rev. D 110~(11) (2024) 114027.
\newblock \href {http://arxiv.org/abs/2406.10681} {\path{arXiv:2406.10681}}, \href {http://dx.doi.org/10.1103/PhysRevD.110.114027} {\path{doi:10.1103/PhysRevD.110.114027}}.

\bibitem{Celiberto:2016zgb}
F.~G. Celiberto, D.~{\relax Yu}. Ivanov, B.~Murdaca, A.~Papa, {Dihadron Production at LHC: BFKL Predictions for Cross Sections and Azimuthal Correlations}, AIP Conf. Proc. 1819~(1) (2017) 060005.
\newblock \href {http://arxiv.org/abs/1611.04811} {\path{arXiv:1611.04811}}, \href {http://dx.doi.org/10.1063/1.4977161} {\path{doi:10.1063/1.4977161}}.

\bibitem{Celiberto:2017uae}
F.~G. Celiberto, D.~{\relax Yu}. Ivanov, B.~Murdaca, A.~Papa, {Inclusive charged light di-hadron production at 7 and 13 TeV LHC in the full NLA BFKL approach}, in: {25th Low-x Meeting}, 2017.
\newblock \href {http://arxiv.org/abs/1709.01128} {\path{arXiv:1709.01128}}.

\bibitem{Celiberto:2017ydk}
F.~G. Celiberto, D.~{\relax Yu}. Ivanov, B.~Murdaca, A.~Papa, {Inclusive dihadron production at the LHC in NLA BFKL}, in: {17th conference on Elastic and Diffractive Scattering}, 2017.
\newblock \href {http://arxiv.org/abs/1709.04758} {\path{arXiv:1709.04758}}.

\bibitem{Caporale:2015vya}
F.~Caporale, G.~Chachamis, B.~Murdaca, A.~Sabio~Vera, {Balitsky-Fadin-Kuraev-Lipatov Predictions for Inclusive Three Jet Production at the LHC}, Phys. Rev. Lett. 116~(1) (2016) 012001.
\newblock \href {http://arxiv.org/abs/1508.07711} {\path{arXiv:1508.07711}}, \href {http://dx.doi.org/10.1103/PhysRevLett.116.012001} {\path{doi:10.1103/PhysRevLett.116.012001}}.

\bibitem{Caporale:2015int}
F.~Caporale, F.~G. Celiberto, G.~Chachamis, A.~Sabio~Vera, {Multi-Regge kinematics and azimuthal angle observables for inclusive four-jet production}, Eur. Phys. J. C 76~(3) (2016) 165.
\newblock \href {http://arxiv.org/abs/1512.03364} {\path{arXiv:1512.03364}}, \href {http://dx.doi.org/10.1140/epjc/s10052-016-3963-6} {\path{doi:10.1140/epjc/s10052-016-3963-6}}.

\bibitem{Caporale:2016soq}
F.~Caporale, F.~G. Celiberto, G.~Chachamis, D.~Gordo~G\'omez, A.~Sabio~Vera, {BFKL azimuthal imprints in inclusive three-jet production at 7 and 13 TeV}, Nucl. Phys. B 910 (2016) 374--386.
\newblock \href {http://arxiv.org/abs/1603.07785} {\path{arXiv:1603.07785}}, \href {http://dx.doi.org/10.1016/j.nuclphysb.2016.07.012} {\path{doi:10.1016/j.nuclphysb.2016.07.012}}.

\bibitem{Caporale:2016vxt}
F.~Caporale, F.~G. Celiberto, G.~Chachamis, A.~Sabio~Vera, {Inclusive four-jet production: a study of Multi-Regge kinematics and BFKL observables}, PoS DIS2016 (2016) 177.
\newblock \href {http://arxiv.org/abs/1610.01880} {\path{arXiv:1610.01880}}, \href {http://dx.doi.org/10.22323/1.265.0177} {\path{doi:10.22323/1.265.0177}}.

\bibitem{Caporale:2016xku}
F.~Caporale, F.~G. Celiberto, G.~Chachamis, D.~Gordo~G\'omez, A.~Sabio~Vera, {Inclusive Four-jet Production at 7 and 13 TeV: Azimuthal Profile in Multi-Regge Kinematics}, Eur. Phys. J. C 77~(1) (2017) 5.
\newblock \href {http://arxiv.org/abs/1606.00574} {\path{arXiv:1606.00574}}, \href {http://dx.doi.org/10.1140/epjc/s10052-016-4557-z} {\path{doi:10.1140/epjc/s10052-016-4557-z}}.

\bibitem{Celiberto:2016vhn}
F.~G. Celiberto, {BFKL phenomenology: resummation of high-energy logs in semi-hard processes at LHC}, Frascati Phys. Ser. 63 (2016) 43--48.
\newblock \href {http://arxiv.org/abs/1606.07327} {\path{arXiv:1606.07327}}.

\bibitem{Caporale:2016djm}
F.~Caporale, F.~G. Celiberto, G.~Chachamis, D.~Gordo~G\'omez, A.~Sabio~Vera, {Inclusive three- and four-jet production in multi-Regge kinematics at the LHC}, AIP Conf. Proc. 1819~(1) (2017) 060009.
\newblock \href {http://arxiv.org/abs/1611.04813} {\path{arXiv:1611.04813}}, \href {http://dx.doi.org/10.1063/1.4977165} {\path{doi:10.1063/1.4977165}}.

\bibitem{Caporale:2016pqe}
F.~Caporale, F.~G. Celiberto, G.~Chachamis, D.~Gordo~Gomez, B.~Murdaca, A.~Sabio~Vera, {High energy effects in multi-jet production at LHC}, in: {24th Low-x Meeting}, Vol.~5, 2017, p.~47.
\newblock \href {http://arxiv.org/abs/1610.04765} {\path{arXiv:1610.04765}}.

\bibitem{Chachamis:2016qct}
G.~Chachamis, F.~Caporale, F.~Celiberto, D.~Gomez~Gordo, A.~Sabio~Vera, {Inclusive three jet production at the LHC for 7 and 13 TeV collision energies}, PoS DIS2016 (2016) 178.
\newblock \href {http://dx.doi.org/10.22323/1.265.0178} {\path{doi:10.22323/1.265.0178}}.

\bibitem{Chachamis:2016lyi}
G.~Chachamis, F.~Caporale, F.~G. Celiberto, D.~G. Gomez, A.~Sabio~Vera, {Inclusive three jet production at the LHC at 7 and 13 TeV collision energies}, PoS DIS2016 (2016) 178.
\newblock \href {http://arxiv.org/abs/1610.01342} {\path{arXiv:1610.01342}}, \href {http://dx.doi.org/10.22323/1.265.0178} {\path{doi:10.22323/1.265.0178}}.

\bibitem{Caporale:2016lnh}
F.~Caporale, F.~G. Celiberto, G.~Chachamis, D.~Gordo~G\'omez, A.~Sabio~Vera, {Probing the BFKL dynamics in inclusive three jet production at the LHC}, EPJ Web Conf. 164 (2017) 07027.
\newblock \href {http://arxiv.org/abs/1612.02771} {\path{arXiv:1612.02771}}, \href {http://dx.doi.org/10.1051/epjconf/201716407027} {\path{doi:10.1051/epjconf/201716407027}}.

\bibitem{Caporale:2016zkc}
F.~Caporale, F.~G. Celiberto, G.~Chachamis, D.~Gordo~G\'omez, A.~Sabio~Vera, {Stability of Azimuthal-angle Observables under Higher Order Corrections in Inclusive Three-jet Production}, Phys. Rev. D 95~(7) (2017) 074007.
\newblock \href {http://arxiv.org/abs/1612.05428} {\path{arXiv:1612.05428}}, \href {http://dx.doi.org/10.1103/PhysRevD.95.074007} {\path{doi:10.1103/PhysRevD.95.074007}}.

\bibitem{Caporale:2017jqj}
F.~Caporale, F.~G. Celiberto, D.~Gordo~Gomez, A.~Sabio~Vera, G.~Chachamis, {Multi-jet production in the high energy limit at LHC}, in: {25th Low-x Meeting}, 2017.
\newblock \href {http://arxiv.org/abs/1801.00014} {\path{arXiv:1801.00014}}.

\bibitem{Chachamis:2017vfa}
G.~Chachamis, F.~Caporale, F.~G. Celiberto, D.~Gordo~Gomez, A.~Sabio~Vera, {Azimuthal-angle Observables in Inclusive Three-jet Production}, PoS DIS2017 (2018) 067.
\newblock \href {http://arxiv.org/abs/1709.02649} {\path{arXiv:1709.02649}}, \href {http://dx.doi.org/10.22323/1.297.0067} {\path{doi:10.22323/1.297.0067}}.

\bibitem{Bolognino:2019cac}
A.~D. Bolognino, F.~G. Celiberto, D.~{\relax Yu}. Ivanov, M.~M.~A. Mohammed, A.~Papa, {High-energy effects in forward inclusive dijet and hadron-jet production}, PoS DIS2019 (2019) 049.
\newblock \href {http://arxiv.org/abs/1906.11800} {\path{arXiv:1906.11800}}, \href {http://dx.doi.org/10.22323/1.352.0049} {\path{doi:10.22323/1.352.0049}}.

\bibitem{Bolognino:2019yqj}
A.~D. Bolognino, F.~G. Celiberto, D.~{\relax Yu}. Ivanov, M.~M.~A. Mohammed, A.~Papa, {Inclusive hadron-jet production at the LHC}, Acta Phys. Polon. Supp. 12~(4) (2019) 773.
\newblock \href {http://arxiv.org/abs/1902.04511} {\path{arXiv:1902.04511}}, \href {http://dx.doi.org/10.5506/APhysPolBSupp.12.773} {\path{doi:10.5506/APhysPolBSupp.12.773}}.

\bibitem{Celiberto:2020rxb}
F.~G. Celiberto, D.~{\relax Yu}. Ivanov, A.~Papa, {Diffractive production of $\Lambda$ hyperons in the high-energy limit of strong interactions}, Phys. Rev. D 102 (2020) 094019.
\newblock \href {http://arxiv.org/abs/2008.10513} {\path{arXiv:2008.10513}}, \href {http://dx.doi.org/10.1103/PhysRevD.102.094019} {\path{doi:10.1103/PhysRevD.102.094019}}.

\bibitem{Celiberto:2020tmb}
F.~G. Celiberto, D.~{\relax Yu}. Ivanov, M.~M.~A. Mohammed, A.~Papa, {High-energy resummed distributions for the inclusive Higgs-plus-jet production at the LHC}, Eur. Phys. J. C 81~(4) (2021) 293.
\newblock \href {http://arxiv.org/abs/2008.00501} {\path{arXiv:2008.00501}}, \href {http://dx.doi.org/10.1140/epjc/s10052-021-09063-2} {\path{doi:10.1140/epjc/s10052-021-09063-2}}.

\bibitem{Celiberto:2021fjf}
F.~G. Celiberto, D.~{\relax Yu}. Ivanov, M.~M.~A. Mohammed, A.~Papa, {High-energy resummation in inclusive hadroproduction of Higgs plus jet}, SciPost Phys. Proc. 8 (2022) 039.
\newblock \href {http://arxiv.org/abs/2107.13037} {\path{arXiv:2107.13037}}, \href {http://dx.doi.org/10.21468/SciPostPhysProc.8.039} {\path{doi:10.21468/SciPostPhysProc.8.039}}.

\bibitem{Celiberto:2021tky}
F.~G. Celiberto, M.~Fucilla, A.~Papa, D.~{\relax Yu}. Ivanov, M.~M.~A. Mohammed, {Higgs-plus-jet inclusive production as stabilizer of the high-energy resummation}, PoS EPS-HEP2021 (2022) 589.
\newblock \href {http://arxiv.org/abs/2110.09358} {\path{arXiv:2110.09358}}, \href {http://dx.doi.org/10.22323/1.398.0589} {\path{doi:10.22323/1.398.0589}}.

\bibitem{Celiberto:2021txb}
F.~G. Celiberto, M.~Fucilla, D.~{\relax Yu}. Ivanov, M.~M.~A. Mohammed, A.~Papa, {Higgs boson production in the high-energy limit of pQCD}, PoS PANIC2021 (2022) 352.
\newblock \href {http://arxiv.org/abs/2111.13090} {\path{arXiv:2111.13090}}, \href {http://dx.doi.org/10.22323/1.380.0352} {\path{doi:10.22323/1.380.0352}}.

\bibitem{Celiberto:2021xpm}
F.~G. Celiberto, M.~Fucilla, D.~{\relax Yu}. Ivanov, M.~M.~A. Mohammed, A.~Papa, {BFKL phenomenology: resummation of high-energy logs in inclusive processes}, SciPost Phys. Proc. 10 (2022) 002.
\newblock \href {http://arxiv.org/abs/2110.12649} {\path{arXiv:2110.12649}}, \href {http://dx.doi.org/10.21468/SciPostPhysProc.10.002} {\path{doi:10.21468/SciPostPhysProc.10.002}}.

\bibitem{Bolognino:2021hxx}
A.~D. Bolognino, F.~G. Celiberto, M.~Fucilla, D.~{\relax Yu}. Ivanov, A.~Papa, {Hybrid high-energy/collinear factorization in a heavy-light dijets system reaction}, SciPost Phys. Proc. 8 (2022) 068.
\newblock \href {http://arxiv.org/abs/2107.12120} {\path{arXiv:2107.12120}}, \href {http://dx.doi.org/10.21468/SciPostPhysProc.8.068} {\path{doi:10.21468/SciPostPhysProc.8.068}}.

\bibitem{Bolognino:2019ouc}
A.~D. Bolognino, F.~G. Celiberto, M.~Fucilla, D.~{\relax Yu}. Ivanov, B.~Murdaca, A.~Papa, {Inclusive production of two rapidity-separated heavy quarks as a probe of BFKL dynamics}, PoS DIS2019 (2019) 067.
\newblock \href {http://arxiv.org/abs/1906.05940} {\path{arXiv:1906.05940}}, \href {http://dx.doi.org/10.22323/1.352.0067} {\path{doi:10.22323/1.352.0067}}.

\bibitem{Bolognino:2022wgl}
A.~D. Bolognino, F.~G. Celiberto, M.~Fucilla, D.~{\relax Yu}. Ivanov, A.~Papa, {Heavy flavored emissions in hybrid collinear/high energy factorization}, PoS EPS-HEP2021 (2022) 389.
\newblock \href {http://arxiv.org/abs/2110.12772} {\path{arXiv:2110.12772}}, \href {http://dx.doi.org/10.22323/1.398.0389} {\path{doi:10.22323/1.398.0389}}.

\bibitem{Bolognino:2022paj}
A.~D. Bolognino, F.~G. Celiberto, M.~Fucilla, D.~{\relax Yu}. Ivanov, M.~M.~A. Mohammed, A.~Papa, {High-energy Signals from Heavy-flavor Physics}, Acta Phys. Polon. Supp. 16~(5) (2023) 17.
\newblock \href {http://arxiv.org/abs/2211.16818} {\path{arXiv:2211.16818}}, \href {http://dx.doi.org/10.5506/APhysPolBSupp.16.5-A17} {\path{doi:10.5506/APhysPolBSupp.16.5-A17}}.

\bibitem{Celiberto:2022qbh}
F.~G. Celiberto, M.~Fucilla, A.~Papa, \href{https://doi.org/10.1051/epjconf/202227000001}{{The high-energy limit of perturbative QCD: Theory and phenomenology}}, EPJ Web Conf. 270 (2022) 00001.
\newblock \href {http://arxiv.org/abs/2209.01372} {\path{arXiv:2209.01372}}, \href {http://dx.doi.org/10.1051/epjconf/202227000001} {\path{doi:10.1051/epjconf/202227000001}}.
\newline\urlprefix\url{https://doi.org/10.1051/epjconf/202227000001}

\bibitem{Celiberto:2022zdg}
F.~G. Celiberto, M.~Fucilla, M.~M.~A. Mohammed, A.~Papa, {Ultraforward production of a charmed hadron plus a Higgs boson in unpolarized proton collisions}, Phys. Rev. D 105~(11) (2022) 114056.
\newblock \href {http://arxiv.org/abs/2205.13429} {\path{arXiv:2205.13429}}, \href {http://dx.doi.org/10.1103/PhysRevD.105.114056} {\path{doi:10.1103/PhysRevD.105.114056}}.

\bibitem{Celiberto:2022kza}
F.~G. Celiberto, M.~Fucilla, {Inclusive $J/\psi$ and $\Upsilon$ emissions from single-parton fragmentation in hybrid high-energy and collinear factorization}, in: {29th International Workshop on Deep-Inelastic Scattering and Related Subjects}, 2022.
\newblock \href {http://arxiv.org/abs/2208.07206} {\path{arXiv:2208.07206}}, \href {http://dx.doi.org/10.5281/zenodo.7237044} {\path{doi:10.5281/zenodo.7237044}}.

\bibitem{Anikin:2009bf}
I.~Anikin, D.~{\relax Yu}. Ivanov, B.~Pire, L.~Szymanowski, S.~Wallon, {QCD factorization of exclusive processes beyond leading twist: $\gamma^\star_T\to\rho_T$ impact factor with twist three accuracy}, Nucl. Phys. B 828 (2010) 1--68.
\newblock \href {http://arxiv.org/abs/0909.4090} {\path{arXiv:0909.4090}}, \href {http://dx.doi.org/10.1016/j.nuclphysb.2009.10.022} {\path{doi:10.1016/j.nuclphysb.2009.10.022}}.

\bibitem{Anikin:2011sa}
I.~Anikin, A.~Besse, D.~{\relax Yu}. Ivanov, B.~Pire, L.~Szymanowski, S.~Wallon, {A phenomenological study of helicity amplitudes of high energy exclusive leptoproduction of the rho meson}, Phys. Rev. D 84 (2011) 054004.
\newblock \href {http://arxiv.org/abs/1105.1761} {\path{arXiv:1105.1761}}, \href {http://dx.doi.org/10.1103/PhysRevD.84.054004} {\path{doi:10.1103/PhysRevD.84.054004}}.

\bibitem{Besse:2013muy}
A.~Besse, L.~Szymanowski, S.~Wallon, {Saturation effects in exclusive rhoT, rhoL meson electroproduction}, JHEP 11 (2013) 062.
\newblock \href {http://arxiv.org/abs/1302.1766} {\path{arXiv:1302.1766}}, \href {http://dx.doi.org/10.1007/JHEP11(2013)062} {\path{doi:10.1007/JHEP11(2013)062}}.

\bibitem{Bolognino:2018rhb}
A.~D. Bolognino, F.~G. Celiberto, D.~{\relax Yu}. Ivanov, A.~Papa, {Unintegrated gluon distribution from forward polarized $\rho$-electroproduction}, Eur. Phys. J. C78~(12) (2018) 1023.
\newblock \href {http://arxiv.org/abs/1808.02395} {\path{arXiv:1808.02395}}, \href {http://dx.doi.org/10.1140/epjc/s10052-018-6493-6} {\path{doi:10.1140/epjc/s10052-018-6493-6}}.

\bibitem{Bolognino:2018mlw}
A.~D. Bolognino, F.~G. Celiberto, D.~{\relax Yu}. Ivanov, A.~Papa, {$\rho$-meson leptoproduction as testfield for the unintegrated gluon distribution in the proton}, Frascati Phys. Ser. 67 (2018) 76--82.
\newblock \href {http://arxiv.org/abs/1808.02958} {\path{arXiv:1808.02958}}.

\bibitem{Bolognino:2019bko}
A.~D. Bolognino, F.~G. Celiberto, D.~{\relax Yu}. Ivanov, A.~Papa, {Leptoproduction of $\rho$-mesons as discriminator for the unintegrated gluon distribution in the proton}, Acta Phys. Polon. Supp. 12~(4) (2019) 891.
\newblock \href {http://arxiv.org/abs/1902.04520} {\path{arXiv:1902.04520}}, \href {http://dx.doi.org/10.5506/APhysPolBSupp.12.891} {\path{doi:10.5506/APhysPolBSupp.12.891}}.

\bibitem{Bolognino:2019pba}
A.~D. Bolognino, A.~Szczurek, W.~Schaefer, {Exclusive production of $\phi$ meson in the $\gamma^*\,p \to \phi\,p$ reaction at large photon virtualities within $k_T$-factorization approach}, Phys. Rev. D 101~(5) (2020) 054041.
\newblock \href {http://arxiv.org/abs/1912.06507} {\path{arXiv:1912.06507}}, \href {http://dx.doi.org/10.1103/PhysRevD.101.054041} {\path{doi:10.1103/PhysRevD.101.054041}}.

\bibitem{Celiberto:2019slj}
F.~G. Celiberto, {Unraveling the Unintegrated Gluon Distribution in the Proton via $\rho$-Meson Leptoproduction}, Nuovo Cim. C42 (2019) 220.
\newblock \href {http://arxiv.org/abs/1912.11313} {\path{arXiv:1912.11313}}, \href {http://dx.doi.org/10.1393/ncc/i2019-19220-9} {\path{doi:10.1393/ncc/i2019-19220-9}}.

\bibitem{Luszczak:2022fkf}
A.~\L{}uszczak, M.~\L{}uszczak, W.~Sch\"afer, {Unintegrated gluon distributions from the color dipole cross section in the BGK saturation model}, Phys. Lett. B 835 (2022) 137582.
\newblock \href {http://arxiv.org/abs/2210.02877} {\path{arXiv:2210.02877}}, \href {http://dx.doi.org/10.1016/j.physletb.2022.137582} {\path{doi:10.1016/j.physletb.2022.137582}}.

\bibitem{Bolognino:2021niq}
A.~D. Bolognino, F.~G. Celiberto, D.~{\relax Yu}. Ivanov, A.~Papa, W.~Sch\"afer, A.~Szczurek, {Exclusive production of $\rho$-mesons in high-energy factorization at HERA and EIC}, Eur. Phys. J. C 81~(10) (2021) 846.
\newblock \href {http://arxiv.org/abs/2107.13415} {\path{arXiv:2107.13415}}, \href {http://dx.doi.org/10.1140/epjc/s10052-021-09593-9} {\path{doi:10.1140/epjc/s10052-021-09593-9}}.

\bibitem{Bolognino:2021gjm}
A.~D. Bolognino, F.~G. Celiberto, D.~{\relax Yu}. Ivanov, A.~Papa, {Exclusive emissions of rho-mesons and the unintegrated gluon distribution}, SciPost Phys. Proc. 8 (2022) 089.
\newblock \href {http://arxiv.org/abs/2107.12725} {\path{arXiv:2107.12725}}, \href {http://dx.doi.org/10.21468/SciPostPhysProc.8.089} {\path{doi:10.21468/SciPostPhysProc.8.089}}.

\bibitem{Bolognino:2022uty}
A.~D. Bolognino, F.~G. Celiberto, M.~Fucilla, D.~{\relax Yu}. Ivanov, A.~Papa, W.~Sch\"afer, A.~Szczurek, {Hadron structure at small-x via unintegrated gluon densities}, Rev. Mex. Fis. Suppl. 3~(3) (2022) 0308109.
\newblock \href {http://arxiv.org/abs/2202.02513} {\path{arXiv:2202.02513}}, \href {http://dx.doi.org/10.31349/SuplRevMexFis.3.0308109} {\path{doi:10.31349/SuplRevMexFis.3.0308109}}.

\bibitem{Celiberto:2022fam}
F.~G. Celiberto, {Phenomenology of the hadronic structure at small-$x$}, 2022.
\newblock \href {http://arxiv.org/abs/2202.04207} {\path{arXiv:2202.04207}}.

\bibitem{Bolognino:2022ndh}
A.~D. Bolognino, F.~G. Celiberto, D.~{\relax Yu.}. Ivanov, A.~Papa, W.~Sch\"afer, A.~Szczurek, {Exclusive emissions of polarized $\rho$ mesons at the EIC and the proton content at low $x$}, in: {29th International Workshop on Deep-Inelastic Scattering and Related Subjects}, 2022.
\newblock \href {http://arxiv.org/abs/2207.05726} {\path{arXiv:2207.05726}}, \href {http://dx.doi.org/10.5281/zenodo.7112750} {\path{doi:10.5281/zenodo.7112750}}.

\bibitem{Bautista:2016xnp}
I.~Bautista, A.~Fernandez~Tellez, M.~Hentschinski, {BFKL evolution and the growth with energy of exclusive $J/\psi$ and $\Upsilon$ photoproduction cross sections}, Phys. Rev. D 94~(5) (2016) 054002.
\newblock \href {http://arxiv.org/abs/1607.05203} {\path{arXiv:1607.05203}}, \href {http://dx.doi.org/10.1103/PhysRevD.94.054002} {\path{doi:10.1103/PhysRevD.94.054002}}.

\bibitem{Garcia:2019tne}
A.~Arroyo~Garcia, M.~Hentschinski, K.~Kutak, {QCD evolution based evidence for the onset of gluon saturation in exclusive photo-production of vector mesons}, Phys. Lett. B 795 (2019) 569--575.
\newblock \href {http://arxiv.org/abs/1904.04394} {\path{arXiv:1904.04394}}, \href {http://dx.doi.org/10.1016/j.physletb.2019.06.061} {\path{doi:10.1016/j.physletb.2019.06.061}}.

\bibitem{Hentschinski:2020yfm}
M.~Hentschinski, E.~Padr\'on~Molina, {Exclusive $J/\Psi$ and $\Psi(2s)$ photo-production as a probe of QCD low $x$ evolution equations}, Phys. Rev. D 103~(7) (2021) 074008.
\newblock \href {http://arxiv.org/abs/2011.02640} {\path{arXiv:2011.02640}}, \href {http://dx.doi.org/10.1103/PhysRevD.103.074008} {\path{doi:10.1103/PhysRevD.103.074008}}.

\bibitem{Peredo:2023oym}
M.~A. Peredo, M.~Hentschinski, {Ratio of $J/\Psi$ and $\Psi(2s)$ exclusive photoproduction cross-sections as an indicator for the presence of non-linear QCD evolution}, Phys. Rev. D 109~(1) (2024) 014032.
\newblock \href {http://arxiv.org/abs/2308.15430} {\path{arXiv:2308.15430}}, \href {http://dx.doi.org/10.1103/PhysRevD.109.014032} {\path{doi:10.1103/PhysRevD.109.014032}}.

\bibitem{Hentschinski:2025ovo}
M.~Hentschinski, R.~R. Ram{\'\i}rez, {The energy dependence of exclusive heavy vector meson photoproduction cross-sections and NLO BFKL evolution}\href {http://arxiv.org/abs/2508.11545} {\path{arXiv:2508.11545}}.

\bibitem{Brzeminski:2016lwh}
D.~Brzeminski, L.~Motyka, M.~Sadzikowski, T.~Stebel, {Twist decomposition of Drell-Yan structure functions: phenomenological implications}, JHEP 01 (2017) 005.
\newblock \href {http://arxiv.org/abs/1611.04449} {\path{arXiv:1611.04449}}, \href {http://dx.doi.org/10.1007/JHEP01(2017)005} {\path{doi:10.1007/JHEP01(2017)005}}.

\bibitem{Motyka:2016lta}
L.~Motyka, M.~Sadzikowski, T.~Stebel, {Lam-Tung relation breaking in $Z^0$ hadroproduction as a probe of parton transverse momentum}, Phys. Rev. D95~(11) (2017) 114025.
\newblock \href {http://arxiv.org/abs/1609.04300} {\path{arXiv:1609.04300}}, \href {http://dx.doi.org/10.1103/PhysRevD.95.114025} {\path{doi:10.1103/PhysRevD.95.114025}}.

\bibitem{Celiberto:2018muu}
F.~G. Celiberto, D.~Gordo~G\'omez, A.~Sabio~Vera, {Forward Drell-Yan production at the LHC in the BFKL formalism with collinear corrections}, Phys. Lett. B786 (2018) 201--206.
\newblock \href {http://arxiv.org/abs/1808.09511} {\path{arXiv:1808.09511}}, \href {http://dx.doi.org/10.1016/j.physletb.2018.09.045} {\path{doi:10.1016/j.physletb.2018.09.045}}.

\bibitem{Chachamis:2015ona}
G.~Chachamis, M.~Deak, M.~Hentschinski, G.~Rodrigo, A.~Sabio~Vera, {Single bottom quark production in k$_{\perp}$-factorisation}, JHEP 09 (2015) 123.
\newblock \href {http://arxiv.org/abs/1507.05778} {\path{arXiv:1507.05778}}, \href {http://dx.doi.org/10.1007/JHEP09(2015)123} {\path{doi:10.1007/JHEP09(2015)123}}.

\bibitem{Chachamis:2013bwa}
G.~Chachamis, M.~Deak, G.~Rodrigo, {Heavy quark impact factor in kT-factorization}, JHEP 12 (2013) 066.
\newblock \href {http://arxiv.org/abs/1310.6611} {\path{arXiv:1310.6611}}, \href {http://dx.doi.org/10.1007/JHEP12(2013)066} {\path{doi:10.1007/JHEP12(2013)066}}.

\bibitem{Chachamis:2009ks}
G.~Chachamis, M.~Hentschinski, A.~Sabio~Vera, C.~Salas, {Exclusive central production of heavy quarks at the LHC}\href {http://arxiv.org/abs/0911.2662} {\path{arXiv:0911.2662}}.

\bibitem{Ball:2017otu}
R.~D. Ball, V.~Bertone, M.~Bonvini, S.~Marzani, J.~Rojo, L.~Rottoli, {Parton distributions with small-x resummation: evidence for BFKL dynamics in HERA data}, Eur. Phys. J. C78~(4) (2018) 321.
\newblock \href {http://arxiv.org/abs/1710.05935} {\path{arXiv:1710.05935}}, \href {http://dx.doi.org/10.1140/epjc/s10052-018-5774-4} {\path{doi:10.1140/epjc/s10052-018-5774-4}}.

\bibitem{Abdolmaleki:2018jln}
H.~Abdolmaleki, et~al., {Impact of low-$x$ resummation on QCD analysis of HERA data}, Eur. Phys. J. C 78~(8) (2018) 621.
\newblock \href {http://arxiv.org/abs/1802.00064} {\path{arXiv:1802.00064}}, \href {http://dx.doi.org/10.1140/epjc/s10052-018-6090-8} {\path{doi:10.1140/epjc/s10052-018-6090-8}}.

\bibitem{Bonvini:2019wxf}
M.~Bonvini, F.~Giuli, {A new simple PDF parametrization: improved description of the HERA data}, Eur. Phys. J. Plus 134~(10) (2019) 531.
\newblock \href {http://arxiv.org/abs/1902.11125} {\path{arXiv:1902.11125}}, \href {http://dx.doi.org/10.1140/epjp/i2019-12872-x} {\path{doi:10.1140/epjp/i2019-12872-x}}.

\bibitem{Silvetti:2022hyc}
F.~Silvetti, M.~Bonvini, {Differential heavy quark pair production at small x}, Eur. Phys. J. C 83~(4) (2023) 267.
\newblock \href {http://arxiv.org/abs/2211.10142} {\path{arXiv:2211.10142}}, \href {http://dx.doi.org/10.1140/epjc/s10052-023-11326-z} {\path{doi:10.1140/epjc/s10052-023-11326-z}}.

\bibitem{Silvetti:2023suu}
F.~Silvetti, {Resummation phenomenology and PDF determination for precision QCD at the LHC}, Phd thesis, Rome U. (2023).
\newblock \href {http://arxiv.org/abs/2403.20315} {\path{arXiv:2403.20315}}.

\bibitem{Rinaudo:2024hdb}
A.~Rinaudo, {Towards the resummation of high-energy next-to-leading logarithms in QCD}, Phd thesis, Genoa U. (2024).
\newblock \href {http://dx.doi.org/10.15167/rinaudo-anna_phd2024-06-14} {\path{doi:10.15167/rinaudo-anna_phd2024-06-14}}.

\bibitem{Celiberto:2025nnq}
F.~G. Celiberto, M.~Bonvini, {Towards Precision Gluon Densities at Small xx: From Resummation to Collider Observables}, PoS EPS-HEP2025 (2026) 222.
\newblock \href {http://arxiv.org/abs/2512.01961} {\path{arXiv:2512.01961}}, \href {http://dx.doi.org/10.22323/1.485.0222} {\path{doi:10.22323/1.485.0222}}.

\bibitem{Celiberto:2021zww}
F.~G. Celiberto, {3D tomography of the nucleon: transverse-momentum-dependent gluon distributions}, Nuovo Cim. C44 (2021) 36.
\newblock \href {http://arxiv.org/abs/2101.04630} {\path{arXiv:2101.04630}}, \href {http://dx.doi.org/10.1393/ncc/i2021-21036-3} {\path{doi:10.1393/ncc/i2021-21036-3}}.

\bibitem{Bacchetta:2021oht}
A.~Bacchetta, F.~G. Celiberto, M.~Radici, P.~Taels, {A spectator-model way to transverse-momentum-dependent gluon distribution functions}, SciPost Phys. Proc. 8 (2022) 040.
\newblock \href {http://arxiv.org/abs/2107.13446} {\path{arXiv:2107.13446}}, \href {http://dx.doi.org/10.21468/SciPostPhysProc.8.040} {\path{doi:10.21468/SciPostPhysProc.8.040}}.

\bibitem{Bacchetta:2021lvw}
A.~Bacchetta, F.~G. Celiberto, M.~Radici, {Toward twist-2 $T$-odd transverse-momentum-dependent gluon distributions: the $f$-type Sivers function}, PoS EPS-HEP2021 (2022) 376.
\newblock \href {http://arxiv.org/abs/2111.01686} {\path{arXiv:2111.01686}}, \href {http://dx.doi.org/10.22323/1.398.0376} {\path{doi:10.22323/1.398.0376}}.

\bibitem{Bacchetta:2021twk}
A.~Bacchetta, F.~G. Celiberto, M.~Radici, {Toward twist-2 $T$-odd transverse-momentum-dependent gluon distributions: the $f$-type linearity function}, PoS PANIC2021 (2022) 378.
\newblock \href {http://arxiv.org/abs/2111.03567} {\path{arXiv:2111.03567}}, \href {http://dx.doi.org/10.22323/1.380.0378} {\path{doi:10.22323/1.380.0378}}.

\bibitem{Bacchetta:2022esb}
A.~Bacchetta, F.~G. Celiberto, M.~Radici, {Towards Leading-twist T-odd TMD Gluon Distributions}, JPS Conf. Proc. 37 (2022) 020124.
\newblock \href {http://arxiv.org/abs/2201.10508} {\path{arXiv:2201.10508}}, \href {http://dx.doi.org/10.7566/JPSCP.37.020124} {\path{doi:10.7566/JPSCP.37.020124}}.

\bibitem{Bacchetta:2022crh}
A.~Bacchetta, F.~G. Celiberto, M.~Radici, {Unveiling the proton structure via transverse-momentum-dependent gluon distributions}, Rev. Mex. Fis. Suppl. 3~(3) (2022) 0308108.
\newblock \href {http://arxiv.org/abs/2206.07815} {\path{arXiv:2206.07815}}, \href {http://dx.doi.org/10.31349/SuplRevMexFis.3.0308108} {\path{doi:10.31349/SuplRevMexFis.3.0308108}}.

\bibitem{Bacchetta:2022nyv}
A.~Bacchetta, F.~G. Celiberto, M.~Radici, A.~Signori, {Phenomenology of gluon TMDs from $\eta_{b,c}$ production}, in: {29th International Workshop on Deep-Inelastic Scattering and Related Subjects}, 2022.
\newblock \href {http://arxiv.org/abs/2208.06252} {\path{arXiv:2208.06252}}, \href {http://dx.doi.org/10.5281/zenodo.7085045} {\path{doi:10.5281/zenodo.7085045}}.

\bibitem{Celiberto:2022omz}
F.~G. Celiberto, {A Journey into the Proton Structure: Progresses and Challenges}, Universe 8~(12) (2022) 661.
\newblock \href {http://arxiv.org/abs/2210.08322} {\path{arXiv:2210.08322}}, \href {http://dx.doi.org/10.3390/universe8120661} {\path{doi:10.3390/universe8120661}}.

\bibitem{Bacchetta:2023zir}
A.~Bacchetta, F.~G. Celiberto, M.~Radici, {Spectator-model studies for spin-dependent gluon TMD PDFs at the LHC and EIC}, PoS EPS-HEP2023 (2024) 247.
\newblock \href {http://arxiv.org/abs/2310.19916} {\path{arXiv:2310.19916}}, \href {http://dx.doi.org/10.22323/1.449.0247} {\path{doi:10.22323/1.449.0247}}.

\bibitem{Bacchetta:2024uxb}
A.~Bacchetta, F.~G. Celiberto, M.~Radici, {Proton 3D reconstruction with T-odd TMD gluon densities}, PoS SPIN2023 (2024) 049.
\newblock \href {http://arxiv.org/abs/2406.04893} {\path{arXiv:2406.04893}}, \href {http://dx.doi.org/10.22323/1.456.0049} {\path{doi:10.22323/1.456.0049}}.

\bibitem{Hentschinski:2021lsh}
M.~Hentschinski, {Transverse momentum dependent gluon distribution within high energy factorization at next-to-leading order}, Phys. Rev. D 104~(5) (2021) 054014.
\newblock \href {http://arxiv.org/abs/2107.06203} {\path{arXiv:2107.06203}}, \href {http://dx.doi.org/10.1103/PhysRevD.104.054014} {\path{doi:10.1103/PhysRevD.104.054014}}.

\bibitem{Mukherjee:2023snp}
S.~Mukherjee, V.~V. Skokov, A.~Tarasov, S.~Tiwari, {Unified description of DGLAP, CSS, and BFKL evolution: TMD factorization bridging large and small x}, Phys. Rev. D 109~(3) (2024) 034035.
\newblock \href {http://arxiv.org/abs/2311.16402} {\path{arXiv:2311.16402}}, \href {http://dx.doi.org/10.1103/PhysRevD.109.034035} {\path{doi:10.1103/PhysRevD.109.034035}}.

\bibitem{Boroun:2023goy}
G.~R. Boroun, {Dipole cross section from the unintegrated gluon distribution at small x}, Phys. Rev. D 108~(3) (2023) 034025.
\newblock \href {http://arxiv.org/abs/2301.01083} {\path{arXiv:2301.01083}}, \href {http://dx.doi.org/10.1103/PhysRevD.108.034025} {\path{doi:10.1103/PhysRevD.108.034025}}.

\bibitem{Boroun:2023ldq}
G.~R. Boroun, {The unintegrated gluon distribution from the GBW and BGK models}, Eur. Phys. J. A 60~(3) (2024) 48.
\newblock \href {http://arxiv.org/abs/2309.04832} {\path{arXiv:2309.04832}}, \href {http://dx.doi.org/10.1140/epja/s10050-024-01255-0} {\path{doi:10.1140/epja/s10050-024-01255-0}}.

\bibitem{Bonvini:2018ixe}
M.~Bonvini, S.~Marzani, {Double resummation for Higgs production}, Phys. Rev. Lett. 120~(20) (2018) 202003.
\newblock \href {http://arxiv.org/abs/1802.07758} {\path{arXiv:1802.07758}}, \href {http://dx.doi.org/10.1103/PhysRevLett.120.202003} {\path{doi:10.1103/PhysRevLett.120.202003}}.

\bibitem{Ball:1995vc}
R.~D. Ball, S.~Forte, {Summation of leading logarithms at small x}, Phys. Lett. B 351 (1995) 313--324.
\newblock \href {http://arxiv.org/abs/hep-ph/9501231} {\path{arXiv:hep-ph/9501231}}, \href {http://dx.doi.org/10.1016/0370-2693(95)00395-2} {\path{doi:10.1016/0370-2693(95)00395-2}}.

\bibitem{Ball:1997vf}
R.~D. Ball, S.~Forte, {Asymptotically free partons at high-energy}, Phys. Lett. B 405 (1997) 317--326.
\newblock \href {http://arxiv.org/abs/hep-ph/9703417} {\path{arXiv:hep-ph/9703417}}, \href {http://dx.doi.org/10.1016/S0370-2693(97)00625-4} {\path{doi:10.1016/S0370-2693(97)00625-4}}.

\bibitem{Altarelli:2001ji}
G.~Altarelli, R.~D. Ball, S.~Forte, {Factorization and resummation of small x scaling violations with running coupling}, Nucl. Phys. B 621 (2002) 359--387.
\newblock \href {http://arxiv.org/abs/hep-ph/0109178} {\path{arXiv:hep-ph/0109178}}, \href {http://dx.doi.org/10.1016/S0550-3213(01)00563-6} {\path{doi:10.1016/S0550-3213(01)00563-6}}.

\bibitem{Altarelli:2003hk}
G.~Altarelli, R.~D. Ball, S.~Forte, {An Anomalous dimension for small x evolution}, Nucl. Phys. B 674 (2003) 459--483.
\newblock \href {http://arxiv.org/abs/hep-ph/0306156} {\path{arXiv:hep-ph/0306156}}, \href {http://dx.doi.org/10.1016/j.nuclphysb.2003.09.040} {\path{doi:10.1016/j.nuclphysb.2003.09.040}}.

\bibitem{Altarelli:2005ni}
G.~Altarelli, R.~D. Ball, S.~Forte, {Perturbatively stable resummed small x evolution kernels}, Nucl. Phys. B 742 (2006) 1--40.
\newblock \href {http://arxiv.org/abs/hep-ph/0512237} {\path{arXiv:hep-ph/0512237}}, \href {http://dx.doi.org/10.1016/j.nuclphysb.2006.01.046} {\path{doi:10.1016/j.nuclphysb.2006.01.046}}.

\bibitem{Altarelli:2008aj}
G.~Altarelli, R.~D. Ball, S.~Forte, {Small x Resummation with Quarks: Deep-Inelastic Scattering}, Nucl. Phys. B 799 (2008) 199--240.
\newblock \href {http://arxiv.org/abs/0802.0032} {\path{arXiv:0802.0032}}, \href {http://dx.doi.org/10.1016/j.nuclphysb.2008.03.003} {\path{doi:10.1016/j.nuclphysb.2008.03.003}}.

\bibitem{White:2006yh}
C.~White, R.~Thorne, {A Global Fit to Scattering Data with NLL BFKL Resummations}, Phys. Rev. D 75 (2007) 034005.
\newblock \href {http://arxiv.org/abs/hep-ph/0611204} {\path{arXiv:hep-ph/0611204}}, \href {http://dx.doi.org/10.1103/PhysRevD.75.034005} {\path{doi:10.1103/PhysRevD.75.034005}}.

\bibitem{Catani:1990xk}
S.~Catani, M.~Ciafaloni, F.~Hautmann, {GLUON CONTRIBUTIONS TO SMALL x HEAVY FLAVOR PRODUCTION}, Phys. Lett. B 242 (1990) 97--102.
\newblock \href {http://dx.doi.org/10.1016/0370-2693(90)91601-7} {\path{doi:10.1016/0370-2693(90)91601-7}}.

\bibitem{Catani:1990eg}
S.~Catani, M.~Ciafaloni, F.~Hautmann, {High-energy factorization and small x heavy flavor production}, Nucl. Phys. B 366 (1991) 135--188.
\newblock \href {http://dx.doi.org/10.1016/0550-3213(91)90055-3} {\path{doi:10.1016/0550-3213(91)90055-3}}.

\bibitem{Collins:1991ty}
J.~C. Collins, R.~Ellis, {Heavy quark production in very high-energy hadron collisions}, Nucl. Phys. B 360 (1991) 3--30.
\newblock \href {http://dx.doi.org/10.1016/0550-3213(91)90288-9} {\path{doi:10.1016/0550-3213(91)90288-9}}.

\bibitem{Catani:1993ww}
S.~Catani, M.~Ciafaloni, F.~Hautmann, {High-energy factorization in QCD and minimal subtraction scheme}, Phys. Lett. B 307 (1993) 147--153.
\newblock \href {http://dx.doi.org/10.1016/0370-2693(93)90204-U} {\path{doi:10.1016/0370-2693(93)90204-U}}.

\bibitem{Catani:1993rn}
S.~Catani, F.~Hautmann, {Quark anomalous dimensions at small x}, Phys. Lett. B 315 (1993) 157--163.
\newblock \href {http://dx.doi.org/10.1016/0370-2693(93)90174-G} {\path{doi:10.1016/0370-2693(93)90174-G}}.

\bibitem{Catani:1994sq}
S.~Catani, F.~Hautmann, {High-energy factorization and small x deep inelastic scattering beyond leading order}, Nucl. Phys. B 427 (1994) 475--524.
\newblock \href {http://arxiv.org/abs/hep-ph/9405388} {\path{arXiv:hep-ph/9405388}}, \href {http://dx.doi.org/10.1016/0550-3213(94)90636-X} {\path{doi:10.1016/0550-3213(94)90636-X}}.

\bibitem{Ball:2007ra}
R.~D. Ball, {Resummation of Hadroproduction Cross-sections at High Energy}, Nucl. Phys. B 796 (2008) 137--183.
\newblock \href {http://arxiv.org/abs/0708.1277} {\path{arXiv:0708.1277}}, \href {http://dx.doi.org/10.1016/j.nuclphysb.2007.12.014} {\path{doi:10.1016/j.nuclphysb.2007.12.014}}.

\bibitem{Caola:2010kv}
F.~Caola, S.~Forte, S.~Marzani, {Small x resummation of rapidity distributions: The Case of Higgs production}, Nucl. Phys. B 846 (2011) 167--211.
\newblock \href {http://arxiv.org/abs/1010.2743} {\path{arXiv:1010.2743}}, \href {http://dx.doi.org/10.1016/j.nuclphysb.2011.01.001} {\path{doi:10.1016/j.nuclphysb.2011.01.001}}.

\bibitem{Brodsky:1996sg}
S.~J. Brodsky, F.~Hautmann, D.~E. Soper, {Probing the QCD pomeron in e+ e- collisions}, Phys. Rev. Lett. 78 (1997) 803--806, [Erratum: Phys.Rev.Lett. 79, 3544 (1997)].
\newblock \href {http://arxiv.org/abs/hep-ph/9610260} {\path{arXiv:hep-ph/9610260}}, \href {http://dx.doi.org/10.1103/PhysRevLett.78.803} {\path{doi:10.1103/PhysRevLett.78.803}}.

\bibitem{Brodsky:1997sd}
S.~J. Brodsky, F.~Hautmann, D.~E. Soper, {Virtual photon scattering at high-energies as a probe of the short distance pomeron}, Phys. Rev. D 56 (1997) 6957--6979.
\newblock \href {http://arxiv.org/abs/hep-ph/9706427} {\path{arXiv:hep-ph/9706427}}, \href {http://dx.doi.org/10.1103/PhysRevD.56.6957} {\path{doi:10.1103/PhysRevD.56.6957}}.

\bibitem{Brodsky:1998kn}
S.~J. Brodsky, V.~S. Fadin, V.~T. Kim, L.~N. Lipatov, G.~B. Pivovarov, {The QCD pomeron with optimal renormalization}, JETP Lett. 70 (1999) 155--160.
\newblock \href {http://arxiv.org/abs/hep-ph/9901229} {\path{arXiv:hep-ph/9901229}}, \href {http://dx.doi.org/10.1134/1.568145} {\path{doi:10.1134/1.568145}}.

\bibitem{Brodsky:2002ka}
S.~J. Brodsky, V.~S. Fadin, V.~T. Kim, L.~N. Lipatov, G.~B. Pivovarov, {High-energy QCD asymptotics of photon-photon collisions}, JETP Lett. 76 (2002) 249--252.
\newblock \href {http://arxiv.org/abs/hep-ph/0207297} {\path{arXiv:hep-ph/0207297}}, \href {http://dx.doi.org/10.1134/1.1520615} {\path{doi:10.1134/1.1520615}}.

\bibitem{Celiberto:2023rtu}
F.~G. Celiberto, A.~Papa, {The high-energy QCD dynamics from Higgs-plus-jet correlations at the FCC}\href {http://arxiv.org/abs/2305.00962} {\path{arXiv:2305.00962}}.

\bibitem{Celiberto:2023uuk}
F.~G. Celiberto, L.~Delle~Rose, M.~Fucilla, G.~Gatto, A.~Papa, {High-energy resummed Higgs-plus-jet distributions at NLL/NLO* with \textsc{Powheg}+\textsc{Jethad}}, in: {57th Rencontres de Moriond on QCD and High Energy Interactions}, 2023.
\newblock \href {http://arxiv.org/abs/2305.05052} {\path{arXiv:2305.05052}}.

\bibitem{Celiberto:2023eba}
F.~G. Celiberto, L.~Delle~Rose, M.~Fucilla, G.~Gatto, A.~Papa, {NLL/NLO$^-$ studies on Higgs-plus-jet production with POWHEG+JETHAD}, PoS RADCOR2023 (2024) 069.
\newblock \href {http://arxiv.org/abs/2309.11573} {\path{arXiv:2309.11573}}, \href {http://dx.doi.org/10.22323/1.432.0069} {\path{doi:10.22323/1.432.0069}}.

\bibitem{Celiberto:2023nym}
F.~G. Celiberto, L.~Delle~Rose, M.~Fucilla, G.~Gatto, A.~Papa, {Towards high-energy Higgs+jet distributions at NLL matched to NLO}, PoS EPS-HEP2023 (2024) 390.
\newblock \href {http://arxiv.org/abs/2310.16967} {\path{arXiv:2310.16967}}, \href {http://dx.doi.org/10.22323/1.449.0390} {\path{doi:10.22323/1.449.0390}}.

\bibitem{Celiberto:2023dkr}
F.~G. Celiberto, M.~Fucilla, D.~{\relax Yu}. Ivanov, M.~M.~A. Mohammed, A.~Papa, {Higgs boson production at next-to-leading logarithmic accuracy}, in: {57th Rencontres de Moriond on QCD and High Energy Interactions}, 2023.
\newblock \href {http://arxiv.org/abs/2305.11760} {\path{arXiv:2305.11760}}.

\bibitem{Celiberto:2023rqp}
F.~G. Celiberto, M.~Fucilla, M.~M.~A. Mohammed, D.~{\relax Yu}. Ivanov, A.~Papa, {High-energy resummation in Higgs production at the next-to-leading order}, PoS RADCOR2023 (2024) 091.
\newblock \href {http://arxiv.org/abs/2309.07570} {\path{arXiv:2309.07570}}, \href {http://dx.doi.org/10.22323/1.432.0091} {\path{doi:10.22323/1.432.0091}}.

\bibitem{Celiberto:2024mdt}
F.~G. Celiberto, L.~Delle~Rose, M.~Fucilla, G.~Gatto, A.~Papa, {Towards Higgs and Z boson plus jet distributions at NLL/NLO+}, PoS DIS2024 (2025) 126.
\newblock \href {http://arxiv.org/abs/2408.08757} {\path{arXiv:2408.08757}}, \href {http://dx.doi.org/10.22323/1.469.0126} {\path{doi:10.22323/1.469.0126}}.

\bibitem{Celiberto:2024bfu}
F.~G. Celiberto, L.~Delle~Rose, M.~Fucilla, G.~Gatto, D.~Y. Ivanov, M.~M.~A. Mohammed, A.~Papa, {Higgs production at NLL accuracy in the BFKL approach}, PoS DIS2024 (2025) 101.
\newblock \href {http://arxiv.org/abs/2408.08731} {\path{arXiv:2408.08731}}, \href {http://dx.doi.org/10.22323/1.469.0101} {\path{doi:10.22323/1.469.0101}}.

\bibitem{Celiberto:2025edg}
F.~G. Celiberto, L.~Delle~Rose, A.~Papa, {Matching NLL to NLO in Higgs and Z plus jet at the LHC and FCC}, PoS EPS-HEP2025 (2026) 352.
\newblock \href {http://arxiv.org/abs/2510.22445} {\path{arXiv:2510.22445}}, \href {http://dx.doi.org/10.22323/1.485.0352} {\path{doi:10.22323/1.485.0352}}.

\bibitem{PhysRevD.18.3998}
W.~A. Bardeen, A.~J. Buras, D.~W. Duke, T.~Muta, Deep-inelastic scattering beyond the leading order in asymptotically free gauge theories, Phys. Rev. D 18 (1978) 3998--4017.
\newblock \href {http://dx.doi.org/10.1103/PhysRevD.18.3998} {\path{doi:10.1103/PhysRevD.18.3998}}.

\bibitem{Yang:2011rp}
Y.~Yang, J.~Ping, C.~Deng, H.-S. Zong, {Possible interpretation of the $Z_b$(10610) and $Z_b$(10650) in a chiral quark model}, J. Phys. G 39 (2012) 105001.
\newblock \href {http://arxiv.org/abs/1105.5935} {\path{arXiv:1105.5935}}, \href {http://dx.doi.org/10.1088/0954-3899/39/10/105001} {\path{doi:10.1088/0954-3899/39/10/105001}}.

\bibitem{Zhou:2025fpp}
H.~Zhou, S.-Q. Luo, X.~Liu, {Triply heavy baryon spectroscopy revisited}, Phys. Rev. D 112~(7) (2025) 074007.
\newblock \href {http://arxiv.org/abs/2507.10243} {\path{arXiv:2507.10243}}, \href {http://dx.doi.org/10.1103/jhr1-ccsw} {\path{doi:10.1103/jhr1-ccsw}}.

\bibitem{Kotikov:2000pm}
A.~V. Kotikov, L.~N. Lipatov, {NLO corrections to the BFKL equation in QCD and in supersymmetric gauge theories}, Nucl. Phys. B 582 (2000) 19--43.
\newblock \href {http://arxiv.org/abs/hep-ph/0004008} {\path{arXiv:hep-ph/0004008}}, \href {http://dx.doi.org/10.1016/S0550-3213(00)00329-1} {\path{doi:10.1016/S0550-3213(00)00329-1}}.

\bibitem{Kotikov:2002ab}
A.~V. Kotikov, L.~N. Lipatov, {DGLAP and BFKL equations in the $N=4$ supersymmetric gauge theory}, Nucl. Phys. B 661 (2003) 19--61, [Erratum: Nucl.Phys.B 685, 405--407 (2004)].
\newblock \href {http://arxiv.org/abs/hep-ph/0208220} {\path{arXiv:hep-ph/0208220}}, \href {http://dx.doi.org/10.1016/S0550-3213(03)00264-5} {\path{doi:10.1016/S0550-3213(03)00264-5}}.

\bibitem{Chekanov:2002rq}
S.~V. Chekanov, {Jet algorithms: A Minireview}, in: {14th Topical Conference on Hadron Collider Physics (HCP 2002)}, 2002, pp. 478--486.
\newblock \href {http://arxiv.org/abs/hep-ph/0211298} {\path{arXiv:hep-ph/0211298}}.

\bibitem{Salam:2010nqg}
G.~P. Salam, {Towards Jetography}, Eur. Phys. J. C 67 (2010) 637--686.
\newblock \href {http://arxiv.org/abs/0906.1833} {\path{arXiv:0906.1833}}, \href {http://dx.doi.org/10.1140/epjc/s10052-010-1314-6} {\path{doi:10.1140/epjc/s10052-010-1314-6}}.

\bibitem{Catani:1993hr}
S.~Catani, Y.~L. Dokshitzer, M.~H. Seymour, B.~R. Webber, {Longitudinally invariant $K_t$ clustering algorithms for hadron hadron collisions}, Nucl. Phys. B 406 (1993) 187--224.
\newblock \href {http://dx.doi.org/10.1016/0550-3213(93)90166-M} {\path{doi:10.1016/0550-3213(93)90166-M}}.

\bibitem{Cacciari:2008gp}
M.~Cacciari, G.~P. Salam, G.~Soyez, {The anti-$k_t$ jet clustering algorithm}, JHEP 04 (2008) 063.
\newblock \href {http://arxiv.org/abs/0802.1189} {\path{arXiv:0802.1189}}, \href {http://dx.doi.org/10.1088/1126-6708/2008/04/063} {\path{doi:10.1088/1126-6708/2008/04/063}}.

\bibitem{Furman:1981kf}
M.~Furman, {Study of a Nonleading \{QCD\} Correction to Hadron Calorimeter Reactions}, Nucl. Phys. B 197 (1982) 413--445.
\newblock \href {http://dx.doi.org/10.1016/0550-3213(82)90452-7} {\path{doi:10.1016/0550-3213(82)90452-7}}.

\bibitem{Aversa:1988vb}
F.~Aversa, P.~Chiappetta, M.~Greco, J.~P. Guillet, {QCD Corrections to Parton-Parton Scattering Processes}, Nucl. Phys. B 327 (1989) 105.
\newblock \href {http://dx.doi.org/10.1016/0550-3213(89)90288-5} {\path{doi:10.1016/0550-3213(89)90288-5}}.

\bibitem{NNPDF:2021uiq}
R.~D. Ball, et~al., {An open-source machine learning framework for global analyses of parton distributions}, Eur. Phys. J. C 81~(10) (2021) 958.
\newblock \href {http://arxiv.org/abs/2109.02671} {\path{arXiv:2109.02671}}.

\bibitem{NNPDF:2021njg}
R.~D. Ball, et~al., {The path to proton structure at 1\% accuracy}, Eur. Phys. J. C 82~(5) (2022) 428.
\newblock \href {http://arxiv.org/abs/2109.02653} {\path{arXiv:2109.02653}}, \href {http://dx.doi.org/10.1140/epjc/s10052-022-10328-7} {\path{doi:10.1140/epjc/s10052-022-10328-7}}.

\bibitem{Forte:2002fg}
S.~Forte, L.~Garrido, J.~I. Latorre, A.~Piccione, {Neural network parametrization of deep inelastic structure functions}, JHEP 05 (2002) 062.
\newblock \href {http://arxiv.org/abs/hep-ph/0204232} {\path{arXiv:hep-ph/0204232}}, \href {http://dx.doi.org/10.1088/1126-6708/2002/05/062} {\path{doi:10.1088/1126-6708/2002/05/062}}.

\bibitem{Bacchetta:2017gcc}
A.~Bacchetta, F.~Delcarro, C.~Pisano, M.~Radici, A.~Signori, {Extraction of partonic transverse momentum distributions from semi-inclusive deep-inelastic scattering, Drell-Yan and Z-boson production}, JHEP 06 (2017) 081, [Erratum: JHEP 06, 051 (2019)].
\newblock \href {http://arxiv.org/abs/1703.10157} {\path{arXiv:1703.10157}}, \href {http://dx.doi.org/10.1007/JHEP06(2017)081} {\path{doi:10.1007/JHEP06(2017)081}}.

\bibitem{Scimemi:2019cmh}
I.~Scimemi, A.~Vladimirov, {Non-perturbative structure of semi-inclusive deep-inelastic and Drell-Yan scattering at small transverse momentum}, JHEP 06 (2020) 137.
\newblock \href {http://arxiv.org/abs/1912.06532} {\path{arXiv:1912.06532}}, \href {http://dx.doi.org/10.1007/JHEP06(2020)137} {\path{doi:10.1007/JHEP06(2020)137}}.

\bibitem{Bacchetta:2019sam}
A.~Bacchetta, V.~Bertone, C.~Bissolotti, G.~Bozzi, F.~Delcarro, F.~Piacenza, M.~Radici, {Transverse-momentum-dependent parton distributions up to N$^{3}$LL from Drell-Yan data}, JHEP 07 (2020) 117.
\newblock \href {http://arxiv.org/abs/1912.07550} {\path{arXiv:1912.07550}}, \href {http://dx.doi.org/10.1007/JHEP07(2020)117} {\path{doi:10.1007/JHEP07(2020)117}}.

\bibitem{Bacchetta:2022awv}
A.~Bacchetta, V.~Bertone, C.~Bissolotti, G.~Bozzi, M.~Cerutti, F.~Piacenza, M.~Radici, A.~Signori, {Unpolarized transverse momentum distributions from a global fit of Drell-Yan and semi-inclusive deep-inelastic scattering data}, JHEP 10 (2022) 127.
\newblock \href {http://arxiv.org/abs/2206.07598} {\path{arXiv:2206.07598}}, \href {http://dx.doi.org/10.1007/JHEP10(2022)127} {\path{doi:10.1007/JHEP10(2022)127}}.

\bibitem{Bury:2022czx}
M.~Bury, F.~Hautmann, S.~Leal-Gomez, I.~Scimemi, A.~Vladimirov, P.~Zurita, {PDF bias and flavor dependence in TMD distributions}, JHEP 10 (2022) 118.
\newblock \href {http://arxiv.org/abs/2201.07114} {\path{arXiv:2201.07114}}, \href {http://dx.doi.org/10.1007/JHEP10(2022)118} {\path{doi:10.1007/JHEP10(2022)118}}.

\bibitem{Moos:2023yfa}
V.~Moos, I.~Scimemi, A.~Vladimirov, P.~Zurita, {Extraction of unpolarized transverse momentum distributions from the fit of Drell-Yan data at N$^{4}$LL}, JHEP 05 (2024) 036.
\newblock \href {http://arxiv.org/abs/2305.07473} {\path{arXiv:2305.07473}}, \href {http://dx.doi.org/10.1007/JHEP05(2024)036} {\path{doi:10.1007/JHEP05(2024)036}}.

\bibitem{Ball:2021dab}
R.~D. Ball, S.~Forte, R.~Stegeman, {Correlation and combination of sets of parton distributions}, Eur. Phys. J. C 81~(11) (2021) 1046.
\newblock \href {http://arxiv.org/abs/2110.08274} {\path{arXiv:2110.08274}}, \href {http://dx.doi.org/10.1140/epjc/s10052-021-09863-6} {\path{doi:10.1140/epjc/s10052-021-09863-6}}.

\bibitem{Salam:1998tj}
G.~P. Salam, {A Resummation of large subleading corrections at small x}, JHEP 07 (1998) 019.
\newblock \href {http://arxiv.org/abs/hep-ph/9806482} {\path{arXiv:hep-ph/9806482}}, \href {http://dx.doi.org/10.1088/1126-6708/1998/07/019} {\path{doi:10.1088/1126-6708/1998/07/019}}.

\bibitem{Ciafaloni:2003rd}
M.~Ciafaloni, D.~Colferai, G.~P. Salam, A.~M. Stasto, {Renormalization group improved small x Green's function}, Phys. Rev. D 68 (2003) 114003.
\newblock \href {http://arxiv.org/abs/hep-ph/0307188} {\path{arXiv:hep-ph/0307188}}, \href {http://dx.doi.org/10.1103/PhysRevD.68.114003} {\path{doi:10.1103/PhysRevD.68.114003}}.

\bibitem{Ciafaloni:2003ek}
M.~Ciafaloni, D.~Colferai, D.~Colferai, G.~P. Salam, A.~M. Stasto, {Extending QCD perturbation theory to higher energies}, Phys. Lett. B 576 (2003) 143--151.
\newblock \href {http://arxiv.org/abs/hep-ph/0305254} {\path{arXiv:hep-ph/0305254}}, \href {http://dx.doi.org/10.1016/j.physletb.2003.09.078} {\path{doi:10.1016/j.physletb.2003.09.078}}.

\bibitem{Ciafaloni:2000cb}
M.~Ciafaloni, D.~Colferai, G.~P. Salam, {On factorization at small x}, JHEP 07 (2000) 054.
\newblock \href {http://arxiv.org/abs/hep-ph/0007240} {\path{arXiv:hep-ph/0007240}}, \href {http://dx.doi.org/10.1088/1126-6708/2000/07/054} {\path{doi:10.1088/1126-6708/2000/07/054}}.

\bibitem{Ciafaloni:1999yw}
M.~Ciafaloni, D.~Colferai, G.~P. Salam, {Renormalization group improved small x equation}, Phys. Rev. D 60 (1999) 114036.
\newblock \href {http://arxiv.org/abs/hep-ph/9905566} {\path{arXiv:hep-ph/9905566}}, \href {http://dx.doi.org/10.1103/PhysRevD.60.114036} {\path{doi:10.1103/PhysRevD.60.114036}}.

\bibitem{Ciafaloni:1998iv}
M.~Ciafaloni, D.~Colferai, {The BFKL equation at next-to-leading level and beyond}, Phys. Lett. B 452 (1999) 372--378.
\newblock \href {http://arxiv.org/abs/hep-ph/9812366} {\path{arXiv:hep-ph/9812366}}, \href {http://dx.doi.org/10.1016/S0370-2693(99)00281-6} {\path{doi:10.1016/S0370-2693(99)00281-6}}.

\bibitem{SabioVera:2005tiv}
A.~Sabio~Vera, {An 'All-poles' approximation to collinear resummations in the Regge limit of perturbative QCD}, Nucl. Phys. B 722 (2005) 65--80.
\newblock \href {http://arxiv.org/abs/hep-ph/0505128} {\path{arXiv:hep-ph/0505128}}, \href {http://dx.doi.org/10.1016/j.nuclphysb.2005.06.003} {\path{doi:10.1016/j.nuclphysb.2005.06.003}}.

\bibitem{Barbieri:1979be}
R.~Barbieri, E.~d'Emilio, G.~Curci, E.~Remiddi, {Strong Radiative Corrections to Annihilations of Quarkonia in QCD}, Nucl. Phys. B 154 (1979) 535--546.
\newblock \href {http://dx.doi.org/10.1016/0550-3213(79)90047-6} {\path{doi:10.1016/0550-3213(79)90047-6}}.

\bibitem{PhysRevLett.42.1435}
W.~Celmaster, R.~J. Gonsalves, Quantum-chromodynamics perturbation expansions in a coupling constant renormalized by momentum-space subtraction, Phys. Rev. Lett. 42 (1979) 1435--1438.
\newblock \href {http://dx.doi.org/10.1103/PhysRevLett.42.1435} {\path{doi:10.1103/PhysRevLett.42.1435}}.

\bibitem{Khachatryan:2016udy}
V.~Khachatryan, et~al., {Azimuthal decorrelation of jets widely separated in rapidity in pp collisions at $ \sqrt{s}=7 $ TeV}, JHEP 08 (2016) 139.
\newblock \href {http://arxiv.org/abs/1601.06713} {\path{arXiv:1601.06713}}, \href {http://dx.doi.org/10.1007/JHEP08(2016)139} {\path{doi:10.1007/JHEP08(2016)139}}.

\bibitem{Mueller:2013wwa}
A.~Mueller, B.-W. Xiao, F.~Yuan, {Sudakov double logarithms resummation in hard processes in the small-x saturation formalism}, Phys. Rev. D 88~(11) (2013) 114010.
\newblock \href {http://arxiv.org/abs/1308.2993} {\path{arXiv:1308.2993}}, \href {http://dx.doi.org/10.1103/PhysRevD.88.114010} {\path{doi:10.1103/PhysRevD.88.114010}}.

\bibitem{Marzani:2015oyb}
S.~Marzani, {Combining $Q_T$ and small-$x$ resummations}, Phys. Rev. D 93~(5) (2016) 054047.
\newblock \href {http://arxiv.org/abs/1511.06039} {\path{arXiv:1511.06039}}, \href {http://dx.doi.org/10.1103/PhysRevD.93.054047} {\path{doi:10.1103/PhysRevD.93.054047}}.

\bibitem{Mueller:2015ael}
A.~Mueller, L.~Szymanowski, S.~Wallon, B.-W. Xiao, F.~Yuan, {Sudakov Resummations in Mueller-Navelet Dijet Production}, JHEP 03 (2016) 096.
\newblock \href {http://arxiv.org/abs/1512.07127} {\path{arXiv:1512.07127}}, \href {http://dx.doi.org/10.1007/JHEP03(2016)096} {\path{doi:10.1007/JHEP03(2016)096}}.

\bibitem{Xiao:2018esv}
B.-W. Xiao, F.~Yuan, {BFKL and Sudakov Resummation in Higgs Boson Plus Jet Production with Large Rapidity Separation}, Phys. Lett. B 782 (2018) 28--33.
\newblock \href {http://arxiv.org/abs/1801.05478} {\path{arXiv:1801.05478}}, \href {http://dx.doi.org/10.1016/j.physletb.2018.04.070} {\path{doi:10.1016/j.physletb.2018.04.070}}.

\bibitem{Hatta:2020bgy}
Y.~Hatta, B.-W. Xiao, F.~Yuan, J.~Zhou, {Anisotropy in Dijet Production in Exclusive and Inclusive Processes}, Phys. Rev. Lett. 126~(14) (2021) 142001.
\newblock \href {http://arxiv.org/abs/2010.10774} {\path{arXiv:2010.10774}}, \href {http://dx.doi.org/10.1103/PhysRevLett.126.142001} {\path{doi:10.1103/PhysRevLett.126.142001}}.

\bibitem{Hatta:2021jcd}
Y.~Hatta, B.-W. Xiao, F.~Yuan, J.~Zhou, {Azimuthal angular asymmetry of soft gluon radiation in jet production}, Phys. Rev. D 104~(5) (2021) 054037.
\newblock \href {http://arxiv.org/abs/2106.05307} {\path{arXiv:2106.05307}}, \href {http://dx.doi.org/10.1103/PhysRevD.104.054037} {\path{doi:10.1103/PhysRevD.104.054037}}.

\bibitem{Andersen:2001kta}
J.~R. Andersen, V.~Del~Duca, S.~Frixione, C.~R. Schmidt, W.~J. Stirling, {Mueller-Navelet jets at hadron colliders}, JHEP 02 (2001) 007.
\newblock \href {http://arxiv.org/abs/hep-ph/0101180} {\path{arXiv:hep-ph/0101180}}, \href {http://dx.doi.org/10.1088/1126-6708/2001/02/007} {\path{doi:10.1088/1126-6708/2001/02/007}}.

\bibitem{Fontannaz:2001nq}
M.~Fontannaz, J.~P. Guillet, G.~Heinrich, {Is a large intrinsic k(T) needed to describe photon + jet photoproduction at HERA?}, Eur. Phys. J. C 22 (2001) 303--315.
\newblock \href {http://arxiv.org/abs/hep-ph/0107262} {\path{arXiv:hep-ph/0107262}}, \href {http://dx.doi.org/10.1007/s100520100797} {\path{doi:10.1007/s100520100797}}.

\bibitem{Ducloue:2014koa}
B.~Duclou\'e, L.~Szymanowski, S.~Wallon, {Violation of energy\textendash{}momentum conservation in Mueller\textendash{}Navelet jets production}, Phys. Lett. B 738 (2014) 311--316.
\newblock \href {http://arxiv.org/abs/1407.6593} {\path{arXiv:1407.6593}}, \href {http://dx.doi.org/10.1016/j.physletb.2014.09.025} {\path{doi:10.1016/j.physletb.2014.09.025}}.

\bibitem{Chatrchyan:2012xg}
S.~Chatrchyan, et~al., {Measurement of the $\Lambda_b$ cross section and the $_{\bar{\Lambda}_b}$ to $\Lambda_b$ ratio with $J/\Psi \Lambda$ decays in $pp$ collisions at $\sqrt{s}=7$ TeV}, Phys. Lett. B 714 (2012) 136--157.
\newblock \href {http://arxiv.org/abs/1205.0594} {\path{arXiv:1205.0594}}, \href {http://dx.doi.org/10.1016/j.physletb.2012.05.063} {\path{doi:10.1016/j.physletb.2012.05.063}}.

\bibitem{Brodsky:1980pb}
S.~J. Brodsky, P.~Hoyer, C.~Peterson, N.~Sakai, {The Intrinsic Charm of the Proton}, Phys. Lett. B 93 (1980) 451--455.
\newblock \href {http://dx.doi.org/10.1016/0370-2693(80)90364-0} {\path{doi:10.1016/0370-2693(80)90364-0}}.

\bibitem{Brodsky:2015fna}
S.~J. Brodsky, A.~Kusina, F.~Lyonnet, I.~Schienbein, H.~Spiesberger, R.~Vogt, {A review of the intrinsic heavy quark content of the nucleon}, Adv. High Energy Phys. 2015 (2015) 231547.
\newblock \href {http://arxiv.org/abs/1504.06287} {\path{arXiv:1504.06287}}, \href {http://dx.doi.org/10.1155/2015/231547} {\path{doi:10.1155/2015/231547}}.

\bibitem{Ball:2022qks}
R.~D. Ball, A.~Candido, J.~Cruz-Martinez, S.~Forte, T.~Giani, F.~Hekhorn, K.~Kudashkin, G.~Magni, J.~Rojo, {Evidence for intrinsic charm quarks in the proton}, Nature 608~(7923) (2022) 483--487.
\newblock \href {http://arxiv.org/abs/2208.08372} {\path{arXiv:2208.08372}}, \href {http://dx.doi.org/10.1038/s41586-022-04998-2} {\path{doi:10.1038/s41586-022-04998-2}}.

\bibitem{Guzzi:2022rca}
M.~Guzzi, T.~J. Hobbs, K.~Xie, J.~Huston, P.~Nadolsky, C.~P. Yuan, {The persistent nonperturbative charm enigma}, Phys. Lett. B 843 (2023) 137975.
\newblock \href {http://arxiv.org/abs/2211.01387} {\path{arXiv:2211.01387}}, \href {http://dx.doi.org/10.1016/j.physletb.2023.137975} {\path{doi:10.1016/j.physletb.2023.137975}}.

\bibitem{NNPDF:2023tyk}
R.~D. Ball, A.~Candido, J.~Cruz-Martinez, S.~Forte, T.~Giani, F.~Hekhorn, G.~Magni, E.~R. Nocera, J.~Rojo, R.~Stegeman, {Intrinsic charm quark valence distribution of the proton}, Phys. Rev. D 109~(9) (2024) L091501.
\newblock \href {http://arxiv.org/abs/2311.00743} {\path{arXiv:2311.00743}}, \href {http://dx.doi.org/10.1103/PhysRevD.109.L091501} {\path{doi:10.1103/PhysRevD.109.L091501}}.

\bibitem{Nocera:2017qgb}
E.~R. Nocera, {Towards a Neural Network Determination of Charged Pion Fragmentation Functions}, in: {22nd International Symposium on Spin Physics}, 2017.
\newblock \href {http://arxiv.org/abs/1701.09186} {\path{arXiv:1701.09186}}.

\bibitem{Bertone:2017xsf}
V.~Bertone, S.~Carrazza, E.~R. Nocera, N.~P. Hartland, J.~Rojo, {Towards a Neural Network determination of Pion Fragmentation Functions}, in: {Parton radiation and fragmentation from LHC to FCC-ee}, 2017, pp. 19--25.

\bibitem{Bertone:2017tyb}
V.~Bertone, S.~Carrazza, N.~P. Hartland, E.~R. Nocera, J.~Rojo, {A determination of the fragmentation functions of pions, kaons, and protons with faithful uncertainties}, Eur. Phys. J. C 77~(8) (2017) 516.
\newblock \href {http://arxiv.org/abs/1706.07049} {\path{arXiv:1706.07049}}, \href {http://dx.doi.org/10.1140/epjc/s10052-017-5088-y} {\path{doi:10.1140/epjc/s10052-017-5088-y}}.

\bibitem{Bertone:2018ecm}
V.~Bertone, N.~P. Hartland, E.~R. Nocera, J.~Rojo, L.~Rottoli, {Charged hadron fragmentation functions from collider data}, Eur. Phys. J. C 78~(8) (2018) 651, [Erratum: Eur.Phys.J.C 84, 155 (2024)].
\newblock \href {http://arxiv.org/abs/1807.03310} {\path{arXiv:1807.03310}}, \href {http://dx.doi.org/10.1140/epjc/s10052-018-6130-4} {\path{doi:10.1140/epjc/s10052-018-6130-4}}.

\bibitem{Khalek:2021gxf}
R.~Abdul~Khalek, V.~Bertone, E.~R. Nocera, {Determination of unpolarized pion fragmentation functions using semi-inclusive deep-inelastic-scattering data}, Phys. Rev. D 104~(3) (2021) 034007.
\newblock \href {http://arxiv.org/abs/2105.08725} {\path{arXiv:2105.08725}}, \href {http://dx.doi.org/10.1103/PhysRevD.104.034007} {\path{doi:10.1103/PhysRevD.104.034007}}.

\bibitem{Khalek:2022vgy}
R.~Abdul~Khalek, V.~Bertone, A.~Khoudli, E.~R. Nocera, {Pion and kaon fragmentation functions at next-to-next-to-leading order}, Phys. Lett. B 834 (2022) 137456.
\newblock \href {http://arxiv.org/abs/2204.10331} {\path{arXiv:2204.10331}}, \href {http://dx.doi.org/10.1016/j.physletb.2022.137456} {\path{doi:10.1016/j.physletb.2022.137456}}.

\bibitem{Soleymaninia:2022qjf}
M.~Soleymaninia, H.~Hashamipour, H.~Khanpour, H.~Spiesberger, {Fragmentation Functions for $\Xi ^-/\bar{\Xi}^+$ Using Neural Networks}, Nucl. Phys. A 2023 (2022) 01.
\newblock \href {http://arxiv.org/abs/2202.05586} {\path{arXiv:2202.05586}}, \href {http://dx.doi.org/10.1016/j.nuclphysa.2022.122564} {\path{doi:10.1016/j.nuclphysa.2022.122564}}.

\bibitem{Soleymaninia:2022alt}
M.~Soleymaninia, H.~Hashamipour, H.~Khanpour, {Neural network QCD analysis of charged hadron fragmentation functions in the presence of SIDIS data}, Phys. Rev. D 105~(11) (2022) 114018.
\newblock \href {http://arxiv.org/abs/2202.10779} {\path{arXiv:2202.10779}}, \href {http://dx.doi.org/10.1103/PhysRevD.105.114018} {\path{doi:10.1103/PhysRevD.105.114018}}.

\bibitem{Caucal:2022ulg}
P.~Caucal, F.~Salazar, B.~Schenke, R.~Venugopalan, {Back-to-back inclusive dijets in DIS at small x: Sudakov suppression and gluon saturation at NLO}, JHEP 11 (2022) 169.
\newblock \href {http://arxiv.org/abs/2208.13872} {\path{arXiv:2208.13872}}, \href {http://dx.doi.org/10.1007/JHEP11(2022)169} {\path{doi:10.1007/JHEP11(2022)169}}.

\bibitem{Taels:2022tza}
P.~Taels, T.~Altinoluk, G.~Beuf, C.~Marquet, {Dijet photoproduction at low x at next-to-leading order and its back-to-back limit}, JHEP 10 (2022) 184.
\newblock \href {http://arxiv.org/abs/2204.11650} {\path{arXiv:2204.11650}}, \href {http://dx.doi.org/10.1007/JHEP10(2022)184} {\path{doi:10.1007/JHEP10(2022)184}}.

\bibitem{Dasgupta:2014yra}
M.~Dasgupta, F.~Dreyer, G.~P. Salam, G.~Soyez, {Small-radius jets to all orders in QCD}, JHEP 04 (2015) 039.
\newblock \href {http://arxiv.org/abs/1411.5182} {\path{arXiv:1411.5182}}, \href {http://dx.doi.org/10.1007/JHEP04(2015)039} {\path{doi:10.1007/JHEP04(2015)039}}.

\bibitem{Dasgupta:2016bnd}
M.~Dasgupta, F.~A. Dreyer, G.~P. Salam, G.~Soyez, {Inclusive jet spectrum for small-radius jets}, JHEP 06 (2016) 057.
\newblock \href {http://arxiv.org/abs/1602.01110} {\path{arXiv:1602.01110}}, \href {http://dx.doi.org/10.1007/JHEP06(2016)057} {\path{doi:10.1007/JHEP06(2016)057}}.

\bibitem{Banfi:2012jm}
A.~Banfi, P.~F. Monni, G.~P. Salam, G.~Zanderighi, {Higgs and Z-boson production with a jet veto}, Phys. Rev. Lett. 109 (2012) 202001.
\newblock \href {http://arxiv.org/abs/1206.4998} {\path{arXiv:1206.4998}}, \href {http://dx.doi.org/10.1103/PhysRevLett.109.202001} {\path{doi:10.1103/PhysRevLett.109.202001}}.

\bibitem{Banfi:2015pju}
A.~Banfi, F.~Caola, F.~A. Dreyer, P.~F. Monni, G.~P. Salam, G.~Zanderighi, F.~Dulat, {Jet-vetoed Higgs cross section in gluon fusion at N$^{3}$LO+NNLL with small-$R$ resummation}, JHEP 04 (2016) 049.
\newblock \href {http://arxiv.org/abs/1511.02886} {\path{arXiv:1511.02886}}, \href {http://dx.doi.org/10.1007/JHEP04(2016)049} {\path{doi:10.1007/JHEP04(2016)049}}.

\bibitem{Liu:2017pbb}
X.~Liu, S.-O. Moch, F.~Ringer, {Threshold and jet radius joint resummation for single-inclusive jet production}, Phys. Rev. Lett. 119~(21) (2017) 212001.
\newblock \href {http://arxiv.org/abs/1708.04641} {\path{arXiv:1708.04641}}, \href {http://dx.doi.org/10.1103/PhysRevLett.119.212001} {\path{doi:10.1103/PhysRevLett.119.212001}}.

\bibitem{Luisoni:2015xha}
G.~Luisoni, S.~Marzani, {QCD resummation for hadronic final states}, J. Phys. G 42~(10) (2015) 103101.
\newblock \href {http://arxiv.org/abs/1505.04084} {\path{arXiv:1505.04084}}, \href {http://dx.doi.org/10.1088/0954-3899/42/10/103101} {\path{doi:10.1088/0954-3899/42/10/103101}}.

\bibitem{Caletti:2021oor}
S.~Caletti, O.~Fedkevych, S.~Marzani, D.~Reichelt, S.~Schumann, G.~Soyez, V.~Theeuwes, {Jet angularities in Z+jet production at the LHC}, JHEP 07 (2021) 076.
\newblock \href {http://arxiv.org/abs/2104.06920} {\path{arXiv:2104.06920}}, \href {http://dx.doi.org/10.1007/JHEP07(2021)076} {\path{doi:10.1007/JHEP07(2021)076}}.

\bibitem{Reichelt:2021svh}
D.~Reichelt, S.~Caletti, O.~Fedkevych, S.~Marzani, S.~Schumann, G.~Soyez, {Phenomenology of jet angularities at the LHC}, JHEP 03 (2022) 131.
\newblock \href {http://arxiv.org/abs/2112.09545} {\path{arXiv:2112.09545}}, \href {http://dx.doi.org/10.1007/JHEP03(2022)131} {\path{doi:10.1007/JHEP03(2022)131}}.

\bibitem{Gelis:2010nm}
F.~Gelis, E.~Iancu, J.~Jalilian-Marian, R.~Venugopalan, {The Color Glass Condensate}, Ann. Rev. Nucl. Part. Sci. 60 (2010) 463--489.
\newblock \href {http://arxiv.org/abs/1002.0333} {\path{arXiv:1002.0333}}, \href {http://dx.doi.org/10.1146/annurev.nucl.010909.083629} {\path{doi:10.1146/annurev.nucl.010909.083629}}.

\bibitem{Kovchegov:2012mbw}
Y.~V. Kovchegov, E.~Levin, {Quantum chromodynamics at high energy}, Vol.~33, Cambridge University Press, 2012.
\newblock \href {http://dx.doi.org/10.1017/CBO9781139022187} {\path{doi:10.1017/CBO9781139022187}}.

\bibitem{Chirilli:2012jd}
G.~A. Chirilli, B.-W. Xiao, F.~Yuan, {Inclusive Hadron Productions in pA Collisions}, Phys. Rev. D 86 (2012) 054005.
\newblock \href {http://arxiv.org/abs/1203.6139} {\path{arXiv:1203.6139}}, \href {http://dx.doi.org/10.1103/PhysRevD.86.054005} {\path{doi:10.1103/PhysRevD.86.054005}}.

\bibitem{Boussarie:2014lxa}
R.~Boussarie, A.~V. Grabovsky, L.~Szymanowski, S.~Wallon, {Impact factor for high-energy two and three jets diffractive production}, JHEP 09 (2014) 026.
\newblock \href {http://arxiv.org/abs/1405.7676} {\path{arXiv:1405.7676}}, \href {http://dx.doi.org/10.1007/JHEP09(2014)026} {\path{doi:10.1007/JHEP09(2014)026}}.

\bibitem{Benic:2016uku}
S.~Benic, K.~Fukushima, O.~Garcia-Montero, R.~Venugopalan, {Probing gluon saturation with next-to-leading order photon production at central rapidities in proton-nucleus collisions}, JHEP 01 (2017) 115.
\newblock \href {http://arxiv.org/abs/1609.09424} {\path{arXiv:1609.09424}}, \href {http://dx.doi.org/10.1007/JHEP01(2017)115} {\path{doi:10.1007/JHEP01(2017)115}}.

\bibitem{Benic:2018hvb}
S.~Beni\'c, K.~Fukushima, O.~Garcia-Montero, R.~Venugopalan, {Constraining unintegrated gluon distributions from inclusive photon production in proton\textendash{}proton collisions at the LHC}, Phys. Lett. B 791 (2019) 11--16.
\newblock \href {http://arxiv.org/abs/1807.03806} {\path{arXiv:1807.03806}}, \href {http://dx.doi.org/10.1016/j.physletb.2019.02.007} {\path{doi:10.1016/j.physletb.2019.02.007}}.

\bibitem{Roy:2019hwr}
K.~Roy, R.~Venugopalan, {NLO impact factor for inclusive photon$+$dijet production in $e+A$ DIS at small $x$}, Phys. Rev. D 101~(3) (2020) 034028.
\newblock \href {http://arxiv.org/abs/1911.04530} {\path{arXiv:1911.04530}}, \href {http://dx.doi.org/10.1103/PhysRevD.101.034028} {\path{doi:10.1103/PhysRevD.101.034028}}.

\bibitem{Roy:2019cux}
K.~Roy, R.~Venugopalan, {Extracting many-body correlators of saturated gluons with precision from inclusive photon+dijet final states in deeply inelastic scattering}, Phys. Rev. D 101~(7) (2020) 071505.
\newblock \href {http://arxiv.org/abs/1911.04519} {\path{arXiv:1911.04519}}, \href {http://dx.doi.org/10.1103/PhysRevD.101.071505} {\path{doi:10.1103/PhysRevD.101.071505}}.

\bibitem{Beuf:2020dxl}
G.~Beuf, H.~H\"anninen, T.~Lappi, H.~M\"antysaari, {Color Glass Condensate at next-to-leading order meets HERA data}, Phys. Rev. D 102 (2020) 074028.
\newblock \href {http://arxiv.org/abs/2007.01645} {\path{arXiv:2007.01645}}, \href {http://dx.doi.org/10.1103/PhysRevD.102.074028} {\path{doi:10.1103/PhysRevD.102.074028}}.

\bibitem{Iancu:2021rup}
E.~Iancu, A.~H. Mueller, D.~N. Triantafyllopoulos, {Probing Parton Saturation and the Gluon Dipole via Diffractive Jet Production at the Electron-Ion Collider}, Phys. Rev. Lett. 128~(20) (2022) 202001.
\newblock \href {http://arxiv.org/abs/2112.06353} {\path{arXiv:2112.06353}}, \href {http://dx.doi.org/10.1103/PhysRevLett.128.202001} {\path{doi:10.1103/PhysRevLett.128.202001}}.

\bibitem{Iancu:2023lel}
E.~Iancu, A.~H. Mueller, D.~N. Triantafyllopoulos, S.~Y. Wei, {Probing gluon saturation via diffractive jets in ultra-peripheral nucleus-nucleus collisions}, Eur. Phys. J. C 83~(11) (2023) 1078.
\newblock \href {http://arxiv.org/abs/2304.12401} {\path{arXiv:2304.12401}}, \href {http://dx.doi.org/10.1140/epjc/s10052-023-12165-8} {\path{doi:10.1140/epjc/s10052-023-12165-8}}.

\bibitem{vanHameren:2023oiq}
A.~van Hameren, H.~Kakkad, P.~Kotko, K.~Kutak, S.~Sapeta, {Searching for saturation in forward dijet production at the LHC}, Eur. Phys. J. C 83~(10) (2023) 947.
\newblock \href {http://arxiv.org/abs/2306.17513} {\path{arXiv:2306.17513}}, \href {http://dx.doi.org/10.1140/epjc/s10052-023-12120-7} {\path{doi:10.1140/epjc/s10052-023-12120-7}}.

\bibitem{Wallon:2023asa}
S.~Wallon, {The QCD Shockwave Approach at NLO: Towards Precision Physics in Gluonic Saturation}, Acta Phys. Polon. Supp. 16~(5) (2023) 26.
\newblock \href {http://arxiv.org/abs/2302.04526} {\path{arXiv:2302.04526}}, \href {http://dx.doi.org/10.5506/APhysPolBSupp.16.5-A26} {\path{doi:10.5506/APhysPolBSupp.16.5-A26}}.

\bibitem{Agostini:2024xqs}
P.~Agostini, T.~Altinoluk, N.~Armesto, {Next-to-eikonal corrections to dijet production in Deep Inelastic Scattering in the dilute limit of the Color Glass Condensate}, JHEP 07 (2024) 137.
\newblock \href {http://arxiv.org/abs/2403.04603} {\path{arXiv:2403.04603}}, \href {http://dx.doi.org/10.1007/JHEP07(2024)137} {\path{doi:10.1007/JHEP07(2024)137}}.

\bibitem{Altinoluk:2024zom}
T.~Altinoluk, G.~Beuf, A.~Czajka, C.~Marquet, {Back-to-back dijet production in DIS at next-to-eikonal accuracy and twist-3 gluon TMDs}, Phys. Rev. D 111~(1) (2025) 014010.
\newblock \href {http://arxiv.org/abs/2410.00612} {\path{arXiv:2410.00612}}, \href {http://dx.doi.org/10.1103/PhysRevD.111.014010} {\path{doi:10.1103/PhysRevD.111.014010}}.

\bibitem{Altinoluk:2025dwd}
T.~Altinoluk, F.~Bergabo, J.~Jalilian-Marian, C.~Marquet, Y.~Shi, {SIDIS at small x at next-to-leading order: Transverse photon}, Phys. Rev. D 112~(5) (2025) 054020.
\newblock \href {http://arxiv.org/abs/2505.04557} {\path{arXiv:2505.04557}}, \href {http://dx.doi.org/10.1103/s561-vqh8} {\path{doi:10.1103/s561-vqh8}}.

\bibitem{Altinoluk:2025tms}
T.~Altinoluk, G.~Beuf, J.~Penttala, {Exploring rapidity regularization schemes at low $x$ with the DIS longitudinal structure function}\href {http://arxiv.org/abs/2510.08550} {\path{arXiv:2510.08550}}.

\bibitem{Caucal:2021ent}
P.~Caucal, F.~Salazar, R.~Venugopalan, {Dijet impact factor in DIS at next-to-leading order in the Color Glass Condensate}, JHEP 11 (2021) 222.
\newblock \href {http://arxiv.org/abs/2108.06347} {\path{arXiv:2108.06347}}, \href {http://dx.doi.org/10.1007/JHEP11(2021)222} {\path{doi:10.1007/JHEP11(2021)222}}.

\bibitem{Kotko:2015ura}
P.~Kotko, K.~Kutak, C.~Marquet, E.~Petreska, S.~Sapeta, A.~van Hameren, {Improved TMD factorization for forward dijet production in dilute-dense hadronic collisions}, JHEP 09 (2015) 106.
\newblock \href {http://arxiv.org/abs/1503.03421} {\path{arXiv:1503.03421}}, \href {http://dx.doi.org/10.1007/JHEP09(2015)106} {\path{doi:10.1007/JHEP09(2015)106}}.

\bibitem{vanHameren:2016ftb}
A.~van Hameren, P.~Kotko, K.~Kutak, C.~Marquet, E.~Petreska, S.~Sapeta, {Forward di-jet production in p+Pb collisions in the small-x improved TMD factorization framework}, JHEP 12 (2016) 034, [Erratum: JHEP 02, 158 (2019)].
\newblock \href {http://arxiv.org/abs/1607.03121} {\path{arXiv:1607.03121}}, \href {http://dx.doi.org/10.1007/JHEP12(2016)034} {\path{doi:10.1007/JHEP12(2016)034}}.

\bibitem{Altinoluk:2020qet}
T.~Altinoluk, R.~Boussarie, C.~Marquet, P.~Taels, {Photoproduction of three jets in the CGC: gluon TMDs and dilute limit}, JHEP 07 (2020) 143.
\newblock \href {http://arxiv.org/abs/2001.00765} {\path{arXiv:2001.00765}}, \href {http://dx.doi.org/10.1007/JHEP07(2020)143} {\path{doi:10.1007/JHEP07(2020)143}}.

\bibitem{Altinoluk:2021ygv}
T.~Altinoluk, C.~Marquet, P.~Taels, {Low-x improved TMD approach to the lepto- and hadroproduction of a heavy-quark pair}, JHEP 06 (2021) 085.
\newblock \href {http://arxiv.org/abs/2103.14495} {\path{arXiv:2103.14495}}, \href {http://dx.doi.org/10.1007/JHEP06(2021)085} {\path{doi:10.1007/JHEP06(2021)085}}.

\bibitem{Boussarie:2021ybe}
R.~Boussarie, H.~M\"antysaari, F.~Salazar, B.~Schenke, {The importance of kinematic twists and genuine saturation effects in dijet production at the Electron-Ion Collider}, JHEP 09 (2021) 178.
\newblock \href {http://arxiv.org/abs/2106.11301} {\path{arXiv:2106.11301}}, \href {http://dx.doi.org/10.1007/JHEP09(2021)178} {\path{doi:10.1007/JHEP09(2021)178}}.

\bibitem{Caucal:2023nci}
P.~Caucal, F.~Salazar, B.~Schenke, T.~Stebel, R.~Venugopalan, {Back-to-back inclusive dijets in DIS at small x: gluon Weizs\"acker-Williams distribution at NLO}, JHEP 08 (2023) 062.
\newblock \href {http://arxiv.org/abs/2304.03304} {\path{arXiv:2304.03304}}, \href {http://dx.doi.org/10.1007/JHEP08(2023)062} {\path{doi:10.1007/JHEP08(2023)062}}.

\bibitem{Cheung:2024qvw}
V.~Cheung, Z.-B. Kang, F.~Salazar, R.~Vogt, {Direct quarkonium production in DIS from a joint CGC and NRQCD framework}, Phys. Rev. D 110~(9) (2024) 094039.
\newblock \href {http://arxiv.org/abs/2409.04080} {\path{arXiv:2409.04080}}, \href {http://dx.doi.org/10.1103/PhysRevD.110.094039} {\path{doi:10.1103/PhysRevD.110.094039}}.

\bibitem{Caucal:2025mth}
P.~Caucal, E.~Iancu, F.~Salazar, F.~Yuan, {Gluon splitting at small $x$: a unified derivation for the JIMWLK, DGLAP and CSS equations}\href {http://arxiv.org/abs/2510.08454} {\path{arXiv:2510.08454}}.

\bibitem{Kang:2013hta}
Z.-B. Kang, Y.-Q. Ma, R.~Venugopalan, {Quarkonium production in high energy proton-nucleus collisions: CGC meets NRQCD}, JHEP 01 (2014) 056.
\newblock \href {http://arxiv.org/abs/1309.7337} {\path{arXiv:1309.7337}}, \href {http://dx.doi.org/10.1007/JHEP01(2014)056} {\path{doi:10.1007/JHEP01(2014)056}}.

\bibitem{Ma:2014mri}
Y.-Q. Ma, R.~Venugopalan, {Comprehensive Description of $J/\psi$ Production in Proton-Proton Collisions at Collider Energies}, Phys. Rev. Lett. 113~(19) (2014) 192301.
\newblock \href {http://arxiv.org/abs/1408.4075} {\path{arXiv:1408.4075}}, \href {http://dx.doi.org/10.1103/PhysRevLett.113.192301} {\path{doi:10.1103/PhysRevLett.113.192301}}.

\bibitem{Ma:2015sia}
Y.-Q. Ma, R.~Venugopalan, H.-F. Zhang, {$J/\psi$ production and suppression in high energy proton-nucleus collisions}, Phys. Rev. D 92 (2015) 071901.
\newblock \href {http://arxiv.org/abs/1503.07772} {\path{arXiv:1503.07772}}, \href {http://dx.doi.org/10.1103/PhysRevD.92.071901} {\path{doi:10.1103/PhysRevD.92.071901}}.

\bibitem{Ma:2018qvc}
Y.-Q. Ma, T.~Stebel, R.~Venugopalan, {$J/\psi$ polarization in the CGC+NRQCD approach}, JHEP 12 (2018) 057.
\newblock \href {http://arxiv.org/abs/1809.03573} {\path{arXiv:1809.03573}}, \href {http://dx.doi.org/10.1007/JHEP12(2018)057} {\path{doi:10.1007/JHEP12(2018)057}}.

\bibitem{Stebel:2021bbn}
T.~Stebel, K.~Watanabe, {$J\psi$ polarization in high multiplicity pp and pA collisions: CGC\,+\,NRQCD approach}, Phys. Rev. D 104~(3) (2021) 034004.
\newblock \href {http://arxiv.org/abs/2103.01724} {\path{arXiv:2103.01724}}, \href {http://dx.doi.org/10.1103/PhysRevD.104.034004} {\path{doi:10.1103/PhysRevD.104.034004}}.

\bibitem{Mantysaari:2021ryb}
H.~M\"antysaari, J.~Penttala, {Exclusive heavy vector meson production at next-to-leading order in the dipole picture}, Phys. Lett. B 823 (2021) 136723.
\newblock \href {http://arxiv.org/abs/2104.02349} {\path{arXiv:2104.02349}}, \href {http://dx.doi.org/10.1016/j.physletb.2021.136723} {\path{doi:10.1016/j.physletb.2021.136723}}.

\bibitem{Mantysaari:2022kdm}
H.~M\"antysaari, J.~Penttala, {Complete calculation of exclusive heavy vector meson production at next-to-leading order in the dipole picture}, JHEP 08 (2022) 247.
\newblock \href {http://arxiv.org/abs/2204.14031} {\path{arXiv:2204.14031}}, \href {http://dx.doi.org/10.1007/JHEP08(2022)247} {\path{doi:10.1007/JHEP08(2022)247}}.

\bibitem{Flore:2020jau}
C.~Flore, J.-P. Lansberg, H.-S. Shao, Y.~Yedelkina, {Large-$P_T$ inclusive photoproduction of $J/\psi$ in electron-proton collisions at HERA and the EIC}, Phys. Lett. B 811 (2020) 135926.
\newblock \href {http://arxiv.org/abs/2009.08264} {\path{arXiv:2009.08264}}, \href {http://dx.doi.org/10.1016/j.physletb.2020.135926} {\path{doi:10.1016/j.physletb.2020.135926}}.

\bibitem{LHCspin:2025lvj}
A.~Accardi, et~al., {LHCspin: a Polarized Gas Target for LHC}\href {http://arxiv.org/abs/2504.16034} {\path{arXiv:2504.16034}}.

\bibitem{Anchordoqui:2021ghd}
L.~A. Anchordoqui, et~al., {The Forward Physics Facility: Sites, experiments, and physics potential}, Phys. Rept. 968 (2022) 1--50.
\newblock \href {http://arxiv.org/abs/2109.10905} {\path{arXiv:2109.10905}}, \href {http://dx.doi.org/10.1016/j.physrep.2022.04.004} {\path{doi:10.1016/j.physrep.2022.04.004}}.

\bibitem{Feng:2022inv}
J.~L. Feng, et~al., {The Forward Physics Facility at the High-Luminosity LHC}, J. Phys. G 50~(3) (2023) 030501.
\newblock \href {http://arxiv.org/abs/2203.05090} {\path{arXiv:2203.05090}}, \href {http://dx.doi.org/10.1088/1361-6471/ac865e} {\path{doi:10.1088/1361-6471/ac865e}}.

\bibitem{Accardi:2012qut}
A.~Accardi, et~al., {Electron Ion Collider: The Next QCD Frontier}: {Understanding the glue that binds us all}, Eur. Phys. J. A 52~(9) (2016) 268.
\newblock \href {http://arxiv.org/abs/1212.1701} {\path{arXiv:1212.1701}}, \href {http://dx.doi.org/10.1140/epja/i2016-16268-9} {\path{doi:10.1140/epja/i2016-16268-9}}.

\bibitem{Acosta:2022ejc}
D.~Acosta, E.~Barberis, N.~Hurley, W.~Li, O.~M. Colin, D.~Wood, X.~Zuo, {The Potential of a TeV-Scale Muon-Ion Collider}, in: {2022 Snowmass Summer Study}, 2022.
\newblock \href {http://arxiv.org/abs/2203.06258} {\path{arXiv:2203.06258}}.

\bibitem{AlexanderAryshev:2022pkx}
I.~Adachi, et~al., {The International Linear Collider: Report to Snowmass 2021}\href {http://arxiv.org/abs/2203.07622} {\path{arXiv:2203.07622}}.

\bibitem{LinearCollider:2025lya}
C.~Balazs, et~al., {The Linear Collider Facility (LCF) at CERN}\href {http://arxiv.org/abs/2503.24049} {\path{arXiv:2503.24049}}.

\bibitem{LinearColliderVision:2025hlt}
H.~Abramowicz, et~al., {A Linear Collider Vision for the Future of Particle Physics}, Eur. Phys. J. ST\href {http://arxiv.org/abs/2503.19983} {\path{arXiv:2503.19983}}, \href {http://dx.doi.org/10.1140/epjs/s11734-026-02153-w} {\path{doi:10.1140/epjs/s11734-026-02153-w}}.

\bibitem{Brunner:2022usy}
O.~Brunner, et~al., {The CLIC project}\href {http://arxiv.org/abs/2203.09186} {\path{arXiv:2203.09186}}.

\bibitem{Arbuzov:2020cqg}
A.~Arbuzov, et~al., {On the physics potential to study the gluon content of proton and deuteron at NICA SPD}, Prog. Part. Nucl. Phys. 119 (2021) 103858.
\newblock \href {http://arxiv.org/abs/2011.15005} {\path{arXiv:2011.15005}}, \href {http://dx.doi.org/10.1016/j.ppnp.2021.103858} {\path{doi:10.1016/j.ppnp.2021.103858}}.

\bibitem{Abazov:2021hku}
V.~M. Abazov, et~al., {Conceptual design of the Spin Physics Detector}\href {http://arxiv.org/abs/2102.00442} {\path{arXiv:2102.00442}}.

\bibitem{Bernardi:2022hny}
G.~Bernardi, et~al., {The Future Circular Collider: a Summary for the US 2021 Snowmass Process}\href {http://arxiv.org/abs/2203.06520} {\path{arXiv:2203.06520}}.

\bibitem{Celiberto:2018hdy}
F.~G. Celiberto, M.~Fucilla, D.~{\relax Yu}. Ivanov, M.~M.~A. Mohammed, A.~Papa, \href{https://inspirehep.net/literature/1841481}{{High-energy QCD at colliders: semi-hard reactions and unintegrated gluon densities}: {Letter of Interest for SnowMass 2021}}, in: {2022 Snowmass Summer Study}, 2021.
\newline\urlprefix\url{https://inspirehep.net/literature/1841481}

\bibitem{Klein:2020nvu}
S.~Klein, et~al., {New opportunities at the photon energy frontier}\href {http://arxiv.org/abs/2009.03838} {\path{arXiv:2009.03838}}.

\bibitem{2064676}
A.~Canepa, M.~D'Onofrio, {Future Accelerator Projects: New Physics at the Energy Frontier (FERMILAB-PUB-22-248-PPD)}.

\bibitem{MuonCollider:2022xlm}
J.~de~Blas, et~al., {The physics case of a 3 TeV muon collider stage}\href {http://arxiv.org/abs/2203.07261} {\path{arXiv:2203.07261}}.

\bibitem{Aime:2022flm}
C.~Aim\`e, et~al., {Muon Collider Physics Summary}\href {http://arxiv.org/abs/2203.07256} {\path{arXiv:2203.07256}}.

\bibitem{MuonCollider:2022ded}
N.~Bartosik, et~al., {Simulated Detector Performance at the Muon Collider}, in: {2022 Snowmass Summer Study}, 2022.
\newblock \href {http://arxiv.org/abs/2203.07964} {\path{arXiv:2203.07964}}.

\bibitem{Black:2022cth}
K.~M. Black, et~al., {Muon Collider Forum report}, JINST 19~(02) (2024) T02015.
\newblock \href {http://arxiv.org/abs/2209.01318} {\path{arXiv:2209.01318}}, \href {http://dx.doi.org/10.1088/1748-0221/19/02/T02015} {\path{doi:10.1088/1748-0221/19/02/T02015}}.

\bibitem{Accettura:2023ked}
C.~Accettura, et~al., {Towards a muon collider}, Eur. Phys. J. C 83~(9) (2023) 864, [Erratum: Eur.Phys.J.C 84, 36 (2024)].
\newblock \href {http://arxiv.org/abs/2303.08533} {\path{arXiv:2303.08533}}, \href {http://dx.doi.org/10.1140/epjc/s10052-023-11889-x} {\path{doi:10.1140/epjc/s10052-023-11889-x}}.

\bibitem{InternationalMuonCollider:2024jyv}
C.~Accettura, et~al., {Interim report for the International Muon Collider Collaboration (IMCC)} 2/2024.
\newblock \href {http://arxiv.org/abs/2407.12450} {\path{arXiv:2407.12450}}, \href {http://dx.doi.org/10.23731/CYRM-2024-002} {\path{doi:10.23731/CYRM-2024-002}}.

\bibitem{MuCoL:2024oxj}
C.~Accettura, et~al., {MuCol Milestone Report No. 5: Preliminary Parameters}\href {http://arxiv.org/abs/2411.02966} {\path{arXiv:2411.02966}}, \href {http://dx.doi.org/10.5281/zenodo.13970100} {\path{doi:10.5281/zenodo.13970100}}.

\bibitem{MuCoL:2025quu}
A.~Chanc{\'e}, et~al., {MuCol Milestone Report No. 7: Consolidated Parameters}\href {http://arxiv.org/abs/2510.27437} {\path{arXiv:2510.27437}}, \href {http://dx.doi.org/10.5281/zenodo.17476875} {\path{doi:10.5281/zenodo.17476875}}.

\bibitem{InternationalMuonCollider:2025sys}
C.~Accettura, et~al., {The Muon Collider}\href {http://arxiv.org/abs/2504.21417} {\path{arXiv:2504.21417}}.

\bibitem{Vignaroli:2023rxr}
N.~Vignaroli, {Charged resonances and MDM bound states at a multi-TeV muon collider}\href {http://arxiv.org/abs/2304.12362} {\path{arXiv:2304.12362}}.

\bibitem{Dawson:2022zbb}
S.~Dawson, et~al., {Report of the Topical Group on Higgs Physics for Snowmass 2021: The Case for Precision Higgs Physics}, in: {2022 Snowmass Summer Study}, 2022.
\newblock \href {http://arxiv.org/abs/2209.07510} {\path{arXiv:2209.07510}}.

\bibitem{Bose:2022obr}
T.~Bose, et~al., {Report of the Topical Group on Physics Beyond the Standard Model at Energy Frontier for Snowmass 2021}, in: {2022 Snowmass Summer Study}, 2022.
\newblock \href {http://arxiv.org/abs/2209.13128} {\path{arXiv:2209.13128}}.

\bibitem{Begel:2022kwp}
M.~Begel, et~al., {Precision QCD, Hadronic Structure \& Forward QCD, Heavy Ions: Report of Energy Frontier Topical Groups 5, 6, 7 submitted to Snowmass 2021}\href {http://arxiv.org/abs/2209.14872} {\path{arXiv:2209.14872}}.

\bibitem{Accardi:2023chb}
A.~Accardi, et~al., {Strong interaction physics at the luminosity frontier with 22 GeV electrons at Jefferson Lab}, Eur. Phys. J. A 60~(9) (2024) 173.
\newblock \href {http://arxiv.org/abs/2306.09360} {\path{arXiv:2306.09360}}, \href {http://dx.doi.org/10.1140/epja/s10050-024-01282-x} {\path{doi:10.1140/epja/s10050-024-01282-x}}.

\bibitem{Gessner:2025acq}
S.~Gessner, et~al., {Design Initiative for a 10 TeV pCM Wakefield Collider}\href {http://arxiv.org/abs/2503.20214} {\path{arXiv:2503.20214}}.

\bibitem{Altmann:2025feg}
J.~Altmann, et~al., {ECFA Higgs, electroweak, and top Factory Study}, Vol. 5/2025 of CERN Yellow Reports: Monographs, 2025.
\newblock \href {http://arxiv.org/abs/2506.15390} {\path{arXiv:2506.15390}}, \href {http://dx.doi.org/10.23731/CYRM-2025-005} {\path{doi:10.23731/CYRM-2025-005}}.

\end{thebibliography}

\end{document}